    \documentclass[aps,showpacs,amsmath,dvips,showkeys]{Tawrevtex4}

\usepackage[english]{babel}
\usepackage{ngerman}
\usepackage{natbib}

\usepackage{graphicx}
\usepackage{natbib}
\usepackage{color}

\def \be  {\begin{equation}}
\def \ee  {\end{equation}}
\def \ee  {\end{equation}}
\def \bea {\begin{eqnarray}}
\def \eea {\end{eqnarray}}

\def \Tr  {\bf{Tr}}

\newcommand{\nn}{\nonumber}

\renewcommand{\figurename}{{\bf Fig.}}
\renewcommand{\tablename}{{\bf Tab.}}
\newcommand{\Dslash}{\ensuremath{D\hspace{-1.5ex} /}}
\newcommand{\Phibar}{\ensuremath{\bar{\Phi}}}
\newcommand{\ua}{\ensuremath{U(1)_A}}
\newcommand{\msig}{\ensuremath{m_{\sigma}}}
\newcommand{\vev}[1]{\ensuremath{\left\langle #1 \right\rangle}}

\begin{document}

\preprint{ECTP-2012-09\hspace*{0.5cm}and\hspace*{0.5cm}WLCAPP-2013-05}

\title{Equilibrium Statistical-Thermal Models in High-Energy Physics}
\author{Abdel Nasser TAWFIK\footnote{http://atawfik.net/}}
\email{atawfik@cern.ch}
\affiliation{Egyptian Center for Theoretical Physics (ECTP), MTI University, 11571 Cairo, Egypt}
\affiliation{World Laboratory for Cosmology And Particle Physics (WLCAPP), Cairo, Egypt}

\begin{abstract}
We review some recent highlights from the applications of statistical-thermal models to different experimental measurements and lattice QCD thermodynamics, that have been made during the last decade. We start with a short review of the historical milestones on the path of constructing statistical-thermal models for heavy-ion physics. We discovered that Heinz Koppe formulated in 1948 an almost complete recipe for the statistical-thermal models. In 1950, Enrico Fermi generalized this statistical approach, in which he started with a general cross-section formula and inserted into it simplifying assumptions about the matrix element of the interaction process that likely reflects many features of the high-energy reactions dominated by density in the phase space of final states. In 1964, Hagedorn systematically analysed the high-energy phenomena using all tools of statistical physics and introduced the concept of limiting temperature based on the statistical bootstrap model. 
It turns to be quite often that many-particle systems can be studied with the help of statistical-thermal methods. The analysis of yield multiplicities in high-energy collisions gives an overwhelming evidence for the chemical equilibrium in the final state. The strange particles might be an exception, as they are suppressed at lower beam energies. However, their relative yields fulfil statistical equilibrium, as well. We review the equilibrium statistical-thermal models for particle production, fluctuations and collective flow in heavy-ion experiments. We also review their reproduction of the lattice QCD thermodynamics at vanishing and finite chemical potential. During the last decade, five conditions have been suggested to describe the universal behavior of the chemical freeze out parameters. The higher order moments of multiplicity have been discussed. They offer deep insights about particle production and to critical fluctuations. Therefore, we use them to describe the freeze out parameters and suggest the location of the QCD critical endpoint. Various extensions have been proposed in order to take into consideration the possible deviations of the ideal hadron gas. We highlight various types of interactions, dissipative properties and location-dependences  (spacial rapidity). Furthermore, we review three models combining hadronic with partonic phases; quasi-particle model, linear $\sigma$-model with Polyakov-loop potentials and compressible bag model.
\end{abstract}

\pacs{12.40.Ee,24.60.-k,05.70.Fh,12.40.-y}
\keywords{Statistical models of strong interactions, Statistical physics of nuclear reactions, Phase transition in statistical mechanics and thermodynamics, Models of strong interactions } 

\maketitle


\tableofcontents
\makeatletter
\let\toc@pre\relax
\let\toc@post\relax
\makeatother

\newpage

 \section{Introduction}

Enormous multiplicity and large transverse energy are likely generated in ultra high-energy nuclear collisions. We study the nuclear collisions for a much more fundamental reason. It intends to understand the extent that the high-energy collision does in order to generate matter in local equilibrium, which can be characterized by thermodynamic parameters, like temperature, pressure,  energy density. The level of equilibrium in produced particles can be tested by analysing the particle abundances or their momentum spectra. The earlier is  established through the chemical composition of the system, while the latter extracts additional information about the dynamical evolution and collective flow. The hadron multiplicities and their correlations are observables which can provide information about nature, composition, and size of the medium from which they are originating.

Furthermore, establishing thermalization is an essential condition in order to raise detailed questions about the equation of state, the underlying dynamics and the change in the effective degrees of freedom. For example,  the extent to which the components of produced mater interact with each others, and finally chemically and thermally freeze out, is an ultimate goal of the high-energy experiments. The properties of strongly interacting matter under extreme conditions of high energy density represent another fundamental goal of the physics program with the ultra-relativistic heavy--ion collisions.  

With the huge progress in operating high-energy experiments, it turns to construct suitable canonical ensembles to incorporate exactly the charge conservation for a given dynamical system as a key tool in the high-energy collisions. In context of quantum statistical mechanics, one implicitly wanted to build Fock spaces of states which are characterized by defined symmetry properties of prescribed charges. In SU(2) symmetry, the energy and total spin dependence of the single--particle density of states in the nuclear Fermi-gas mode has been calculated. About six decades ago, micro--canonical calculations taking into consideration four-momentum conservation have been presented. A general approach to calculate the phase volume was implemented in order to study the consequences of including momentum conservation in the relativistic collisions. 

The statistical description of particle production in high--energy nuclear interactions dates back to 1948, when Heinz Koppe published his paper entitled: ''Die Mesonenausbeute beim Beschuss von leichten Kernen mit Alpha--Teilchen [German]'' or ''Meson Yields from Bombarding Light Nuclei with Alpha Particles [English]'' in Zeitschrift f\"ur Naturforschung. Koppe introduced an almost complete recipe for statistical-thermal models. Particle production (mesons or lightest Goldtone modes), formation and decay of resonances, temporal and thermal evolution of the interacting system, approaches applying either Fermi-Dirac or Bose-Einstein or Stefan-Boltzmann statistics in thermal medium and under equilibrium condition in the final state of the nuclear interaction were implemented. Furthermore, Koppe estimated the equilibrium concentrations of each type of the produced particles

In 1950, Fermi introduced a statistical theory based on the assumption that in multiple processes occurring in high-energy collisions there is a localization of the energy in a small spatial volume which then decays into various possible final states. The decay is conjectured to be compatible with the constant motion relative a priori probabilities proportional to their statistical weights. Thus, the Fermi model is preferably applicable in energy range compared with that of  the cosmic rays. At the lower end of this energy range, the model becomes unreliable, because only a few final states are accessible and their details are important at the higher end. Because of insufficient consideration of peripheral interactions, deviations likely occur at low energy. This should be restricted due to conservation laws. Conservation of the charges, baryon number and strangeness was studied within the thermodynamical model of high-energy hadronic collisions. 

As {\it dictated} in Koppe model, the particle production in Fermi model is treated by means of statistical tools, here weights. Furthermore,  Fermi model gives a generalization of the {\it ''statistical model''}, in which one starts with a general cross-section formula and inserts into it a simplifying assumption about the matrix element of the process, which reflects that many features of the high-energy reactions dominated by the density in phase space of the final states. The final states are accessible from the given initial state under the consideration of all conservation laws. Therefore, the formulation of Fermi model  is sufficiently general to allow an application to many different processes such as nucleon-antinucleon annihilation, and the production of pions and strange particles in meson-nucleon and nucleon-nucleon ({\it elementary}) collisions. 

A general description for the quantum statistics in  high-energy collisions with a variable number of ideal (non-interacting) particles was introduced by Magalinski and Terletskii. The generalization extends to an aggregate of oppositely charged particles, for which the conservation laws are fulfilled. Expressions for total number of particles and for the total energy are deduced. They are distinguishable from the ordinary quantum statistics. In almost all implications of the statistical-thermal methods, charge conservation, the underlying symmetry corresponds to one or several unrelated U(1) groups should be respected. This apparently means that in order to obtain the total charge of a many-particle state, the charges of the constituents can be simply added. There is an increasing interest in gauge theories based on various non-Abelian internal symmetries, which was introduced by Yang and Mills.

The phase transition of strongly interacting matter has been intensively studied for many years. Already in fifties of last century, long before the discovery of the Quantum Chromodynamics (QCD), there have been first speculations about a possible new phase of strongly interacting matter, based on studies of the thermodynamics of a hadron resonance gas. Pomeranchuk came up with an assumption that a finite hadron size would imply a critical density above which the hadronic matter cannot be in the compound state, known as hadrons. In 1964, Hagedorn introduced the mass spectrum in order to characterize the abundant formation of hadron resonances with increasing masses and rotational degrees of freedom. Accordingly,  the mass spectrum  formulated the concept of limiting temperature based on the statistical bootstrap model. 

In 2003, the statistical-thermal model has been confronted to the lattice QCD simulations. Details about the critical conditions near deconfinement were possible. The lattice QCD calculations suggest that this transition should be a true transition only in small quark mass intervals. In a broad intermediate quark masses regime the transition is not related to any singular behavior of the QCD partition function. Nonetheless, it still is well localized and  characterized by rapid changes of thermodynamic quantities in a narrow temperature interval. The transition temperature thus is well defined and determined in the lattice QCD calculations through the location of maxima in response functions such as the chiral susceptibility. 

At critical temperature, the additional degrees of freedom carried by the quark-gluon plasma, the new state of matter, are to be released resulting in an increase in the thermodynamic quantities like energy and pressure densities. This was physical picture before 2003. The success of statistical-thermal model in reproducing lattice QCD results at various quark flavors and masses (below $T_c$) changed this picture drastically. Instead of releasing additional degrees of freedom at $T>T_c$, the hadronic system increases its effective degrees of freedom, namely at $T<T_c$. In other words, the hadron resonance gas (HRG) has much more degrees of freedom than QGP.

Since the pioneering work of 2003, a wide spectrum of applications has absolved the statistical-thermal models. In this review, we will discuss recent progress in constructing  statistical-thermal models and their applications to phenomenological descriptions of particle production, correlations and higher order moments of multiplicity in nucleus-nucleus collisions. The importance of conservation laws and their different implementations in the statistical-thermal approach is emphasized.

The present review is organized as follows. In section \ref{sec:histr}, a short history of the statistical-thermal models shall be outlined. The historical milestones on the path of constructing statistical-thermal models for the  heavy-ion collisions are shortly summarized. The early birth of statistical description dates back to early days of atomic physics. Thomas-Fermi statistical theory, section \ref{sec:tf}, is one of the earliest tractable schemes for solving the many-body problem. The theory was suggested to describe atoms as uniformly distributed electrons (negatively charged clouds) around nuclei in a six-dimensional phase space (momentum and coordinates). Apparently, this is an oversimplification of the actual many-body problem.

Koppe introduced an almost complete recipe for statistical-thermal models, section \ref{sec:hk}, including particle production, formation and decay of resonances, temporal and thermal evolution of the interacting system under equilibrium condition in the final state of the nuclear interaction.

Fermi model, section \ref{sec:firmim}, implements statistical weights and gives a generalization of the {\it ''statistical model''}. One starts with a general cross-section formula and inserts into it a simplifying assumption about the matrix element of the process (high-energy reactions dominated by the density in phase space). From the given initial state under all conservation laws, the final states are accessible. 

In section \ref{lsmt}, a general approach to calculate the phase volume shall be introduced. This method was suggested to study the consequences of including momentum conservation in the relativistic collisions. A general description for the quantum statistics with a variable number of ideal (non-interacting) particles was presented.  Another general feature of statistical description in high-energy collisions is to determine the phase space or Fock space. This should be restricted due to conservation laws. Conservation of the charges, baryon number and strangeness was studied within the thermodynamical model of high-energy hadronic collisions.
 
The Hagedorn approach shall be outlined in section \ref{sec:hagd}. The abundant formation of resonances of increasing mass and rotational degrees of freedom was
one of the most striking features of strong interaction physics, which has attracted intense theoretical attention for about six decades. Hagedorn utilized all tools of statistical physics. He assumed that {\it higher and higher resonances of strongly interacting particle occur and take part in the thermodynamics as if they were particles} and introduced the concept of the limiting temperature based on the statistical bootstrap model. That was the origin of multiphase structure of hadronic matter. The limiting temperature  arose as a consequence of the mass spectrum which relates the number of hadronic resonances to their masses as an exponential. 

In order to use the resonance gas model for further comparison with lattice QCD results, Karsch, Redlich and Tawfik suggested to rescale the masses of hadron resonances, section \ref{sec:krt1}. The dependence of critical temperature as estimated in the statistical-thermal model with varying masses according to the quark mass and number of flavors in lattice QCD simulations can be deduced.  The dependence of the critical energy density on the quark mass and number of flavors shall be studied. The resulting hadronic states with changing the masses of the lightest Goldstone mesons shall be elaborated. These three topics characterize the resulting statistical-thermal model and its matching with the lattice QCD simulations

The statistical-thermal models in non-ideal hadron gas shall be discussed in section \ref{sec:nonidealG1}. Three categories of hadronic interactions shall be discussed in section \ref{sec:intr}. We start with van der Waals repulsive interactions, section \ref{sec:exclvV}, then we introduce the Uhlenbeck and Gropper statistical interactions, section \ref{sec:UG} and general $S$-matrix and strong interactions \ref{sec:smatrx}.

The dissipative properties shall be studied in section \ref{sec:visc}. This shall be divided into a single-component fluid, section \ref{sec:viscone}, and multiple-component fluid, section \ref{sec:viscmult}. Chapman-Enskog and Relaxation time approximation shall be utilized. Topics like, $K$-matrix parametrization of hadronic cross-sections, relation of $T$-matrix to $K$-matrix and the relaxation time shall be discussed.

Section \ref{sec:rapd} shall be devoted to the statistical-thermal models in rapidity space.
The goal is the study of the dependence of thermal parameters on the location, in particular on the spatial rapidity. A single freeze-out temperature model shall be given in section \ref{sec:sftm1}. The rapidity dependence of thermal parameters shall be discussed in section \ref{sec:rptp1}.  An approximation of  beam rapidity dependency of  particle ratios can be derived in the Regge model. Accordingly, the baryon-pair production at very high energy is governed by Pomeron exchange, section \ref{sec:rpbr1}. The asymmetry between baryons and anti-baryons can be expressed by the string-junction transport and by an exchange with negative $C$-parity. The proton ratios versus Kaon ratios shall be reviewed in section \ref{sec:kkpp}. 

Section \ref{sec:hic1} is devoted to the equilibrium statistical-thermal models in high-energy physics. The conservation laws shall be presented in section \ref{sec:conserv}. 
The extensive comparison with heavy-ion collisions shall be reviewed in section \ref{sec:compHIC}. The particle multiplicity shall be outlined in section \ref{sec:multp}. First, we shortly introduce Bjorken model, in which early thermalization, vanishing baryon chemical potential in the fluid and one-dimensional expansion besides boost symmetry of the initial conditions, the initial independence on rapidity and the isotropic (homogenises) rapidity over the spacetime are the main assumptions. Second, the experimental results shall be confronted to the statistical-thermal models. The particle abundances at Alternating Gradient Synchrotron (AGS), Super Proton Synchrotron (SPS) and RHIC, and recently at LHC energies are found consistent with equilibrium populations. This makes it possible to extract both freeze-out parameters over a wide range of Nucleus-Nucleus center-of-mass energies $\sqrt{s_{NN}}$ from fits of measured particle ratios with the thermal models. The event generators  PYTHIA $6.4.21$ and Heavy-Ion Jet INteraction Generator (HIJING) shall be shortly introduced. The dynamical fluctuations of particle ratios shall be the other feature to compare with, section \ref{sec:dynmFlct}.

Section \ref{sec:fo1} shall be devoted to the chemical freeze-out. In order to deduce  a universal  relation between chemical freeze-out parameters, $T$ and $\mu_b$ and nucleus-nucleus center-of-mass energy, $\sqrt{s_{NN}}$, a common method used is to fit the experimental hadron ratios. Starting with a certain value of baryon chemical potential $\mu_b$, temperature $T$ is increased very slowly. At this value of $\mu_b$ and at each raise in $T$, the strangeness chemical potential $\mu_S$ is determined under the condition that the strange quantum numbers should remain conserved in heavy-ion collisions. Having three values of $\mu_b$, $T$  and $\mu_S$, then all thermodynamic quantities including the number density $n$ of each spices are calculated. When the ratio of two particles reaches the experimental, then the temperature $T$ and chemical potential $\mu_b$ are registered. This procedure is repeated for all particle ratios measured in different high-energy experiments.

Six universal conditions are supposed to describe the freeze-out diagram:
\begin{itemize}
\item Cleymans and Redlich: constant energy per particle, section \ref{sec:cr_en1},
\item Braun-Munzinger and Stachel: constant-total baryon density, section \ref{sec:totalb},
\item Magas and Satz: percolation theory in heavy-ion collisions, section \ref{sec:stzm},
\item Tawfik: constant-normalized entropy, $s/T^3$, section \ref{sec:Tawst3},
\item Tawfik: vanishing product of kurtosis and susceptibility, $\kappa\; \sigma^2$, section \ref{sec:fo-hm}, and
\item Tawfik: constant-trace anomaly, $(\epsilon-3 p)/T^4$, section \ref{taw:It4}.
\end{itemize}

The comparison with the  lattice QCD calculations shall be introduced in section \ref{sec:lqcd}. The lattice QCD simulations turn to be compatible with experiments. About ten years ago, the challenge of confronting HRG with the lattice QCD results was risen. It intended to learn about the critical conditions near deconfinement and the underlying degrees of freedom. The comparison shall be divided into thermodynamics at vanishing chemical potential, section \ref{sec:QCDmu0}, thermodynamics at vanishing chemical potential, section \ref{sec:QCDmun0}, chiral phase transition, section \ref{sec:chiTr}, and confinement-deconfinement phase diagram, section \ref{sec:QCDpd}.

The higher order moments of charged-particle multiplicity distribution were predicted four decades ago. The empirical relevance to the experimental measurements is now clear. Section \ref{sec:hm} shall be devoted to the higher order moments of particle multiplicities. The measurement of the correlation length $\xi$ seems to be very much crucial. It has been pointed out that the contribution of the critical fluctuations to higher order moments of multiplicities is proportional to a positive power of $\xi$. The latter is conjectured to diverge at QCD critical endpoint and such an assumption is valid in the thermodynamic limit.

The topic is divided into non-normalized, section \ref{nonnorlTc}, normalized higher order moments, section \ref{sec:norm}, and products of higher order moments, section \ref{sec:mult}. The comparison with the experimental results shall be discussed in section \ref{sec:HMexp}. The comparison with the lattice QCD results shall be discussed in section \ref{sec:HMlqcd}.

Three models combining hadronic ($<T_c$) with partonic ($>T_c$) phases shall be presented in section \ref{sec:ABTc}. They are examples about other models for thermodynamics of hadronic matter. Their analogy to the statistical-thermal models is obvious, i.e. possessing comparable effective mass and coupling. The effective degrees of freedom is remarkably reduced and therefore the ideal gas behavior of lattice QCD at high temperatures can not be explained. The hadronic and partonic models as:
\begin{itemize}
\item quasi-particle model, section \ref{hrg_quasiprtcl},
\item linear $\sigma$-model with Polyakov-loop potentials, section \ref{hrg_schaefer}, and 
\item compressible bag model, section \ref{hrg_koch}.
\end{itemize}

\section{A short history}
\label{sec:histr}

One of the earliest reviews \cite{despers} on statistical-thermal methods in high-energy physics \cite{kretz} assumed that the dispersion relations was the first approach for a general description of the interaction of $\pi$-mesons and nuclei. With increasing energies, especially when the reactions with three or more particles are taken into consideration, the complexity of the mathematical formalism seems to grow rapidly.

Based on Gibbs condition, the equilibrium behavior of thermodynamical observables is to be evaluated as an average over statistical ensembles. This is essentially different than the time average for a particular state, which in turn would be essential when dealing with out-of-equilibrium processes \cite{Tawfik:2010pt,Tawfik:2010kz,Tawfik:2010uh}. Therefore, the equilibrium distribution could be an average over the accessible phase space. The statistical ensemble corresponding to the thermodynamic equilibrium is the one in which the phase space density should be uniform over the whole accessible phase space. The level of equilibrium is determined by the level of uniform occupying the accessible phase space.  The agreement between observables and predictions using the statistical operator likely imply equilibrium \cite{redb}.

A short review of the historical milestones on the path of constructing statistical-thermal models for the  heavy-ion physics is outlined in undermentioned sections. Recently, a brief history of strong interaction thermodynamics in quark matter and nuclear collisions was presented by Satz \cite{satzBHistry}.
 
\subsection{(1927-1928) Thomas-Fermi: a statistical approach for atomic physics}
\label{sec:tf}

The Thomas-Fermi statistical theory is one of the earliest tractable schemes for solving the many-body problem.  The theory was suggested to describe atoms as uniformly distributed electrons (negatively charged clouds) around nuclei in a six-dimensional phase space (momentum and coordinates). Obviously, this is an oversimplification of the actual many-body problem.

Shortly after Schr\"odinger introduced his quantum mechanical wave equation, the Thomas-Fermi statistical theory was invented by L.H. Thomas in January 1927 \cite{thomas27} and independently by E. Fermi a year later \cite{efermi28A,efermi28B}. The basic idea is a follows. Based on the Pauli exclusion principle, the degenerate gas of fermions will be enforced to occupy the phase space with two particles (two fermions) per volume $\hbar^3$ of this space \cite{myers}. The fermions move in an effective potential field which is determined by nuclear charge and by assumed uniform distribution of the particles. For a large number of atoms or molecules, i.e. large $N$ of electrons, this theory approximately describes the electron density, $\rho(\textbf{x})$, where $\textbf{x}\in R^3$ and ground state entropy, $E(N)$.

Although, Schr\"odinger equation gives the exact density and energy, it cannot be easily worked out for large $N$. The particle density $\rho(\textbf{x})$ is the central variable rather than the wavefunction and the total energy of a system is given as a function of the particle (fermion) density. As pointed out by Lieb in an unpublished work \cite{Lieb}, the ordinary matter like atoms and molecules are in their ground state most of the time. This would highlight the importance of Thomas-Fermi model in providing a framework to estimate the properties of matter in its ground. In other words, the statistical Thomas-Fermi model is to be applied to a comprehensive survey for macroscopic nuclear properties. The statistical approximation enters the game, when dealing with nuclei. Various improvements have been applied to the phase space in the atomic models, mainly coming up with density-gradient corrections and other improvements \cite{corr1,corr2,corr3}. The widely used one is based on a formal expansion in powers of $\hbar$ \cite{corr4}.

The functional energy is the starting point for the theory. For example, for a molecule with $k$ positively charged atoms, i.e. nuclei with $z_i>0$ at fixed locations $R_i\in R^3$, where $i$ runs over $k$, the functional energy is to be divided into {\it four} contributions as given below, Eqs. (\ref{eq:ETM1}) and (\ref{eq:ETM2}). The total energy of a nucleus is given by integral over all space of an energy density, which mainly consists of Thomas-Fermi kinetic energy and Coulomb and nuclear interaction. The nuclear interaction $E$ can be divided into repulsive and attractive interactions \cite{myers},
\bea
E_{nn\; \text{or}\; pp} &=& \frac{1}{2} \int d^3 r_1\; \int d^3 r_2\;  \int_{n\, \text{or}\, p} d^3 p_1\; \int_{n\, \text{or}\, p} d^3 p_2\; \left(\frac{2}{\hbar^3}\right)^2\; u_{12}^{(\text{rep})}, \label{eq:ETM1}\\
E_{np} &=& \frac{1}{2} \int d^3 r_1\; \int d^3 r_2\;  \int_{n} d^3 p_1\; \int_{p} d^3 p_2\; \left(\frac{2}{\hbar^3}\right)^2\; u_{12}^{(\text{attr})}.  \label{eq:ETM2}
\eea 
The proton or neutron density is $2/\hbar^3$, which apparently assumed $2$ degenerate particles. In this case fermions occupy an element $\hbar^3$ of the available phase space. For $i$-th particle, the phase space is characterized by coordinates $\vec{r}_i$ and $\vec{p}_i$.

The effective interaction energy has been worked out in literature \cite{myers,Lieb},
\bea
E_{\rho} &=& A_k \int \rho(\mathbf{x})^{5/2}\; d \mathbf{x}  
+ \int \rho(\mathbf{x})\, u(\mathbf{x})\; d\, \mathbf{x} 
+ \frac{1}{2} \int \int \frac{\rho(\mathbf{x}) \; \rho(\mathbf{x^{\prime}})}{\left|\mathbf{x} - \mathbf{x}^{\prime} \right|}\; d\, \mathbf{x} \, d\, \mathbf{x}^{\prime}. \label{eq:ETM3}
\eea
{\bf The first term} gives the Thomas-Fermi kinetic energy, which is associated with a system of non-interacting particles in a homogeneous gas. This functional equation could be applied to electrons in atoms encountering the most important idea of the modern density functional theory and local density approximation. This is obtained by integrating the kinetic energy density of a homogeneous gas, 
\bea \label{eq:t0}
\int t_0[\rho(\mathbf{x}]\; d\, \mathbf{x} &=& \int \left(\frac{2}{(2 \pi)^3} \int \frac{k^2}{2}\, \rho_k\, d\, k\right)\; d\, \mathbf{x} = \int \left(\frac{1}{2 \pi^2} \int_0^{k_F} k^4 \, d\, k\right)\; d\, \mathbf{x},
\eea
which in turn is obtained  by summing all of the free-particle energy states $k^2/2$  up to the Fermi wavevector, $k_F=(3\, \pi^2\, \rho(\mathbf{x}))^{1/3}$. The coefficient $A_k$ is given as $(3/10) (3\, \pi^2)^{2/3}$.

{\bf The second term in Eq. (\ref{eq:ETM3})} is  the classical electrostatic energy of attraction and repulsion between the nuclei and the electrons, where $u$ is the static Coulomb potential energy arising from the nuclei
\bea
u(\mathbf{x})) &=& - \sum_{j=1}^N \frac{z_j}{\left|\mathbf{r} - \mathbf{R}_j \right|}.
\eea

{\bf The third term in Eq. (\ref{eq:ETM3})} represents the repulsive interactions of the system, concretely, the classical Coulomb repulsion or  Hartree energy, which is a physical constant (two times the binding energy of the electron on the ground state of the 
hydrogen atom).
 
It is worthwhile to mention that the ground state density and energy of a system can be obtained, when Eq. (\ref{eq:t0}) is minimized, under the condition of charge conservation. This type of constrained minimization problem, which occurs frequently within many-body methods can be performed using the technique of Lagrange multipliers.

The Thomas-Fermi theory has been before to study potential fields and charge density in metals and the  equation of states of elements \cite{feymns49}. However, this method suffers from many deficiencies  \cite{dfec1}. Probably, the most serious defect is that the theory does not predict bonding between atoms \cite{dfec1,dfec2,dfec3}. Therefore, molecules and solids are not allowed to be formed in this theory. The large approximation in the kinetic energy is the main reason. The kinetic energy represents a substantial portion of the total energy of a system and so even small errors prove disastrous. Another shortcoming is the over-simplified description of the particle interactions, which are treated classically and so do not take account of quantum phenomenon such as the exchange interaction. 

In view of  modern density functional theory, Thomas-Fermi theory can be considered as an approximation to the more accurate theory. Nevertheless, Thomas-Fermi theory remains a great milestone on the path of construction methods to deal with many-body problem, especially in six-dimensional phase space.

\subsection{(1948) Heinz Koppe: statistical methods for particle production}
\label{sec:hk}

Heinz Koppe formulated the abstract of his pioneer paper \cite{hkoppeA,hkoppeB} as follows. \\
{\bf German:} [{\it Mittels des neuen Berkely-Betatrons ist es möglich gewesen, durch Beschuß von leichten Kernen (insbesondere C) mit $\alpha$-Teilchen von etwa $380~$MeV Mesonen zu erzeugen. Im folgenden soll eine einfache Methode angegeben werden, nach der sich die dabei zu erwartende Ausbeute absch\"atzen l\"a\3t.}] \\
{\bf English:} [{\it By means of the new Berkely-Betatrons, it turns to be possible through bombarding light nuclei (especially $C$) with $\alpha$-particle of about $380~$MeV, mesons will be produced. We present a simple method that can be used to determine the produced particle.}]

Apparently, Koppe formulated an almost complete recipe for statistical-thermal models. The Koppe's method includes particle production (mesons or lightest Goldtone modes), formation and decay of resonances, temporal and thermal evolution of the interacting system, approaches applying either Fermi-Dirac, or Bose-Einstein or Stefan-Boltzmann statistics in thermal medium and equilibrium condition in the final state of the nuclear interaction. Utilizing statistical methods, Koppe made estimation for the equilibrium concentrations of each type of the produced particles \cite{hkoppeA,hkoppeB}. 

\subsubsection{Particle production in thermal medium}

Starting with the excitation energy per nucleon 
\bea
U &=& \frac{m_2}{(m_1+m_2)^2}\; E,
\eea
where $m_1 (m_2)$ and $E$ are the mass of projectile (target) nucleus and the kinetic energy, respectively. The temperature of {\it excited} daughter  nucleus $T_0$ is related to $U$; $T_0=3.8\sqrt{N}$, where $N$ refers to the number of excited nuclei (resonances). In Berkely-Betatrons, $T_0$ was measured as $\sim 10~$MeV. At such very high temperature, the projectile ($\alpha$-particle) can not remain stable \cite{hkoppeA,hkoppeB}. Under these circumstances, pair production is likely \cite{hkoppe2}. The meson production is apparently not possible. {\it ''Vacuum dissociation''} \cite{refffff2} or {\it ''pair-degeneracy''} \cite{hkoppe2} were mainly investigated from electron-positron pair production. Koppe assumed that the same considerations make it possible to apply this in the production of meson pairs \cite{hkoppeA,hkoppeB}. The thermal production of a particle begins once the temperature approaches the order of the rest-energy of the particle of interest. Because the rest-energy of mesons is about $150~$MeV, Koppe expected a very small number of mesons and in this case the equations of {\it low} temperatures \cite{hkoppe2} can be utilized. For electrons with a rest-energy of $0.5~$MeV, a situation is formed, in which pair-degeneracy apparently becomes dominant. 

Koppe assumed that the electron gas can be treated as cavity radiation with energy density
\bea
u &=& \frac{7}{4} \frac{\pi^2}{15\, c^3\, \hbar^3}\; T^4.
\eea
The radiation loss is simply related to the cross-section of the excited daughter nuclei $\sigma$. The time dependence of temperature loss reads
\bea \label{eq:Tt1}
T(t) &=& \left(T_0^{-2} + 2 \, B \, t\right)^{-1/2},
\eea
where $B={\cal O}(\sigma/(m_1+m_2))$. Equation (\ref{eq:Tt1}) gives the effects of the radiation loss and the expansion of the interacting system. Both control the temperature loss, which apparently continues till no more mesons is produced.

\subsubsection{Statistical methods}

For determining the possibility of available mesons, statistical tools come into play as follows \cite{hkoppe2}. Because the speed of electrons shall have almost the same order of $c$, then the energy flux, which is caused by the electrons, relative to light quantum radiation will be increased by the same factor ($7/8$) so that one can relate both to each other and calculate using the Stefan-Boltzmann law with the factor $(1+7/4)=11/4$ (the factor $7/8$ stands for the difference between Bose-Einstein and Fermi-Dirac statistic).
\bea
n(\epsilon) &=& \frac{(2 m_b)^{3/2}}{\pi^2\, \hbar^3}\; \epsilon^{1/2}\, \left[1+e^{\frac{m_b\, c^2 + \epsilon}{T}}\right]^{-1},
\eea
where $m_b$ and $\epsilon$ are meson mass and density of the kinetic energy, respectively. The rate of produced particle (number of meson per unit time) is
\bea
\nu(T(t)) &=& \frac{m_b\, \sigma}{\pi^2\, \hbar^3}\, T(t)^2\, e^{-\frac{m_b\, c^2}{T(t)}}.
\eea
Then, the integration results in
\bea
n &=& a (m_1+m_2) \, T_0 \, e^{-\frac{m_b\, c^2}{T_0}},
\eea
where $a=0.031$. Substituting with given values, then the number of mesons which will be produced in $\alpha-A$ collisions at $380~$MeV per unit time is $\sim1.7\times10^{-4}$.

\subsection{(1950) Enrico Fermi: statistical weights for particle production}
\label{sec:firmim}

As {\it dictated} in Koppe model \cite{hkoppeA,hkoppeB}, the particle production in Fermi model \cite{fermi1,fermi2,fermi3,fermi4,fermi5} is treated by means of statistical tools, here weights. Furthermore,  Fermi model \cite{fermi1,fermi2,fermi3,fermi4,fermi5} gives a generalization of the {\it ''statistical model''}, in which one starts with a general cross-section formula and inserts into it a simplifying assumption about the matrix element of the process, which likely reflects  many features of the high-energy reactions dominated by the density in phase space of the final states. The final states are accessible from the given initial state under the consideration of all conservation laws. Therefore, the formulation of Fermi model \cite{fermi1,fermi2,fermi3,fermi4,fermi5} is sufficiently general to allow applications to many different processes such as nucleon-antinucleon annihilation, and the production of pions and strange particles in meson-nucleon and nucleon-nucleon ({\it elementary}) collisions. 

Fermi model is based on the assumption that in multiple processes occurring in high-energy collisions, there is a localization of the energy in a small spatial volume which then decays into various possible final states. The decay is conjectured to be compatible with the constant of motion with relative a priori probabilities proportional to their statistical weights. Thus, the Fermi model is preferably applicable in energy range compared with that of the cosmic rays \cite{kretz}. At lower end of this energy range, the model becomes unreliable, because only a few final states are accessible and their details are important at higher end. Because of insufficient consideration of peripheral interactions, deviations likely occur at low energy.

\subsubsection{Cross-section in general scattering theory}

From $S$-matrix theory in natural units and for  initial $i$ and  final $f$ state describing a collision of two ({\it fundamental}) particles with masses $m^{\prime}_k$, where $k=1,2$ and four-momenta $p^{\prime}_k=(\omega^{\prime}_k, p^{\prime}_k)$, where $\omega^{\prime}_{k^2}=m^{\prime}_{k^2}+p^{\prime}_{k^2}$, the cross-section of that reaction which leads to any state out of $\Phi$, is given as
\bea
\sigma_{\Phi} &=& \frac{(2 \pi)^2 \; \omega^{\prime}_1 \; \omega^{\prime}_2 }{\sqrt{(p^{\prime}_1\, p^{\prime}_2)^2 -  {m^{\prime}}_1^2\, {m^{\prime}}_2^2}} \; \sum_{f\in\Phi} \delta^4 (P_f - p^{\prime}_1 - p^{\prime}_2)\; \left| M_{f i}\right|^2, \label{eq:sgmscrt1}
\eea
where $P_f$ is the total four-momentum in the final state $f$. The matrix element $M_{f i}$ can  rigorously be related to the scattering matrix $S$
\bea
S_{f i} &=& I_{f i} + i\, .\, \delta^4 (P_f - p^{\prime}_1 - p^{\prime}_2)\;  M_{f i},
\eea
where $I_{f i}$ is a unitary operator. The problem of determining the matrix element $M_{f i}$ still unsolved. This is one of the basic problems of quantum field theory. The invariance of  $M_{f i}$ under certain groups of transformations like translation, Lorentz transformation and rotation in isotropic spin space, is essential. 

The strong interactions, which are simply sufficiently isolated, obey the local conservation laws of total four-momentum $P$, total angular momentum $J$, total isotropic spin $T$, total baryon number $N$, total strangeness $S$, and total electric charge. The total electric charge is the conservation of the three-component $T_3$.
\bea
M_{f i} &=& \delta_{N_f\, N_i} \, \delta_{S_f\, S_i} \, \delta_{T_{1 f}\, N_{1 i}} \times  \nonumber \\  &&\sum_{\alpha,\, J,\, T} \; \sum_{\alpha^{\prime},\, J^{\prime},\, T^{\prime}} \langle f | J, T, \alpha \rangle \, \delta_{J J^{\prime}} \delta_{T T^{\prime}} \, M(\alpha, \alpha^{\prime}, J, T) \langle J^{\prime}, T^{\prime}, \alpha^{\prime}| i \rangle. 
\eea 
The unknown matrix element $M(\alpha, \alpha^{\prime}, J, T)$ contains all dynamical effects. Some remarks in dealing with $M(\alpha, \alpha^{\prime}, J, T)$ are elaborated in Ref. \cite{kretz}.

\subsubsection{Statistical aspects of particle production in final state}

Any assumption for the dependence of  $M(\alpha, \alpha^{\prime}, J, T)$ on conserved quantities like mass, momenta and quantum numbers can be checked for usefulness by the statistical theory. Ignoring spins, charges and all conservations except the total four-momentum, then $\sigma_{\Phi}$ is unity, i.e. Lorentz invariant and 
\bea
M_{f i} &=& \left(\frac{m^{\prime}_1 . m^{\prime}_2 . m_1 .  m_2 \cdots m_n}{\omega^{\prime}_1 . \omega^{\prime}_2 . \omega_1 .  \omega_2 \cdots \omega_n}\right)^{1/2} .\, M_{f i}^{\prime},
\eea
where $m_i$ and $\omega_i$ are masses and energies of $i$-th state. It is obvious that $M_{f i}^{\prime}$ depends on the Lorentz-invariant and $M_{f i}=M_{f i}^{\prime}$ in the non-relativistic limit.

The matrix element can be given as \cite{fermi1,fermi2}
\bea \label{eq:mfi}
\left|M_{f i}^{\prime}\right|^2 \propto (2 \pi)^{-3(n-1)} \, \Omega^{n-1},
\eea
where $\Omega$ does not depend on momentum of the particle in the final state and has the volume dimension. Furthermore, $\Omega$ might depend on the total center-of-mass energy $E$. Concretely, $\Omega$ is assumed to be the volume of the larger out of the two Lorentz-contracted spheres. Therefore, Eq. (\ref{eq:sgmscrt1}) can be simplified as follows.
\bea \label{eq:sgm1}
\sigma(E; m_1,\cdots,m_n) &\propto & P(E; m_1, \cdots, m_n) \nonumber \\
&=& \left(\frac{\Omega}{V}\right)^{n-1} \,  \left(\frac{V}{8\, \pi^3}\right)^{n-1} \; \rho(E,0; m_1, \cdots, m_n),
\eea
where $V$ is the volume. Expressions (\ref{eq:mfi}) and (\ref{eq:sgm1}) can be interpreted as follows. In point-of-view of an observer in the center-of-mass system, each of the two incoming particles appears as a Lorentz-contracted sphere. Then, the interacting spheres penetrate each other. The total energy $E$ will be concentrated in a small space region of the volume $\Omega$. Because of the strong interactions, a rapid succession of transitions likely sets in. Particles of various kinds will be created and annihilated. With increasing the center-of-mass energy, the number of produced particles rapidly increases. Such a circumstance apparently suggests that some sort of statistical equilibrium will be attained. Therefore, each accessible state will finally be excited to its average statistical strength. In other words, the system reaches a state, in which the particles can be considered as a gas of a free constituents confined to the large normalization volume $V$. The cross-section of each final state is proportional to $(\Omega/V)^{n-1}$. There is another indirect proportionality  to the number of produced particles $n$. The latter is likely proportional to the volume. Furthermore, the cross-section of each final state is proportional to the density of the final state. It is apparent that this approach is likely the extreme opposite of the perturbation theory \cite{fermi1}.

The phase space integral \cite{sriv,neum1} reads 
\bea
\rho_S(E,0; m_1, \cdots, m_n) &=&\int dp_1\cdots\int dp_n\, \frac{m^{\prime}_1 . m^{\prime}_2 . m_1 .  m_2 \cdots m_n}{\omega^{\prime}_1 . \omega^{\prime}_2 . \omega_1 .  \omega_2 \cdots \omega_n} 
\delta\left(E-\sum_{i=1}^n\varepsilon_i\right)\, \delta^3 \left(E-\sum_{i=1}^n p_i\right), \label{eq:rhoS}
\eea
where the dispersion relation $\varepsilon_i=\sqrt{m_i^2+p_i^2}$. There is another theory \cite{lepor} assuming the replacement of $(\omega_1 \cdot \omega_2 \cdots \omega_n)^{-1}$  by $(\omega_1 \cdot \omega_2 \cdots \omega_n)^{-3/2}$, then
\bea
\rho_F(E, P; m_1, \cdots, m_n) &=&\int dp_1\cdots\int dp_n\,  \delta\left(E-\sum_{i=1}^n\varepsilon_i\right)\, \delta^3 \left(E-\sum_{i=1}^n p_i\right). \label{eq:rhoF}
\eea
As pointed out in Ref. \cite{kretz}, a systematic comparison between Eq. (\ref{eq:rhoS}) and (\ref{eq:rhoF}) is not possible. Because the Lorentz invariance in Eq. (\ref{eq:rhoS})  apparently permits the use of simple recurrence formulas, a numerical evaluation is comparatively easy. The property of Lorentz invariance in Eq. (\ref{eq:mfi}) likely does not necessarily guarantee
a better agreement with the experiment.

The statistical model of Fermi \cite{fermi1} was successfully applied to large angle elastic and exchange scattering \cite{hgd2,hgd3,hgd4}.  For non-invariant statistical model and  probabilities $P_j$ for all channels $j$ on the reaction \hbox{$p+p \rightarrow j'$}, then in natural units,
\bea
\left(\frac{P_0}{\sum_j P_j}\right)_{p p} &=& e^{{\cal O}(E)},
\eea
where $P_0$ is the probability in the elastic channel. The asymptotic behavior out of the sums over phase space integrals \cite{hgd5,hgd7} of  out-of-thermodynamic methods \cite {hgd8,hgd9}, 
\bea
\left(\frac{P_0}{\sum_j P_j}\right)_{p p} &=& e^{-a\, E^{\alpha}},
\eea
where $a$ and $\alpha$ are constants. To keep the energy per particle and consequently the temperature constant, increasing $E$ is followed by increasing the number of possible kinds of particles \cite{hgd9}. This is the heart of Hagedorn model, section \ref{sec:hagd}.

\subsection{(1954-1957) Lepore, Stuart, Magalinski and Terletskii: statistics of charge-conserving systems} 
\label{lsmt}

With the huge progress in conducting high-energy experiments, it turns to construct suitable canonical ensembles in order to exactly  incorporate the charge conservation for a given dynamical system as a key tool in the high-energy collisions. In context of quantum statistical mechanics, one implicitly wanted to build Fock spaces of states which are characterized by defined symmetry properties or prescribed charges \cite{Elze}.  In SU(2) symmetry, the energy and total spin dependence of the single-particle density of states in the nuclear Fermi-gas mode has been calculated \cite{magl2,magl1A,magl1B}. About six decades ago, micro-canonical calculations taking into consideration four-momentum conservation have been presented \cite{lepor,magl3A,magl3B}.

\subsubsection{Phase space volume and momentum conservation}

Lepore and Stuart \cite{lepor} introduced a general approach to calculate the phase volume. This method is used to study the consequences of including momentum conservation in the relativistic collisions. The statistical weight  of a state that leads to $n$ particles with masses $m_1, m_2, \cdots, m_n$ has to fulfil the rule 
\bea
S_n &=& {\cal R}\, \frac{\Omega^{n-1}}{(2 \pi)^{3(n-1)}} \, \int \prod_{i=1}^{n} d^{3} \, p_i, 
\eea
where $\Omega$ is the configurational space volume, $\Omega=(2 m/E) \Omega_0= (2 m/E) (4 \pi / 3)/r^3$, with energy $E$, mass $m$ and proper coordinate $r$. The ratio ${\cal R}={\cal S}/{\cal G}$ stands for weight factor ${\cal S}$ due to conservation of spin and isotopic spin and ${\cal G}$ which converts the specific phase space volume into a generic one appropriate to indistinguishable particles. 

The requirements of momentum and energy conservation is to be imposed on the statistical weight,
\bea
S_n &=& {\cal R}\, \frac{\Omega^{n-1}}{(2 \pi)^{3(n-1)}}\, \frac{d}{d (i E)} \, \int_{-\infty-i \epsilon}^{\infty-i \epsilon}\, \frac{d \alpha}{\alpha}\, \int_{-\infty}^{\infty}\, d^{3} \, \lambda \, \prod_{i} d^{3} \, p_i 
\exp\left\{i \left[\lambda \cdot \sum_i p_i + \alpha \left(E-\sum_i \varepsilon_i \right)\right]\right\}, \hspace*{10mm}
\eea
where $\varepsilon_i$ is the dispersion relation of $i$-th particle, while $\epsilon$ has a infinitesimal positive value. 

Each of the momentum integral can be written as \cite{lepor}
\bea
I &=& \int d^3\, p \; \exp\left[i \left(\lambda \cdot p - \alpha\, \varepsilon \right)\right].
\eea
With the variable change $p=m \sinh \theta$, 
\bea
I &=& - \frac{2 \pi m}{\lambda} \frac{d}{d \lambda} \int_{-\infty}^{\infty} d \theta \, \cosh \theta \; \exp\left[i\, m\left(\lambda\, \sinh \theta- \alpha\, \cosh \theta \right)\right],
\eea
which can reduced to the standard forms, if $\alpha>\lambda>0$ is valid
\bea
I &=& - \frac{2 \pi m}{\lambda} \frac{d}{d \lambda} \int_{-\infty}^{\infty} d \theta \, \cosh \theta \; \exp\left[i\, m\, \left(\alpha^2-\lambda^2\right)^{1/2}\, \cosh\left(\theta-\phi_1\right)\right].
\eea 
With another variable change $\theta^{\prime}=\left(\theta-\phi_1\right)$, then, the integrals can be expressed in Hankel functions
\bea
I &=& \frac{2 \pi^2 m^2 \alpha}{\alpha^2-\lambda^2} \, H_2^{(2)}\left(m (\alpha^2-\lambda^2)^{1/2}\right).
\eea

In relativistic limit, the density of state reads
\bea
S_n &=& {\cal R}\, \frac{\Omega^{n-1}}{(2 \pi)^{2(n-1)}}\, \frac{(4 n -3)!\, E^{3n -4}}{2^{4(n-1)}\,(2 n -1)!\, (2n -2)! (3n-4)!}.
\eea 
Comparing this expression with the one given in Ref. \cite{bookFermiEq} (page 576, Eq. (13)), it can be found that ${\cal R}$ is absent. As discussed above, ${\cal R}$ counts for momentum conservation.  For large $n$,
\bea
S_n &=& {\cal R}\, \frac{\Omega^{n-1}}{\pi^{2n}}\, \frac{E^{3n -4}}{(3n-4)!}\, \left(\frac{2 \pi^2}{n}\right)^{1/2},
\eea  
where Stirling's approximation is applied.

\subsubsection{Quantum statistics with a variable number of idea particles}

Magalinski and Terletskii introduced \cite{magl3A,magl3B} a general description for the quantum statistics in systems (like high-energy collisions) with a variable number of ideal (non-interacting) particles. The generalization extends to an aggregate of oppositely charged particles, for which the conservation laws are fulfilled. Expressions for total number of particles and the total energy are deduced. They are distinguishable from the ordinary quantum statistics. 

A general feature of statistical description in high-energy studies is to determine the phase space or Fock space. Both should be restricted due to conservation laws. Conservation of the charges, baryon number and strangeness was studied within the thermodynamical model of high-energy hadronic collisions \cite{elze5,elze6}. Finite-size effects were found.

In almost all applications of statistical-thermal methods applied to study systems with charge conservation, the underlying symmetry corresponds to one or several unrelated U(1) groups \cite{Elze}. This apparently means that in order to obtain the total charge of a many-particle state, the charges of the constituents can be simply added. There is an increasing interest in gauge theories based on various non-Abelian internal symmetries \cite{elze9}. This has been  introduced by Yang and Mills \cite{elze8}.

\subsection{(1964) Hagedorn: statistical-thermodynamical approach to strong interactions}
\label{sec:hagd}

The abundant formation of resonances of increasing mass and rotational degrees of freedom is
one of the most striking features of strong interaction physics, which has attracted intense theoretical attention for about six decades \cite{hgdrnA,hgdrnB}. Hagedorn was the first who systematically analysed high-energy phenomena using all tools of statistical physics. It is assumed that {\it higher and higher resonances of strongly interacting particle occur and take part in the thermodynamics as if they were particles}. Hagedorn introduced the concept of limiting temperature based on the statistical bootstrap model. This was the origin of the  multiphase structure of the hadronic matter. The limiting temperature arose as a consequence of the mass spectrum $\rho(m)$ \cite{hgdrnA,hgdrnB}, which relates the number of hadronic resonances to their masses as an exponential. The density of states is given as \cite{hgdrnA,hgdrnB} 
\bea
\sigma(E,V) &=& \sum_{n=1}^{\infty} \frac{V_0^n}{n!} \int \delta \left(m-\sum_{i=1}^{n} E_i\right) \prod_{i=1}^{n} \rho(m_i) d m_i d^3 p_i,
\eea
In general, all thermodynamical quantities are sensitive to $\rho(m)$. For up-to-date undermined resonance states, a parametrization for total spectral weight has been introduced  \cite{brnt}
\bea
\rho(m) &=& f(m)\, e^{m/T_0},
\eea
where $f(m)$ strongly depends on the hadron resonances. Hegedorn \cite{hgdrnA,hgdrnB} estimated its asymptotic behavior as follows.
\bea
\rho(m) &=& \text{const.~} m^{-5/2}.
\eea

The Hagedorn model assumes that a certain volume $V_0$, which is a thermodynamical equilibrium, is established. The latter is also described {\it ''by statistical  thermodynamics of an unlimited and undetermined number of more or less excited hadrons''}. Then, these hadronic states should leave the region of interaction and strongly decay through {\it ''a number of steps''} into stable hadrons. The Hagedorn model defines the excited hadrons as a highly excited hadron. In other words, the highly excited hadron is nothing but a fireball consists of undistinguished resonances. The thermodynamical system is the fireball itself.

The number of excited hadrons with masses between $m$ and $m+d m$ is given by the mass spectrum $\rho(m)$. In thermodynamical language, $\rho(E)$ denotes the number of states between $E$ and $E+d E$. The partition function corresponding to $\rho(m)$ is conjectured to diverge for $T\rightarrow T_0$. In the present review, we only include known resonance states with mass $\leq 2~$GeV instead of the Hagedorn mass spectrum \cite{hgdrnA,hgdrnB}. 

In the Hagedorn model \cite{hgdrnA,hgdrnB}, it is assumed that the excluded-volume correction should be  proportional to the pointlike energy density. Furthermore, it is assumed that the density
of states of the finite-size particles in total volume $V$ can be taken as precisely the same
as that of pointlike particles in the available volume $V-\sum_i V_0^{(i)}$, where $V_0^{(i)}$ is the eigenvalue of the $i$-th particle in hadron resonance ensemble.

\subsubsection{Hagedorn model: quantum statistics}
\label{sec:hrg1}

The hadron resonances treated as a free gas~\cite{Karsch:2003vd,Karsch:2003zq,Redlich:2004gp,Tawfik:2004sw,Tawfik:2004vv} are conjectured to add to the thermodynamical pressure in the hadronic phase (below $T_c$). This statement is valid for free as well as for interacting resonances. 
It has been shown that the thermodynamics of strongly interacting  system can also be approximated to an ideal gas composed of hadron resonances with masses $\le 2~$GeV ~\cite{Tawfik:2004sw,Vunog}. Therefore, the confined phase of QCD, the hadronic phase, was modelled as a non-interacting gas of resonances ~\cite{Karsch:2003vd,Karsch:2003zq,Redlich:2004gp,Tawfik:2004sw,Tawfik:2004vv}. The grand canonical partition function reads
\bea
Z(T, \mu, V) &=&\Tr\left[ \exp^{\frac{\mu\, N-H}{T}}\right],
\eea
where $H$ is the Hamiltonian of the system and $T$ ($\mu$) being temperature (chemical potential). The Hamiltonian is given by sum of the kinetic energies of the relativistic Fermi and Bose particles. The main motivation of using this Hamiltonian is that
\begin{itemize} 
\item it contains all relevant degrees of freedom of confined and  {\it strongly interacting} matter. 
\item It  implicitly includes the interactions that result in formation of resonances. 
\item In addition, it has been shown that this model can offer a quite satisfactory description of the particle production in the heavy-ion collisions, section \ref{sec:compHIC}. 
\end{itemize}
With the above assumptions, dynamics of the partition function can be calculated exactly and apparently expressed as a sum over {\it single-particle partition} functions $Z_i^1$ of all hadrons and their resonances.
\bea \label{eq:lnz1}
\ln Z(T, \mu_i ,V)&=&\sum_i \ln Z^1_i(T,V)=\sum_i\pm \frac{V g_i}{2\pi^2}\int_0^{\infty} k^2 dk \ln\left\{1\pm \exp[(\mu_i -\varepsilon_i)/T]\right\}, \hspace*{1cm}
\eea
where $\varepsilon_i(k)=(k^2+ m_i^2)^{1/2}$ is the $i-$th particle dispersion relation, $g_i$ is
spin-isospin degeneracy factor and $\pm$ stands for bosons and fermions, respectively.

Before the discovery of QCD, a probable phase transition of a massless pion gas to a new phase of matter was speculated \cite{lsm1}. Based on statistical models like Hagedorn \cite{hgdrn1A,hgdrn1B} and Bootstrap model \cite{boots1,boots2}, the thermodynamics of such an ideal pion gas has been studied, extensively. After the QCD, the new phase of matter is now known as the quark gluon plasma (QGP). 

The physical picture was that at $T_c$ the additional degrees of freedom carried by QGP are to be released resulting in an increase in the thermodynamical quantities like energy and pressure densities. The success of HRG model in reproducing lattice QCD results at various quark flavors and masses (below $T_c$) changed this physical picture drastically. {\it Instead of releasing additional degrees of freedom at $T>T_c$, the hadronic system increases its effective degrees of freedom, namely at $T<T_c$. In other words, the hadron gas has much more degrees of freedom than QGP}.

At finite temperature $T$ and baryon chemical potential $\mu_i $, the pressure of the $i$-th hadron or resonance species reads 
\begin{equation}
\label{eq:prss} 
p(T,\mu_i ) = \pm \frac{g_i}{2\pi^2}T \int_{0}^{\infty}
           k^2 dk  \ln\left\{1 \pm \exp[(\mu_i -\varepsilon_i)/T]\right\}.
\end{equation} 
As no phase transition is conjectured in the HRG model, summing over all hadron resonances results in the final thermodynamic pressure (of the hadronic phase). 

The switching between hadron and quark chemistry is given by the relations between  the {\it hadronic} chemical potentials and the quark constituents;  $\mu_i =3\, n_b\, \mu_q + n_s\, \mu_S$, where $n_b$($n_s$) being baryon (strange) quantum number. The chemical potential assigned to the light quarks is given as $\mu_q=(\mu_u+\mu_d)/2$ and the one assigned to strange quark reads $\mu_S=\mu_q-\mu_s$. It is worthwhile to notice that the strangeness chemical potential $\mu_S$ should be calculated as a function of $T$ and $\mu_i $. In doing this, it is assumed that the overall strange quantum number has to remain conserved in the heavy-ion collisions~\cite{Tawfik:2004sw}.  

\subsubsection{Hagedorn model: Boltzmann statistics}
\label{sec:lzBessl1}

The relativistic momentum integrals can be replaced by summation over modified Bessel functions \cite{boots2}, 
\bea
K_2\left(\frac{m}{T}\right) &=& \frac{\sqrt{\pi}}{\Gamma(5/2)} \frac{1}{4} \left(\frac{m}{T}\right)^2 \int_0^{\infty} p^4 dp \frac{\exp(-\varepsilon/T)}{\varepsilon}, \\
&=& \frac{\sqrt{\pi}}{\Gamma(5/2)} \frac{1}{4} \left(\frac{m}{T}\right)^2 \int_1^{\infty} \exp(-2 m/T) \, (t^2-1)^{3/2}\, d t,
\eea
where $\Gamma(5/3)=(3/2) \Gamma(3/2)=(3/2) \sqrt{\pi}/2$. In non-relativistic limit,
\bea
K_2\left(\frac{m}{T}\right) &=& \sqrt{\frac{\pi T}{2 m}}\; \exp(-m/T) \left[1+\frac{4 m^2/T^2-1}{8 m/T} + \cdots\right].
\eea
In relativistic limit,
\bea
K_2\left(\frac{m}{T}\right) &=& \frac{2 T^2}{m^2}- \frac{1}{2} - \left[ \ln\left(\frac{m}{2 T}\right) +\gamma - \frac{3}{4}\right] \frac{m^2}{8 T^2} + \cdots,
\eea
where $\gamma \simeq 0.577$ is Euler constant.

Then, the partition function reads
\bea
\ln\, z_{cl}(T,\mu,V) &=& \frac{V}{2\, \pi^2}\, T^3 \sum_{i=1}^N \,  \exp\left(\frac{\mu_i}{T}\right)\, \left(\frac{m_i}{T}\right)^2 \, K_2\left(\frac{m_i}{T}\right).
\eea
The logarithm can be expressed as a series. Each term comprises contributions from all contributing bosons and fermions. Then, from Eqs. (\ref{eq:prss}) and (\ref{eq:lnz1}), we get  
\bea
\ln {\cal Z}_{qs}(T,\mu,V) &=& \frac{T^3\, V}{2 \, \pi^2} \sum_{n=1}^{\infty} \frac{g_n}{n^4} \, \exp\left(n\frac{\mu}{T}\right) \left(n\frac{m}{T}\right)^2 \; K_2\left(n\frac{m}{T}\right),
\eea
which is restricted to $m-\mu<0$ and the fermionic factor $(-)^{n+1}$ should be combined in the degeneracy factor.

\subsubsection{Mass spectrum: statistical bootstrap and dual resonance model}

The statistical bootstrap method introduced in 1979 by Bradley Efron \cite{afron4} measures the accuracy in a sample. This was inspired by earlier work on the jackknife \cite{jack5,jack6}. A Bayesian extension was developed in 1981 \cite{bar10}.  

But, in early sixties of last century, the statistical bootstrap model was suggested by Hagedorn to describe hadron production \cite{boots137,boots140}. This approach was refined as the understanding of hadronic structure advanced, and ultimately it has been modified to allow for the possibility that individual, confined hadron-gas particles dissolve into quark-gluon plasma (QGP).

For multi-hadron system, as in strong interactions, the statistical bootstrap and dual resonance model estimate an asymptotic density of the states
\bea \label{eq:boots}
\sigma (m^2) &=& c\; m^2\; e^{b\; m},
\eea
where $a$, $b$ and $c$ are constants and $m$ is the center-of-mass energy. Obviously, the constant $b$ gets the interpretation as the inverse temperature. Adding up new resonance masses leads to an increase in the density of states rather in the kinetic energy. 

The resonance excitation spectrum is given as \cite{satzold7}
\bea
\rho(p^2) &=& \delta (p^2-\mu^2) + \frac{1}{B} \left[\sigma(p^2) - \delta^{(\nu)}(p) - B \rho(p^2)\right],
\eea
where $\nu-1$ plays the role of $d$, the dimensionality $B$ is a ($\nu-2$)-D space volume, $B=2\, \mu\, V$, and $\mu$ is the mass of the basic hadron, the pion mass. In non-relativistic limit, $B$ is responsible for the connection to a 3D coordinate space volume. It is worthwhile to mention that
\bea
p^2 &\equiv & p^2_0 - \sum_{i=1}^{\nu-1}  p^2_i.
\eea

The quantity which carries dynamical information in the bootstrap scheme is the mass spectrum $\rho(m)$ of the fireballs. In a generic form, $\rho(m)$ satisfies an integral equation so that \cite{boots4a}
\bea
\rho(m) &=& \delta(m-m_0) + \sum_{n=2}^{\infty} \frac{1}{n !} \; \delta\left(m-\sum_{n=1}^{n} m_i \right) \, \prod_{i=1}^{n} \rho(m_i)\, d m_i, 
\eea
where $m_0$ is the mass of hadron resonance and $m_i$, $i=1,2,\cdots,\infty$ stands for the fireball masses in an ascending order of complexity. $\rho(m)$ determines the number of hadron resonances or fireballs with rest-frame volume $V$. This volume should reside inside the momentum region $d^3 p$ around $\vec{p}$, so that
\bea
\frac{V}{\hbar^3} d^3 p \int \rho(m)\, d m &\rightarrow & 2 \frac{V^{\mu}\, p_{\mu}}{2 \pi^3} \int \delta_p (p^2 - m^2) d^4 p \cdot \tau(m^2) \, d m^2,
\eea
where the four-volume $V^{\mu}$ is adjusted parallel to the four-momentum  $p^{\mu}$ so that
\bea
\rho(m)\, d m &=& \tau(m^2)\, d m^2, \\
V^{\mu} &=& \frac{V}{m} \, p^{\mu}.
\eea

The similarity between the two models is due to similar underlying partition features. The level density of one-dimensional dual resonance model is the number of root systems of the equation
\bea
\sum_{r=1}^N r\, \lambda_r = N,
\eea
where $r$ and $\lambda$ are positive and $N=m^2$. The asymptotic solution reads \cite{satzold}
\bea
Z_1(x) &=& \sum_{r=1}^{\infty} \sigma_1(N)\; x^N = \prod_{k=1}^{\infty} (1-x^k)^{-1},
\eea
which can  lead to Eq. (\ref{eq:boots}).

The dimensionality $d$ is related to $a$ \cite{satzold}
\bea
a &=& - \frac{1}{2} \; (d-1).
\eea
The inverse temperature parameter is related to the slope $\alpha^{\prime}$ of the Regge trajectories \cite{reggetr2010}
\bea
b &=& 2\, \pi\; \left(\frac{\alpha^{\prime}\; d}{6}\right)^{1/2},
\eea
For $d=4$ and $\alpha^{\prime}=1~$GeV$^{-2}$, the resulting critical temperature reads  $\sim 174~$MeV.

\subsection{(2003) Karsch, Redlich and Tawfik: scaling of resonance masses for lattice simulations}

\label{sec:krt1}

In order to use the resonance gas model for further comparison with lattice QCD results we should take into account that the lattice QCD  calculations are generally performed with quark masses heavier than those realized in nature \cite{Karsch:2003vd}. This was the case 10 years ago. In fact, we should take advantage of this by comparing lattice results obtained for different quark masses with resonance gas model calculations based on a modified resonance spectrum depending on the quark masses. Rather than converting the bare quark masses used in lattice QCD calculation into a renormalized mass, it is much more convenient to use directly the pion mass ($m_\pi \sim \sqrt{m_q}$), i.e. the mass of the lightest Goldstone particle, as a control parameter for the quark mass dependence of the hadron spectrum.

To leading order, the quark mass dependence of all other hadron masses may then be parametrized as $m_H \sim m_{H,0} + c_H \sqrt{m_\pi}$, while in the heavy quark mass limit the relation is determined by the parton model, $m_H = n m_\pi$, with $n=2$ and 3
for mesons and baryons, respectively.

For thermodynamic considerations, we want to extract information on features of the quark mass dependence of a large set of resonances \cite{Karsch:2003vd,Karsch:2003zq,Redlich:2004gp}. In order to study the quark mass dependence of hadron masses in the intermediate  region between the chiral and heavy quark mass limits, an approach is adopted which is based on the Hamiltonian of the MIT bag-model \cite{bag1}. MIT stands for Massachusetts Institute of Technology.

\subsubsection{MIT bag model}
\label{sec:mitBAG}

Although, the original Hamiltonian breaks explicitly chiral symmetry and implies non-conservation of the axial-vector current, it still provides a satisfactory description of the hadron mass spectrum that can be used for our thermodynamic considerations. In the limit of a static, spherical cavity the energy of the bag of radius R is given by
\bea
E&=&E_V+E_0+E_K+E_M+E_E. \label{eq1}
\eea
The first two terms are due to quantum fluctuations and are assumed to  depend only on the bag radius. The volume and the zero-point energy terms have a generic form
\bea
 E_V &=& \frac{4}{3}\pi BR^3, \\
  E_0 &=& -\frac{Z_0}{R},\label{eq2}
\eea
where  $B$ is the bag constant and  $Z_0$ is a phenomenological parameter attributed to the surface energy.

The quarks inside the bag contribute with their kinetic and rest energy. Assuming $N$ quarks of mass $m_i$, the  quark  kinetic energy is determined from
 \bea
 E_K &=& \frac{1}{R}\sum_{i=1}^N[x_i^2+(m_iR)^2]^{1/2}, \label{eq4}
 \eea
where $x_i(m_i,R)$ enters the expression on the frequency $\omega =[x^2+(mR)^2]^{1/2}/R$  of the lowest quark mode and is obtained \cite{bag1} as the smallest positive root of the following
equation
 \bea
 \tan(x_i) &=& \frac{x_i}{1-m_iR-\sqrt{x_i^2+(m_iR)^2}}. \label{eq5}
 \eea
The last two terms in Eq.~(\ref{eq1}) represent the color--magnetic and electric  interaction of quarks.  It is described by the exchange of a single gluon between two quarks
inside the bag. The color electric energy was found  very small \cite{bag1} and will be neglected in our further discussion. The color magnetic exchange term is given by
\bea
E_M &=& {8}k\alpha_c\sum_{i<j}\frac{M(m_i R, m_j R)}{R} (\vec{\sigma_i}\cdot\vec{\sigma_j}). \label{eq6}
\eea
Here $\alpha_c$ is the strong coupling constant and $k=1$ for baryons and $2$ for mesons. For a given spin configuration of the bag, the scalar spin product in Eq.~(\ref{eq6}) can easily be calculated. The function $M(x,y)$ depends on the quark modes magnetic moment and is described in detail in Ref. \cite{bag1}. 

The dependence of the energy on the bag radius can be eliminated by the condition that the quark and gluon field pressure balance the external vacuum pressure. For a static spherical bag this condition is equivalent to minimizing $E$ with respect to $R$. The true radius $R_0$ of the bag is determined from the condition $\partial E/\partial R=0$ and the hadron bag mass is then obtained from Eq. (\ref{eq1})  with $R=R_0$.

To extract the physical mass spectrum from the MIT bag model one
still needs to fix a set of five parameters that determine  the
bag energy  \cite{Karsch:2003vd}. Following the original fit to experimental data made
in Ref. \cite{bag1}, the parameters have been fixed
\bea
B^{1/4} &=& 0.145\;\, \text{GeV}\nn \\ 
Z_0 &=& 1.84, \nn  \\
\alpha_c &=& 0.55,  \\
m_u=m_d &=& 0, \nn \text{and} \\ 
m_s &=& 0.279\; \text{GeV}. \nn
\eea 
This provides a quite satisfactory description of hadron masses belonging to the octet and decuplet of baryons and the octet of vector mesons. 

\begin{figure}[t]
\begin{center}
\vskip -2cm
\includegraphics[width=25.5pc, height=28.5pc,angle=180]{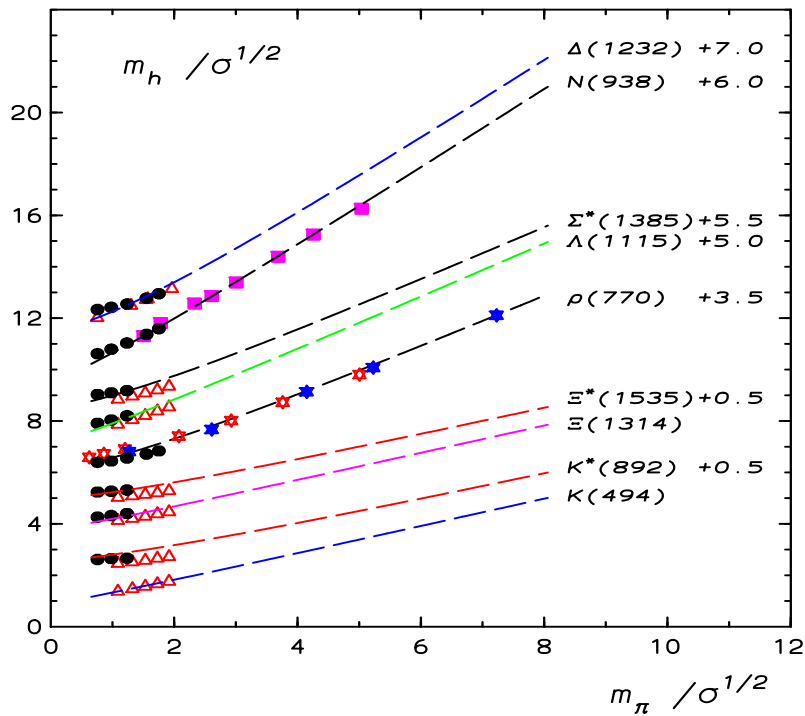}
\end{center}
\vskip -1.5truecm \caption{Dependence of different hadron masses
$m_h$ on the pion mass $m_\pi$.  Both $m_h$ and $m_\pi$ are
expressed in units of the string tension $\sqrt\sigma\simeq 420~$MeV.
Curves are the MIT bag model results. The
filled circles represent the PC-PACS lattice results 
\cite{data1}. The filled diamonds are $n_f=3$ whereas the
open-diamonds are $n_f=2$ quark flavor results \cite{peikert1}.
The filled-boxes are quenched QCD results \cite{data3}. All
other points are from Ref. \cite{data6}. Both lattice
data and bag model results are shifted in $m_h$-direction by
a constant factors (indicated). The graph taken from Ref.  \cite{Karsch:2003vd}.}
\label{qm10}
\end{figure}

The MIT bag model provides an explicit dependence  of hadron
masses  on the constituent quark mass  \cite{Karsch:2003vd}. This dependence is entirely
determined by the kinetic and magnetic energy of the quarks. To
compare bag model calculations with lattice calculations, which do
not provide values for constituent quark masses, it is best to
express the quark mass dependence in terms of the pion mass, which
is most sensitive to changes of the quark masses. In
Fig.~\ref{qm10}, the resulting dependence of different
hadron masses on the pion mass is presented with the bag parameters described
above but with varying $m_u$. The masses are expressed in units of
the square root of the string tension for which we use
$\sqrt{\sigma} = 420$~MeV. We also keep the strange and light quark masses to be indistinguishable. The model predictions are
compared with recent lattice data on hadron masses calculated for
different current quark masses \cite{peikert1,data1,data3,data6}.

For large quark masses the bag model description of hadron masses
reproduces the naive parton model picture and consequently all
hadron masses are almost linearly increasing with the pion mass as
seen in  Fig.~\ref{qm10} . This is to be expected as in this case
the energy of the bag is entirely determined by the quark rest
mass. As seen in  Fig.~\ref{qm10} the slope increases with the
number of non--strange constituent  quarks inside the bag.
Consequently, the slops of ($\Xi^*,\Xi$) and ($K,K^*$) or
($\Sigma^* ,\Lambda$) and $\rho$ coincide at large $m_\pi$. A phenomenological parametrization of the quark mass dependence of resonances, expressed in terms of the pion mass \cite{Karsch:2003vd}
\begin{equation}
\frac{M(x)}{\sqrt\sigma} \simeq N_u a_1 x+ \frac{m_0}{1+a_2x+a_3x^2+a_4x^3+a_5x^4}, \label{parA}
\end{equation}
which provides a good description of the MIT bag model result for
non-strange hadron masses calculated for different values of
$m_\pi$. Here $x\equiv {m_\pi /\sqrt\sigma}$, $m_0\equiv
{m_{hadron} /\sqrt\sigma}$, $N_u$ is the number of light quarks
inside the hadron; $N_u=2$ for mesons $N_u=3$ for baryons.

\begin{table}[h] 
\centering{
\begin{tabular}{|c|c|c|c|c|} \hline
\multicolumn{5}{|c|}{ Parametrization of the non--strange hadron
masses from Eq.~(12)}
\\\hline
 $a_1$    & ~ $ a_2$ ~  & $a_3 $  &  $a_4$ & $a_5$   \\ \hline
0.51$\pm$ 0.1 &   ~  $a_1 \frac{N_u}{m_0}$ ~~& 0.115$\pm$ 0.02
&-0.0223$\pm$ 0.008 &
 0.0028$\pm$ 0.0015 \\\hline
\end{tabular}
}
\caption{Parameters entering the interpolation formula for non-strange
hadron masses  \cite{Karsch:2003vd}. } \label{tab:fitt}
\end{table}

The fitting parameters, Eq.~(\ref{parA}),  are summarized in Tab. \ref{tab:fitt}.  Accordingly, Eq.~(\ref{parA}) reproduces the quark mass dependence of all  non-strange hadron masses obtained from the bag model within a relative error of $\leq$6$\%$.
\begin{eqnarray}
\left.m_h(m_{\pi})\right|_{n_s=0} &=& 0.5 n_q m_{\pi} + \frac{m_h^0}{1+ 0.16 m_{\pi}^2 - 0.0517 m_{\pi}^3+0.008m_{\pi}^4} \label{eq4}\\
\left.m_h(m_{\pi})\right|_{n_s=\pm 1} &=& 0.55 n_q m_{\pi} +
  \frac{m_h^0}{0.468+ 0.058 m_{\pi} - 0.001 m_{\pi}^2}  \label{eq5}
\end{eqnarray}
where $n_q$ is the number of light quarks.

\begin{figure}[htb]
\begin{center}
\includegraphics[width=90mm]{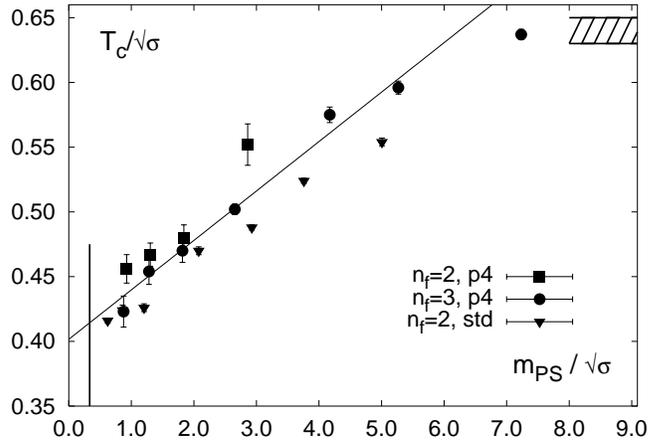}
\end{center}
\caption{The transition temperature in 2 (filled squares) and 3 (circles) flavor  QCD versus $m_{PS}/\sqrt{\sigma}$ using an improved staggered fermion action (p4-action). Also shown are results for 2-flavor QCD obtained with the standard staggered fermion action (open squares). The dashed band indicates the uncertainty on $T_c/\sqrt{\sigma}$ in the quenched limit. Graph taken from Ref.  \cite{Karsch:2003vd}.}
\label{fig:tc_pion}
\end{figure}

Section \ref{sec:TcPionm} is devoted to the dependence of critical temperature as estimated in the statistical-thermal model with varying masses on the quark mass and number of flavors.  The dependence of the critical energy density on the quark mass and number of flavors shall be studied in section \ref{sec:ecPionm}. The resulting hadronic states with changing the masses of the lightest Goldstone mesons shall be elaborated in section \ref{sec:Npions}. These three topics characterize the resulting statistical-thermal model and its matching with the lattice QCD simulations \cite{Karsch:2003vd,Karsch:2003zq,Redlich:2004gp}.

 \subsubsection{Dependence of critical temperature on quark mass and flavor}
\label{sec:TcPionm}
 
A collection of transition temperatures obtained in calculations with $2$ and $3$ quark flavors with degenerate masses is shown in Fig.~\ref{fig:tc_pion}. The main feature of the numerical results  is that the transition temperature varies rather slowly with the quark mass. The linear behavior has been described by the fit \cite{karsch1}, Fig.~\ref{fig:tc_pion},
\begin{equation}
\biggl(\frac{T_c}{\sqrt{\sigma}}\biggr)_{m_{PS}/\sqrt{\sigma}} =
\biggl(\frac{T_c}{\sqrt{\sigma}} \biggr)_0
0.4 + 0.04(1)\; \biggl(\frac{m_{PS}}{\sqrt{\sigma}} \biggr).
\label{tcfit}
\end{equation}
For pion masses $m_{PS} \sim (6-7)\sqrt{\sigma} \simeq 2.5$~GeV, the transition temperature reaches the pure gauge value, $T_c/\sqrt{\sigma} \simeq 0.632 (2)$ \cite{qgp3}.

We use $T_c = 175\;(15)$~MeV for 2-flavor QCD and $T_c =
155\;(15)$~MeV for 3-flavor QCD, respectively. For the energy
densities at the transition point we then find
\begin{equation}
\biggl( \frac{\epsilon}{T^4} \biggr)_{T=T_c} \simeq 
\begin{array}{ll}
4.5 \pm 1.5, & 2-\text{flavor}, \\
7.5 \pm 2,  & 3-\text{flavor}.
\end{array}
\label{epsilonres}
\end{equation}

\begin{figure}[htb!]
  \begin{center}
    \includegraphics[width=10cm]{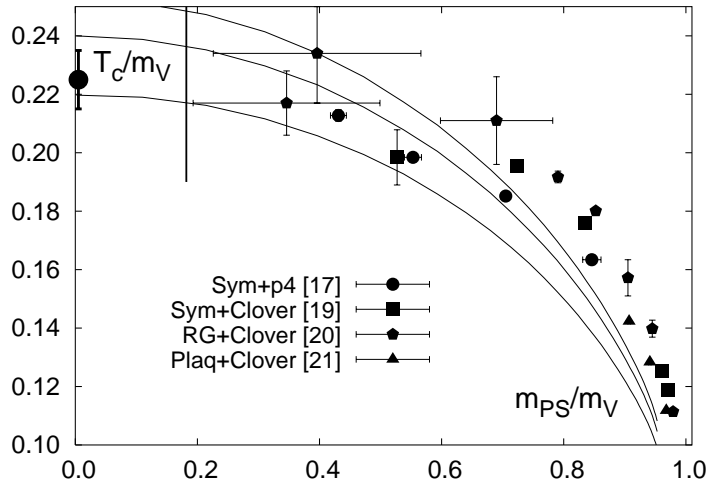}
    \caption{The  transition temperature vs. pion mass obtained
in lattice calculations. A comparison of constant energy density lines at $1.2$ (upper),
$0.8$ (middle) and $0.4$ (lower) GeV/fm$^3$ with lattice results for
2-flavor QCD obtained with improved staggered \cite{karsch1}
as well as improved
Wilson \cite{Bernard,Ali,Edwards} fermion formulations. $T_c$ as
well as $m_{PS}$ are expressed in terms of the corresponding
vector meson mass. The graph and references therein taken from Ref. \cite{karsch1}. }
    \label{fig:conditions}
  \end{center}
\end{figure}

In Fig.~\ref{fig:conditions}, results from HRG at constant energy density are compared to $T_c$ obtained in lattice calculations. As can be
seen the agreement is quite good up to masses, $m_{PS}\simeq
3\; \sqrt{\sigma}$ or $m_{PS}\simeq 1.2$~GeV.
The reason for the deviations at larger values of the quark mass,
of course, is due to the fact that  the glueball sector is not taken into consideration. We use gb as an abbreviation for glueball(s).

 \begin{figure}[htb!]
   \begin{center}
     \includegraphics[width=10cm]{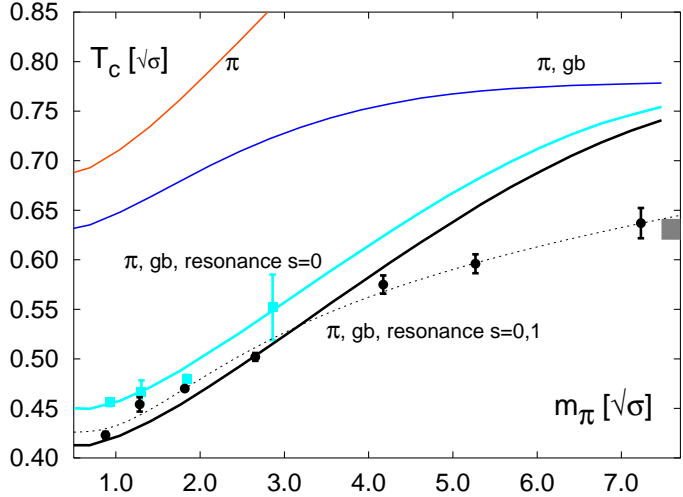}
     \caption{The results for $2$ and $3$ flavor QCD compared to lines of constant energy density of  $\sim 0.8$ GeV/fm$^3$ with varying pion-masses $m_{\pi}$. Here $T_c$ and the lightest pseudo-scalar mass $m_{\pi}$ are expressed in units of $\sqrt{\sigma}$. The different curves are related to different constituents of HRG.}
 \label{fig:tcpion1}
\end{center}
 \end{figure}

When the lightest hadron mass becomes comparable to typical
glueball masses, also the glueball sector will start to contribute
a significant fraction to the energy density. Using a set of $15$
different glueball states was identified in lattice
calculations \cite{glu1,glu2}, we have calculated their contribution to
the total energy density. At $m_{PS}/\sqrt{\sigma} \simeq 6.5$,
they contribute as much as the entire hadronic sector \cite{Karsch:2003vd}.
However, the contribution of these $15$ states only leads to a small shift in the lines of constant energy density, Fig.~\ref{fig:tcpion1}.
Further support for this comes from a calculation of the
energy density of the $15$ known glueball states at the transition
temperature of the pure gauge theory, $T=0.63 \sqrt{\sigma}$ \cite{Karsch:2003vd}.
For this we
obtain $\epsilon (T=0.63 \sqrt{\sigma}) \simeq 0.06$ GeV/fm$^3$ or
equivalently $\epsilon/T_c^4 \simeq 0.1$, which is about 20\% of
the overall energy density at $T_c$.
The situation is similar to that in the chiral limit if one neglects the contribution of heavy resonances and includes only a few low-lying hadronic states. The contribution of the $15$ glueball states thus does not seem to be sufficient. In fact, the transition temperature in $d$-dimensional
$SU(N_c)$ gauge theories is well understood in terms of the
critical temperature of string models,
$T_c/\sqrt{\sigma}=\sqrt{3/\pi(d-2)}$, which also is due to an
exponentially rising mass spectrum for string excitations
\cite{Pisarski}.

For even larger $m_\pi$ the contribution of the hadron mass spectrum to the QCD equation of state is negligible since it is suppressed by the Boltzmann factor. Here, thermodynamics is entirely determined  by glueball states. To compare the model prediction with lattice data in the broad $m_\pi$-range  one still needs to implement these important degrees of freedom in the partition function. We have included $15$ different glueball states \cite{glu1,glu2}. The corresponding lines of constant energy density are compared to the transition temperatures in 3-flavor QCD in Fig.~\ref{fig:tcpion1}.

It is conceivable that extending the glueball mass spectrum to all
higher excited states will improve the results shown in
Fig.~\ref{fig:tcpion1}. On the other hand, one also should stress that
the glueball states were obtained in quenched QCD and at zero temperature \cite{glu1,glu2}. There are indications from lattice calculations that glueball masses could be modified
substantially in the presence of dynamical quarks \cite{g25} as
well as at finite temperature \cite{g20}. The analysis of glueball
states at high temperature \cite{g20} suggests that their masses
can drop by $\sim (20-40)\%$. As all glueballs are heavy on
the temperature scale of interest, shifts in their masses influence
the thermodynamics much more strongly than in the
light quark mass regime, where the lowest state has already a mass
which is of the order of the transition temperature. In fact, we
find that taking into account a possible decrease of the glueball
masses close to $T_c$ seems to be more important than adding further
heavy states to the spectrum. We thus have
included a possible reduction of glueball masses in the equation
of state. The resulting $T_c$ with this modification is also shown
in Fig.~\ref{fig:tcpion1}. Decreasing the glueball masses increases
the thermal phase space available for particles. Consequently,
the temperature required to get $\epsilon =0.8$ GeV/fm$^3$ is
decreasing \cite{Karsch:2003vd}. As can be seen in Fig.~\ref{fig:tcpion1}, a reduction of
glueball masses by 40$\%$ is sufficient to reproduce lattice
results in the whole $m_\pi$ range. However, to make this comparison
more precise, it clearly is important to get a more detailed
understanding of the glueball sector in the future.

\subsubsection{Dependence of critical energy density on quark mass and flavor}
\label{sec:ecPionm}

\begin{figure}[htb]
\begin{center}
\includegraphics[width=7cm,angle=-90]{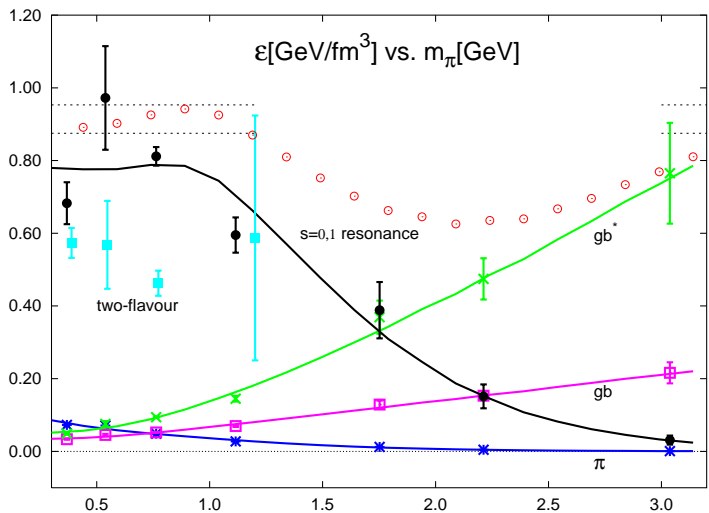}
\caption{The dependence of critical energy density on the mass of the lightest pseudo-scaler boson is given in physics units for $2$ and $3$-flavors. HRG with pions or $15$ states of glueballs with and without scaled masses are not able to reproduce the lattice results. Only HRG in which strange and non-strange fermion and bosons in additional to the $15$ states of glueballs are included describes very well the lattice results.}
 \label{fig:epsilonpion1}
\end{center}
 \end{figure}
 
The dependence of energy density on the mass of the lightest pseudo-scalar meson (both in physical units) is presented in Fig. \ref{fig:epsilonpion1}. The lattice calculations for $2$ and $3$-flavors are given as solid rectangles and circles, respectively.  The lattice results for $3$ flavors indicate that the critical energy density seems to remain constant, while the mass of pseudo-scalar meson gets at large as $\sim 1~$GeV. With further increasing mass, the critical energy density almost exponentially decreases. The available results for $2$ flavors are not as broaden as that for $3$ flavors. The first region ($0.3<m_{\pi}<1.2~$GeV) seems to confirm constant critical energy density.

\begin{figure}[htb]
   \begin{center}
     \includegraphics[width=7cm,angle=-90]{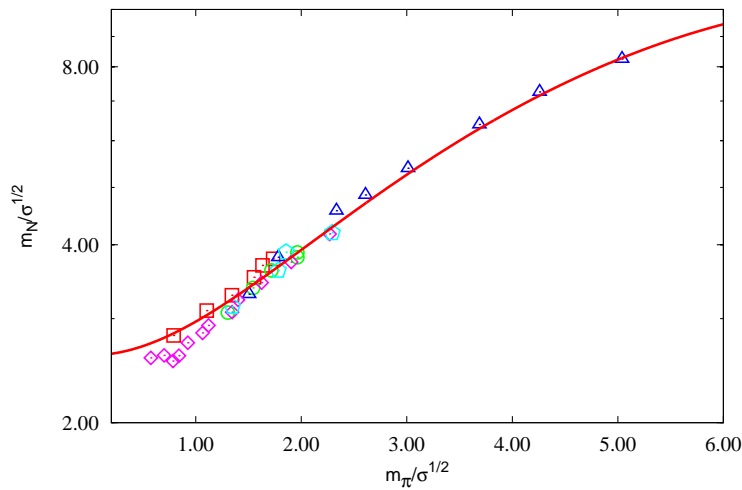}
     \caption{Nucleon mass is given in dependence on the mass of the lightest pseudo-scaler boson (both in physics units) for $3$-flavors. HRG in which strange and non-strange fermion and bosons in additional to the $15$ states of glueballs are included describes very well the lattice results. The lattice results are taken from \cite{ddatt1} (rectangle), \cite{ddatt2} (circle), \cite{data3} (triangle), \cite{ddatt4} (quadrilateral) and \cite{ddatt5} (pentacle).}
 \label{fig:mnpion1}
\end{center}
 \end{figure}
 
It is obvious that the HRG model reproduces very well the lattice results, especially for $3$ flavors. In order to analysis the effects on different degrees of freedom, different components are included in the HRG model. It is obvious that the pions alone are not able to reproduce the critical energy density. Also the $15$ states of glueballs with and without scaled masses are not able, as well. In both cases, raising the mass of lightest pseudo-scalar boson leads an increase in the resulting critical energy density, so that very heavy pseudo-scalar boson would results in critical energy density comparable to the one measured in lattice gauge theory.

Using the results presented in Fig. \ref{fig:tcpion1} that $T_c$ dependence on $m_{PS}$, the corresponding energy density is calculated and drawn as astride, open rectangle and stars for pions, glueballs without and with scaled masses, respectively. The three curves represent the calculations for critical energy density in dependence on the mass of the lightest pseudo-scalar boson, where the input critical temperatures have been taken from lattice QCD calculations \cite{Karsch:2003vd}. The agreement is obviously excellent. 

The open circles represent the HRG results, in which  both strange and non-strange fermion and bosons in additional to the $15$ states of glueballs are included. The resulting critical energy density remains constant ($\sim 0.93\pm0.05~$GeV/fm$^{-3}$) for $m_{PS}<1.2~$GeV. A further raise in $m_{PS}$ decreases the critical energy density. There is a minimum value at $m_{PS}\simeq 2.0~$GeV \cite{Karsch:2003vd}. Afterwards, the critical energy density increases till it retrains back the value it started with. 

\subsubsection{Dependence of hadron masses on quark mass and flavor}
\label{sec:Npions}

\begin{figure}[htb]
\centering{
\includegraphics[width=7cm,angle=-90]{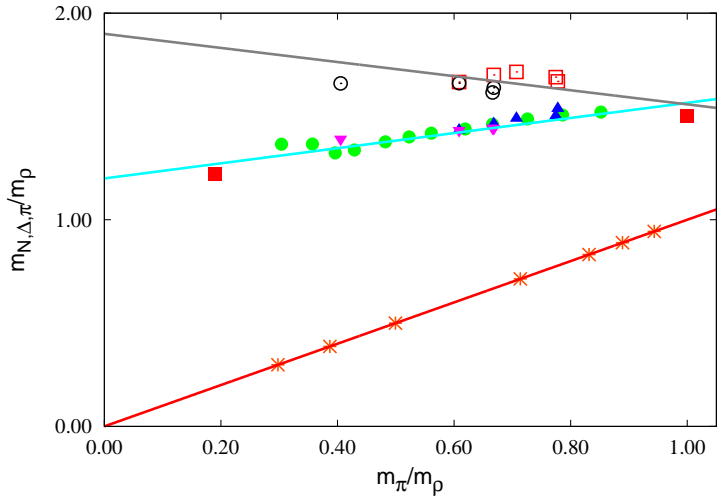}
\caption{The dependence of the masses of different hadronic states on the lightest pseudo-scaler boson (in physics units)  for $3$-flavors. HRG in which strange and non-strange fermion and bosons in additional to the $15$ states of glueballs are included describes very well the lattice results, which are taken from \cite{ddatt2} (solid up and down triangle and empty square and circle), \cite{ddatt4} (solid circle) and \cite{Karsch:2000kv} (astride).}
\label{fig:hadronspion1}
}
 \end{figure}

The nucleon masses calculated in different lattice QCD simulations in dependence on the mass of the lightest pseudo-scaler boson, $m_{PS}$ are compared with each others. Both quantities are given in $\sqrt{\sigma}$ units. HRG model, in which strange and non-strange fermion and bosons in additional to the $15$ states of glueballs are included describes very well the lattice results. As shown in Fig. \ref{fig:mnpion1}, the dependence reads
\bea
n_N &=& 2.6 + 0.4\, m_{PS}^2 - 0.035 \, m_{PS}^3.
\eea
It  is obvious that the agreement between lattice QCD and HRG calculations is excellent.

In units of $\rho$-meson, the masses of $N$, $\Delta$ and $\pi$ are given in dependence on the mass of the lightest pseudo-scalar boson, Fig. \ref{fig:hadronspion1}. Different lattice calculations are used,  \cite{ddatt2} (solid up and down triangle and empty square and circle), \cite{ddatt4} (solid circle) and \cite{Karsch:2000kv} (astride). Open symbols represent $m_{\Delta}/m_{\rho}$. Solid symbols stand for $m_N/m_{\rho}$, compare with Fig. \ref{fig:hadronspion1}. The astride gives $m_{\pi}/m_{\rho}$. The HRG with strange and non-strange fermion and bosons in additional to the $15$ states of glueballs is apparently able to reproduce the correct dependence on the three masses.

\section{Statistical-thermal models in non-ideal hadron gas}
\label{sec:nonidealG1}

Since, analytic or perturbative solutions in low-energy QCD are hard or even impossible due to the highly nonlinear nature of the strong force in this limit, the formulation of QCD in discrete rather than continuous spacetime was suggested four decades ago \cite{wlsonn74}.  The discritization naturally introduces a momentum cut-off at the order $a^{-1}$, where $a$ is the lattice spacing. With the lattice QCD, it turns to be possible to have a {\it first-principle} framework for the investigation of non-perturbative phenomena, like thermodynamics of hadronic and partonic matter and confinement-deconfinement phase transition, which are intractable in analytic field theories. Lattice QCD benefits from the huge progress in soft and hard ware facilities. Furthermore, lattice QCD has made successful contact to many experiments, for instant, in estimating the mass of the proton. There is a deviation of  less than $2\%$ \cite{durr}.

Various types of interactions shall be introduced in section \ref{sec:intr}. We start with 
van der Waals repulsive interactions, section \ref{sec:exclvV}, then we introduce the Uhlenbeck and Gropper statistical interactions, section \ref{sec:UG} and general $S$-matrix and strong interactions \ref{sec:smatrx}.

The dissipative properties shall be studied in section \ref{sec:visc}. This shall be divided into a single-component fluid, section \ref{sec:viscone}, and multiple-component fluid, section \ref{sec:viscmult}. Chapman-Enskog and Relaxation time approximation shall be utilized. Topics like, $K$-matrix parametrization of hadronic cross-sections, relation of $T$-matrix to $K$-matrix and the relaxation time shall be discussed.

Section \ref{sec:rapd} shall be devoted to the statistical-thermal models in rapidity space.
The goal is the study of the dependence of thermal parameters on the location, in particular on the spatial rapidity. A single freeze-out temperature model shall be given in section \ref{sec:sftm1}. The rapidity dependence of thermal parameters shall be discussed in section \ref{sec:rptp1}.  An approximation of  beam rapidity dependency of  particle ratios can be derived in the Regge model. Accordingly, the baryon-pair production at very high energy is governed by Pomeron exchange, section \ref{sec:rpbr1}. The asymmetry between baryons and anti-baryons can be expressed by the string-junction transport and by an exchange with negative $C$-parity. The proton ratios versus Kaon ratios shall be reviewed in section \ref{sec:kkpp}.

\subsection{Interactions in hadron gas}
\label{sec:intr}

In literature, there are at least three types of interactions to be implemented to the hadron gas: 
\begin{itemize}
\item The van der Waals repulsive interactions shall be studied in section \ref{sec:exclvV}. 
\item The Uhlenbeck and Gropper statistical interactions shall be introduced in section \ref{sec:UG}. 
\item Section \ref{sec:smatrx} is devoted to strong interactions represented by the generic $S$-matrix. 
\end{itemize}

\subsubsection{Approaches for excluded volume: van der Waals repulsive interactions}
\label{sec:exclvV}

\begin{figure}[htb]
\centering{
\includegraphics[angle=-90,width=10.cm]{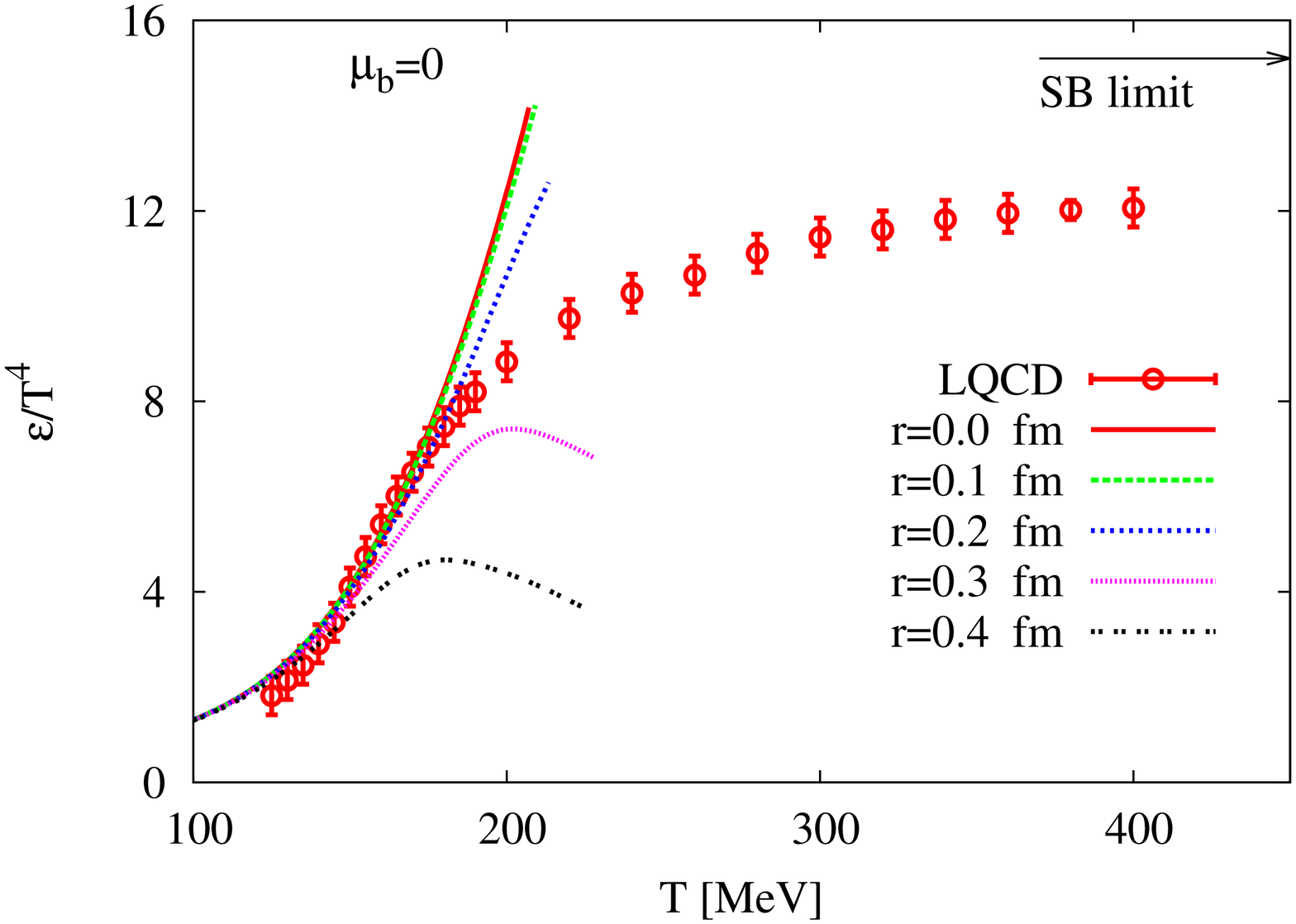} \\
\includegraphics[angle=-90,width=10.cm]{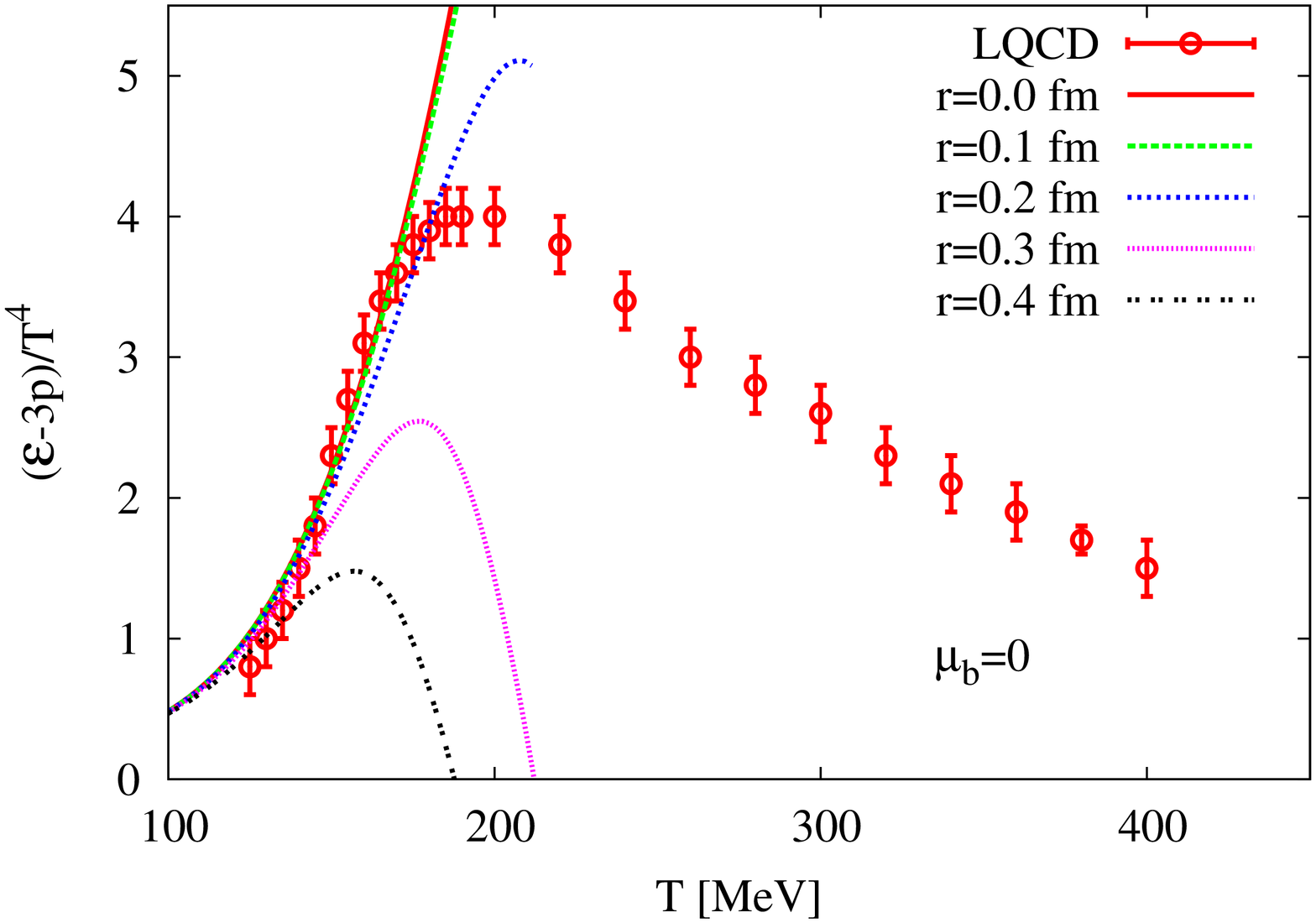}
\caption{Top panel: normalized energy density is given in dependence on the temperature at vanishing chemical potential. The symbols represent the lattice QCD simulations \cite{latFodor}. The curves are the HRG calculations at different hard-core radii of the hadron. Bottom panel presents the same but for the normalized trace anomaly. The results at $T>T_c$ are out of scope of the HRG model. }
\label{fig:vdW} 
}
\end{figure}

 The repulsive interactions between hadrons are considered as a phenomenological extension, which would be exclusively based on van der Waals excluded volume \cite{exclV1,exclV2,exclV3,exclV4}.  Accordingly, considerable modifications in thermodynamics of hadron gas including energy, entropy and number densities are likely. There are intensive theoretical works devoted to estimate the excluded volume and its effects on the particle production and fluctuations \cite{exclV5}, for instance. It is conjectured that the hard-core radius of hadron nuclei can be related to the multiplicity fluctuations \cite{exclV6}. Assuming that hadrons are spheres and  all have the same radius, we compare between different  radii in Fig. \ref{fig:vdW}. On the other hand, the assumption that the radii would depend on the hadron masses and sizes could come up with a very small improvement.

The first principle lattice QCD simulations for various thermodynamic quantities offer an essential framework to check the ability of extended {\it ideal} hadron gas, in which the excluded volume is taken into consideration \cite{apj}, to describe the hadronic matter in thermal and dense medium. Figure \ref{fig:vdW} compares normalized energy density and trace anomaly as calculated in lattice QCD and HRG model. The symbols with error bars represent the lattice QCD simulations for $2+1$ quark flavors with physical quark masses in continuum limit, i.e. vanishing lattice spacing \cite{latFodor}. The curves are the HRG calculations at different hard-core radii of hadron resonances, $r$. We note that increasing the hard-core radius reduces the ability to reproduce the lattice QCD results. Two remarks are now in order. At $0\leq r<0.2~$fm, the ability of HRG model to reproduce the lattice energy density or trace anomaly is apparently very high. Furthermore, we note that varying $r$ in this region makes almost no effect, i.e. the three radii, $r=[0.0,0.1,0.2]~$fm, have almost the same results. At $r>0.2~$fm, the disagreement  becomes obvious and increases with increasing $r$. At higher temperatures, the resulting thermodynamic quantities, for instance energy density and trace anomaly become {\it non}-physical. For example, the energy density and trance anomaly nearly tends toward vanishing. 

So far, we conclude that the excluded volume is practically irrelevant. It comes up with a negligible effect, at $r\leq 0.2~$fm. On the other hand, a remarkable deviation from the lattice QCD calculations appears, especially when relative large values are assigned to $r$. With this regard, it has to be taken into consideration that the excluded volume itself is conjectured to assure the thermodynamic consistency in the HRG model \cite{exclV1,exclV2,exclV3,exclV4,exclV5,exclV6}. 

It is obvious that the thermodynamic quantities calculated in HRG model are likely to diverge at $T_c$ \cite{hgdrnA,hgdrnB,hgdrn1A,hgdrn1B}. It is a remarkable finding that despite the mass cut-off at $2~$GeV,  the energy density remains finite even when $T$ exceeds $T_c$. Apparently, this is the main reason why the trace anomaly gets negative values. The correction to pressure is tiny or negligible  \cite{exclV1,exclV2,exclV3,exclV4,exclV5,exclV6}.  Nevertheless, the finite hard-core should not be believed to reproduce the lattice QCD simulations at $T>T_c$. The validity of HRG model is strictly limited to $T<T_c$.

The excluded-volume approach \cite{extndd}, section \ref{sec:exclvV}, assumes that the energy normalized by $4{\cal B}$ equals the excluded-volume and the intensive quantity $T_{pt}$ in the point-type particle approach and the other thermodynamic quantities \cite{extndd2}, respectively, have to be {\it corrected} as follows.
\begin{eqnarray}
T &=& \frac{T_{pt}}{1-\frac{P_{pt}(T_{pt})}{4{\cal B}}}, \\
p(T) &=& \frac{p_{pt}(T_{pt})}{1-\frac{P_{pt}(T_{pt})}{4{\cal B}}},
\end{eqnarray}
where ${\cal B}^{1/4}=0.34\,$GeV stands for the MIT bag constant.
  
\subsubsection{Uhlenbeck and Gropper statistical interactions}
\label{sec:UG}

The second type of extensions has been introduced by Uhlenbeck and Gropper \cite{uhlnb}. This is mainly a statistical interaction. The {\it non}-ideal (correlated) hadron statistics is given by the classical integral if the Boltzmann factor $\exp(-\phi_{ij}/T)$ is corrected as follows. 
\bea \label{eq:uhlb}
\exp\left(-\frac{\phi_{ij}}{T}\right)\left[1\pm e^{-m\, T\, r_{ij}^2}\right],
\eea
where $r_{ij}$ is the average correlation distance between $i$-th and $j$-th particle. $\phi_{i j}$ is the interaction potential between $i$-th and $j$-th particle pairs. It is not an exception to apply this to any type of interactions. But, the potential must be such that the
Boltzmann factor precisely equals to the pair correlation function \cite{ug4}. Apparently, the summation over all pairs gives the total potential energy. This kind of modifications takes into account correlations and also the {\it non-ideality} of the hadron gas. The latter would among others refer to the discreteness of the energy levels. Uhlenbeck and Gropper  introduced an additional correction but concluded that it is only valid at very low temperatures \cite{uhlnb}. It should be noticed that the correction, expression (\ref{eq:uhlb}), is belonging to {\it generic} types of correlation interactions.

The quantum correction introduced by statistics can appear as an attractive potential for Bose-Einstein statistics and as a repulsive potential for Fermi-Dirac statistics. Pathria \cite{ug4} developed a mathematical expression for the effective interaction between fermions or bosons. There are several examples of the statistical interactions. The virial correction to the pressure of an ideal gas is most likely the origin of this idea about the effective interaction. Applications as in physics of white dwarf stars (classic example
of Fermi repulsion) and of triplet atomic hydrogen atom (effective repulsion
between like-spin electrons due to the Paul principle) are well-known  \cite{mb2003}. The diatomic hydrogen atom is bound in electron singlet state, while the triplet is not. 
Another example can be found in a system of rare gas atoms. When one gas approaches  another, there is an exponential repulsion between the atoms, which often is explained by the electron statistical repulsion. Similarly, the trapped bosons condense, like boson stars, which likely collapse to a smaller region in the center of the trap, gives the impression of an effective boson statistical attraction \cite{mb2003}.

\subsubsection{$S$-matrix and strong interactions}
\label{sec:smatrx}

As introduced in Ref. \cite{Tawfik:2004sw}, the third type of interactions to be implemented in the ideal hadron gas is the attraction. The Hagedorn states are considered as the framework to study the physics of strongly interacting matter for temperatures below $T_c$. The Hagedorn interaction finds its description in the hadron mass spectrum, $\rho(m)$, Eq. (\ref{eq:Imuel1}). Using the hadrons and resonances which are verified experimentally, the limits of the exponential $\rho(m)$ can be determined. For instance, the mass cut-off may vary from strange to non-strange states \cite{brnt}, for strange $1.5~$GeV and for non-strange $2.0~$GeV or from bosons to fermions, etc. According to the Bootstrap model \cite{boots1,boots2}, the fireballs are treated as hadronic massive states possessing all conventional hadronic properties. It is apparent that the fireball mass is determined by the mass spectrum. Furthermore, fireballs are consisting of further fireballs. This is only valid at the statistical equilibrium of an ensemble consisting of an undetermined number of states (fireballs).

On one hand, the whole spectrum of possible interactions can be represented by $S$-matrix, which can re-write the partition function, Eq.~(\ref{eq:lnz1}), as an expansion of the fugacity term \cite{Tawfik:2004sw}  
\begin{eqnarray}
\ln {\cal Z}^{(int)}(V,T,\mu) &=& \ln {\cal Z}^{(id)}(V,T,\mu) +
\sum_{\nu=2}^{\infty} a_{\nu}(T) \exp(\mu_{\nu}/T). \label{eq:za3}
\end{eqnarray}
The $S$-matrix describes the scattering processes in the thermodynamical system~\cite{Dashen:1969mb}. $a_{\nu}(T)$ are the so-called virial coefficients and the subscript $\nu$ refers to the order of the
multiple-particle interactions
\begin{eqnarray}
a_{\nu}(T) &=& \frac{g_r}{2\pi^3}
\int_{M_{\nu}}^{\infty}dw\;e^{-\varepsilon_r(w)/T}\;
\sum_l(2l+1)\frac{\partial}{\partial w}\delta_l(w). \label{eq:za4}
\end{eqnarray}
The sum runs over the spatial waves. The phase shift $\delta_l(w)$ of two-body inelastic interactions, for instance, depends on the resonance half-width $\Gamma_r$, spin and mass of produced resonances,  
\bea
\ln {\cal Z}^{(int)}(V,T,\mu) &=& \ln {\cal Z}^{(id)}(V,T,\mu) +
\frac{g_r}{2\pi^3}\int_{M_{\nu}}^{\infty}dw
      \frac{\Gamma_r\;e^{(-\varepsilon_r(w)+\mu_r)/T}
      }{(M_r-w)^2+\left(\frac{\Gamma_r}{2}\right)^2}.
      \hspace*{10mm} \label{eq:za5}  
\eea
 For narrow width and/or at low $T$, the virial term reduces, so that we will get the {\it non-relativistic} ideal partition function of the hadron resonances with effective masses $M_{\nu}$. In other words, the \underline{\it resonance contributions to the partition function are the same as that of free particles} \underline{\it with effective masses}. At temperatures comparable to $\Gamma_r$, the effective mass approaches the physical one. Thus, at high temperatures, the strong interactions are  taken into consideration via including heavy resonances,
Eq.~(\ref{eq:za3}). We therefore suggest to use the canonical partition function Eq.~(\ref{eq:za5}) without any corrections. 

In such a way, the $S$-matrix would give the plausible scattering processes taking place in the system of interest. The strong interactions are taken into consideration via heavy resonances. These conclusions suggest that the grand canonical partition function is able to simulate various types of interactions, when hadron resonances with masses up to $2~$GeV are included. As discussed, this sets the limits of Hagerdorn mass spectrum, Eq. (\ref{eq:Imuel1}). The predefined mass cut-off is supposed to avoid Hagedorn's singularity.  A conclusive convincing proof has been presented through confronting HRG to lattice QCD results  \cite{Karsch:2003vd,Karsch:2003zq,Redlich:2004gp,Tawfik:2004sw,Tawfik:2004vv}. Fig. \ref{fig:vdW} illustrates the excellent agreement between HRG with $r=0~$fm and the lattice QCD calculations. It should be noticed that the results at $T>T_c$ are out of scope of the HRG model. \\


We conclude that the attraction interaction is very sufficient to overcome the hard-core repulsion interaction. In light of this, we comment on the conclusion of Ref. \cite{apj}. In framework of  interacting hadron resonance gas, a thermal evaluation of thermodynamic quantities has been proposed. The interactions to be implemented in the HRG model are mainly van der Waals repulsion which are included through correction for the finite size of hadrons. Different values for the hadron radii can be assigned to the baryons and mesons. The authors studied the sensitivity of the modified HRG model calculations to the hadron radii. The results on different thermodynamic quantities were confronted with predictions from lattice QCD simulations. Therefore, the conclusion would be understood, as  hadron resonances with masses up to $3~$GeV are taken into consideration. At this mass cut-off, the exponential description of the mass spectrum would be no longer valid. Furthermore, it is straightforward to deduce that heavier masses are connected with lower thermodynamical quantities. It is correctly emphasized \cite{lFodor10} that including all known hadrons up to $2.5$ or even $3.0$ GeV would increase the number of hadron resonances by a few states with masses $>2~$GeV. An attempt to improve the HRG model by including an exponential mass spectrum for these very heavy resonances has been proposed \cite{brnt}. In Refs. \cite{lFodor10,apj} only known states and not the mass spectrum are taken into account. For this reason, the authors of Ref. \cite{apj} concluded that the HRG model with small or even vanishing radii gives thermodynamic quantities which apparently are less steeply than in case of a {\it ideal} HRG.

\subsection{Dissipative properties of hadron gas: viscous coefficients}
\label{sec:visc}

The dissipative properties in hadronic matter can approximately be studied under a simplification that the hadron gas has only one component. The Relaxation time and Chapman-Enskog approximation shall be applied \cite{wirananta09} in section \ref{sec:viscone}. Section \ref{sec:viscmult} is devoted to the case of multiple components. $K$-matrix parametrization of hadronic cross-sections, $T$-matrix and its relation to $K$-matrix and relaxation time shall be elaborated.

\subsubsection{Dissipative quantities in a single-component fluid}
\label{sec:viscone}

\paragraph{\bf Chapman-Enskog approximation \\}

The dependence of bulk viscosity on the speed of sound, the thermodynamic properties of the system and the interaction cross-sections between the constituents of a hadronic system was studied \cite{wirananta09,wirananta12}. An approximation to bulk viscosity in physical units gives
\bea
\eta_{\nu} &=& k\, T\, \frac{\alpha_2^2}{2\, w_0^{(2)}},
\eea
where the parameters read
\bea
\alpha_2 &=& \frac{3}{2}\, \left[z \, \tilde{h}\left(\gamma-\frac{5}{3}\right)+\gamma\right], \label{eq:alf21}\\
z &=& \frac{m\, c^2}{k\, T}, \\
\tilde{h} &=& \frac{K_3(z)}{K_2(z)},
\eea
$\gamma$ is the ratio of specific heats at constant pressure and volume, respectively, $c_p/c_v$, and $K_i$ is the modified Bessel function of $i-$th order. The  omega integral, $w_0^{(2)}$, gives information about the cross-section of scattering particles,
\bea
w_i^{(2)} &=& \frac{2\, \pi\, c\, z^3}{K_2(z)^2} \int_0^{\infty} d\, \psi\, \sinh^7(\psi) \cosh^i(\phi)\, K_j(2\, z\, \cosh(\psi)) 
\int_0^{\pi} d\, \Theta \, \sin(\Theta)\, \sigma(\psi,\Theta)\, \left(1-\cos^2(\Theta)\right),  
\eea
where the order $j=5/3+(-1)^i/2$ and $\sigma(\psi,\Theta)$ is the differential cross section. The hyperbolic functions read 
\bea
\sinh(\psi) &=& \frac{p_1-p_2}{2\, m\, c}, \\
\cosh(\psi) &=& \frac{\sqrt{- p_{\alpha}\, p^{\alpha}}}{2\, m\, c}.
\eea
The ratio of specific heats can be related to the adiabatic speed of sound, $c_s=\sqrt{\partial (-p_{\alpha}\, p^{\alpha})^{1/2}/\partial \epsilon|_{S}}$ at constant entropy $S$,
\bea
\gamma &=&1+ \left.\frac{\partial (-p_{\alpha}\, p^{\alpha})^{1/2}}{\partial \epsilon}\right|_{S}=1+c_s^2.
\eea
Also, the parameter $\alpha_2$, Eq. (\ref{eq:alf21}), can be expressed in terms of the speed of sound,
\bea
\alpha_2 &=& \frac{3}{2}\, \left[-\left(z \, \tilde{h}+1\right)\left(\frac{1}{3}-c_s^2\right)-\frac{1}{3}\, z\, \tilde{h}+\frac{4}{3}\right].
\eea
Then, the bulk viscosity reads
\bea
\eta_{\nu} &=& k\, T\, \frac{a^2 \left(\frac{1}{3}-c_s^2\right)^2 + 2\, a\, b \left(\frac{1}{3}-c_s^2\right)+b^2}{2\, w_0^{(2)}},
\eea
where $a=-3(z\,\tilde{h}+1)/2$ and $b=-(z\,\tilde{h}-4)/2$. \\

\paragraph{\bf Relaxation time approximation \\}

\begin{figure}
\centering
\includegraphics[width=7cm]{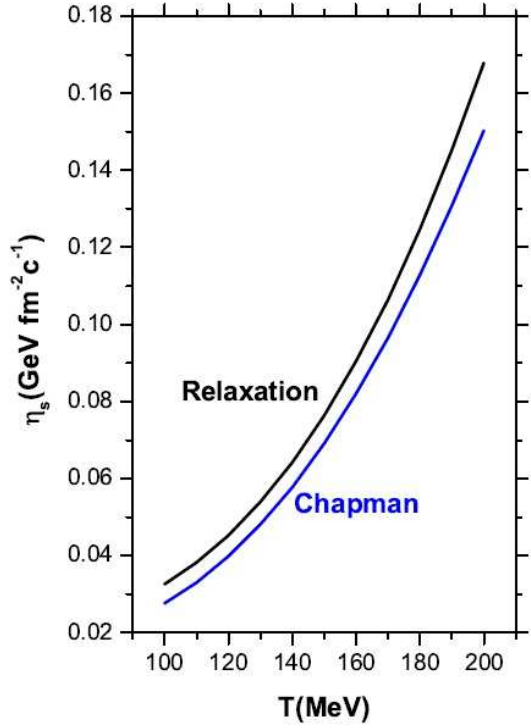}
\caption{Shear viscosity coefficients of pion gas as calculated from Chapman-Enskog (bottom curve) and relaxation time (top curve) approximations as function of temperature. Both quantities are given in physical units. Graph taken from Ref. \cite{wirananta12}.}
\label{wiranaS}
\end{figure}

We start with energy-stress tensor \cite{xiovereta1a,wirananta09,wirananta12},
\bea
T^{i\, j} &=& T_0^{i\, j} + \frac{1}{(2\, \pi)^3} \int d^3\,p\,  \frac{p^i}{p^0} \;\; v_p^j\, \delta\, f_p, \\
\delta\, f_p &=& -\tau \left(\frac{\partial\, f_p^0}{\partial\, t} + v_p \cdot \nabla\, f_p^0\right), \label{eq:tij_W}
\eea
where $f_p^0$ is the equilibrium distribution function and $v_p$ is the velocity of a particle with momentum $p$ and energy $p^0=\epsilon_p$. The dissipative part of the energy-stress tensor reads
\bea
T^{i\, j}_{diss} &=& - \eta \left(\frac{\partial\, u^i}{\partial\, x^j}+\frac{\partial\, u^j}{\partial\, x^j}\right) - \left(\eta_{\nu}-\frac{2}{3}\right)\; \nabla \cdot u \, \delta^{i\, j},  \label{eq:tij_Wdiss}
\eea
where $u$ is the fluid velocity field. Comparing Eq. (\ref{eq:tij_W}) with Eq. (\ref{eq:tij_Wdiss}), the bulk viscosity can be deduced from 
\bea
\eta_{\nu} &=& \frac{\tau}{9\, T} \frac{1}{(2\, \pi)^3}\int d^3\,p\,  \frac{f^0}{p^0} \left[\left(1-3\frac{h}{c_s\, T}\right) p^2 - \left(m\, \frac{p}{\epsilon_p}\right)^2\right], \\
 &=& \frac{\tau}{9\, T} \frac{1}{(2\, \pi)^3} \int d^3\,p\,  \frac{f^0}{p^0} \left[\left(1-3 c_s^2\right) \epsilon_p - \frac{m^2}{\epsilon_p}\right]^2,
\eea
where $\tau$ is a momentum-independent relaxation time. The bulk viscosity would require an energy-dependent relaxation time
\bea
\eta_{\nu} &=& \frac{1}{9\, T} \frac{1}{(2\, \pi)^3} \sum_i \int d^3\,p\,  \frac{\tau_i(\epsilon_i)}{\epsilon_i^2} \left[\left(1-3 c_s^2\right) \epsilon_i^2 - m^2\right]^2 \, f_i^0,
\eea

As concluded in Ref. \cite{wirananta09}, two extreme limits apparently leads to vanishing bulk viscosity: 
\begin{itemize} 
\item $m\rightarrow 0$ and $c_s^2\rightarrow 1/3$ and 
\item $m\sim \epsilon_p$ and $c_s^2\rightarrow 2/3$. 
\end{itemize}  
It was concluded that the intermediate mass particles are the ones contributing the most to the bulk viscosity. To derive expressions for the shear viscosity, almost the same procedure was applied \cite{wirananta12}. Fig. \ref{wiranaS} presents the shear viscosity coefficients of a pion gas calculated using Chapman-Enskog (bottom curve) and relaxation time (top curve) approximations. In both approximations,  increasing $T$ leads to a rapid increase in the shear viscosity.

\subsubsection{Dissipative quantities in a multiple-component fluid}
\label{sec:viscmult}

\paragraph{\bf Relaxation time approximation \\}

The relativistic kinetic theory gives the transport equations for classic and colored particles in a non-Abelian external field. The transport coefficients for {\it confined} QCD matter composed of known particles and resonances are studied at vanishing chemical potential. The relaxation time approximation of Boltzmann equation is applied. The hadronic density states in {\it confined} QCD are taken into consideration. The transport equations describe the evolution of the phase space distribution function of the particles of interest. The transport properties are defined as the coefficients of spatial component of the difference between the energy-momentum tensors out of and at equilibrium corresponding to the Lagrangian density.  

The Hagedorn fluid of QCD {\it confined} phases can be modelled as a non-interacting gas composed hadron resonances \cite{Tawfik:2010mb}. The main motivation of doing this was discussed in previous sections. This refers to all relevant degrees of freedom of the {\it confined} strongly interacting matter and implicitly including the interactions that likely result in resonance formation \cite{hgdrnA,hgdrnB}, i.e. the strong interaction.

In spherical polar coordinates, the energy-momentum tensor of a single particle with $p$- and $T$-independent mass $m$ is defined as  
\begin{eqnarray}
T^{\mu\nu}_1 &=& \frac{g}{2\pi^2} \rho(m)\int p^2\, dp \;
\frac{p^{\mu}p^{\nu}}{\varepsilon}\;  n(p,T),   
\end{eqnarray}
where $p^{\mu}=(\varepsilon,\vec{p})$ is momentum four-vector and $g$ is degeneracy factor of the hadron resonances. The single particle energy is given by the dispersion relation 
$\varepsilon=(\vec{p}\,^2+m^2)^{1/2}$.  The Hagedorn mass spectrum $\rho(m)$, section \ref{sec:hagd}, implies growth of the hadron mass spectrum with increasing the resonance masses. 
\begin{eqnarray}\label{eq:rhom}
\rho(m) &=& c\left(m_0^2+m^2\right)^{k/4} \exp(m/T_H),
\end{eqnarray}
with $k=-5$, $c=0.5\,$GeV$^{3/2}$, $m_0=0.5\,$GeV and $T_H=0.195\,$GeV. 

With the above assumptions on Hagedorn viscous fluid, the overall energy-momentum tensor can be calculated as a sum over energy-momentum tensors $T^{\mu\nu}_1$ of all hadrons resonances,
\begin{eqnarray}
T^{\mu\nu} & =& \sum_i T^{\mu\nu}_i . 
\end{eqnarray}
This reflects the algebraic properties, here the addition, of the energy-momentum tensor.
In momentum phase space and assuming that the system is in a state with vanishing chemical potential but near equilibrium, the distribution function $n(p,T)$ reads
\begin{eqnarray}\label{eq:noTp}
n(p,T) & =& \frac{1}{\exp\left(\frac{\varepsilon - \vec{p}\cdot\vec{u}}{T}\right)\pm1}, 
\end{eqnarray}
where $\pm$ stands for fermion and boson statistics, respectively. The local flow velocity $\vec{u}$ is compatible with the Eckart fluid \cite{eckrt}, implying that $T^{\mu\nu}u_{\mu}u_{\nu}=\varepsilon$. It is obvious that $n(p,T)$ satisfies the kinetic theory \cite{V-hadron2a,V-hadron2b} and second law of thermodynamics. The solution of kinetic equation is obtainable by deviating the distribution function from its local equilibrium.

The deviation of energy-momentum tensor from its local equilibrium is corresponding to the difference between the distribution function near and at equilibrium, $\delta n=n-n_0$. The latter can be determined by {\it relaxation time approximation} with vanishing external and self-consistent forces \cite{V-hadron2a,V-hadron2b,relaxx}
\begin{eqnarray}
\delta n(p,T) &=& - \tau \frac{p^{\mu} }{\vec{p}\cdot\vec{u}}\; \partial_{\mu} n_0(p,T).
\end{eqnarray}

\begin{figure}
\centering
\includegraphics[width=10cm]{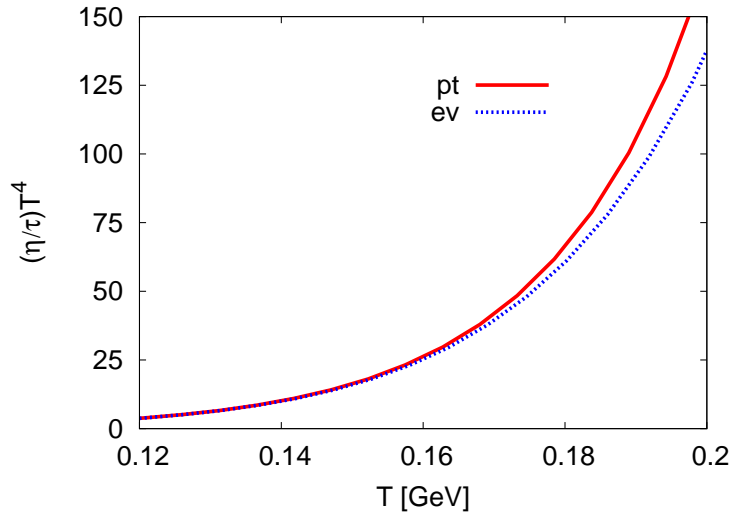}
\caption{Bulk  viscosity coefficients of Hagedorn fluid are depicted as function of the heat bath temperature. The effects of excluded-volume approach are illustrated (dashed lines).}
\label{2fig1a}
\end{figure}

\begin{figure}
\centering
\includegraphics[width=10cm]{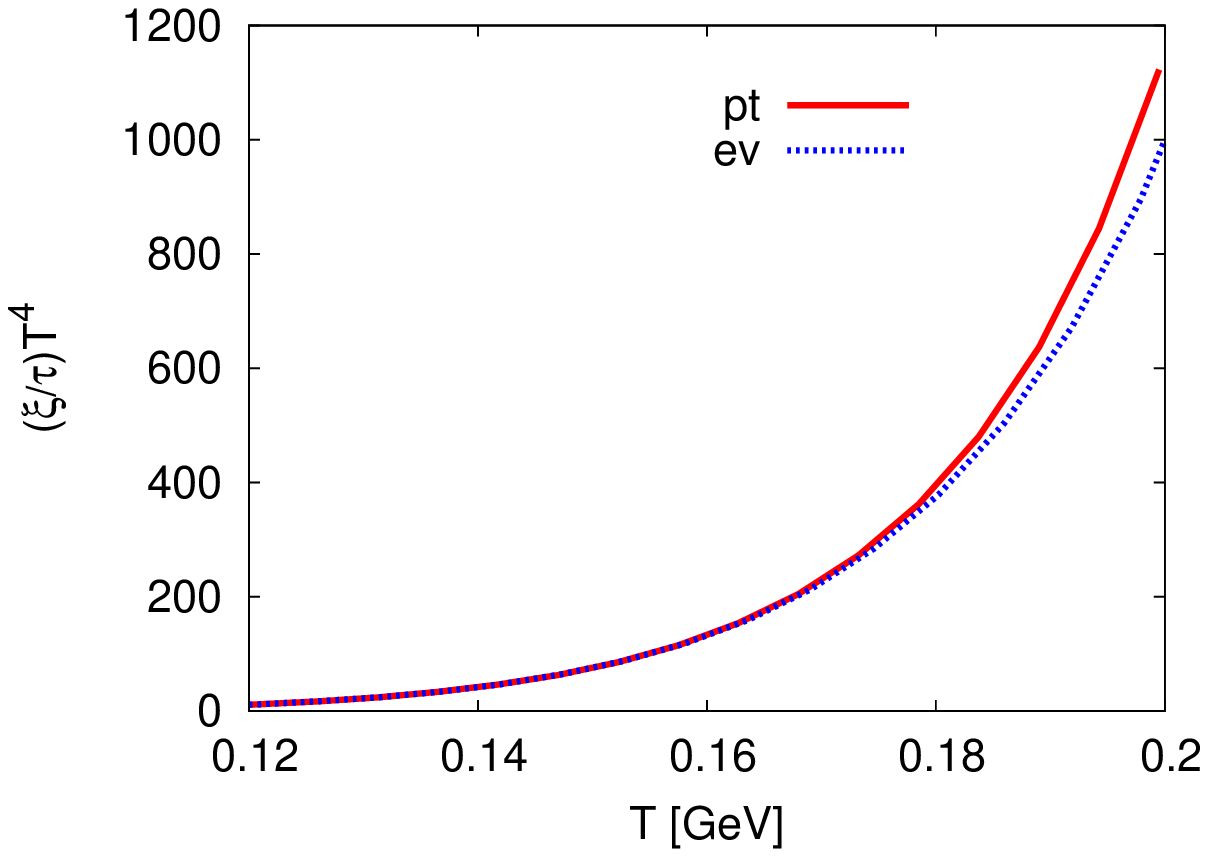}
\caption{Shear viscosity coefficients of Hagedorn fluid are depicted as function of the heat bath temperature. The effects of excluded-volume approach are illustrated (dashed lines).}
\label{2fig1b}
\end{figure}

\begin{figure}[htb]
\centering
\includegraphics[width=7cm]{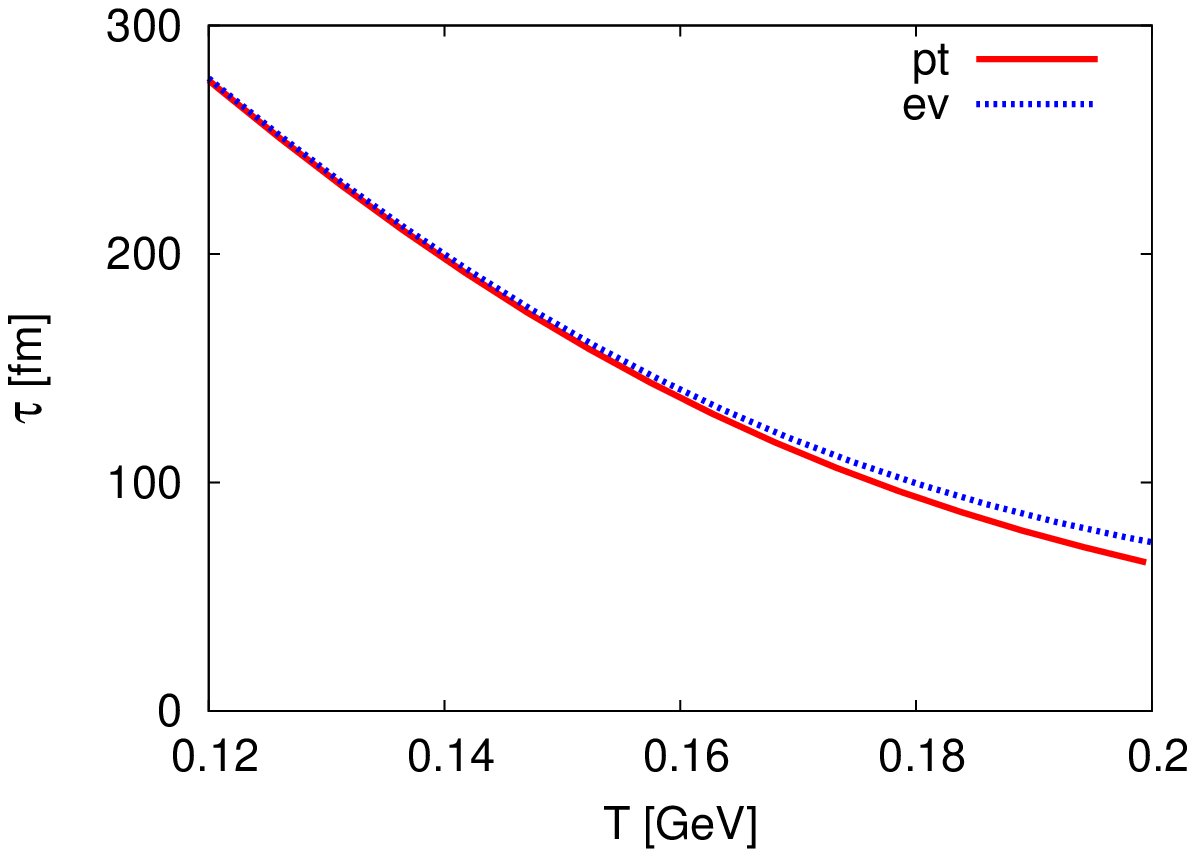}
\includegraphics[width=7cm]{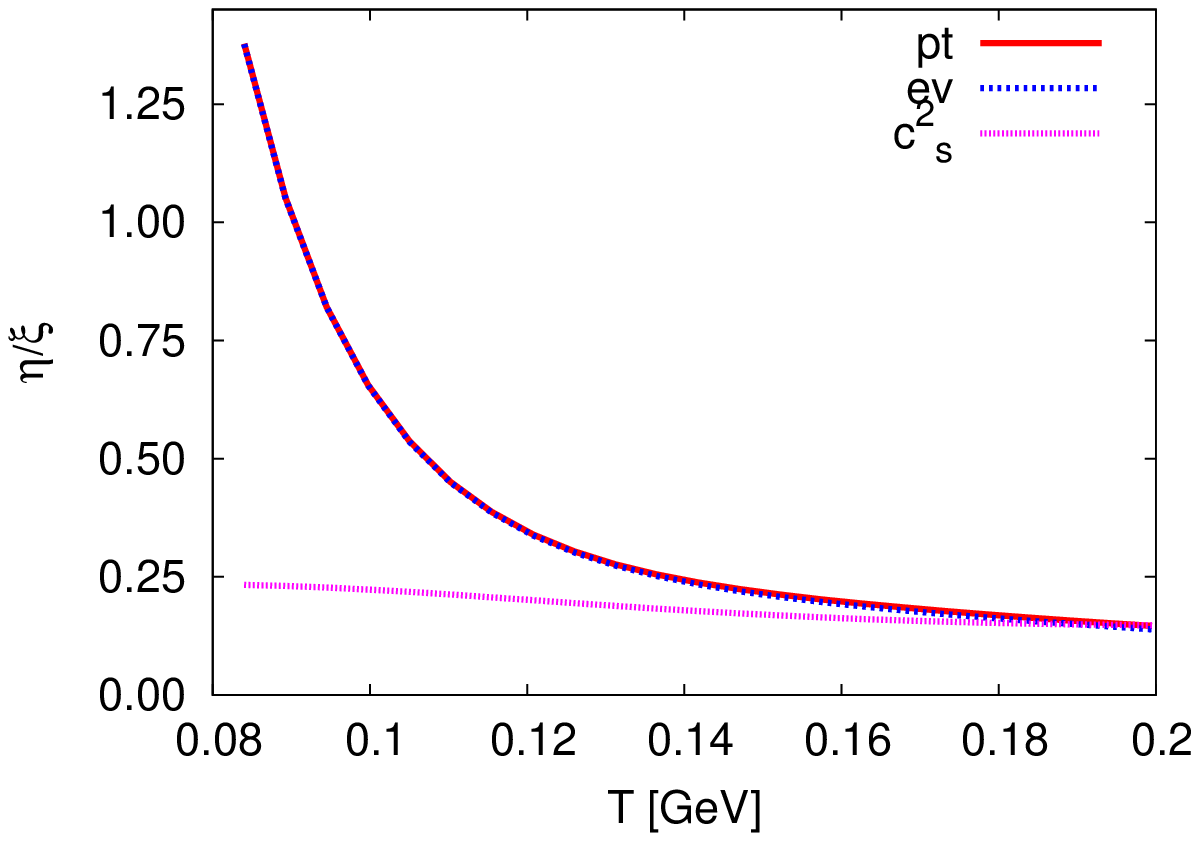}
\caption{Left panel: The thermal change of the relaxation time in Hagedorn fluid. The point-type and excluded-volume approaches are compared. Dashed-dotted line gives the speed of sound.. Right panel: The ratio $\eta/\xi$ is given in dependence on temperature $T$. The point-type and excluded-volume approaches are compared. Dashed-dotted line gives the speed of sound.}
\label{2fig2a}
\end{figure}

Then, the difference between near and equilibrium energy-momentum tensor reads
\begin{eqnarray}
\delta T^{\mu\nu}_1 &=& -\frac{g}{2\pi^2} \rho(m)\int_{0}^{\infty} p^2\, dp \;
\frac{p^{\mu}p^{\nu}}{\varepsilon^2}\; \tau\; p^{\alpha}\partial_{\alpha}\; n_0(p,T).   
\end{eqnarray}
Using the symmetric projection tensor $h_{\alpha\beta}$ \cite{maartns}, the components of the derivative $\partial_{\alpha}$ can split to parallel and orthogonal to $u^{\mu}$. $h_{\alpha\beta}$ generates a 3-matric and projects each point into the instantaneous rest space of the fluid.  
\begin{eqnarray}
\partial_{\mu} &=& D u^{\mu} + \nabla_{\mu},
\end{eqnarray}
where $D=u^{\alpha} \partial_{\alpha}=(\partial_{t},0)$ gives the temporal derivative and $\nabla_{\mu}=\partial_{\mu}-u_{\mu} D =(0,\partial_{i})$ are spacial derivative \cite{V-hadron2a,V-hadron2b}. Such a splitting has to guarantee the conservation of equilibrium energy-momentum tensor; $\partial_{\mu} T^{\mu\nu}=0$ and fulfil the laws of thermodynamics at equilibrium \cite{V-hadron2a,V-hadron2b,maartns}. In ref \cite{V-hadron1}, the non-equilibrium $n(p,T)$ has been decomposed using the {\it relaxation time approach} $n=n_0+\tau n_1+\cdots$. Alternatively, as $n(p,T)$ embeds the 1$^{st}$-rank tensor $u$, $\delta T^{\mu\nu}_1$ can be decomposed into $u$ \cite{V-hadron2a,V-hadron2b} in order to deduce its spatial components
\begin{eqnarray}\label{eq:decomp}
\partial_k u^l &=& \frac{1}{2}\left(\partial_k u^l + \partial_l u^k - \frac{2}{3}\delta_{kl}\partial_i u^i\right) + \frac{1}{3}\delta_{kl}\partial_i u^i \equiv \frac{1}{2} W_{kl} + \frac{1}{3}\delta_{kl}\partial_i u^i.
\end{eqnarray}
Applying the equation of hydrodynamics, then the deviation from equilibrium 
\begin{eqnarray}
\delta T^{\mu\nu}_1 &=& \frac{g}{2\pi^2}\rho(m)\tau \int_{0}^{\infty} 
\frac{p^{\mu}p^{\nu}}{T} n_0(1+n_0) 
\left[\vec{p}\cdot\vec{u} c_s^2 \nabla_{\alpha} u^{\alpha} + p^{\alpha}\left(\frac{\nabla_{\alpha} p}{\epsilon+p} - \frac{\nabla_{\alpha} T}{T}\right)+\frac{p^{\alpha}p^{\beta}}{\vec{p}\cdot\vec{u}} \nabla_{\alpha} u_{\beta}\right] p^2 dp, \hspace*{10mm}
\end{eqnarray}
can be re-written as 
\begin{eqnarray} \label{finaldT}
\delta T^{i j}_1 &=& \frac{g}{2\pi^2}\frac{\tau}{T}\rho(m)\int_{0}^{\infty} 
p^{i}p^{j} n_0(1+n_0) 
\left[\left(\varepsilon c_s^2-\frac{\vec{p}\, ^2}{3\varepsilon}\right)\partial_i u^i - 
\frac{p^kp^l}{2\varepsilon} W_{kl}\right] p^2 dp,
\end{eqnarray}
where $c_s^2=\partial p/\partial \epsilon$ is the speed of sound in this viscous fluid. The bulk $\xi$ and shear $\eta$ viscosity can be deduced from Eq. (\ref{finaldT}) by comparing it with  
\begin{eqnarray} \label{eq:dltT}
\delta T^{i j}_1 &=& -\xi \delta_{ij} \partial_k u^k - \eta W_{ij}.
\end{eqnarray}
To find shear viscosity $\eta$, we put $i\neq j$ in Eqs.  (\ref{eq:decomp}) and (\ref{eq:dltT}). To find bulk viscosity $\xi$, we substitute $i$ with $j$ and $T^{\mu\nu}_0$ with $3P$. The subscript $0$ (as in the distribution function $n$) refers to the equilibrium state. Although we keep the gradients of velocity, we put $\vec{u}=0$ in the final expressions. 
The intensive quantities $\eta$ and $\xi$ of Hagedorn fluid \footnote{As we assume a vanishing chemical potential, the heat conductivity vanishes as well.}  in comoving frame, respectively, read
\begin{eqnarray}
\xi(T) &=& \frac{g}{2\pi^2} \frac{\tau}{T} \sum_i\rho(m_i)\int_{0}^{\infty} n_0(1+n_0)\left(c_s^2 \varepsilon_i^2 - \frac{1}{3}\vec{p}\, ^2 \right)^2 p^2 dp,\\
\eta(T) &=& \frac{g}{30\pi^2} \frac{\tau}{T} \sum_i\rho(m_i)\int_{0}^{\infty} n_0(1+n_0) \frac{\vec{p}\, ^4}{\varepsilon_i^2} p^2 dp.
\end{eqnarray}
The $T$--dependence of dimensionless ratios $\xi T^4/\tau$ and $\eta T^4/\tau$ is depicted in Figs. \ref{2fig1a} and \ref{2fig1b}, respectively. With increasing $T$, bulk and shear viscosities increase, significantly. We note that $\xi$ seems to be about one order of magnitude larger than $\eta$. Right panel of Fig. \ref{2fig2a} illustrates such a comparison. At low $T$, $\eta$ starts with larger values than $\xi$'s. But with increasing $T$, $\xi$ gets larger \cite{xiovereta1a,xiovereta1b}. 

The ratio $\xi/\eta$ can be related to the speed of sound $c_s^2$ in a gas of massless pions. Apparently, there are essential differences between this system and the one of Hagedorn fluid. According to Refs. \cite{xiovereta1a,xiovereta1b,etaxiratio1a,etaxiratio1b}, the ratio of $\xi/\eta$ in $N=2^*$ plasma is conjectured to remain finite across the second--order phase transition. This behavior seems to be illustrated in Figs. \ref{2fig1a} and \ref{2fig1b}. In the Hagedorn fluid, the system is assumed to be drifted away from equilibrium and it should relax after a characteristic time $\tau$. Should we implement a phase transition in the Hagedorn fluid, then $\tau\propto \xi^{z}$, where $z$ is the critical exponents, likely diverges near $T_c$. \\

\paragraph{\bf $K$-matrix parametrization of hadronic cross-sections \\}

The shear viscosity and entropy density of HRG are calculated using the Chapman-Enskog and virial expansion methods \cite{wirananta13}. The latter is known as $K$-matrix parametrization of hadronic cross-sections which preserves the unitarity of the $T$-matrix. The magnitude of shear viscosity is strongly determined  by the strength of interactions between the constituent particles in a system. For example, the shear viscosity is inversely proportional to the differential cross-section of the interacting particles. In light of this, it is likely that large cross-sections, which apparently a characteristic feature of  strongly interacting systems, lead to small viscosities. 

\begin{figure}[htb]
\centering
\includegraphics[width=8cm]{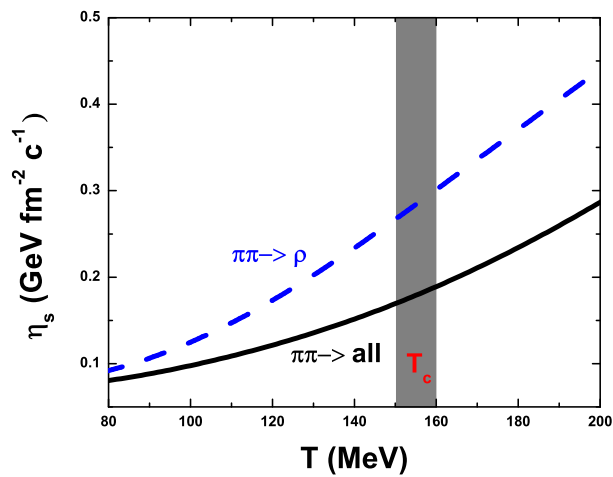}
\caption{Using $K$-Matrix formalism, the shear viscosity is given as function of temperature. The dashed curve shows results when only $\rho$-resonance was considered. The results from all hadron resonances, the ones that formed through $\pi - \pi$ interactions, are shown by the solid curve. The graph taken from Ref. \cite{wirananta13}. }
\label{fig:wiranaEta1}
\end{figure}

\begin{figure}[htb]
\centering
\includegraphics[width=10cm]{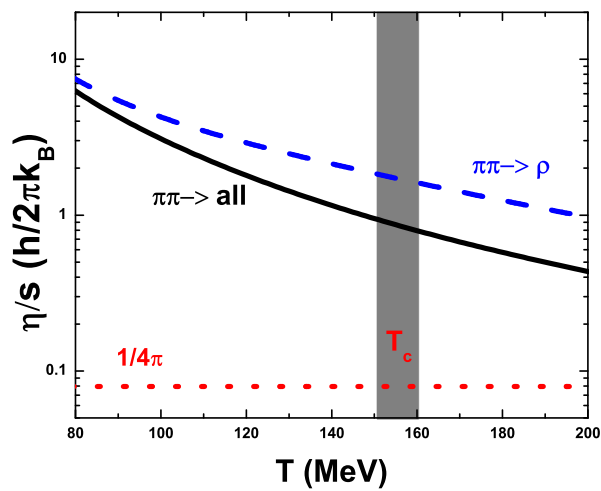}
\caption{Using $K$-Matrix formalism, the ratio of shear viscosity to entropy density is given as function of temperature. The dashed curve shows results when only $\rho$-resonance was considered. The results from all hadron resonances, the ones that formed through $\pi - \pi$ interactions, are shown by the solid curve. For reference, the AdS/CFT lower bound is given by the red dashed-line \cite{etaTOs2001}. The graph taken from Ref. \cite{wirananta13}. }
\label{fig:wiranaEta1}
\end{figure}

Resonant interactions, including the widths of the various resonances, are incorporated consistently in calculations of both the shear viscosity and the entropy density. The $K$-matrix parametrization of the hadronic cross-sections \cite{wira52,wira53,wira49} are employed. This accommodates multiple resonances and preserves the unitarity of the $T$-matrix in all channels. The various types of interactions, that would take place in the hadronic matter, have been discussed in section \ref{sec:intr}. Furthermore, the interaction or $T$-matrix for hadronic interaction through a single resonance, $a+b \rightarrow R \rightarrow a+b$ can  be parametrized  in Breit-Wigner equation \cite{bw2006}
\bea
T &=& \frac{m_R\, \Gamma_{R\rightarrow a\, b}(\sqrt{s})}{(m_R^2 - s) - i\, m_R\, \Gamma_{R}^{tot}(\sqrt{s}) },
\eea
where $m_R$ is the mass of the resonance, the total width $\Gamma_{R}^{tot}(\sqrt{s}) =\sum_{c, d} \Gamma_{R\rightarrow c d}$ and the partial width $\Gamma_{R\rightarrow c d}$ is the one for the channel $R\rightarrow c d$ of the resonance $R$. The differential cross-section for such an interaction at center-of-mass energy $\sqrt{s}$ reads
\bea
\sigma(\sqrt{s},\theta) &=& \frac{C(I,l)}{q_{a b}^2} \; \frac{m_R^2\, \Gamma^2_{R\rightarrow a\, b}(\sqrt{s})}{(m_R^2 - s)^2 + m^2_R\, \left(\Gamma_{R}^{tot}\right)^2(\sqrt{s}) }\,   P_l(\cos(\theta)),  \label{eq:sgmstheta1}
\eea
where $C(I,l)$ is the symmetry factor including the spin-isospin multiplicities for the corresponding resonance, $\theta$ is the polar angle, $I$ denotes the isospin and $I_3$ is the third component of the isospin. $l$ is the orbital angular momentum of the exit channel or decay channel and $P_l(\cos(\theta))$ are Legendre polynomials, which account for the angular momentum dependence of the exit channel.  The center-of-mass momentum of the incoming particle reads
\bea
q_{a b}(\sqrt{s}) &=& \frac{1}{2 \sqrt{s})} \, \sqrt{(s-(m_a+m_b)^2)\, (s-(m_a-m_b)^2)}.
\eea

By integrating over the polar angle and summing over all resonances, the total cross-section for the reaction $a+b\rightarrow c+d$ can be obtained
\bea
\sigma(\sqrt{s})_{a+b\rightarrow c+d} &=& \sum_R \frac{2\, S_R + 1}{(2\, S_a + 1) (2\, S_b + 1)} \, \frac{\pi}{q_{a b}} \, \frac{\Gamma_{R\rightarrow a b} \, \Gamma_{R\rightarrow c d}}{(m_R-\sqrt{s})^2 + \left(\Gamma_{R}^{tot}\right)^2(\sqrt{s})/4},\hspace*{10mm}
\eea
where $S$ is the spin for hadrons and their resonances.

The energy dependence on the partial width of the channel $R\rightarrow a b$ of resonance $R$
\bea
\Gamma_{R\rightarrow c d}(\sqrt{s}) &=& \Gamma^0_{R\rightarrow a b} \frac{m_R}{\sqrt{s}} \left(\frac{q_{a b}(\sqrt{s})}{q_{a b}(m_R)}\right)^{2 l + 1} \, \frac{1.2}{1+0.2\, \left(\frac{q_{a b}(\sqrt{s})}{q_{a b}(m_R)}\right)^{2 l} },
\eea
where $\Gamma^0_{R\rightarrow a b}$ is the  width for the channel $R \rightarrow a b$ at the pole. The last fraction is related to the Blatt-Weisskopf $B-$factor \cite{bw1952}, which will be defined later on. The differential cross-section for the process $a b \rightarrow c d$, Eq. (\ref{eq:sgmstheta1}), can be given in terms of the invariant amplitude ${\cal M}$ or the scattering amplitude $f(\sqrt{s},\theta)$ \cite{wira49}
\bea
\sigma(\sqrt{s},\theta) &=& \frac{1}{(8 \pi)^2 \, s} \left(\frac{q_{a b}}{q_{c d}}\right)\, \left|{\cal M}\right|^2 = \left|f\left(\sqrt{s},\theta\right)\right|^2,\\
\left|f\left(\sqrt{s},\theta\right)\right|^2 &=& \frac{1}{q_{a b}} \sum_l (2\, l+1)\, T^d(s)\, P_l(\cos(\theta)), 
\eea
where $q_{a b} (q_{c d})$ is the breakup momentum in the initial (final) state and $T^d(s)$ is the interaction term.

Using $K$-Matrix formalism, the thermal evolution of shear viscosity is given in Fig. \ref{fig:wiranaEta1}. The dashed curve shows results when only $\rho$-resonance was considered. The results from all hadron resonances, the ones that formed through $\pi - \pi$ interactions, are shown by the solid curve. The temperature dependence of shear viscosity normalized to the entropy density is presented in Fig. \ref{fig:wiranaEta1}.  For reference, the AdS/CFT lower bound is given by the red dashed-line \cite{etaTOs2001}. \\

\paragraph{\bf $T$-matrix and its relation to $K$-matrix \\} 

The overlap matrix between the initial and final state of the collision can be used to define the $T$-matrix,
\bea
S_{a\, b \rightarrow c\, d} &=& \langle c\, d |S|a\, b\rangle.
\eea
The scattering operator (matrix) is in turn defined
\bea
S &=& I + 2\, i\, T,
\eea
where $I\, (T)$ is the identity operators related to collision-free case, $S\, S^{\dagger} = S^{\dagger}\, S=I$. Then, we get $(T^{-1}+i\, I)^{\dagger} = T^{-1}+i\, I$. Accordingly, the Hermitian of $K$-matrix can be given as
\bea
K^{-1} &=& T^{-1} + i\, I,  \\
K &=& K^{\dagger}.
\eea
To prove that $K$-matrix is indeed symmetric, one can study the time reversal symmetry of both $S$- and $T$-matrix. Then, real and imaginary parts of $T$-matrix can be expressed in $K$-matrix
\bea
\text{Re}\; T &=& (I+K^2)^{-1}\, K = K\, (I+K^2)^{-1}, \\
\text{Im}\, T &=& (I+K^2)^{-1} \, K^2 = K^2\, (I+K^2)^{-1}.
\eea

The hadron resonances with masses $m_R$ can be represented as a sum of poles in $K$-matrix
\bea
K_{a\, b \rightarrow c\, d} &=& \sum_R \frac{g_{R\rightarrow a b}(\sqrt{s})\; g_{R\rightarrow c d}(\sqrt{s})}{(\sqrt{s})m_R^2 - s},
\eea
where the decay coupling is given as
\bea
g^2_{R\rightarrow a b}(\sqrt{s}) &=& m_R\; \Gamma_{R\rightarrow a b}(\sqrt{s}) = \Gamma^0_{R\rightarrow a b}(\sqrt{s})\, \frac{m_R^2}{\sqrt{s}} \, \frac{q_{a b}}{q_{a b 0}} \left[B^l(q_{a b},\, q_{a b 0})\right]^2,
\eea
and $q_{a b 0}$ being the breakup momentum at energy $\sqrt{s}=m_R$. The Blatt-Weisskopf barrier factors can be given as
\bea
B^l(q_{a b},\, q_{a b 0}) &=& \frac{F_l(q_{a b})}{F_l(q_{a b 0})},
\eea
where for $0\leq l \leq 4$ and $z=(q/q_R)^2$ with $q_R=0.1973~$GeV/c  \cite{wirananta13}
\bea
F_0(q) &=& 1, \\
F_1(q) &=&  \sqrt{\frac{2\, z}{z+1}},\\
F_2(q) &=&  \sqrt{\frac{13\, z^2}{(z-3)^2+ 9  z}},\\ 
F_3(q) &=&  \sqrt{\frac{277\, z^3}{z\, (z-5)^2 + 9\, (2 z - 5)}},\\
F_4(q) &=&  \sqrt{\frac{12746\, z^4}{(z^2-45 z + 105)^2 + 25 z (2 z - 21)^2}}.
\eea
\newline

\paragraph{\bf The relaxation time \\}

The relaxation time depends on the relative cross-section as
\begin{eqnarray}
\tau(T) &=& \frac{1}{n_f(T)\langle v(T)\sigma(T)\rangle},
\end{eqnarray}
where $v(T)$ is the relative velocity of two particles in a binary collision and $n_f(T)$ is the density of each of the two species. The thermal-averaged transport rate or cross-section is $\langle v(T)\sigma(T)\rangle$. The transport equation of single-particle distribution function in momentum space, $n(r,p,t)$ \cite{GSt}, 
\begin{eqnarray}\label{BUU}
\frac{\partial}{\partial t} n +\vec{v}\cdot\vec{\nabla}_{r} - \vec{\nabla}_{r}U\cdot \vec{\nabla}_{p} n &=& -\int\frac{d^3 p_2 d^3p_1^{\prime}, d^3\, p_2^{\prime}}{(2\pi)^6} \sigma v \times \nonumber \\ 
& & \hspace*{4mm} \left[n\, n_2(1-n_1^{\prime})(1-n_2^{\prime}) - n_1^{\prime} n_2^{\prime}(1-n)(1-n_2)\right]   \times \hspace*{10mm} \\ 
& & \hspace*{4mm}    
	\delta^4(p+p_2-p_1^{\prime}-p_2^{\prime}).   \nonumber
\end{eqnarray}
First line in r.h.s. of Eq. (\ref{BUU}) gives the Boltzmann collision term. The second line adds the Uehling-Uhlenbeck factors \cite{uu33,vuu845}. The third line accounts for the Pauli-blocking of the final states. The total derivative of $n$ is given by the collision integral. To solve Eq. (\ref{BUU}), several gradients must be take into account. Real and imaginary parts of the $G$-Matrix \cite{corr4} are taken to describe the potential of nuclear interactions and the cross-section of the binary interaction, respectively. The in-medium effects in final (Pauli blocking) and also in intermediate states have to be taken into consideration. The kinetic Boltzmann-Uehling-Uhlenbeck equations \cite{uu33} for pure nucleon system have been analysed \cite{ref:tau1}. The relaxation time in non-relativistic approximation has been deduced as
\begin{eqnarray}\label{eq:NN}
\tau(T) & \approx & \frac{850}{T^2}\left(\frac{n}{n_0}\right)^{1/3} \left[1+0.04 T\frac{n}{n_0}\right] + \frac{38}{\sqrt{T}(1+\frac{160}{T^2})} \frac{n_0}{n},
\end{eqnarray}
where $n$ is the baryon density and $n_0$ is the nuclear saturation density $n_0\approx 0.145~$fm$^{-3}$. 

When fitting the decay widths $\Gamma_i$ of $i$-th hadron resonance, then the decay relaxation time $\tau_i$ in GeV$^{-1}$ reads $\tau_i\equiv\Gamma_i^{-1}=\left(0.151 m_i -0.058\right)^{-1}$ \cite{extndd2,eta-hadrons2,eta-hadrons5}. As the resonance mass $m$ is conjectured to remain constant in thermal and dense medium, this linear fit apparently implies that $\tau$ remains unchanged, as well. In the Hagedorn fluid, where the inter-particle collisions as in Eq. (\ref{eq:NN}) are minimized, we are left with specific processes to estimate $\tau$ (decay and repulsion for instance). Formation from {\it free space} vacuum and decay to stable resonances; $P_1+P_2\leftrightarrow P_3$ \cite{newRafls} are examples. 

In rest frame of particle $P_3$ boosting from the laboratory frame, the kinetic equation for the time evolution of the number density $n_3(T)$ reads
\begin{eqnarray}\label{eq:noT}
\frac{d}{d t} n_3(T) = \frac{d}{dVdt}\left(W_{12\rightarrow3} - W_{3\rightarrow12}\right).
\end{eqnarray}
The backward (inverse) direction is also valid. Note that $n(T)$, Eq. (\ref{eq:noT}), and $n(p,T)$, Eq. (\ref{eq:noTp}), are related to each other via $n(T)=N(T)/V=g/(2\pi^2)\sum_i\rho(m_i)\int p^2 dp\, n_i(p,T)$. Therefore, $n(p,T)$ is a Lorentz scalar, whereas $n(T)$ not. The thermal decay and production rate $dW/dVdt$ have been discussed \cite{newRafls}. 
In Boltzmann limit and assuming that the repulsive interaction does not contribute meaningfully to the overall relaxation time, the decay time in rest frame is given in textbooks 
\begin{eqnarray}\label{eq:finalTau}
\tau &=& \frac{8\pi m_3^2 g_3 I}{p\sum_{spin} |\langle\vec{p},-\vec{p}|{\cal M}|m_3\rangle|^2} \left\langle\frac{\varepsilon_3}{m_3}\right\rangle,
\end{eqnarray}
where $p$ is the pressure. $I$ is a step functions for particle distinguishability; $I=2$ for indistinguishable and $I=1$ for distinguishable particles. ${\cal M}$ is the hadronic reaction matrix. 

In left panel of Fig. \ref{2fig2a}, the relaxation time given in fm-units, Eq. (\ref{eq:finalTau}), is given as function of $T$ in GeV. We note that increasing the temperature $T$ leads to reducing the relaxation time $\tau$. It might mean that the decay processes get faster when $T$ increases. Near $T_c$, the effect of excluded-volume approach is considerable. 

As an application of these results, we mention the cosmological {\it viscous} models \cite{Tawfik:2011sh,Tawfik:2011mw,Tawfik:2010pm,Tawfik:2010ht,Tawfik:2010bm,Tawfik:2009mk}, which require a complete set of thermo- \cite{Karsch:2003vd,Karsch:2003zq,Redlich:2004gp,Tawfik:2004sw,Tawfik:2005qn,Tawfik:2010uh,Tawfik:2008ii} and hydro-dynamic equations of state in order to solve the evolution equation in early Universe and study the nucleosynthesis.

\subsection{Statistical-thermal models in rapidity space}
\label{sec:rapd}

About three decades ago, it was found that the ratios of particles to antiparticles depend on the longitudinal momentum in $pp$ and $pA$ collisions \cite{greiner84}. This has been confirmed in many collisions \cite{alice2013rpidity}. At least since $2006$, we have models for the rapidity representation of the statistical-thermal models \cite{sm06,rapidTM07a,rapidTM07b,rapidTM07c}. The analysis of rapidity distributions of identified particles has been suggested by Stiles and Murray \cite{sm06}.  The proposal to analyse the freeze-out parameters on dependence of rapidity was motivated by RHIC data at $200~$GeV \cite{rhicRpdt,brahms1,brahms2}. The BRAHMS data \cite{rhicRpdt,brahms1,brahms2} shows that the antiparticle to particle ratios have a maximum at mid-rapidity and slowly decrease as increasing rapidity. It has been suggested that the particle ratios at RHIC energies at large rapidity are consistent with those measured at the SPS energies \cite{rohlch07}. The earlier are related to mid-rapidity for approximately boost-invariant systems, while the latter are related to $4\, \pi$ studies.  

The ultimate goal is studying the dependence of thermal parameters on the location, in particular on the spatial rapidity. Based on a single freeze-out temperature, approaches of the rapidity dependence of particle ratios have been suggested \cite{rapidTM07a,rapidTM07b}, section \ref{sec:sftm1}. Based on rapidity dependence of the thermal parameters,  a systematic behavior towards an increase in chemical potential with increasing rapidity has been observed \cite{cleymRpd06}, section \ref{sec:rptp1}. Implementing Regge phenomenological approach \cite{aliceRapA,alice2010}, a parametrization relating the baryon ratios with the rapidity was suggested  \cite{alice2013rpidity}, section \ref{sec:rpbr1}.

\subsubsection{A single freeze-out temperature model}
\label{sec:sftm1}

\begin{figure}[tb]
\centerline{\includegraphics[width=8.cm]{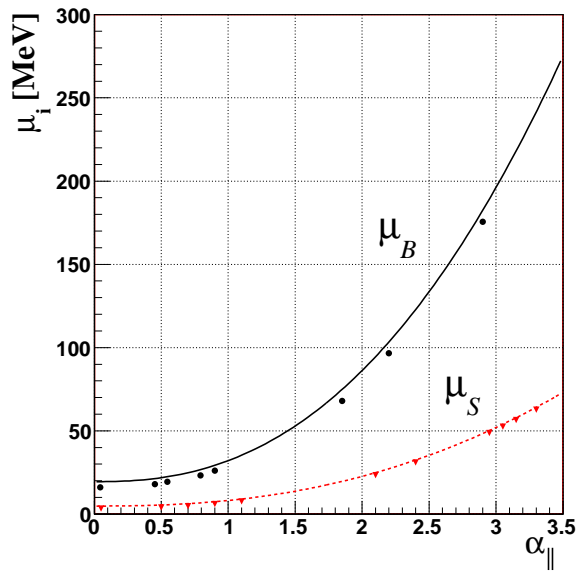}}
\caption{Baryon and strange chemical potentials are plotted as functions of $y_{||}$. Parameters of Eq. (\ref{eq:muOFy2}) are obtained from the fit to the BRAHMS data \cite{brahms1,brahms2}. The points represent a naive calculation based of Eqs. (\ref{eq:pap2})-(\ref{eq:pippim2}). The graph taken from Ref. \cite{rapidTM07b}.} 
  \label{Fig:mueps1}   
\end{figure}

A single-freeze-out model with thermal and geometric parameters dependent on the position within the fireball was proposed \cite{rapidTM07a,rapidTM07b}. This was used to describe the rapidity and transverse-momentum spectra of pions, kaons, protons and antiprotons measured at RHIC energy \cite{brahms1,brahms2}. The extension to boost-non-invariant systems consists of two elements: the choice of the freeze-out hypersurface (and collective flow) and the incorporation of the dependence of thermal parameters on the location \cite{rapidTM07a,rapidTM07b}.
\begin{itemize}
\item The freeze-out hypersurface can be parametrized as
\bea
x^{\mu} &=& \tau \left(\cosh(y_{\perp})  \cosh(y_{||}), \sinh(y_{\perp}) \cos(\phi), \sinh(y_{\perp}) \sin(\phi), \cosh(y_{\perp}) \sinh(y_{||})\right), \hspace*{10mm}
\eea
where $y_{\perp}$ is related to the transverse radius and $y_{||}$ is the spacial rapidity. The parameter $\tau$ stands for the proper time at freeze-out. The  four momentum of the particle can be given in terms of the transverse mass $m_{\perp}$ and rapidity $y$
\bea
p^{\mu} &=& \left(m_{\perp} \cosh(y), p_{\perp} \cos(\psi), p_{\perp} \sin(\psi), m_{\perp} \sinh(y)\right).
\eea
Then,
\bea
d^3 \sum \cdot p &=& \tau^3\, d y_{||}\, d \phi\, \sinh(y_{\perp}) \, d y_{\perp}\, p \cdot u, 
\eea
where
\bea
p \cdot u &=& m_{\perp} \cosh(y_{\perp}) \cosh(y_{||}-y) - p_{\perp} \sinh(y_{\perp}) \cos(\phi-\psi), 
\eea
and $d^3 \sum$ is the volume element of the hypersurface. The Cooper-Frye formula for the momentum density of a given species \cite{rapidTM07a,rapidTM07b,cf74} was used
\bea
\frac{d^2\, N}{2 \pi\, p_T\, d p_T\, d y} &=& \tau^3\, \int_{-\infty}^{\infty} d\, y_{||} \int_0^{y^{max}_{\perp}(y_{||})} d y_{\perp} \int_0^{2 \pi} d \phi \, p\cdot u\, f\left(\frac{p\cdot u -  \mu(y_{||})}{T}\right), \hspace*{.7cm}
\eea
where $f$ is Bose-Einstein or Fermi-Dirac distribution function. The chemical potential counts for all charges
\bea
\mu(y_{||}) &=& B\, \mu_b(y_{||}) + S\, \mu_s(y_{||}) + I_3\, \mu_{I_3}(y_{||}) + \cdots, \label{eq:muOFy1}
\eea
where $B$, $S$ and $I_3$ are baryon and strangeness quantum numbers and the third component of  isospin, respectively. Then, the rapidity dependence of the chemical potential is obvious. The thermal dependence can be deduced from the various universal conditions that will be elaborated in section \ref{sec:fo1}.
\item The functional form of the chemical potentials given in Eq. (\ref{eq:muOFy1}), can be parametrized as follows \cite{rapidTM07a,rapidTM07b}.
\bea
\mu_i(y_{||}) &=& \mu_i(0) \left(1+A_i\, y_{||}^{2.4}\right), \label{eq:muOFy2}
\eea
where $A_i$ is a fitting parameter for the given charge and $\mu_i(0)$ is the $i$-th charge chemical potential at mid-rapidity (the most central collisions).
\end{itemize}

The baryon and strange chemical potentials  as functions of $y_{||}$ are plotted in Fig. \ref{Fig:mueps1}. The parameters of Eq. (\ref{eq:muOFy2}) are obtained from the fit to the BRAHMS data \cite{brahms1,brahms2} and given in Ref. \cite{rapidTM07b}. 
  
We may replace Bose-Einstein or Fermi-Dirac distribution functions by Boltzmann one. According, $y_{||}\simeq y$ and the ratios of particle to antiparticle can be directly related to the chemical potential, for instance,
\bea
\frac{p}{\bar{p}} \simeq \exp\left(2 \frac{\mu_b}{T}\right), \label{eq:pap1}\\
\frac{K^+}{K^-} \simeq \exp\left(2 \frac{\mu_s}{T}\right), \label{eq:kpkm1}\\
\frac{\pi^+}{\pi^-} \simeq \exp\left(2 \frac{\mu_{I_3}}{T}\right), \label{eq:pippim1}
\eea
or
\bea
\mu_b(y) \simeq \frac{1}{2}\, T\, \log\left(\frac{p}{\bar{p}}\right),  \label{eq:pap2} \\
\mu_s(y) \simeq \frac{1}{2}\, T\, \log\left(\frac{K^+}{K^-}\right), \label{eq:kpkm2} \\
\mu_{I_3}(y) \simeq \frac{1}{2}\, T\, \log\left(\frac{\pi^+}{\pi^-}\right). \label{eq:pippim2}
\eea
The symbols in Fig. \ref{Fig:mueps1} represent a naive calculation based of Eqs. (\ref{eq:pap2})-(\ref{eq:pippim2}). 

\subsubsection{Rapidity dependence of thermal parameters}
\label{sec:rptp1}

Assuming that the rapidity axis is populated with fireballs following a  Gaussian distribution function given by $\rho(y_{FB})$, where $y_{FB}$ is the rapidity of the fireball \cite{rapidTM07c}
\bea
\rho(y_{FB}) &=& \frac{\exp\left(-\frac{y_{FB}^2}{2\, \sigma^2}\right)}{\sqrt{2} \, \pi \, \sigma},
\eea
and the width of the distribution $\sigma$ which can be estimated from the experimental data \cite{rhicRpdt}. This distribution of fireballs can be used to calculate the momentum distribution of $i$-th particle and the distribution of particles emitted from a single fireball $E_i \, d^3\, N_1^i/d^3\, p$. Along the rapidity axis  
\bea
E_i \frac{d^3\, N_i}{d^3\, p} &=& \int_{-\infty}^{\infty} \rho(y_{FB}) \, E_i \frac{d^3\, N_1^i}{d^3\, p} \, \left(y-y_{FB}\right) \, d\, y_{FB}.
\eea 
The rapidity distribution of particles with masses $m_0$ emitted from a single fireball is given as 
\bea
\frac{d\, N_1^i}{d\, y} &=& 2\, \pi\, g_i \, \exp\left(\frac{\mu_i}{T}\right)\, T\, \left[m_0^2+\frac{2\, m_0\, T}{\cosh(y)} + \frac{2\, T^2}{\cosh^2(y)}\right] \, \exp\left(-\frac{m_0\, \cosh(y)}{T}\right). 
\eea
The freeze-out parameters, $T$ and $\mu$, vary with changing the rapidity. The idea of Ref. \cite{rapidTM07c} is that $T$ and $\mu$ are always related to each other. The universal conditions describing the freeze-out curve have been suggested by many authors \cite{prcl22,prcl24,prcl25,Tawfik:2004ss,Tawfik:2005qn,Tawfik:2012si,Tawfik:2013dba}. In order words, changing rapidity changes $T$, which is turn change $\mu$ and vice versa. The relation between $\mu$ and the fireball rapidity can be given as
\bea 
\mu &=& \left\{\begin{array}{ll}
0.0245 + 0.011\, y_{FB}^2 & \hspace*{1cm} \text{RHIC}\\
& \\
0.237 + 0.011\, y_{FB}^2 & \hspace*{1cm}  \text{SPS}\end{array}\right..
\eea

\subsubsection{Rapidity dependence of baryon ratios}
\label{sec:rpbr1}

\begin{figure}[tb]
\centerline{\includegraphics[width=8.cm]{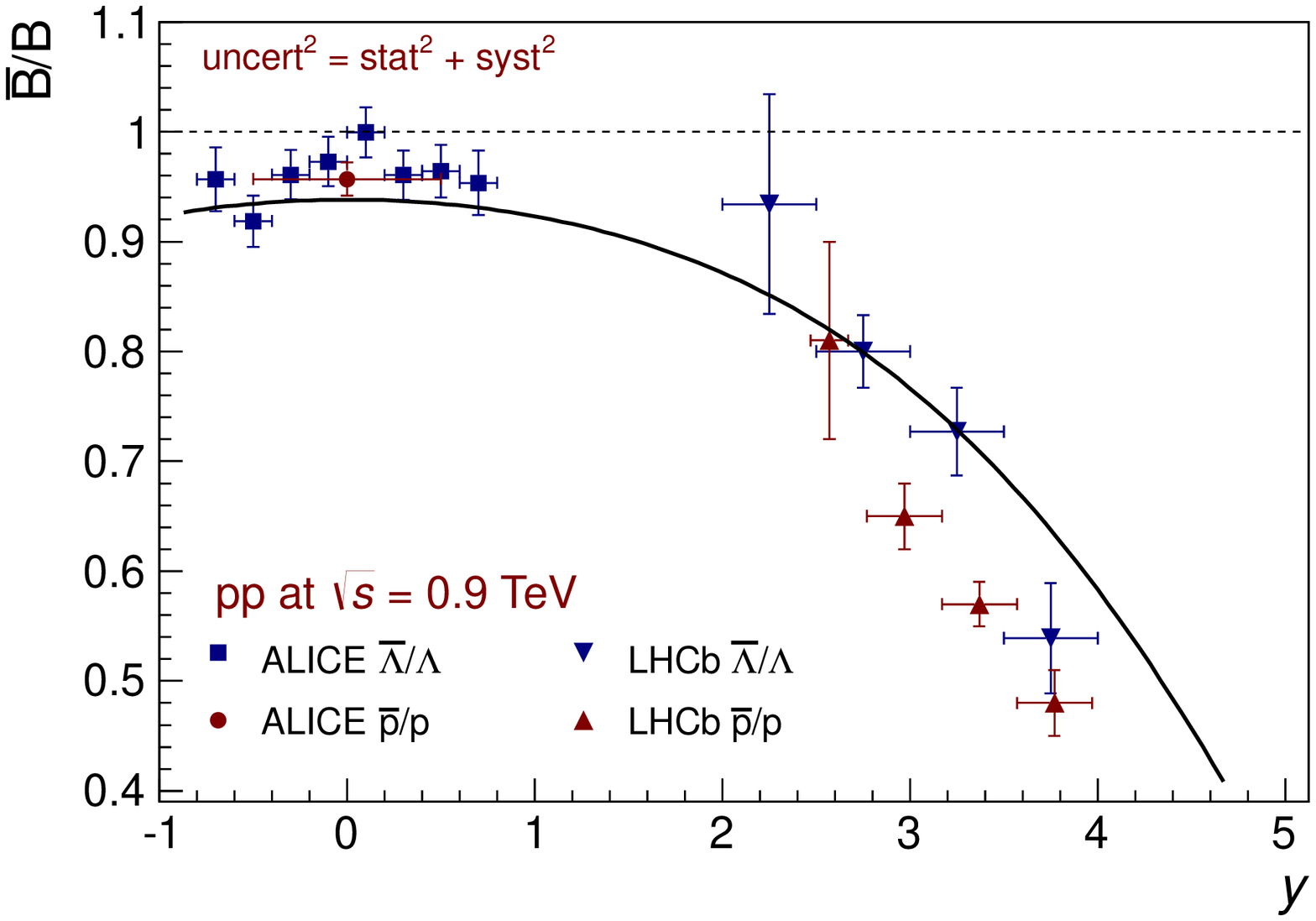}} 
\centerline{\includegraphics[width=8.cm]{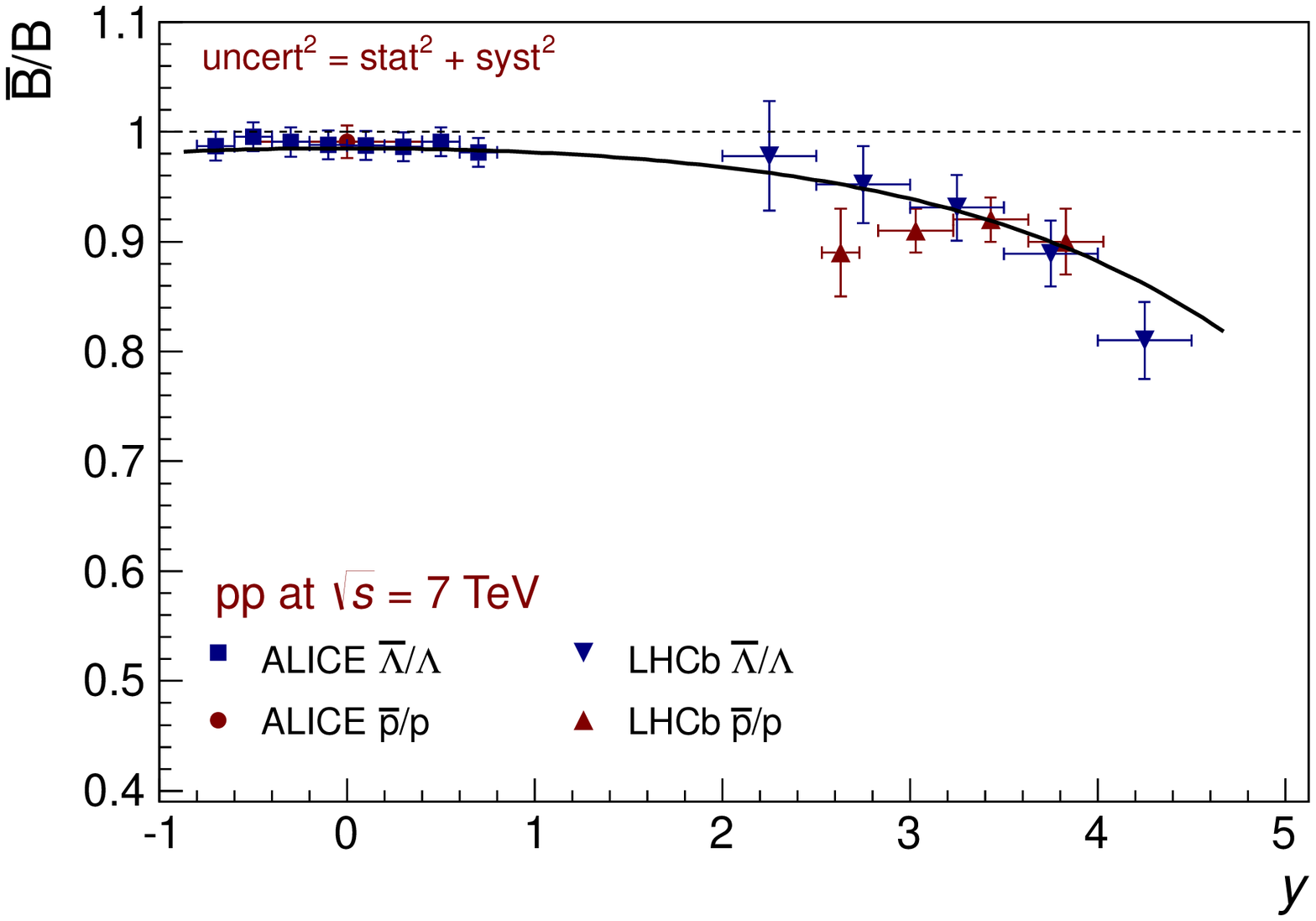}}
\caption{Ratios of $\bar{p}/p$ and $\bar{\Lambda}/\Lambda$ as functions of rapidity $y$ at $\sqrt{s}=0.9~$TeV (top panel) and $7~$TeV (bottom panel). The graphs taken from Ref. \cite{alice2013rpidity} combining ALICE and LHCb results.} 
  \label{Fig:ratiosalicerpdt1}   
\end{figure}

An approximation for  $y_{beam}-$ and $y-$dependencies of particle ratios can be derived in Regge model \cite{regge77}. Accordingly, the baryon-pair production at very high energy is governed by Pomeron exchange. The asymmetry between baryons and anti-baryons can be expressed by the string-junction transport and by an exchange with negative $C$-parity, e.g. $\omega$ exchange. A parametrization for particle ratio was suggested  \cite{aliceRapA,alice2010,alice2013rpidity} 
\bea
R(y_{beam}, y) &=& \frac{1+C_1\, \exp(\alpha_j-\alpha_p)\, y_{beam} \, \cosh(\alpha_j-\alpha_p) \, y}{1+C_2\, \exp(\alpha_j-\alpha_p)\, y_{beam} \, \cosh(\alpha_j-\alpha_p) \, y}, \label{eq:alicePrm}
\eea
where the Pomeron intercept $\alpha_p=1.2$ \cite{aliceP32a,aliceP32b} and the string-junction intercept $\alpha_j=0.5$ \cite{aliceP2}. It is assumed to equal the intercept of the secondary Reggions.

If $C_1=0$, then the given parametrization, Eq. (\ref{eq:alicePrm}), counts only the contribution of string-junction and/or for the case when in anti-proton spectrum the secondary Reggeons with positive C-parity (e.g. $f$ exchange) should have the same contributions as the secondary Reggeons with negative C-parity \cite{alice2013rpidity} have. The parameters $C_2=−C_1=3.9$ give the curves in Fig. \ref{Fig:ratiosalicerpdt1}. A Reggeon with negative C-parity and $\alpha_j=0.5$ is sufficient to describe the experimental data, especially ALICE data. Any significant contribution to the ratio of anti-baryon to baryon at mid-rapidity is likely due to an exchange, which is not suppressed with increasing rapidity interval would be disfavoured \cite{alice2013rpidity}.

\subsubsection{Proton ratios versus Kaon ratios}
 \label{sec:kkpp}
 
 \begin{figure}[tbh]
\includegraphics[width=10.cm]{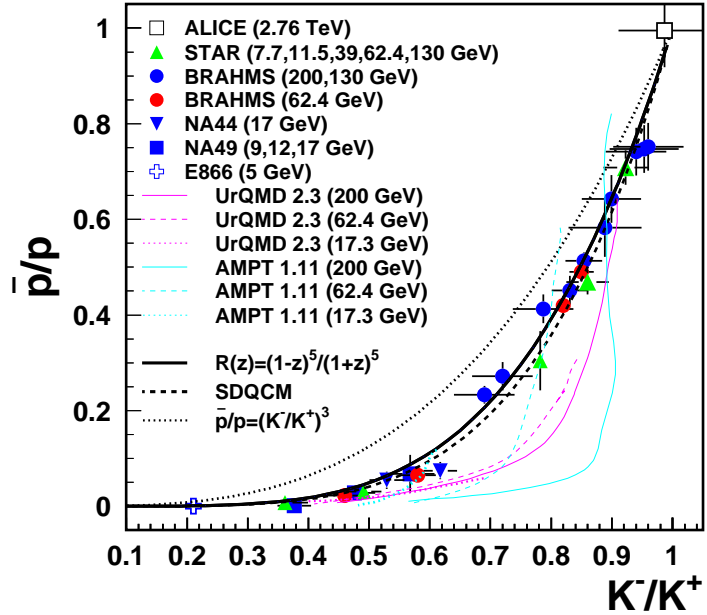}
\caption{The $\bar{p}/p$ ratio is given in dependence on $K^{+}/K^{-}$. Both are measured in central nucleus-nucleus collisions. The experimental results are compared with lattice QCD and even-generator calculations. The graph  taken from Ref. \cite{pap_kkFig}. 
\label{Fig:KPratios1}
}   
\end{figure}

The ratios of antiprotons-to-protons and that of $K^{-}$-to-$K^{+}$ are very significant observables measuring the hadron-antihadron asymmetry in central $AA$-collisions \cite{pap_kkFig}. Obviously, the proton ratios in the final state carry essential information about the production of baryons and antibaryons, while the Kaon ratio $K^{-}/K^{+}$ likely cancels the effect of strangeness production and therefore reflects the asymmetry between charged mesons and their antiparticles \cite{pap_kkFig}. As given in Fig. \ref{Fig:KPratios1}, the experimental measurements seem to follow a kind of a universal curve \cite{arsene:2010}. The comparison with the event generators, A Multiphase Transport Model (AMPT) \cite{LinAMPT05} and the Ultra-Relativistic Quantum Molecular Dynamics (UrQMD) \cite{urqmd98review} allows the conclusion that the hadronization with String fragmentation would not explain such a universal phenomenon \cite{arsene:2010}.  The symbols refer to the RHIC results (BRAHMS and STAR measurements) \cite{brahms1,bearden:2001,arsene:2010,Abelev:2008ab,kuma11bes}. The lower energy results are deduced from Refs. \cite{ahle:1999A,ahle:1999B,na49:2002A,na49:2002B,IGbearden:2002}, while the LHC measurements at $\sqrt{s}= 2.76$ TeV from Ref. \cite{RPregh11PbPbLhc}. AMPT $1.11$ and UrQMD $2.3$ were presented in Ref. \cite{arsene:2010}.

In the quark combination model \cite{ccm1,ccm2}, the fraction of two ratios can be parametrized with the global production of baryons, antibaryons and mesons \cite{pap_kkFig}
\bea
R(z) &=& \frac{\bar{B}(x,z)}{B(x,z)}=\frac{\mathcal{B}(-z)}{\mathcal{B}(z)}=1-\frac{z}{3\,\mathcal{B}(z)}, \label{eq:Rz}
\eea
where $\mathcal{B}(z)$ is defined in Ref. \cite{pap_kkFig}. $x$ gives the total number of quarks and antiquarks. This should characterize the bulk property of the system related to size or energy of the system. $z$ stands for the asymmetry between quarks and antiquarks, $|z|\le 1$. Apparently, this plays the role of measuring the baryon number density,
\bea
x &=& N_q + N_{\bar{q}}, \\
z &=& \frac{N_q - N_{\bar{q}}}{N_q + N_{\bar{q}}}.
\eea
In Fig. \ref{Fig:KPratios1}, the baryons, the antibaryons and the mesons are studied as function of the numbers of constituent quarks and antiquarks before hadronization. They should follow a universal correlation, for instance the relation between $\bar{p}/p$ and $K^{-}/K^{+}$ ratios, Eq. (\ref{eq:Rz}). It is obvious that this dependence can be well explained in the framework of quark combination model \cite{ccm1,ccm2}. 
\bea
z &=& \frac{1-\frac{K^+}{K^-}}{1+\left(1+0.14\frac{1+R_{VP}}{R_{VP}}\right)\frac{K^+}{K^-}},
\eea
where $R_{VP}=0.45$.  As pointed out in Ref. \cite{pap_kkFig}, understanding the low $p_T$ hadron production in the relativistic heavy-ion collisions is an essential framework to describe the particle production.


\section{Statistical-thermal models in high-energy physics}
\label{sec:hic1}

As outlined in section \ref{sec:histr}, the implication of statistical-thermal model to high-energy collisions dates back to about six decades \cite{hkoppe2}. In 1951, Pomeranchuk \cite{aa:pom} came up with the conjecture that a finite hadron size would imply a critical density above which the hadronic matter cannot be in the compound state, known as hadrons. In 1964, Hagedorn introduced the mass spectrum to describe the abundant formation of resonances with increasing masses and rotational degrees of freedom \cite{hgdrnA,hgdrnB} which relate the number of hadronic resonances to their masses as an exponential. Accordingly,  Hagedorn formulated the concept of limiting temperature based on the statistical bootstrap model. 

The ultimate goal of the physics program of high-energy collisions is the study of the properties of strongly interacting matter under extreme conditions of temperature and/or compression \cite{Tawfik:2010pt,Tawfik:2010kz,Tawfik:2010uh,Tawfik:2008ii,Tawfik:2006zh,Tawfik:2006yq,Tawfik:2006dk,Adamovich:2001hi}. The particle multiplicities and their fluctuations and correlations are experimental tools to analyse  the nature, composition, and size of the medium from which they are originating. Of particular interest is the extent to which the measured particle yields are showing equilibration. Based on analysing the particle abundances \cite{rfRED6,rfRED9} or momentum spectra \cite{rfRED9,rfRED20,rfRED46,rfRED47}. The degree of equilibrium of the produced particles can be estimated. The particle abundances can help to establish the chemical composition of the system. The momentum spectra can give additional information on the dynamical evolution and the collective flow. Based on Gibbs approach, the equilibrium behavior of thermodynamical observables can be evaluated as an average over statistical ensembles rather than as a time average for a particular state. The latter would be essential when dealing with out-of-equilibrium processes \cite{Tawfik:2010pt,Tawfik:2010kz,Tawfik:2010uh}.

In this review, we discuss the implications of statistical-thermal model on deducing a phenomenological description for the particles production and their fluctuations and correlations. The statistical-thermal approach provides a very satisfactory description for the experimental results covering a wide range of center-of-mass energies. 
The conservation laws shall be introduced in section \ref{sec:conserv}.
A detailed comparison with the heavy-ion collisions shall be given in section \ref{sec:compHIC}. Section \ref{sec:fo1} will be devoted to the chemical freeze-out. We list out argumentation for several unified descriptions for the chemical freeze-out of the produced particles.  In section \ref{sec:lqcd}, we compare between lattice QCD thermodynamics and the statistical-thermal model. The accuracy of lattice QCD simulation in thermal and dense medium makes it possible to estimate essential quantities, like higher order moments and  chemical freeze-out. 

\subsection{Conservation laws}
\label{sec:conserv}

\begin{figure}[tb]
\centerline{\includegraphics[width=10.cm]{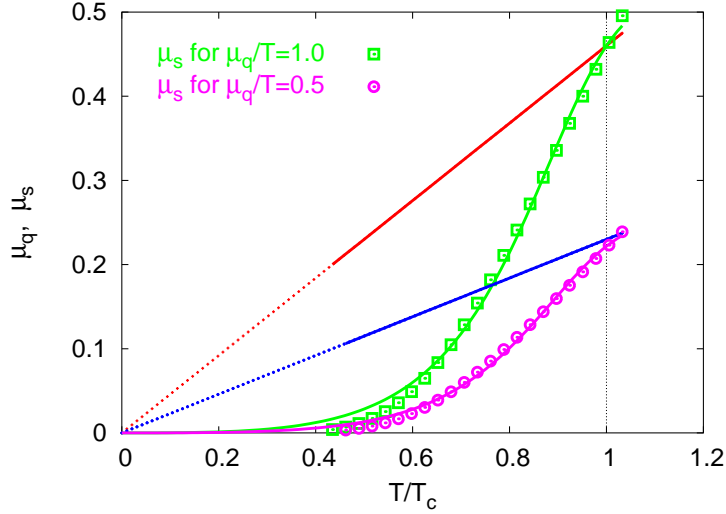}}
\caption{The strange quark chemical potential $\mu_s$ vs. $T/T_c$ for $\mu_q/T=1$ and $\mu_q/T=0.5$ (straight lines). The results are fitted according to Eq.~(\ref{eq:muFit}). At $T=0$, we find that $\mu_q=\mu_s=0$. As $T\rightarrow T_c$, the strangeness chemical potential approaches the baryon chemical potential, $\mu_s=\mu_q$. The units used here are $\sqrt{\sigma}\sim420\;$MeV. The graph taken from Ref. \cite{Tawfik:2004sw}.} 
  \label{Fig:musmub1}   
\end{figure}

The {\it hadron}-based chemical potentials $\mu_B$ and $\mu_S$ are related to the {\it quark}-based ones, $\mu_q$ and $\mu_s$ 
\begin{eqnarray}
\mu_{B} &=&   3\,\mu_{q},\\ 
\mu_{S} &=&  \mu_{q} - \mu_{s}, \label{eq:mu1}
\end{eqnarray}
Assuming that the isospin and charge chemical potentials are vanishing, we use the following combination for the hadron resonances 
\begin{eqnarray}
\mu     &=& 3b\mu_{q} + s\mu_{s}, \label{eq:mu2}
\end{eqnarray}
where $b$ and $s$ are the baryon and strange quantum numbers, respectively. 

The initial conditions in heavy-ion collisions apparently include zero net strangeness. This is expected to remain the case during the whole interaction unless an asymmetry in the production of strange particles happens during the hadronization. As we are interested in the hadron thermodynamics and location of QCD phase transition, we suppose that the net strangeness is entirely vanishing. The average strange particle number reads    
\begin{eqnarray}
<n_s> &=& \frac{1}{N}\sum_i^N\lambda^{(i)}_s\frac{\partial \ln {\cal
  Z}^{(i)}(V,T,\mu)}{\partial \lambda^{(i)}_s}.
  \label{eq:ns1} 
\end{eqnarray}
$\ln {\cal Z}$ is given in Eq.~(\ref{eq:za3}) and $\mu_S$ is given in Eq.~(\ref{eq:mu1}). $\lambda_s=\exp(\mu_S/T)$ is the fugacity factor of $s$ quark. The procedure used to calculate the {\it quark}-based $\mu_s$ is the following: For given $T$ and $\mu_q$ (or $\mu_B$), we iteratively increase $\mu_s$ and in each iteration, we calculate the difference $<n_s>-<n_{\bar{s}}>$, Eq.~(\ref{eq:ns1}). The value of $\mu_s$ which disposes zero net-strangeness is the one we read out and shall use in calculating the thermodynamic quantities. As in Eq.~(\ref{eq:mu1}), the relation between $\mu_s$ and $\mu_S$ is given by taking into consideration the baryonic property of $s$ quark. The resulting $\mu_s$ for different $\mu_q$ (or $\mu_B$) and $T$ are depicted in Fig.~\ref{Fig:musmub1}. From this {\it numerical} method, we can fit $\mu_s$ as a function of $T$ and $\mu_B$, 
\begin{eqnarray}
\mu_s & \approx &
  \frac{0.138\;\vartheta\; \theta^3}{1-2.4\;\theta^2
  +2.7\;\theta^3},  \label{eq:muFit} 
\end{eqnarray}
where $\vartheta\equiv\mu_q/T$ and $\theta\equiv T/T_c$.  {\it Re-scaled}  resonance masses can be implemented \cite{Karsch:2003zq, Karsch:2003vd}, section \ref{sec:krt1}. 
 
In order to guarantee vanishing net strange particle numbers, it is not enough to simply set $\mu_s=0$ and consequently, \hbox{$\mu_S=\mu_q=\mu_B/3$} in Eq.~(\ref{eq:ns1}). However, there are publications in which the authors have assigned $\mu_s$ to zero in hadron matter and afterwards applied the Gibbs condition for the {\it first order} phase transition to QGP. Aside the baryons, the strange mesons with different contents of $s$ quarks play determining roles at different temperatures and therefore, affect the final results,
Eq.~(\ref{eq:mu1}). Setting $\mu_s=0$, leads to violating the strange quantum numbers. Nevertheless, we will show here calculations in which we set $\mu_s=0$. We do this in order to extensively compare with lattice results~\cite{Tawfik:2005qh}. 

For completeness, we recall the situation in the plasma regime. For conserving strangeness at $T>T_c$, we have to suppose that $\mu_S=0$. $\mu_S$ consists of one baryonic part $\mu_B/3$ and another part coming from the strangeness quantum number $-\mu_s$. From Eq.~(\ref{eq:mu1}), we then get 
\begin{eqnarray}
\mu_s &=& \mu_q=\mu_B/3.
\end{eqnarray}
This result is numerically confirmed in Fig.~\ref{Fig:musmub1}. For $\mu_q=0$ (or $\mu_B=0$), we find that $\mu_s=0$ for all temperatures. $\mu_s$ increases with increasing both $\mu_q$ and $T$. At $T_c$, we find that $\mu_q\approx\mu_s$. Therefore, we can suggest to set $\mu_q=\mu_s$  for all temperatures above $T_c$. 

We  summarize that $\mu_S$ in the hadronic matter has to be calculated in dependence on $\mu_B$ and $T$ under the assumption that the net strangeness is vanishing. In the QGP phase, one might fulfil this assumption by setting $\mu_s = \mu_B/3$.

\subsubsection{Strangeness chemical potential in lattice QCD}
\label{sec:2b}

In the Euclidian path integral formulation, the partition function of lattice QCD at finite $T$ and $\mu$ reads 
\begin{eqnarray}
{\cal Z}(T,\mu) &=& \Tr \; e^{-(H-\mu N)/T} = \int {\cal D}\psi\;{\cal D}\bar{\psi}\;{\cal D}A\;e^{{\cal
 S}_f(V,T,\mu)+{\cal S}_g(V,T)},  \hspace*{10mm}
\end{eqnarray}
where $(\psi,\bar{\psi})$ and $A$ are the fermion and gauge fields, respectively. The chemical potential $\mu$ is given in Eq.~(\ref{eq:mu2}). By Legendre transformation of  Hamiltonian $H$, we get the Euclidian action \hbox{${\cal S}=\int_0^{1/T}dt \int_V d^3x {\cal L}$}. The fermionic action is given as 
\begin{eqnarray}
{\cal S}_f &=& a^3\sum_x\left[m a \bar{\psi}_x\psi_x + \frac{1}{2}
\sum_{k=1}^{4} \left(\bar{\psi}_x\gamma_{k}\psi_{x+\hat{k}} -
  \bar{\psi}_{x+\hat{k}} \gamma_{k}\psi_x \right)+
  \mu a \bar{\psi}_x\gamma_4\psi_x \right], \hspace*{10mm}
\end{eqnarray}
where $a$ is the lattice spacing. As given in Eq.~(\ref{eq:ns1}), the number density of  quarks with flavor number $x$ is obtained by the derivation with respect to $\mu_x$,  
\begin{eqnarray}
n_x &=& \frac{\partial}{\partial \mu_x} \ln {\cal Z}(T,\mu_x). \label{quarkn}
\end{eqnarray}
For checking the dependence of $\mu_s$ on $\mu_q$ and consequently on $\mu_B$, it is enough to approximate the fermionic part of lattice QCD Lagrangian for three quark flavors as    
\begin{eqnarray}
{\cal L} &\approx&
\mu_q\left(\sum_{x\in\{u,d\}}\bar{\psi}_x\gamma_4\psi_x  \right) + 
       \mu_s \bar{\psi}_s\gamma_4\psi_s  \approx  \mu_q n_u + \mu_q n_d + \mu_s n_s. \label{musLatt}
\end{eqnarray}
To taken into account the conservation of the baryon and strange quantum numbers, the summation in last expression has to run over $s$ quarks, too. By doing this, last term turns to be $(\mu_q-\mu_s)\,n_s$. Then we expect that the strangeness on lattice vanishes at $\mu_s=\mu_q$. But from Eq.~(\ref{musLatt}), which reflects the situation in current lattice simulations, we find that $n_s=0$ for $\mu_s=0$. 

As discussed in previous section, the strangeness in QGP is  conserved at $\mu_s=\mu_q=\mu_B/3$. In the hadronic regime, especially at large $\mu_B$, $\mu_s$ (or $\mu_S$) has to be calculated as a function of  $T$ and $\mu_B$, Fig.~\ref{Fig:musmub1}. In spite of these considerations, the  reliable lattice QCD simulations are still limited to \hbox{$\mu_B\approx  3 T_c$}. At this small value, there is practically no big difference between $\mu_s=0$ and \hbox{$\mu_s=f(T,\mu_B)$}.

\subsection{Comparing with heavy-ion collisions}
\label{sec:compHIC}

\subsubsection{Particle abundances and their ratios}
\label{sec:multp}

Understanding the dynamical properties of hot and dense hadronic matter is one
of the main motivations of heavy-ion experiments, which in turn offer unique
possibilities to study hadronic matter under extreme conditions \cite{pp1a,pp1b,s200k,pp3a,pp3b} and to compare with the lattice QCD simulations \cite{pp4a}. RHIC has shown that the bulk matter created in such collisions can be quantitatively described by hydrodynamic models \cite{pp3b}. The created hot and dense partonic matter, QGP or colored quarks and gluons, where  quarks and gluons can move freely over large volumes comparing to the typical size of a hadron, rapidly expands and cools down. Over this path, it likely undergoes phase transition(s) back to the hadronic matter. Different thermal models can very well reproduce the particle abundances, which are governed in chemical equilibrium by two parameters, the chemical freeze-out temperature $T_{ch}$ and the baryon chemical potential $\mu_b$, where the latter reflects the net baryon content of the system, directly, and the center-of-mass energy, indirectly, section \ref{sec:fo1}. A schematic description of the successive stages of a heavy-ion collision is given by the Bjorken model \cite{bjorken83}. \\

\paragraph{\bf Bjorken model \\}

\begin{figure}[tbh]
\begin{center}
\includegraphics[width=8.cm]{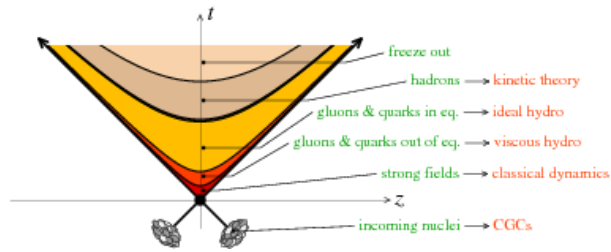}
\caption{Various stages of a heavy-ion collision are given in plane of the proper time $t$ and the longitudinal coordinate $z$. The stages (green) are related to the underlying physics (red). Graph taken from Ref. \cite{edmund}. } 
\label{Figjbrk}   
\end{center}
\end{figure}

The Bojorken model \cite{bjorken83} is illustrated in Fig. \ref{Figjbrk}. Early thermalization, vanishing baryon chemical potential in the fluid and one-dimensional expansion besides boost symmetry of the initial conditions, the initial independence on rapidity and the isotropic (homogenises) rapidity over the spacetime are the main assumptions of this model.  A schematic representation of the various stages of a heavy-ion collision is given as a function of time $t$ and the longitudinal coordinate $z$ (the collision axis) in Fig \ref{Figjbrk}. The {\it time} variable is related to the {\em proper time} $\tau\equiv\sqrt{t^2-z^2}$, which has a Lorentz-invariant meaning. This should remain constant along the hyperbolic curves separating various stages \cite{edmund}, Fig. \ref{Figjbrk}.

Accordingly, the boost-invariant four-velocity is defined as
\bea
u^{\mu} &=& \frac{\tilde{X}^{\mu}}{\tau} = \frac{t}{\tau}\, \left(1,0,0,\frac{z}{t}\right) =\frac{(t,0,0,z)}{\sqrt{t^2-z^2}}.
\eea
Due to Lorentz symmetry and the initial conditions for energy density, pressure and temperature, respectively
\bea
\epsilon &=& \epsilon(\tau,y) \rightarrow \epsilon(\tau), \\
P &=& P(\tau,y) \rightarrow P(\tau), \\
T &=& T(\tau,y) \rightarrow T(\tau),
\eea
then the equation of motion reads
\bea
\frac{d \epsilon}{d \tau} &=& -\frac{\epsilon + P_s}{\tau} \left(1- \frac{4}{3\, \tau\, T} \frac{\eta}{s} -  \frac{1}{\tau\, T} \frac{\xi}{s}\right),
\eea
where $s$ is the entropy and $\eta\, (\xi)$ being bulk (shear) viscosity coefficient. Reliable EoS of hadronic matter has been elaborated in Ref. \cite{Tawfik:2011sh}. 

\begin{figure}[tbh]
\begin{center}
\includegraphics[width=7.cm,height=5cm]{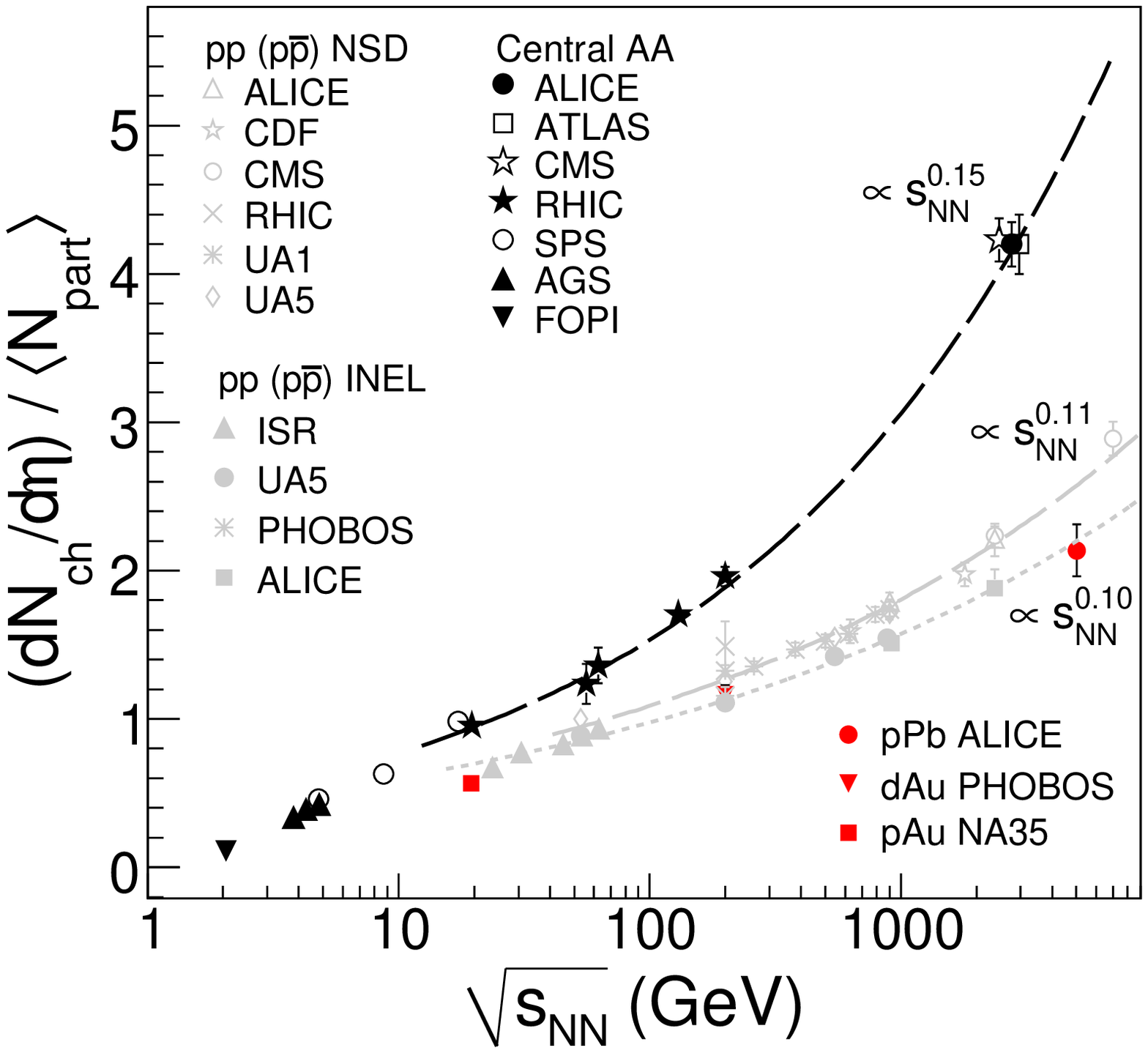}
\includegraphics[width=7.cm,height=5cm]{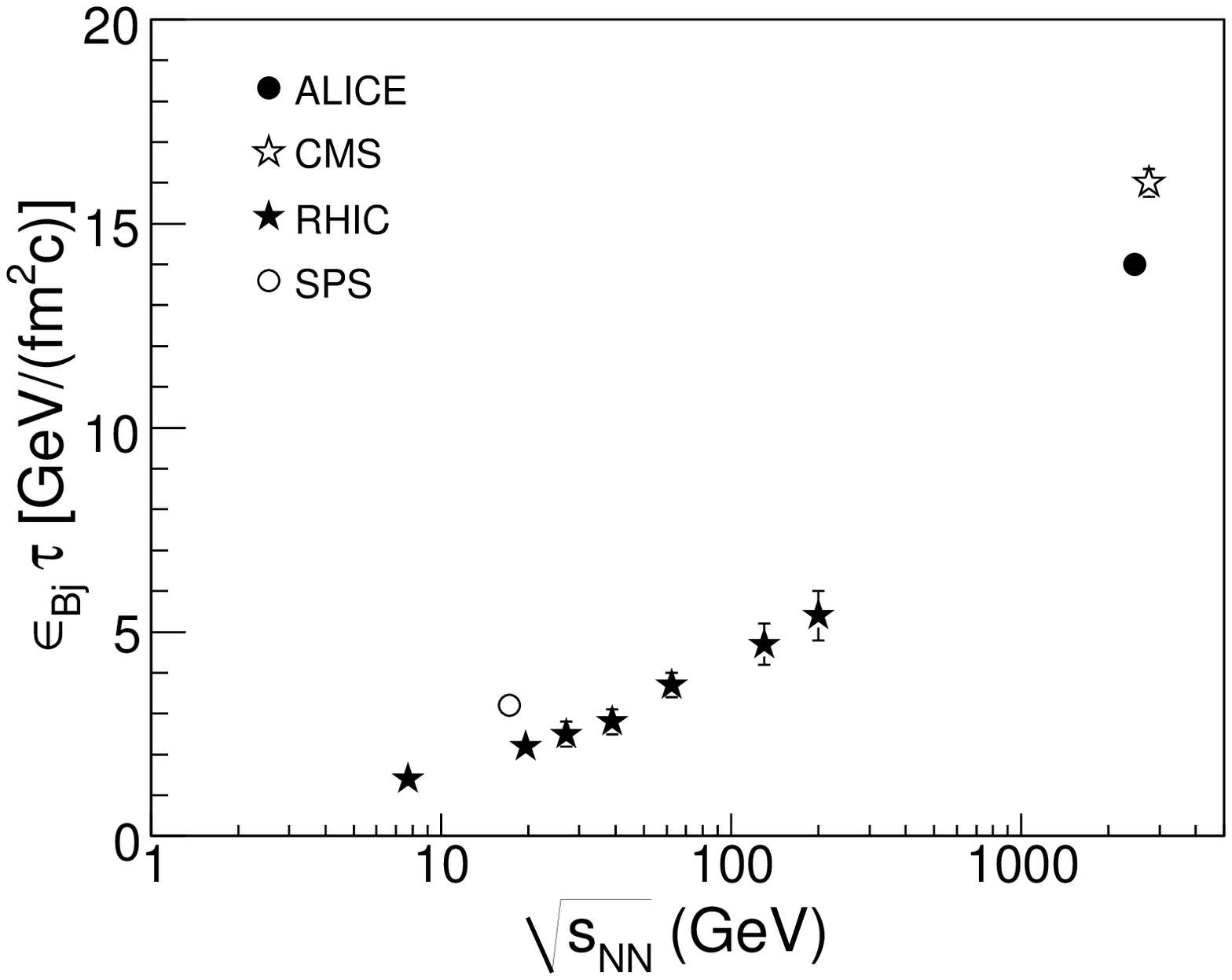}
\caption{Left panel: a comparison of $dN_{ch}/d\eta$ per participating nucleon at
mid-rapidity in central heavy-ion collisions ~\cite{Back:2003hx,Abreu:2002fx,Adler:2001yq,Bearden:2001xw,Bearden:2001qq,Adcox:2000sp,Back:2000gw,Back:2002wb,Aamodt:2010pb,ATLAS:2011ag,Chatrchyan:2011pb} to corresponding results from $p$+$p$($\bar{p}$)~\cite{ua1,ua5A,ua5B,Abelev:2008ab,Abe:1989td,Aamodt:2010pp,Khachatryan:2010xs,isr,Aamodt:2010ft,Alver:2010ck}  and $p(d)$+A collisions~\cite{ALICE:2012xs,Alber:1997sn,Back:2003hx}.
Right panel: the product of Bjorken energy density, $\epsilon_{Bj}$~\cite{bjorken83}, and  the formation time $\tau$ in central heavy-ion collisions at mid-rapidity is drawn in dependence on the center-of-mass energy. The quantities are given in physical units. Graphs taken from Ref. \cite{epsl0tau}. }
\end{center} 
\label{FigjbrkEpslTau}  
\end{figure}

In the simplest case, i.e. $\eta=\xi=0$ and $P=\epsilon/3$, the energy change per unit rapidity, $y$, is given as
\bea
\frac{d E}{d y} &=& \frac{d^3 V}{d y} \epsilon(\tau_f) = \pi\, R^2\, \tau_f\, \epsilon_0\, \tau_0 \left(\frac{\tau_0}{\tau_f}\right)^{\gamma}. 
\eea
Assuming that no expansion takes place at $\tau>\tau_t$, then
\bea \label{eq:epsl0}
\epsilon_0 &=& \frac{1}{\pi\, R^2\, \tau_0}\, \frac{d E}{d y} = \left.\frac{<m_t>}{\pi\, R^2} \frac{d\, E}{d y}\right|_{z=0} \frac{d\, N_{ch}}{d y}.
\eea
Eq. (\ref{eq:epsl0}) allows to estimate the initial energy density, which apparently is related to experimental quantities, like $d E/d y$ and $d N_{ch}/d y$. The dependence on other quantities is also possible \cite{trainer13,Bylinkin13}. A recent estimation for the quantity $\epsilon_0\, \tau$ at various center-of-mass energies is given in right panel of Fig. \ref{FigjbrkEpslTau}. The pseudo-rapidity $\eta$ is related $y$. At very high energy, $\eta=y$. An extensive comparison between the particle multiplicity $dN_{ch}/d\eta$ per participating nucleon at mid-rapidity in central heavy-ion collisions ~\cite{Back:2003hx,Abreu:2002fx,Adler:2001yq,Bearden:2001xw,Bearden:2001qq,Adcox:2000sp,Back:2000gw,Back:2002wb,Aamodt:2010pb,ATLAS:2011ag,Chatrchyan:2011pb} and the corresponding results from $p$+$p$($\bar{p}$)~\cite{ua1,ua5A,ua5B,Abelev:2008ab,Abe:1989td,Aamodt:2010pp,Khachatryan:2010xs,isr,Aamodt:2010ft,Alver:2010ck}  and $p(d)$+A collisions~\cite{ALICE:2012xs,Alber:1997sn,Back:2003hx} is presented in left panel of Fig. \ref{FigjbrkEpslTau}. \\

\paragraph{\bf Experimental particle abundances and their ratios\\}

The particle abundances at Alternating Gradient Synchrotron (AGS), Super Proton Synchrotron (SPS) and RHIC, and recently at LHC energies are consistent with equilibrium populations \cite{therm1,Tawfik:2012zp} (compare with left panel of Fig. \ref{FigjbrkEpslTau}). This makes it possible to extract both freeze-out parameters over a wide range of Nucleus-Nucleus center-of-mass energies $\sqrt{s_{NN}}$ from fits of measured particle ratios with the thermal models. The various types of  hadrons interactions in the final state can - in the best cases - partly taken into account. Details on how to extract the freeze-out parameters are elaborated in section \ref{sec:fo1}. Nevertheless, the formation of resonances {\it themselves} can only be materialized through strong interactions, since the resonances (fireballs) are composed of further resonances (fireballs), which in  turn consist of resonances (fireballs) and so on \cite{boots137,Tawfik:2012zp}. Taking into consideration all kinds of resonance interactions by means of the S-matrix, which describes the scattering processes in the thermodynamical system, reduces the resulting virial term, so that the partition function turns to be reduced to the non-relativistic limit, especially at narrow width and/or low temperature $T$ \cite{Tawfik:2004sw}.  All possible interactions modifying the relative abundances are found negligible in the hadronic phase \cite{Tawfik:2004sw,intr1}, especially at low chemical potentials or high energies  \cite{Karsch:2003vd,Karsch:2003zq,Redlich:2004gp,Tawfik:2004vv}. Details on all types of interactions in the hadronic phase are given in section \ref{sec:intr}.  This discussion serves as an introduction on how the experimental particle abundances are related to statistical-thermal models.

\begin{figure}[tbh]
\begin{center}
\includegraphics[width=10.cm]{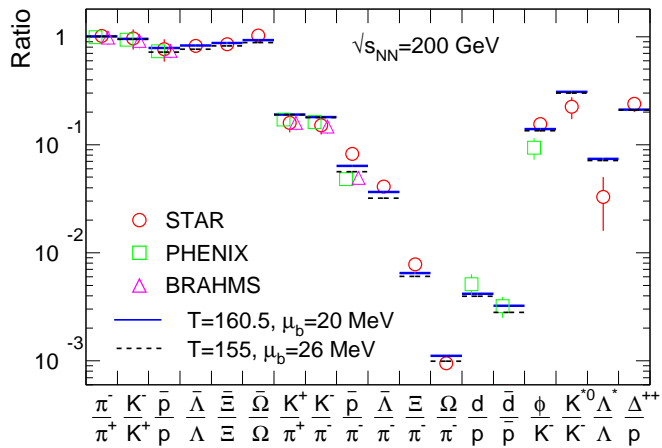}
\caption{The ratios of various hadron yields measured at $\sqrt{s_{NN}}$=200 GeV are compared with the results from the statistical-thermal models. Graph taken from Ref. \cite{dataCR}. } 
\label{Fig:Ratio200}   
\end{center}
\end{figure}

The statistical operator of HRG formulated in grand canonical ensemble in central heavy-ion collisions from top AGS up to RHIC \cite{dataCR}, Fig. \ref{Fig:Ratio200}, gives an good description for the particle yields. For example,  at $\sqrt{s_{NN}}=200~$GeV, the ratios of identified hadron yields are compared with the best fits using statistical-thermal models. With identical ratios, we mean particle-antiparticle ratios. The yields of $\pi^\pm$, $K^\pm$, $p$, and $\bar{p}$ are measured in PHENIX \cite{p200d}, STAR \cite{s200}, and BRAHMS \cite{b200} experiments. The comparison with statistical-thermal models using two sets of $T$ and $\mu$ parameters is quite good. The comparison for available yields of $\phi$ \cite{s200phi,p200phi}, $K(892)^{*}$ \cite{s200k}, $d$ and $\bar{d}$ \cite{p200d} is also illustrated. The measured hadron ratios with corresponding uncertainties are also included \cite{s200phi,p200phi,s200k}. In case of missing measurements, the ratios are calculated from the available - published - yields \cite{dataCR}. In additional, the ratios of strange hyperons \cite{s200h}, $\Delta^{++}/p$ \cite{s200d}, $\bar{p}/\pi^-$, $\bar{\Lambda}/\pi^-$,  $\Xi/\pi^-$, $\Omega/\pi^-$, and $\Lambda^*/\Lambda$\cite{s200s} are also illustrated.

\begin{figure}[tbh]
\begin{center}
\includegraphics[width=8.cm,angle=-90]{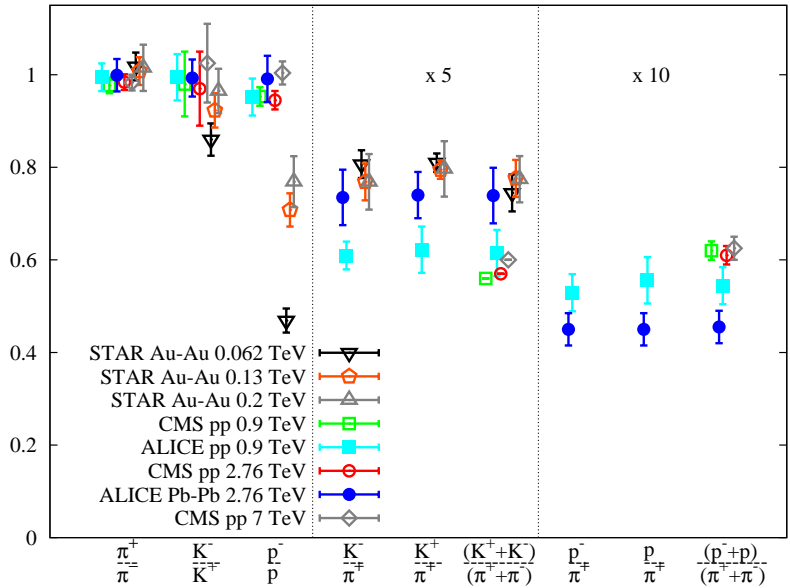}
\caption{The ratios of various hadron yields measured at RHIC and LHC energies in different collisions are compared with to each others. A detailed comparison between these ratios and the statistical-thermal fits are elaborated in Ref. \cite{Tawfik:2012zp}. Even their evolution with $\sqrt{s_{NN}}$ is also given. Graph taken from Ref. \cite{Tawfik:2012zp}. } 
\label{Fig:RatioALLs}   
\end{center}
\end{figure}

\begin{figure}[tbh]
\begin{center}
\includegraphics[width=3.cm,angle=-90]{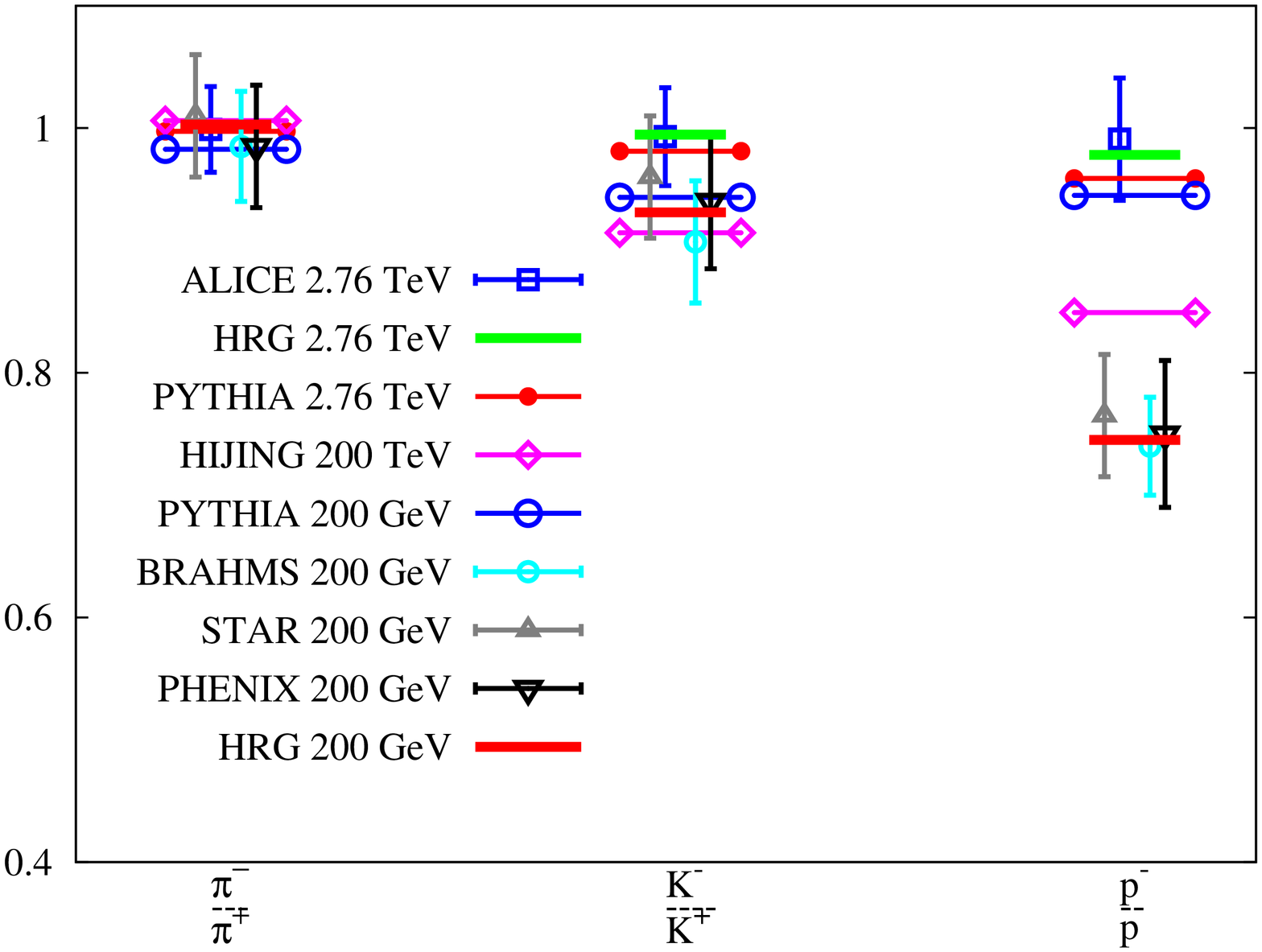}
\includegraphics[width=3.cm,angle=-90]{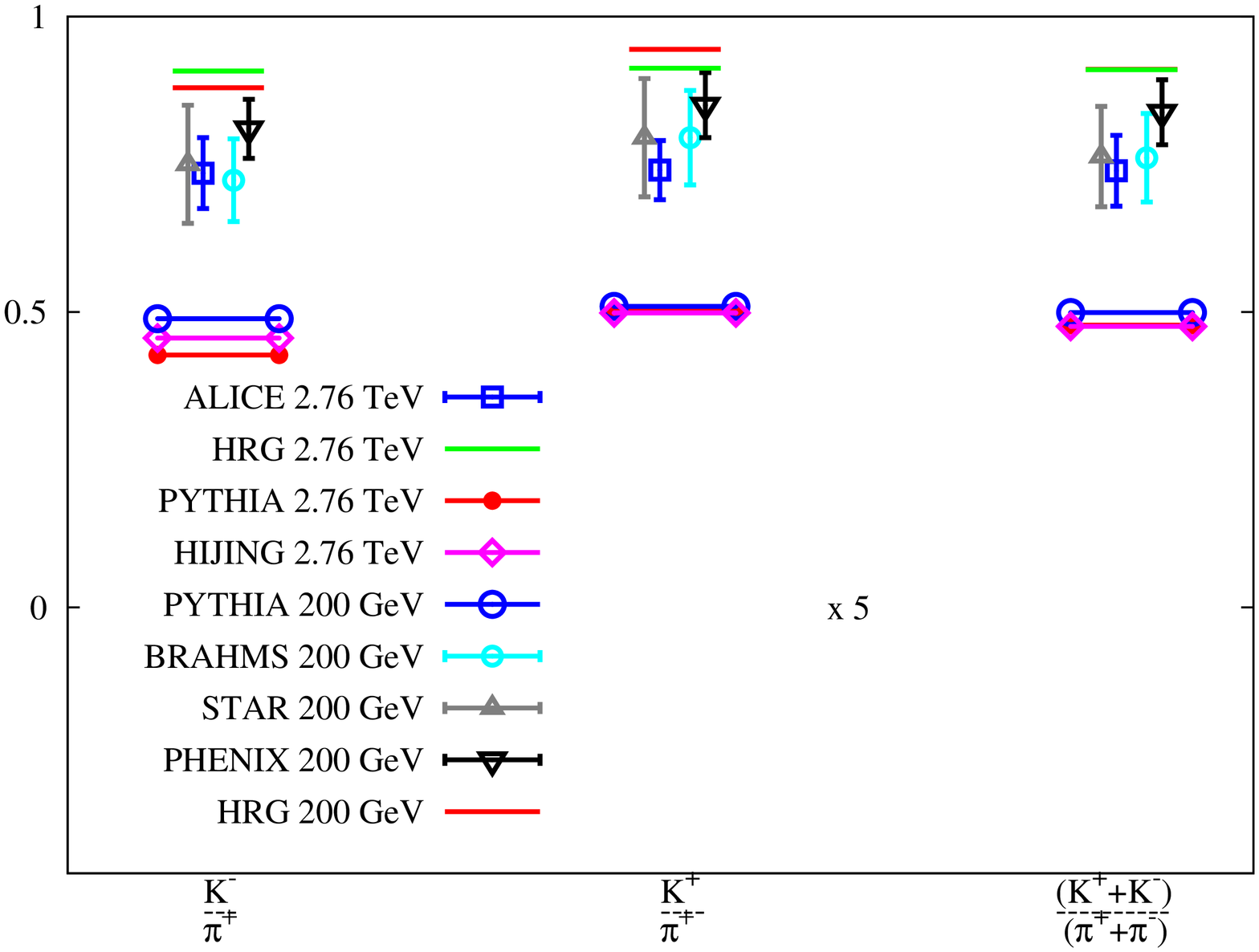}
\includegraphics[width=3.cm,angle=-90]{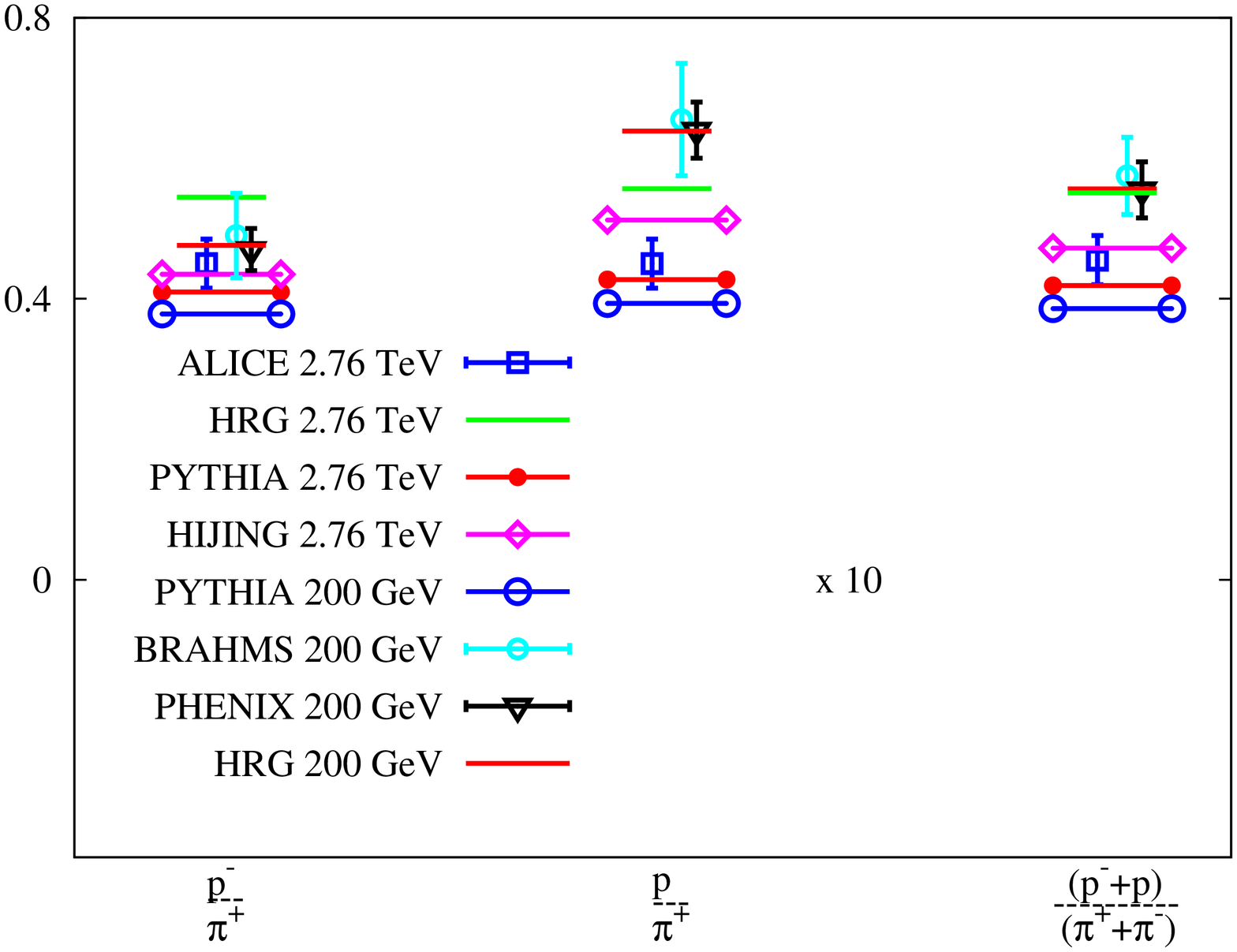}
\caption{Different particle ratios measured at RHIC and LHC energies (symbols with errors) are confronted to statistical-thermal models (horizontal lines) and event generators PYTHIA $6.4.21$ and HIJING. Graphs taken from Ref. \cite{Tawfik:2012zp}. } 
\label{Fig:Ratio3ALLs}   
\end{center}
\end{figure}

In Fig. \ref{Fig:RatioALLs}, the ratios $\pi^-/\pi^+$, $K^-/K^+$ and $\bar{p}/p$ in Au-Au collisions at RHIC \cite{Abelev:2008ab,phenix2003,brahms2005} and Pb-Pb collisions at LHC \cite{alice2012} energies (symbols with error bars) are compared with each others.
The HRG  (thick horizontal lines) and event generators PYTHIA $6.4.21$ and HIJING (symbols with horizontal lines) results are confronted to the experimental data (symbols) \cite{Tawfik:2012zp}, Fig. \ref{Fig:Ratio3ALLs}.

The PYTHIA $6.4.21$ simulations are performed for $pp$ collisions, while the HIJING simulations are performed for Pb-Pb collisions in the rapidity range $|y|<=0.5$  with centrality ($0-5\%$) and  \hbox{$4.5 >p_t>0.1~$GeV}.
As introduced, the difference between the particle production in $pp$ and $AA$ collisions almost disappears at very high energy \cite{Tawfik:2010pt,Tawfik:2010kz}. Thus, $pp$ and $AA$ results are conjectured to be non-distinguishable. It is obvious, that the experimental measurements for homogeneous particles to antiparticles ratios, i.e. same particle species, are well reproduced by means of the HRG model and both event generators. At $\sqrt{s_{NN}}=200~$GeV, it is found that $T_{ch}=171~$MeV,  and $\mu_b=8.4~$MeV. The corresponding strange chemical potential reads $\mu_S=5.9~$MeV.  At LHC energy $\sqrt{s_{NN}}=2760~$GeV, it is found that $T_{ch}=172~$MeV, $\mu_b=6.3~$MeV and $\mu_S=4.6~$MeV.  
Comparing to HIJING, PYTHIA $6.4.21$ describes very well the ALICE experiments. We note that PYTHIA $6.4.21$ overestimates the RHIC results for $\bar{p}/p$. This might be originated to the initial conditions related to the $pp$ collisions. In comparison with both event generators, the HRG model reproduces very well the experimental results at both energies.

\begin{figure}[tbh]
\begin{center}
\includegraphics[width=5.cm,angle=-90]{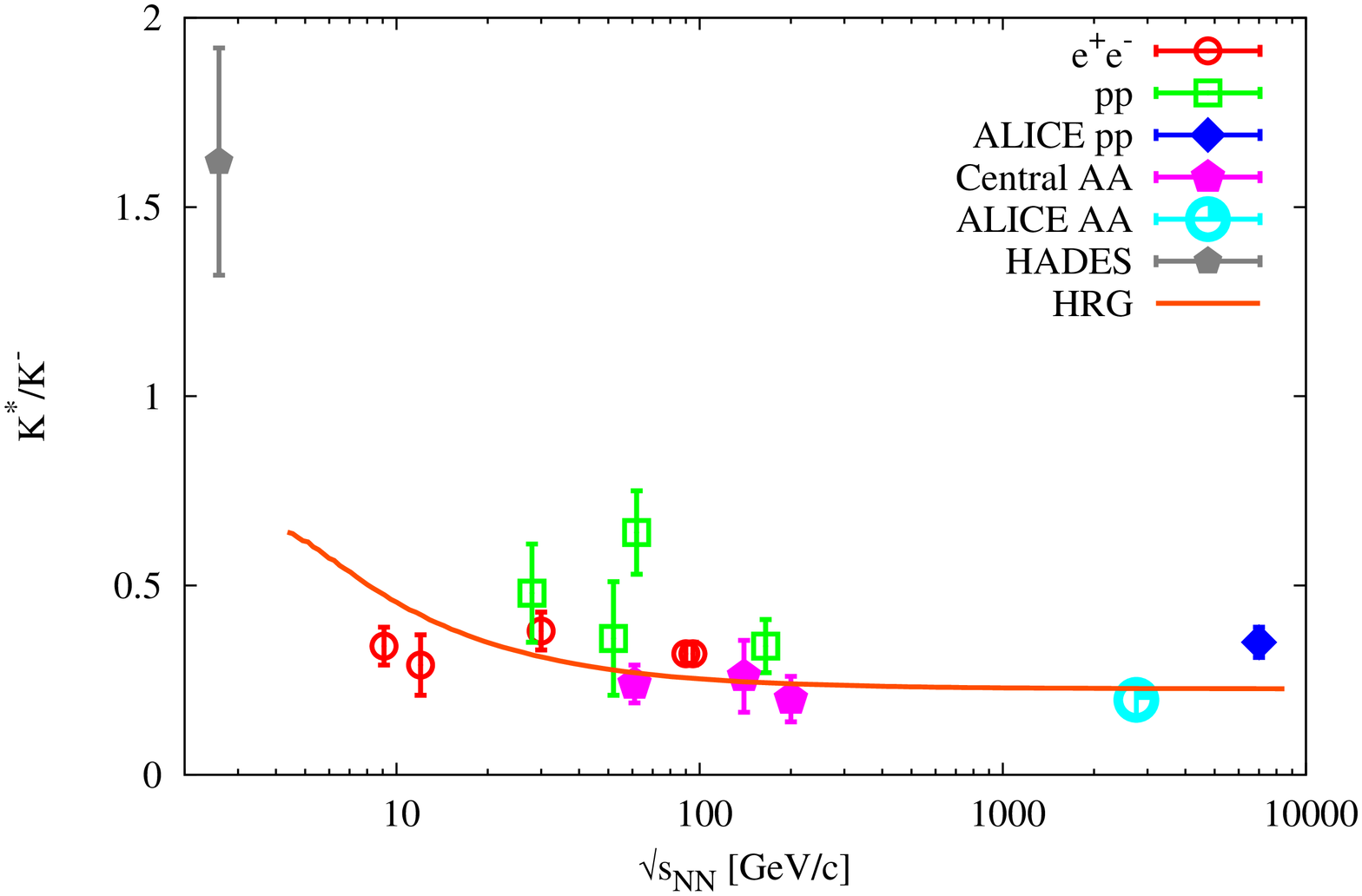}
\includegraphics[width=5.cm,angle=-90]{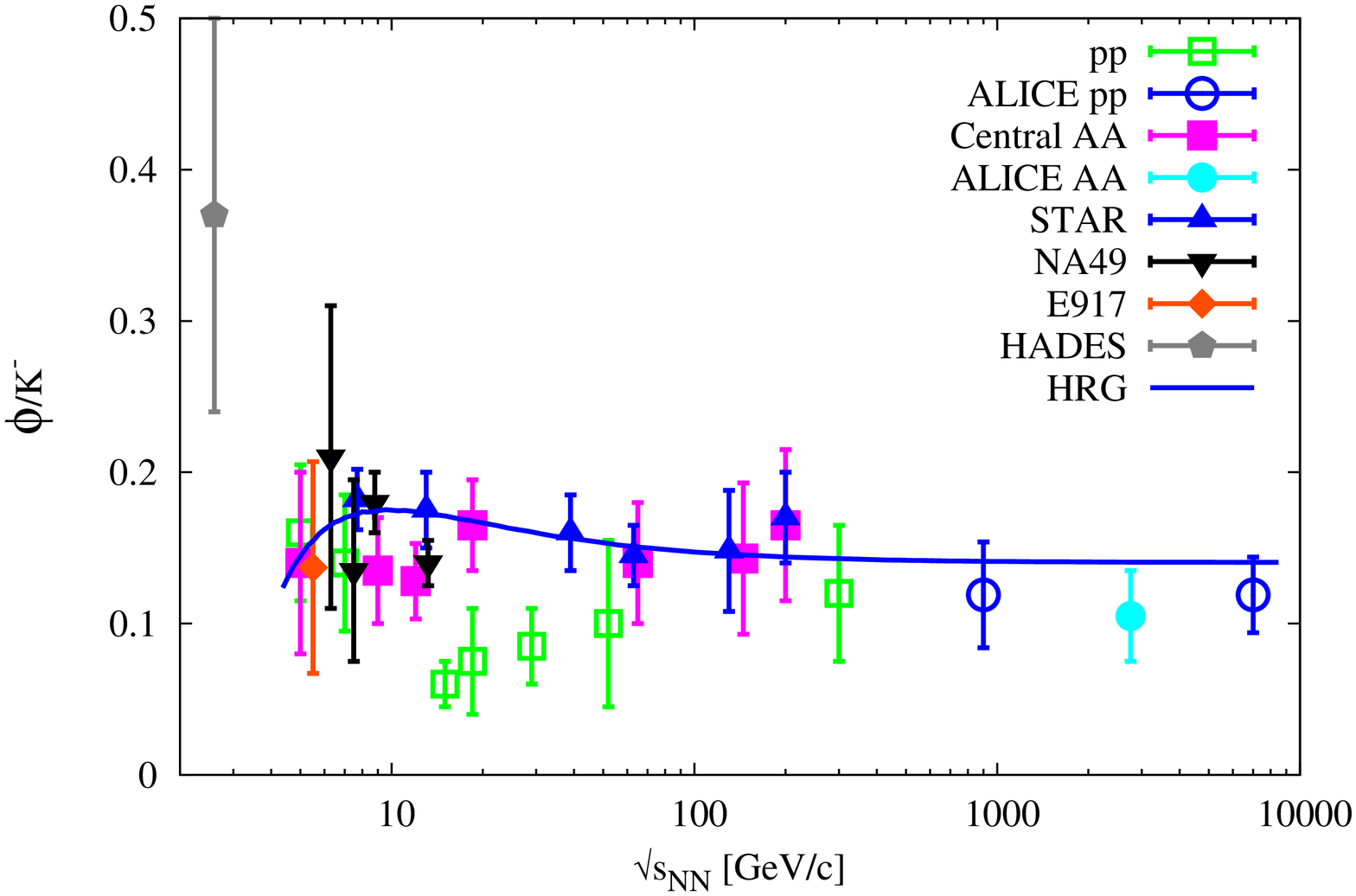}
\end{center}
\caption{The strange ratios $K^*/K^-$ and $\phi/K^-$ are given in dependence on $\sqrt{s_{NN}}$ for different collisions (symbols with errors). The ratios from HRG are given by the curves.} 
\label{Fig:KstrPhi}   
\end{figure}

Fig. \ref{Fig:KstrPhi} shows the energy dependence of the ratios $K^*/K^{-}$ and $\phi/K^-$ as measured in different systems and compared with the HRG results. The experimental data includes, $e^+ e^-$ \cite{ee1,ee2,ee3,ee4,ee5}, $pp$ \cite{pp1a,pp1b,s200k,pp3,pp4b,pp5,pp6} and ALICE $pp$ \cite{ALICEpp}, Central $AA$ including STAR \cite{s200k,pp3,pp6,pp5,pp6} and ALICE \cite{ALICEAA2012} besides HADES \cite{hades1,hades2} results. Left panel includes data from NA49 \cite{na49:2002A,na49b} and E917 \cite{e917} collaborations.   It is obvious that the statistical-thermal model reproduces well both ratios.  The deviation at low energies (very large chemical potential) may refer to the hadron interactions, especially van der Waals repulsion, section \ref{sec:intr}.

\begin{figure}[hbt]
\centering{
\includegraphics[width=10cm]{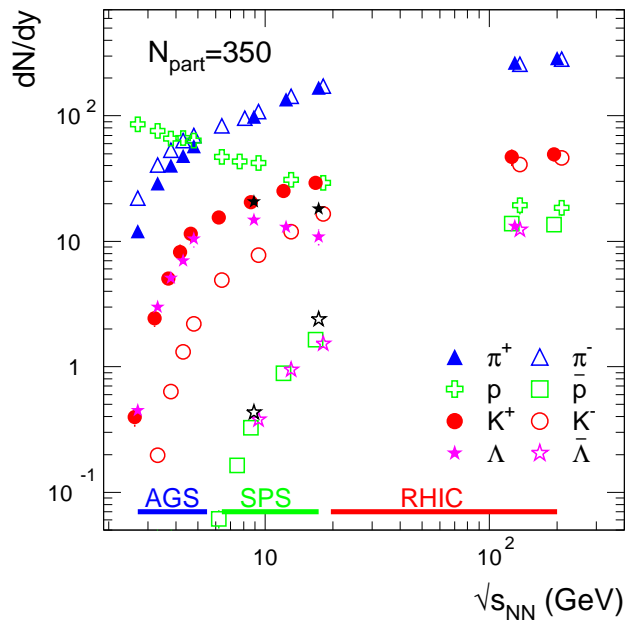}
\caption{The energy dependence of the experimental hadron yields at the mid-rapidity for various particle species produced in central collisions. Graph taken from Ref. \cite{dataCR}.} 
}
\label{figmy0}
\end{figure}

An extensive comparison between the measurements of yields at the mid-rapidity for abundant particle species  in  central collisions is shown in Fig.~\ref{figmy0}.  
The energy regimes for the different accelerators are sketched. For the SPS energies, two independent measurements are available for the $\Lambda$ hyperona.
At lower energies (AGS at $\sqrt{s_{NN}}\le5~$GeV), it is conjectured that the fireball is dominated by the incoming nucleons, while the produced pion yields have  a strong energy dependence, especially in the energy regime \cite{dataCR}. Furthermore, the importance of the isospin is apparent in the different yields of $\pi^+$ and $\pi^-$. It was observed that the decreasing yield of protons should be related to an increasing transparency of the incoming nuclei as a function of the incident energy. At energies larger than $\sqrt{s_{NN}}\simeq$100 GeV, it seems that the produced protons are dominant. Detailed discussion about the results can be found in Ref. \cite{dataCR}. \\

\paragraph{\bf PYTHIA $6.4.21$ \\}

Although the experimental data which can be compared with statistical-thermal models is taken from heavy-ion collisions, the comparison with PYTHIA $6.4.21$, which is designed to generate multi-particle production in collisions between elementary particles, $e^+ e^-$, $pp$ and $ep$, turns to be possible at very high energies. As discussed, the measured particle production is conjectured as an indicator for the formation of QGP, especially in heavy-ion collisions. In $pp$-collisions, the spatial and time evolution of the system is too short to assure initial conditions required to drive hadronic matter into partonic QGP. The potential difference between pp- and AA-collisions is supposed to disappear at LHC energies \cite{Tawfik:2010pt,Tawfik:2010kz}. We use PYTHIA $6.4.21$ \cite{pythia6421A,pythia6421B} with the Perugia-$0$ tune \cite{perugia} in the framework of AliRoot \cite{aliroot}. The bulk of PYTHIA multiplicities is formed in jets, i.e. in collimated bunches of hadrons or resonances decaying into further hadrons produced by the hadronization of partons \cite{pythia6421A,pythia6421B}. The relative proportion of strange particles is as expected small comparing with non-strange hadrons \cite{se2}. PYTHIA is capable of simulating for different processes including hard and soft interactions, parton distributions, initial/final-state parton showers, multiple interactions, fragmentation and decay.

In Refs. \cite{Tawfik:2010pt,Tawfik:2010kz}, we have noticed that the collective flow of strongly interacting matter in heavy-ion collisions makes the HRG model underestimating the particle ratios measured in $pp$-collisions, especially at low energies. It was found that the differences between the particle ratios in $pp$- and $AA$-collisions almost disappear  at the LHC energies. In light of this, the comparison with PYTHIA remains an enlightening feature. Although, it gives comparable high-energy results as the ones from the heavy-ion collisions, its initial conditions would be reflected in the collective properties in the final state. This would include - among others - strangeness suppression, as the mass of strange quark is heavier than that of up and down quarks. Therefore, the production of strange hadrons should be generally suppressed relative to hadrons containing up and down quarks only.  

The data sample presented in present work consists of $500$ thousands minimum bias events for $pp$-collisions at $200~$GeV and $2.76~$TeV with \hbox{$0.1~$MeV $<p_t<4.5~$MeV} and \hbox{$|y|<=0.5$}.   \\

\paragraph{\bf Heavy-Ion Jet INteraction Generator (HIJING)\\}

Taking into consideration the role of minijets in $pp$, $pA$ and $AA$ interactions, the HIJING Monte Carlo model was developed in early 1990's \cite{hijing1A,hijing1B}. It combines a QCD-inspired model for jet production using Lund model for jet fragmentation \cite{jetsetA,jetsetB}. It is  expected that hard or semi-hard parton scatterings with transverse momenta of a few GeV dominate the  high-energy heavy-ion collisions. In light of this, HIJING provides a qualitative understanding of the interplay between soft string dynamics and hard QCD interactions. In particular, HIJING reproduces many inclusive spectra, two-particle correlations, as well as the observed flavour and multiplicity dependence of the average transverse momentum.   The nuclear shadowing and jet quenching are two important features of HIJING event generator. 
\begin{itemize}
\item The shadowing describes the modification of the free nucleon parton density. The parton shadowing is taken into account using a parametrization of the modification. It has been observed that at low-momentum fractions, the nuclear shadowing results in a decrease in the multiplicity. 
\item With jet quenching it is meant the energy of partons in nuclear matter responsible for an increase of the particle multiplicity at central rapidity. Jet quenching is taken into account by an expected energy loss of partons traversing dense matter. 
\end{itemize}
A simple color configuration is assumed for the multi-jet system and the Lund fragmentation model is used for the hadronization. It is worthwhile to mention that HIJING does not consider the secondary interactions.

The HIJING event generator uses some PYTHIA subroutines of tune $5.3$ in order to generate the kinetic variables for hard scattering and associated radiations. Also, it implies JETSET $7.2$ for jet fragmentation \cite{jetsetA,jetsetB}. The data sample presented in this work has the same characteristics as the ones of PYTHIA $6.4$. $500$ thousands minimum bias events for Pb-Pb collisions at $200~$GeV and $2.76~$TeV with \hbox{$0.1~$MeV $<p_t<4.5~$MeV} and \hbox{$|y|<=0.5$} are analysed.

\begin{figure}[tbh]
\begin{center}
\includegraphics[width=8.cm,angle=0]{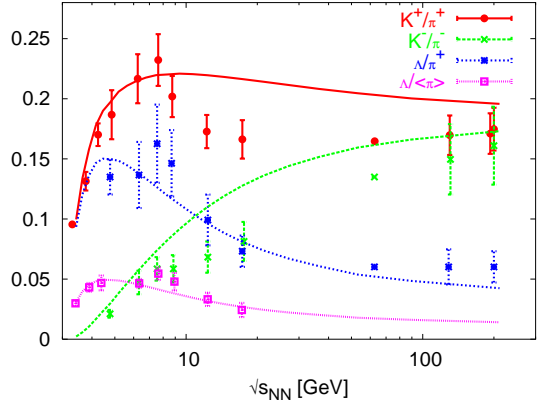}
\caption{The $K^+/\pi^+$, $K^-/\pi^-$, $\Lambda/\pi^+$ and \hbox{$\Lambda/<\pi>$}
ratios at AGS ($\sqrt{s_{NN}}\leq4.84\;$GeV), SPS
($6.26\leq\sqrt{s_{NN}}\leq17.27\;$GeV) and RHIC energies
($62.4\leq\sqrt{s_{NN}}\leq200\;$GeV). The quark occupation parameters $\gamma_q$ and $\gamma_s$ are equal and both assigned to units.  Graph taken from Ref. \cite{Tawfik:2005gk}. } 
\label{Fig:varusRts2}   
\end{center}
\end{figure}

\begin{figure}[tbh]
\begin{center}
\includegraphics[width=8.cm,angle=0]{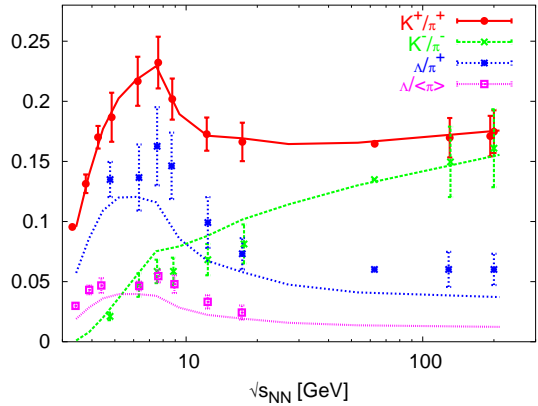}
\caption{The same as in Fig. \ref{Fig:varusRts2} but $\gamma_q\neq1$ while $\gamma_s=1$.  Graph taken from Ref. \cite{Tawfik:2005gk}. } 
\label{Fig:varusRts3}   
\end{center}
\end{figure}

\begin{figure}[tbh]
\begin{center}
\includegraphics[width=8.cm,angle=0]{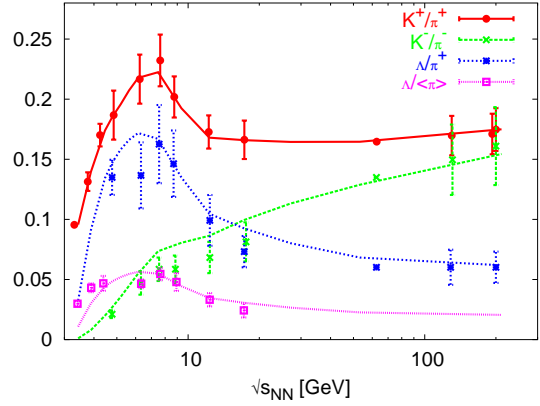}
\caption{The same as in Fig. \ref{Fig:varusRts2} but $\gamma_q=\gamma_s\neq1$.  Graph taken from Ref. \cite{Tawfik:2005gk}. } 
\label{Fig:varusRts1}   
\end{center}
\end{figure}

The energy dependence of particle ratios may depend on the quark occupation parameters 
\cite{Tawfik:2005gk}. For finite isospin fugacity $\lambda_{I_3}$, we get 
\begin{eqnarray}
\frac{n_{K^+}}{n_{\pi^+}} \equiv \frac{K^+}{\pi^+} &\propto&
                     \lambda_s^{-1}
                     \left(\frac{\lambda_q}{\lambda_{I_3}}\right)\; 
                      \frac{\gamma_q}{\gamma_s},  \label{K20piplus} \\
\frac{n_{K^-}}{n_{\pi^-}} \equiv \frac{K^-}{\pi^-} &\propto&
                     \lambda_s\,
                     \left(\frac{\lambda_{I_3}}{\lambda_q}\right)\; 
                     \frac{\gamma_s}{\gamma_q},  \label{K20piminus}\\
\frac{n_{\Lambda}}{n_{\pi^+}} \equiv \frac{\Lambda}{\pi^+} &\propto&
                     \lambda_s\left(\frac{\lambda_q}{\lambda_{I_3}^2}\right)^2 
                     \; \gamma_q^2 \gamma_s.
                     \label{K20pilambdapiplus}
\end{eqnarray}
The particle numbers at zero chemical potential, \hbox{$n_j(T)\simeq
  T\,m_j^2K_2(m_j/T)$}, represent the proportional factors in these 
  expressions. $n_j(T)$ is a smooth function of $T$. The fugacity
  $\lambda$ is also a smooth function of $T$.  Correspondingly, a 
  monotonic increase in the particle ratios is expected with  energy. 

The statistical parameter $\gamma$ appearing in the front of Boltzmann exponential, $\exp(-\varepsilon/T)$ gives the averaged occupancy of the phase space relative to equilibrium limit. Therefore, in the equilibrium limit $\gamma=1$. Assuming the time evolution of the system, we can describe 
$\gamma_i$ as the ratio between the change in particle number before
and after the chemical freeze-out, i.e. $\gamma_i=n_i(t)/n_i(\infty)$. The chemical freeze-out is defined as a time scale, at which there is no longer particle production and the collisions
become entirely elastic. In case of phase transition, $\gamma_i$ is expected to
be larger than one, because of the large degrees of freedom,
weak coupling and expanding phase space at temperatures larger than the critical temperature, i.e. QGP.

\subsubsection{Dynamical fluctuations of particle ratios}
\label{sec:dynmFlct}

The particle number density is given by the derivative of partition function with respect to the chemical potential of interest. The fluctuations in  the particle number are given by the susceptibility density, which is the second derivative with respect to the  chemical potential $\mu$ \cite{Tawfik:2008ii}
\bea 
\langle n\rangle &=& \sum_i \frac{g_i}{2\pi^2} \int dk k^2
\frac{e^{(\mu_i-\varepsilon_i)/T}}{1\pm e^{(\mu_i-\varepsilon_i)/T}}, \label{eq:n1}  \\
\langle (\Delta n)^2\rangle &=& \sum_i \frac{g_i}{2\pi^2} \int dk k^2
           \frac{e^{(\varepsilon_i-\mu_i)/T}}
	   {\left(e^{(\varepsilon_i-\mu_i)/T}\pm1\right)^2}.
        \label{eq:dn1}
\eea

\begin{figure}[tbh]
\begin{center}
\includegraphics[width=7.cm,angle=0]{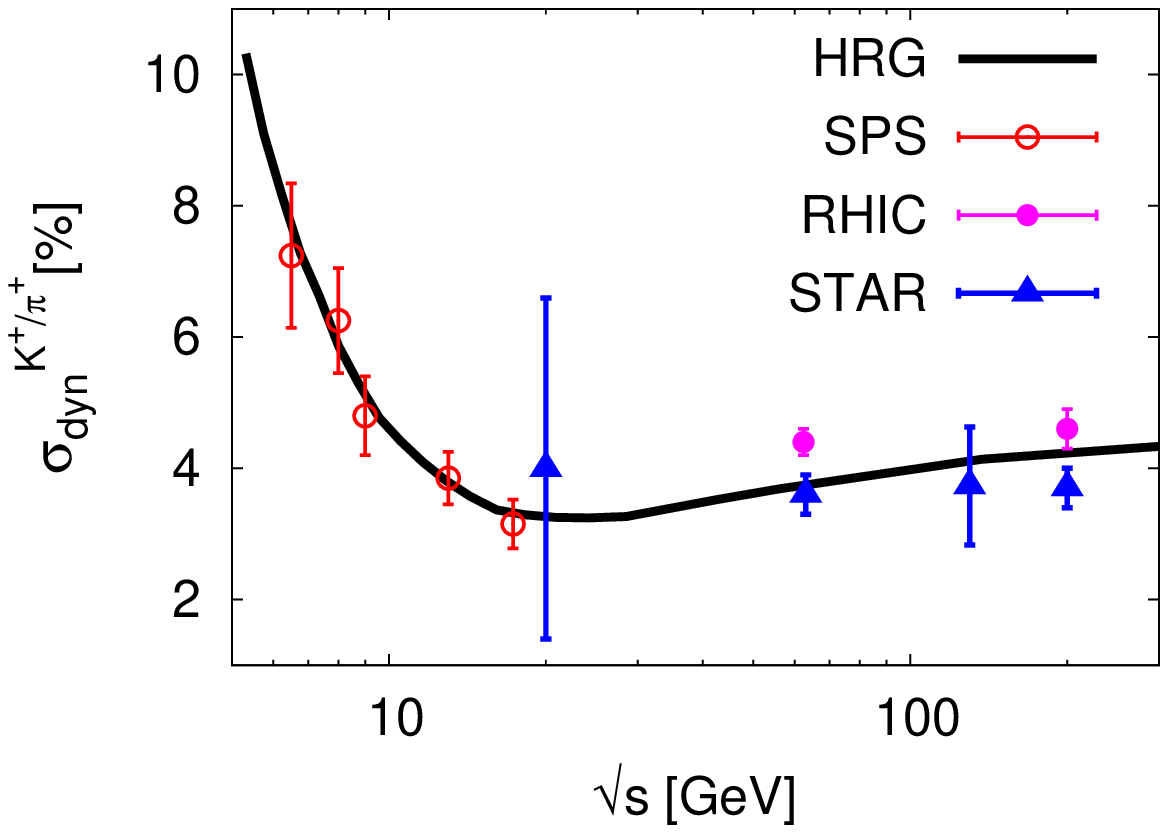}
\includegraphics[width=7.cm,angle=0]{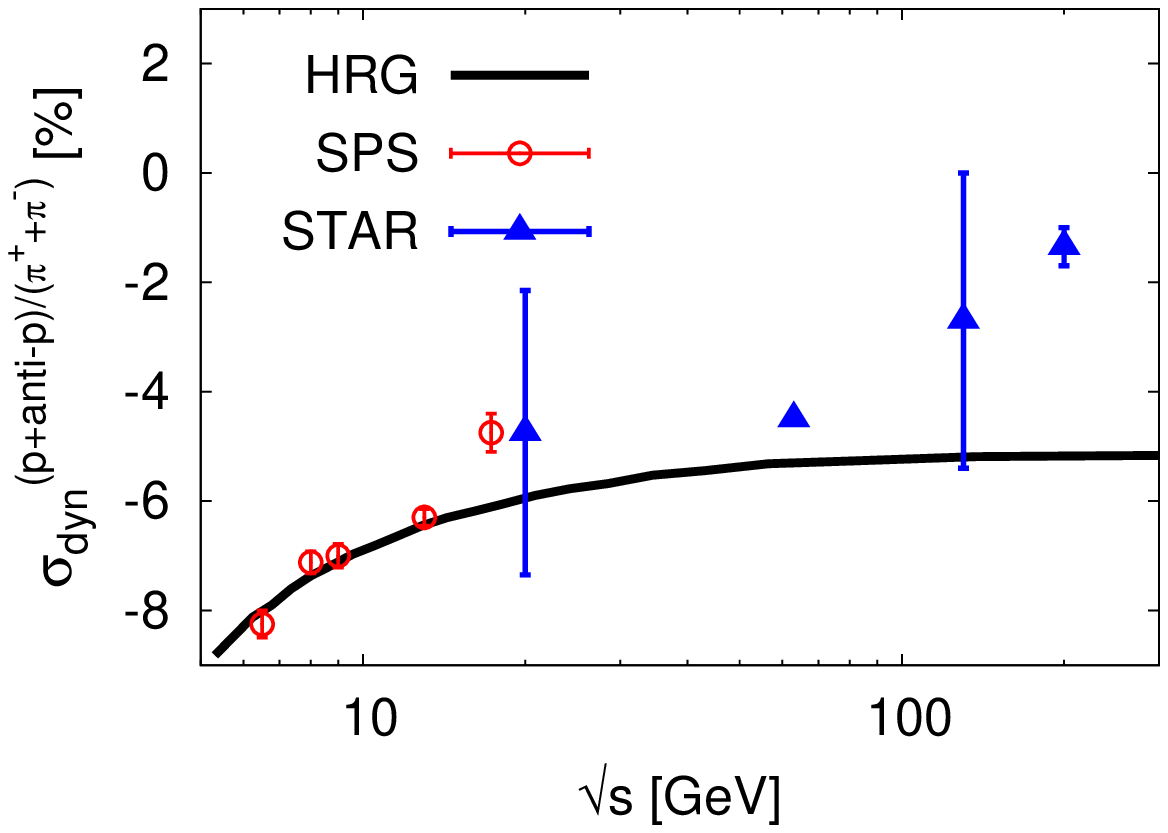}
\caption{The dynamical fluctuations of the particle ratios $K^+/\pi^+$ and $(p+\bar{p})/(\pi^++\pi^-)$ is given as function of center-of-mass energy.  Graphs taken from Ref. \cite{Tawfik:2008ii}. } 
\label{Fig:sgmPKpi}   
\end{center}
\end{figure}

When the system absolved the chemical freeze-out process, the hadron resonances are conjectured to decay either to stable particles or
to other resonances. The particle number and fluctuation densities in final state
have to take into account this chemical process 
\bea 
\langle n_i^{final}\rangle &=& \langle n_i^{direct}\rangle + \sum_{j\neq
i} b_{j\rightarrow i} \langle n_j\rangle, \label{eq:n2} \\ 
\langle (\Delta n_{j\rightarrow i})^2\rangle &=& b_{j\rightarrow i} (1-b_{j\rightarrow
i}) \langle n_j\rangle + b_{j\rightarrow i}^2 \langle (\Delta
n_{j})^2\rangle,  \label{eq:dn2} 
\eea
where $b_{j\rightarrow i}$ being branching ratio for the decay of $j$-th resonance
to $i$-th particle. To characterize when the chemical freeze-out takes place, we assumed that ratio $s/T^3$, where $s$ is the entropy density, get a constant value \cite{Tawfik:2004vv,Tawfik:2004ss,Tawfik:2005qn}. Other conditions describing the freeze-out curve have been reported \cite{Tawfik:2013pd,Tawfik:2013dba,Tawfik:2013eua}, section \ref{sec:fo1}.

The fluctuations in the particle ratio $n_1/n_2$ are~\cite{koch}
\bea  
\sigma^2_{n_1/n_2} &=& \frac{\langle (\Delta n_1)^2\rangle}{\langle n_1\rangle^2} + 
                       \frac{\langle (\Delta n_2)^2\rangle}{\langle n_2\rangle^2} - 
                     2 \frac{\langle \Delta n_1 \; \Delta
		     n_2\rangle}{\langle n_1\rangle \; \langle
		     n_2\rangle},  \label{eq:sigma}
\eea
which include dynamical as well as statistical fluctuations. The third term of
Eq.~(\ref{eq:sigma}) counts for the fluctuations from the hadron resonances which 
decay into particle $1$ and particle $2$, simultaneously. In such a
mixing channel, all correlations including quantum statistics ones are
taken into account.  Obviously, 
this decay channel results in strongly correlated particles. To extract
statistical fluctuation, we apply Poisson scaling in the mixed decay
channels,
\bea 
(\sigma^2_{n_1/n_2})_{stat} &=& \frac{1}{\langle n_1\rangle} +
\frac{1}{\langle n_2\rangle}. \label{eq:sigmaStat}
\eea
Subtracting Eq.~(\ref{eq:sigmaStat}) from Eq.~(\ref{eq:sigma}), we get
dynamical fluctuations of particle ratio $n_1/n_2$. 
\bea
(\sigma^2_{n_1/n_2})_{dyn} &=& 
          \frac{\langle n_1^2\rangle}{\langle n_1\rangle^2} +
          \frac{\langle n_2^2\rangle}{\langle n_2\rangle^2} -
         \frac{\langle n_1\rangle+\langle n_2\rangle +
	 2\langle n_1n_2\rangle}{\langle n_1\rangle\langle n_2\rangle}. \label{eq:sigma2}
\eea

The values of dynamical fluctuations of strangeness particle ratios are
greater than that of non-strangeness ones. Their dependence on
$\sqrt{s}$ is also non-monotonic. While fluctuations from $K^-/\pi^-$
exponentially decrease with increasing $\sqrt{s}$, fluctuations from
$\Lambda/\pi$ and $\Xi^+/\pi$ have minimum values at $\sqrt{s}\sim 10\;$GeV \cite{Tawfik:2006dk}. 

Comparing strangeness with non-strangeness fluctuations, we find that 
replacing pion by its anti-particle has almost no influence on
strangeness dynamical fluctuations. Non-strangeness dynamical
fluctuations are dramatically changed, when pion in
denominator assuming that the particle ratios can
mathematically be seen as fractions. has been replaced by its
anti-particle.  For instance, for $p/\pi$ ratio \cite{Tawfik:2006dk}
\bea
\langle \left(\Delta n_{p/\pi^+}\right)^2 \rangle &\rightarrow&  
    \langle \left(\Delta n_u\right)^2 \rangle  + 2 \langle \left(\Delta
    n_d\right)^2\rangle,  \label{hq1}\\
\langle \left(\Delta n_{p/\pi^-}\right)^2 \rangle &\rightarrow&  
    3 \langle \left(\Delta
    n_u\right)^2\rangle. \label{hq2}
\eea   
As in right panel in Fig.~\ref{Fig:sgmPKpi}, while fluctuations of first
    ratio move from negative to positive values, second ratio remains
    negative at all energies.  
In hadronic phase, i.e. particle ratios, quarks are strongly confined
    into hadronic states. It 
    is believed that quarks in the unconfined phase may be strongly
    correlated~\cite{Tawfik:2006dk}. It would be 
    interesting to verify above expressions. In doing this, we have to
    take into consideration volume fluctuations 
    on lattice. Resonance gas model can not be applied at temperatures
    higher than critical one. \\

The responsibility of non-equilibrium quark occupancy of phase space is utilized for particle production \cite{Tawfik:2005gk}. We want clarify the physical reason behind the non-monotonic behavior. The dependence of single-particle
entropy on the collision energy is related to the averaged phase
space density. In Boltzmann limit and for one particle
\begin{eqnarray}
\frac{s}{n} &=& \frac{\varepsilon}{T} + 1 - \frac{\mu}{T},
 \label{sOvern} 
\end{eqnarray}
where $\varepsilon$ is the single-particle energy. In this
expression, $s/n$ is related to $\varepsilon$. Apparently, $s/n$ 
becomes maximum when the chemical potential gets 
as large as the single-particle energy $\varepsilon$ \cite{Tawfik:2013pd}. As we assumed
Boltzmann limit, the maximum is unity in this case. Depending on the
chemical potential $\mu$, we can insert particles into the phase space. The
maximum occupation is reached at $\mu=\varepsilon$. This is an upper
limit. Then, it becomes prohibited to insert more particles. On the other
hand, we can expect - at least theoretically - occupation values larger than
this classical upper limit, only if the phase space itself is changed. This
situation is most likely provoked by the phase transition. From this
discussion we can apparently realize that $s/n$ might play the same role
as the statistical parameter 
$\gamma$ does. 

\begin{figure}[htb] 
\begin{center}
\includegraphics[width=10.cm]{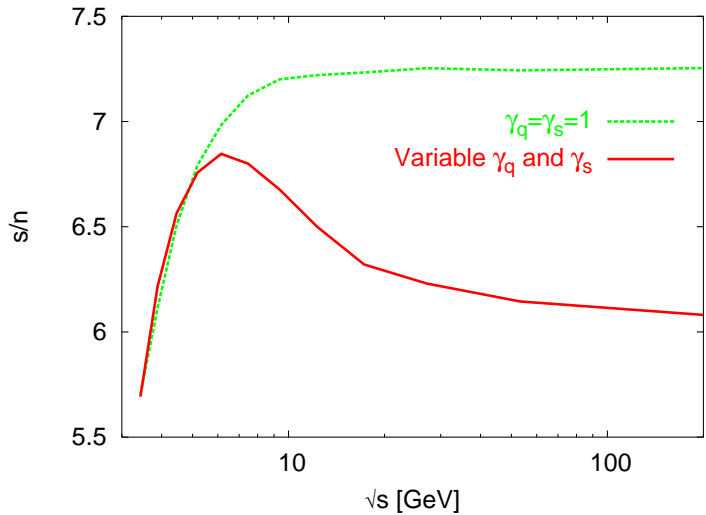}
\caption{The entropy per particle $s/n$ as function of $\sqrt{s_{NN}}$. Only for
variable  quark occupancy parameters $\gamma_q$ and $\gamma_s$ there is a singularity
in $s/n$ ratio. It is important to notice that the singularity is located at
almost the same energy as the peaks of particle ratios.  Graph taken from Ref. \cite{Tawfik:2013pd}.
\label{Fig:5a} }
\end{center}
\end{figure}

The results on  $s/n$ vs. $\sqrt{s_{NN}}$ are depicted in
Fig.~\ref{Fig:5a}. Again we use here many hadron resonances and full grand-canonical statistical
set of the thermodynamic parameters. In this case, the complete dependence of $s/n$ on
$T$ and $\mu$ and consequently on $\sqrt{s_{NN}}$, can straightforwardly be obtained. For $\gamma_q=\gamma_s=1$, we find that
$s/n$ increases as the energy raises from AGS to low SPS
($\sqrt{s_{NN}}\leq10\;$GeV). As shown in Figs.~\ref{Fig:varusRts2}-\ref{Fig:varusRts1}, a
mild maximum  in $K^+/\pi^+$ is located at the same value of
$\sqrt{s_{NN}}$. For higher energies, $s/n$ remains constant.

For the non-equilibrium case, i.e. variable $\gamma_q$ and $\gamma_s$, we
find almost the same behavior up to $\sqrt{s_{NN}}\simeq6.5\;$GeV. At
this energy, there is a singularity in $s/n$. The
singularity might be related to a certain critical
phenomenon. For the  rest of SPS energies, $s/n$ rapidly decreases. Although the energy is
increased and consequently the produced particles, the single-particle
entropy decreases. This can only be understood, by assuming that the phase
space shrinks. At RHIC, $s/n$ decreases slowly with the energy. The
shrinking in phase space becomes slow. If this
model would give the correct description, we now might have for the first time a theoretical
explanation for the dependence of phase space on energy. The phase
space at SPS energy is apparently larger than the phase space at RHIC
and LHC. The consequences are that the quark-gluon plasma might be produced at
SPS. And detecting its signatures at RHIC might be non-trivial.



\subsection{Chemical freeze-out}
\label{sec:fo1}

Reducing the temperature, the partonic matter \cite{satz94} shall hadronize. At some temperature $T_{ch}$, the produced hadrons entirely freeze out. Both freeze-out parameters, $\mu_B$ and $T_{ch}$, can be determined in statistical-thermal models by combining various ratios of integrated particle yields~\cite{Sollfrank:1999cy,Cleymans:1998yf}. Such a way, we obtain a window in the $T_{ch}-\mu_B$ phase-diagram compatible to the  experimental values. $\mu_B$ and $T$ are free parameters in these fits. 

The statistical-thermal models provide a systematic study of many important properties of hot and  dense hadron gas, especially at the chemical freeze-out. In order to deduce  a universal  relation between chemical freeze-out parameters, $T$ and $\mu_b$ and nucleus-nucleus center-of-mass energy, $\sqrt{s_{NN}}$, a common method is used to fit the experimental hadron ratios as follows. Starting with a certain value of baryon chemical potential $\mu_b$, for instance, the temperature $T$ is increased very slowly. At this value of $\mu_b$ and at each raise in $T$, the strangeness chemical potential $\mu_S$ is determined under the condition that the strange quantum numbers should remain conserved in the heavy-ion collisions. Having the three values of $\mu_b$, $T$  and $\mu_S$, then all thermodynamic quantities including the number density $n$ of each spices are calculated. When the ratio of two particles, like $\pi^+/\pi^-$ reaches the experimental values, then the temperature $T$ and chemical potential $\mu_b$ are registered. This procedure is repeated for all particle ratios measured in different high-energy experiments. Details on this procedure are reviewed in Ref. \cite{jean2006,dataCR}. 

In the evolution of the hadronic system, the chemical freeze-out point is the stage, in which the inelastic collisions cease and the relative particle ratios become fixed. The search for common properties of the freeze-out parameters in heavy-ion collisions has a long tradition~\cite{redlich2,rfRED6}.  After chemical freeze-out, the particle composition inside the fireball should be fixed. On the other hand, the elastic collisions still keep the system together until {\it thermal  freeze-out}. At this stage, the momentum distributions of particles no longer changes. Therefore, the transverse momentum spectra determine the thermal freeze-out parameters.

\begin{figure}[thb]
\includegraphics[width=12.cm,angle=0]{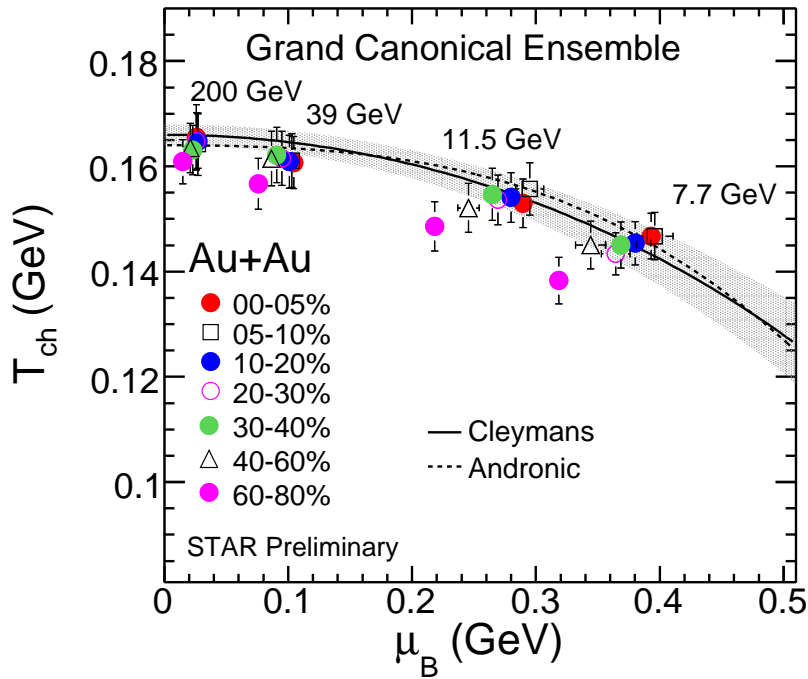}
\caption{The variation of $T_{\rm{ch}}$ with the baryon chemical potential $\mu_{B}$ for
different energies and centralities. The graph taken from Ref. \cite{Kumar2012}. } 
\label{fig:frzSTAR}
\end{figure}

Fig. \ref{fig:frzSTAR} illustrates the variation of the extracted chemical freeze-out parameters using the Grand-Canonical Ensemble (GCE) approach for different energies and centralities \cite{dataCR}. It is obvious that at lower energies, $T_{\rm{ch}}$ shows a variation with $\mu_{B}$ as a function of centrality \cite{Kumar2012}. When moving from central to peripheral collisions, both $T_{\rm{ch}}$ and $\mu_{B}$ values decreas. It is noted that
use of Strangeness Canonical Ensemble (SCE) has an opposite trend, i.e. the $T_{\rm{ch}}$ values seem to increase when moving from central to
peripheral collisions. However, it is found that the $\chi^{2}/NDF$ is high in SCE of peripheral collisions \cite{dataCR}.

The two curves as results from statistical-thermal models using two different conditions for the freeze-out, section \ref{sec:cr_en1} and \ref{sec:totalb}, respectively.

\subsubsection{Cleymans and Redlich: constant energy per particle}
\label{sec:cr_en1}

A criterion of constant energy per particle was proposed as a universal curve unifying freeze-out results obtained at SIS, top AGS and SPS \cite{prcl22,redlich2}. This criterion is frequently cited literature and used to make predictions \cite{prcl46} for freeze-out parameters at SPS energies of $40$ and $80~$AGeV for Pb–Pb collisions long before the data were taken. Since then, the criterion has proven its validity also at RHIC \cite{prcl5,redlich1} and even at the lower SPS energies \cite{prcl1,prcl7,prcl46}.
These predictions turned out to be in agreement with the statistical model analysis of recent experimental data obtained at these collision energies \cite{prcl7}.

In a non-relativistic system, constant energy per particle would be expressed as the summation of average thermal mass of the fireball plus some function of freeze-out temperature $T$ \cite{prcl47}. At SIS, the thermal mass is of the order of the nucleon mass, whereas the freeze-out temperature $T\simeq 50~$MeV. At SPS and RHIC, the leading particles in the final state are pions. However, at chemical freeze-out most of the pions are still hidden in the mesonic and baryonic resonances. Thus, here the average thermal mass corresponds approximately to the $\rho$-meson mass. Consequently, with $T\simeq 160-170~$MeV a comparable value of energy per particle  is obtained at SPS and RHIC as in a much lower energy at SIS \cite{redlich1,redlich2,redlich3,redlich4}.

In UrQMD, the correlation between $1~$GeV energy per particle and inelasticity was investigated in central Pb–Pb collisions at SPS energy \cite{prcl48}. It was shown that there is a clear correlation between the chemical break-up in terms of inelastic scattering rates and the rapid decrease in energy per particle. If the value of energy per particle $\rightarrow 1~$GeV, the inelastic scattering likely rates drop. Further evolution of the system is then due to the elastic collisions that substantially preserve the chemical composition of the collision in the fireball. The results from RHIC seems to favor a higher value for the energy per particle than that at SPS and lower energies \cite{prcl12}. This can be interpreted as a change in the baryonic composition of a two-component thermal source.

\subsubsection{Braun-Munzinger and Stachel: constant-total baryon density}
\label{sec:totalb}

\begin{figure}[thb]
\centering{
\includegraphics[width=6.cm,angle=-90]{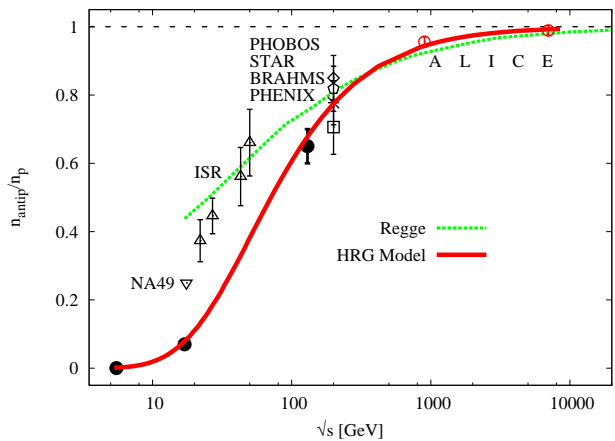}
\caption{\normalsize $n_{\bar{p}}/n_p$ ratios depicted in the whole available range of $\sqrt{s}$. Open symbols stand for the results from various $pp$ experiments (labeled). The solid symbols give the heavy-ion results from AGS, SPS and RHIC, respectively. The fitting of $pp$ results according to Regge model is given by the dashed curve \cite{alice2010}. The solid curve is the HRG results. Extending it to the LHC energy obviously shows that the ratio itself is very much close to unity. Contrary to the dashed curve, the solid line is not a fitting to experimental data. The graph taken from Ref. \cite{Tawfik:2010pt}. } 
\label{fig:pap2010}
}
\end{figure}

\begin{figure}[htb]
\centering{
\includegraphics[width=.6\textwidth,height=.4\textwidth]{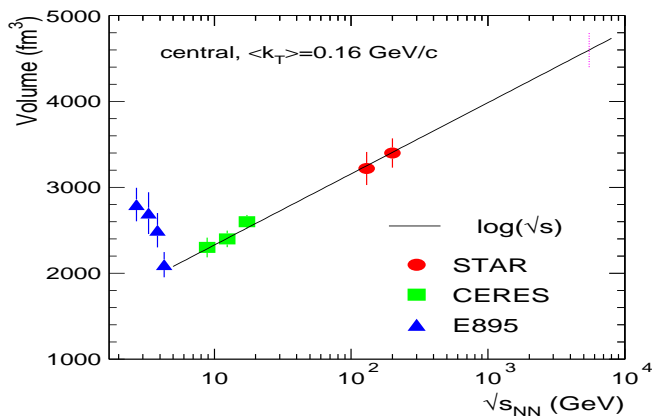}
\caption{Excitation function of the freeze-out volume extracted from  
pion HBT correlations is given in dependence on the center-of-mass energy. The line is a $\log(\sqrt{s_{NN}})$ dependence arbitrarily normalized. The graph taken from Ref. \cite{prcl40a}. }
} 
\label{aa:vol} 
\end{figure}

The net baryon density, i.e. the difference between the density of baryons $n_b$ and anti-baryons $n_{\bar{b}}$ is conjectured to characterize the contributions of baryons to thermodynamics \cite{Tawfik:2010pt}.  In Fig. \ref{fig:pap2010}, the results of $p/\bar{p}$  calculated in HRG are represented by solid line, which  seems to be a kind of a universal curve. In heavy-ion collisions, the antiproton-to-proton ratio varies strongly with the collision energy. HRG describes very well the heavy-ion results. Also, ALICE $pp$ results are reproduced by means of HRG model. The ratios from $pp$- and $AA$-collisions runs very close to unity implying almost vanishing matter-antimatter asymmetry. On the other hand, it can also be concluded that the statistical-thermal models including HRG seem to  perfectly describe the hadronization at very large energies and the condition deriving the chemical freeze-out at the final state of hadronization, the constant degrees of freedom or $S(\sqrt{s},T)=7 (4/\pi^2) V T^3$, seems to be valid at all center-of-mass-energies spanning between AGS and LHC.

On the other hand, the conservation of baryon number, which is mainly given by the number of participants $A_{part}$ and the variation of net baryon number $n_b-n_{\bar{b}}$, which is related to the density number of participants ($A_{part}/V$) both are conjectured to reflect the corresponding
change in the volume of the fireball. 

Comparing total yields of different particle species in the heavy-ion collisions with the calculations in the statistical-thermal model has shown that the volume has a minimum between AGS and the lowest SPS energy \cite{prcl40a,prcl40b} and exhibits a dependence on the center-of-mass energy opposite to that of $n_b$. In Fig.~\ref{aa:vol}, the energy dependence of volume of the fireball as extracted from pion Hanbury~Brown-Twiss (HBT) correlations \cite{tp2} is given in dependence on the center-of-mass energy. The behavior at the  lowest energies is different than that at top AGS and above, for which a  $\log(\sqrt{s_{NN}})$ dependence describe well the measurements \cite{aa:hbt}. This non-monotonic behavior can be understood  quantitatively as the result of an universal pion freeze-out at a critical  mean free path $\lambda_f\simeq$1~fm, independent of energy \cite{aa:hbtcer}.  

Despite this large variation, it was first noticed \cite{prcl24} that the sum of baryon and anti-baryon densities remains remarkably constant. Thus, it was proposed \cite{prcl24} that the chemical freeze-out curve can be obtained from the condition of fixed density of total number of baryons and anti-baryons
\bea\label{eq:nbnab}
n_b+n_{\bar{b}} &\simeq & 0.12~\text{fm}^{-3}.
\eea
At low energy, the production of anti-baryons is suppressed in high-ion collisions. Therefore, the net baryon number density would be even equivalent to the total density of baryons. At high energy, the condition, Eq. (\ref{eq:nbnab}), gives a phenomenological explanation for the enhancement of the low mass dilepton yield observed by the CERES Collaboration \cite{prcl40a,aa:hbtcer}.

\subsubsection{Magas and Satz: percolation theory in heavy-ion collisions}
\label{sec:stzm}

The percolation theory successfully gives a description for the critical properties of QCD matter \cite{prcl37A,prcl37B}. In particular, color deconfinement in pure gauge theory can be treated as a percolation phenomenon \cite{prcl38A,prcl38B}. Furthermore, models based on percolation theory are used to provide a qualitative description for charmonium production and even the possible suppressions of various particle yields in heavy-ion collisions \cite{prcl39A,prcl39B}.

Based on geometric estimates using percolation theory, a self-consistent equation for the densities is suggested to reproduce the freeze-out parameters \cite{prcl25}. The percolation theory was used to formulate and quantify the chemical freeze-out conditions in the heavy-ion collisions. It is assumed that at vanishing baryon density, the hadronic matter is conjectured to freeze out by vacuum percolation. At finite baryon density, the freeze-out takes place according to baryon percolation. The condition that describes the freeze-out line in heavy-ion collisions was formulated as \cite{prcl25}
\bea \label{eq:prcl}
n(T,\mu) &=& \frac{1.24}{V_h} \left[1-\frac{n_b(T,\mu)}{n(T,\mu)}\right] + \frac{0.34}{V_h} \left[1-\frac{n_b(T,\mu)}{n(T,\mu)}\right],
\eea
where $V_h$ is the hadronic volume corresponding to an average radius $r_h\simeq 0.8~$fm, the numbers $1.24$ and $0.34$ have been calculation in the percolation theory  \cite{prcl44}. When the size of the largest cluster falls below the size of the overall spatial volume, the number $0.34$ appears. The disappearance of any large-scale vacuum and strongly interacting medium become the one that spans the entire space, both are related $1.24$.

Equation (\ref{eq:prcl}) has a unique solution at $T=T(\mu)$. This is characterized by combining the total density of all hadrons $n(T, \mu_b)$ and baryon number density $n_b(T, \mu_b)$ from the statistical-thermal model of HRG, for instance.

\subsubsection{Tawfik: constant-normalized entropy, $s/T^3$}
\label{sec:Tawst3}

\begin{figure}[thb] 
\centerline{\includegraphics[width=10.cm]{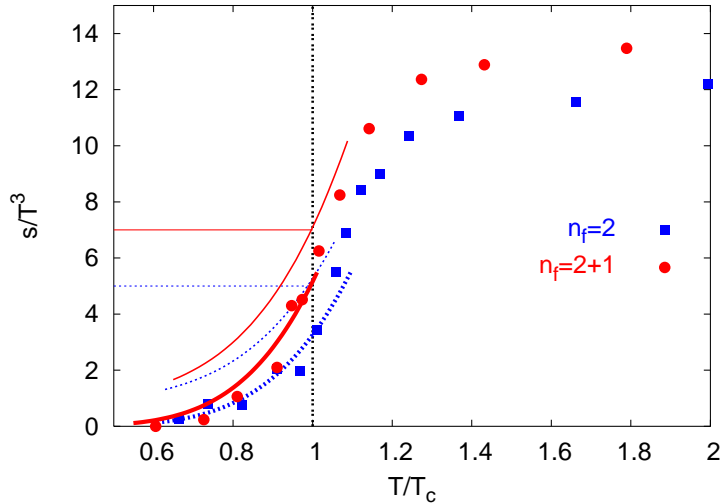}} 
\caption{Lattice QCD results on the entropy  density normalized to $T^3$ for $n_f=2$ (full circles) and $n_f=2+1$  (full squares) quark flavors at $\mu_B=0$~\cite{Karsch:2000kv,Karsch:2003zq} on the top of the results from the HRG model (curves). The thick curves represent the results from HRG with rescaled masses. We find a well agreement with the lattice QCD simulations. The HRG  calculations with the physical resonance masses are given by the thin curves. The horizontal lines indicate to $s/T^3$-values for the physical resonance masses at the different quark flavors and critical temperature $T_c$. The graph taken from Ref. \cite{Tawfik:2004sw}. \label{Fig:3}
}    
\end{figure}

\begin{figure}[thb] 
\centerline{\includegraphics[width=10.cm]{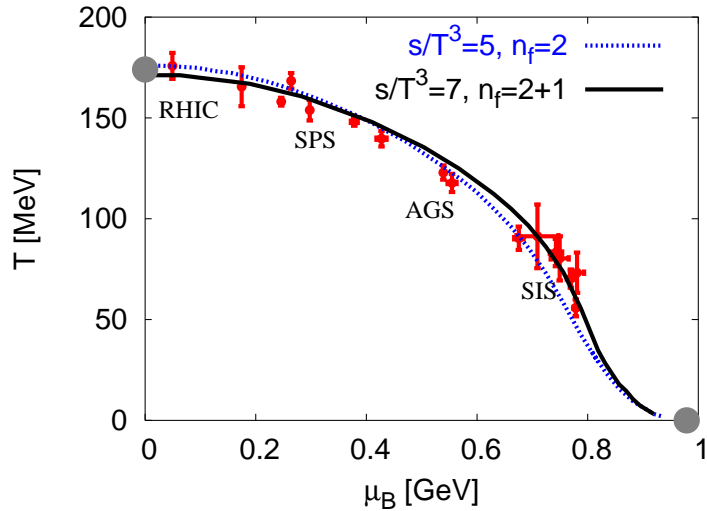}} 
\caption{\label{Fig:4}
  The freeze-out parameters calculated according to constant $s/T^3$ on the top of {\it
  phenomenologically} estimated freeze-out parameters
  (small full circles with errors). For non-strange 
  hadron resonances, $s/T^3=5$ is assumed. For all hadron resonances, $s/T^3=7$ is used. Both values taken taken from lattice QCD
  simulations, Fig.~\ref{Fig:3}. We find
  that strangeness degrees of freedom are essential for reproducing the
  freeze-out parameters at low incident energies. The condition of constant
  $s/T^3$ have been applied for $T>5~$MeV. At smaller temperatures, the
  HRG is no longer applicable. The graph taken from Ref. \cite{Tawfik:2004sw}. }
\end{figure}

For vanishing free energy, i.e. at the chemical freeze-out, the {\it  equilibrium} entropy gives the amount of energy which can't be used to produce additional work. We can in this context define the entropy as the degree of sharing and spreading the energy inside the equilibrium system. Furthermore, we find that the strangeness degrees of freedom are essential at low collision energies, where the strangeness chemical potential $\mu_S$ is as large as $\mu_B$. According to the strangeness conservation in the heavy-ion collisions, we find that the higher is the collision energy, the smaller is $\mu_S$~\cite{Tawfik:2004sw}.  

We plot in Fig.~\ref{Fig:3}, the lattice results on $s/T^3$ vs. $T/T_c$ at  $\mu_B=0$ for different quark flavors $n_f$~\cite{Karsch:2000kv,Karsch:2003zq}. The quark masses used in the lattice calculations are heavier than their physical masses in vacuum. For a reliable comparison with the lattice QCD, the hadron resonance masses included in HRG have to be re-scaled to values heavier than the physical ones~\cite{Karsch:2003vd,Karsch:2003zq}, section \ref{sec:krt1}. As shown in Fig.~\ref{Fig:3}, HRG
 can very well reproduce the lattice results for the different quark  flavors under this re-scaling condition, section \ref{sec:krt1}. In the same figure, we draw the  results for physical resonance masses as thin curves, i.e. the case if  lattice QCD simulations were done for physical quark masses. The two  horizontal lines point at the values of $s/T^3$ at the critical temperature $T_c$. As mentioned above, the  critical $T_c$ and the freeze-out temperature $T_{ch}$ are assumed to be the same at small $\mu_B$. We find that  $s/T^3=5$ for $n_f=2$ and $s/T^3=7$ for $n_f=2+1$. These  two values will be used in order to describe the freeze-out parameters for different quark flavors. The normalization with   respect to $T^3$ should not be connected with massless particles. Either the resonances in HRG or the quarks on lattice are massive.

In HRG, temperature $T_{ch}$ was calculated at different $\mu_B$
according to constant $s/T^3$. $\mu_B$ can be related to the collision
energy. On the other hand, $\mu_S$, the strangeness chemical potential, has
been calculated in dependence on $\mu_B$ and $T$. The resulting $T$ and $\mu_B$
are plotted in Fig.~\ref{Fig:4} on the top of the freeze-out parameters
($T_{ch}$ and $\mu_B$) which, as mentioned above, have been estimated by
thermal fits of various ratios of the particle yields produced in different
heavy-ion collisions. The dotted curve represents our results for
$n_f=2$. In this case, only the non-strange hadron resonances are included in
the partition function. The entropy is given by
$\partial T\ln{Z}(T,\mu_B,\mu_S)/\partial T$. The condition applied  
in this case is $s/T^3=5$. The solid curve gives $2+1~$results (two light
quarks plus one heavy strange quark). Here all resonances are included in
the partition function and the condition reads $s/T^3=7$. We find that both
conditions can satisfactory describe the freeze-out parameters at high
collision energy. We notice that the $n_f=2$ curve does not go through
the SIS data points. Therefore, we can conclude that the non-strange
degrees of freedom alone might be non-sufficient at the SIS energy.

It is worthwhile to notice that both entropy density $s$ and the corresponding
temperature $T_{ch}$ decrease with increasing $\mu_B$. The entropy density
is much faster than $T$, so that the ratio $s/T^3$ becomes greater than $7$
at very large $\mu_B$. In this limit, {\it thermal} entropy density $s$ is
expected to vanish, since it becomes proportional to $T$. The {\it quantum}
entropy~\cite{Miller:2003ha,Miller:2003hh,Miller:2003ch,Miller:2004uc,Hamieh:2004ni,Miller:2004em} is entirely disregarded in these calculations.

In rest frame of produced particle, the hadronic matter can be determined by constant degrees of freedom, for instance, $s/T^3 (4/\pi^2)=const$ \cite{Tawfik:2005qn,Tawfik:2004ss}. The quantity $const$ is assigned to $5$ and $7$ for two and three quark flavors, respectively. The chemical freeze-out is related to the particle creation. Therefore, the abundances of different particle species are controlled by the chemical potential, which obviously depends on $T$. With the beam energy, $T$ is increasing, while the baryon densities at mid-rapidity decreases. The estimation of the macroscopic parameters of the chemical freeze-out can be extracted from particle ratio. These parameters collected over the last three decades seem to fellow regular patterns as the beam energy increases \cite{jean2006,Tawfik:2005qn,Tawfik:2004ss}. In section \ref{sec:fo-hm}, the higher order moments have been suggested to control the chemical freeze-out, so that several conditions have been proposed \cite{HM_FO1,HM_FO2,HM_FO3,HM_FO4}.

\subsubsection{Tawfik: vanishing product of kurtosis and susceptibility, $\kappa\; \sigma^2$}
\label{sec:fo-hm}

In section \ref{sec:hm}, the higher order moments of multiplicity shall be studied. The normalization is done with respect to the standard deviation $\sigma$, which is related to the susceptibility $\chi$ and correlation length $\xi$ \cite{Tawfik:2013dba}. Therefore, the normalization provides with a tool to relate moments to experimental measurements.  The normalization of $4-$th order moment known as heteroskedacity or kurtosis $\kappa$ means varying volatility or more accurately, varying variance.  Actually, the kurtosis is given by normalized $4-$th order moment minus $3$. The subtraction of $3$, which arises from the Gaussian distribution will be elaborated in section \ref{sec:fo-hm}.

To scale the correlation functions, there are several techniques, for instance, the survey system's optional statistics module including product moment correlation. This module includes the so-called partial correlation which seems to be useful when the relationship between two variables is to be highlighted, while effect of one or two other variables can be removed \cite{Tawfik:2013dba}. That certain products can be directly connected to the corresponding susceptibilities in Lattice QCD simulation and related to long range correlations, we utilize product moment, section \ref{sec:mult}. 
 
 \begin{figure}[htb]
\includegraphics[angle=-90,width=10.cm]{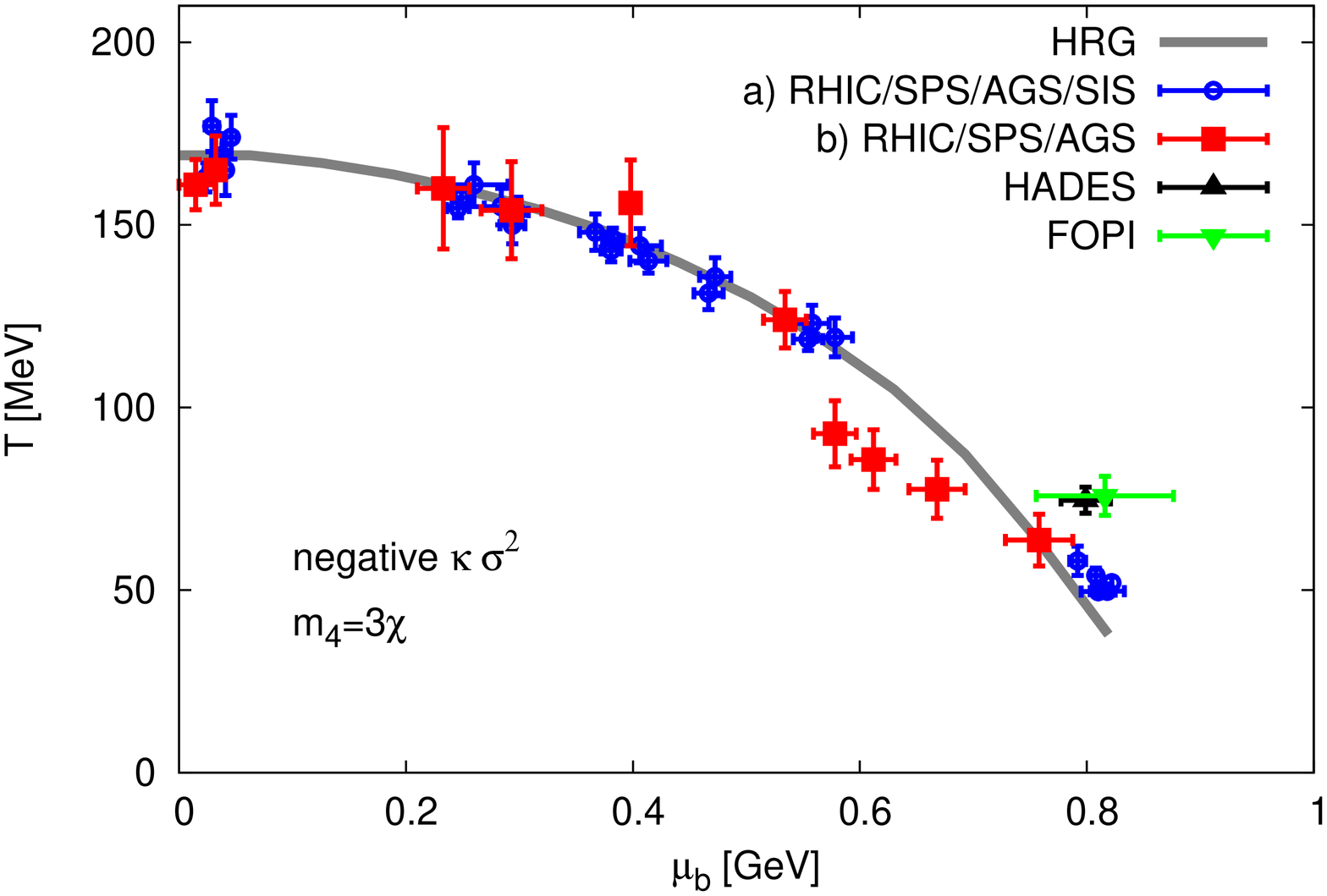}
\caption{The chemical freeze-out boundary is given in dependence on the temperature $T$. The experimental data are presented by the solid circles. The curves represent the results from the HRG model, which are determined when the sign of ${\kappa}\, \sigma^2$ is flipped. The filled circles are taken from \cite{jean2006} while filled squares are quoted from \cite{dataCR}. The upwards and downwards triangle give HADES \cite{hades} and FOPI \cite{fopi} results, respectively. Graph taken from Ref. \cite{Tawfik:2012si,Tawfik:2013dba}.}
\label{fig:fezeout-sT3} 
\end{figure}

As will be shown in Fig. \ref{fig:kS2muu}, the thermal evolution of ${\kappa}\,\sigma^2$ slowly decreases. It vanishes and even changes its sign. This result seems to update previous studies \cite{lqcd1a,HM_FO2,r42}, where ${\kappa}\,\sigma^2$ was assumed to remain finite and positive with increasing $\mu$. In the present work, we find that the sign of ${\kappa}\,\sigma^2$ is flipped at high $T$ \cite{HM_FO1}. Furthermore, we find that the $T$  and $\mu$ parameters, at which the sign is flipped are amazingly coincide with the ones of the chemical freeze-out. 
Vanishing $\kappa\, \sigma^2$ for boson and fermion, respective, reads
\bea 
\int_0^{\infty}  \left\{\text{cosh}\left[\frac{\varepsilon_i
      -\mu_i }{T}\right] + 2\right\}\; \text{csch}\left[\frac{\varepsilon_i
      -\mu_i }{2\, T}\right]^4 \; k^2\,dk 
&=& \frac{3\, g_i}{\pi^2}\, \frac{1}{T^3}\, \left[\int_0^{\infty} \left(1-\text{cosh}
  \left[\frac{\varepsilon_i -\mu_i }{T}\right]\right)^{-1}\; k^2\,dk\right]^2, \hspace*{7mm}\label{eq:lsigma2g} \\
\int_0^{\infty}  \left\{\text{cosh}\left[\frac{\varepsilon_i
      -\mu_i }{T}\right] - 2\right\}\; \text{sech}\left[\frac{\varepsilon_i
      -\mu_i }{2\, T}\right]^4 \; k^2\,dk 
&=& \frac{3\, g_i}{\pi^2}\, \frac{1}{T^3}\, \left[\int_0^{\infty} \left(\text{cosh}
  \left[\frac{\varepsilon_i -\mu_i }{T}\right]+1\right)^{-1}\; k^2\,dk\right]^2. \hspace*{10mm}\label{eq:lsigma2fc}
\eea
The rhs and lhs in both expressions can be re-written as 
\bea
16  \frac{\pi^2}{g_i} T^3\, m_4(T,\mu) &=& 48 \frac{\pi^2}{g_i} T^3\, \chi^2(T,\mu),
\eea
which is valid for both bosons and fermions. Then, the chemical freeze-out is defined, if the condition
\bea
m_4(T,\mu) &=& 3\, \chi^2(T,\mu),
\eea 
is fulfilled. At the chemical freeze-out curve, a naive estimation leads to $\xi\sim 3^{1/3}~$fm. In doing this, it is assumed that the proportionality coefficients of  $\kappa \sim \xi^7$ and $\chi\sim\xi^2$, are equal.  An estimation for $\xi$  in the heavy-ion collisions has been reported \cite{xxii2}. Near a critical point, the experimental value $\sim 2-3\,$fm (only factor 3 larger) agrees well with our estimation.

At the chemical freeze-out curve, the intensive parameters $T$ and $\mu$ which are related to the extensive properties entropy and particle number, respectively, have to be determined over a wide range of beam energies. Fig. \ref{fig:fezeout-sT3} collects a large experimental data set. For a recent review, we refer to \cite{jean2006} (filled squares) and the references therein. The filled circles are taken from Ref. \cite{dataCR}. The upwards and downwards triangle represent HADES \cite{hades} and FOPI \cite{fopi} results, respectively. The solid curve represents a set of $T$ and $\mu$, at which ${\kappa}\,\sigma^2$ vanishes as calculated in HRG \cite{Tawfik:2012si,Tawfik:2013dba}. It is obvious that this curve reproduces very well the experimental data. As given above, at this curve the normalized fourth order moment $\kappa$ is equal to three times the squared susceptibility $\chi$.  This new condition seems to guarantee the condition introduced in \cite{Tawfik:2005qn,Tawfik:2004ss}; $s/T^3=const.$ over the range $0<\mu<0.8~$GeV.

\subsubsection{Tawfik: constant-trace anomaly, $(\epsilon-3 p)/T^4$}
\label{taw:It4}

The QCD trace anomaly $(\epsilon-3 p)/T^4$ also known as the interaction measure  can be derived from the trace of the energy-momentum tensor, $T_{\mu}^{\nu}=\epsilon-3 p$ and apparently is conjectured to be sensitive to the presence of massive hadronic state. For instance, for non-interacting hadron gas at vanishing chemical potential \cite{Tawfik:2013eua}
\bea
\frac{\epsilon-3 p}{T^4} &=& \int_0^{\infty} dm \, \rho(m) \int \frac{d^3 p}{(2 \pi)^3} \frac{E^2-p^2}{E} \exp(-E/T) \simeq \frac{1}{2 \pi^2} \int_0^{\infty} \left(\frac{m}{T}\right)^3 \, \rho(m)  \, K_1\left(\frac{m}{T}\right)\, dm, \label{eq:Imuel1}
\eea
where $\rho(m)$ is the mass spectrum \cite{hgdrnA,hgdrnB}. In the classical limit, the trace anomaly at finite chemical potential $\mu$ reads
\bea
\frac{\epsilon-3 p}{T^4} &\sim &  \frac{g}{2 \pi^2} \, e^{\mu/T} \, \left(\frac{m}{T}\right)^2 \left[\frac{m}{T}\, K_1\left(\frac{m}{T}\right)\right].
\eea

The QCD equation of state can be deduced from the energy-momentum tensor, for instance, the normalized pressure can be obtained for the integral of $(\epsilon-3 p)/T^5$.  For completeness, we mention that the trace anomaly gets related the QCD coupling constant so that $I(T)/T^4\propto T^4\, \alpha_s^2$ \cite{peter}, where $I(T)=\epsilon(T)-3 p(T)$. In light of this, essential information about weakly coupled systems would be provided through trace anomaly. 

A universal parametrization for the QCD trace anomaly at $\mu=0$ was proposed \cite{lgtFdr12}
\bea \label{eq:parmQCD}
\frac{I(T,\mu_b=0)}{T^4} &=& e^{-\frac{h_1}{t}-\frac{h_2}{t^2}} \left\{h_0+\frac{f_0 \left[\tanh(f_1 t+f_2)\right]+1}{1+g_1 t+g_2 t^2}\right\},
\eea
where $t=T/200$. The fitting parameters are listed in Ref. \cite{lgtFdr12}. This gives the thermal evolution of $I(T)$. A parametrization at finite  $\mu$ was suggested \cite{OweRev}
\bea
\label{eq:OweP1}
\frac{I(T,\mu_b)}{T^4} &=& \frac{I(T,\mu_b=0)}{T^4} + \frac{1}{2} \frac{\mu_b^2}{T} \frac{\partial \chi_2(T,\mu_b)}{\partial T},
\eea
where the susceptibility $\chi_2(T,\mu_b)$  is given by the second derivative of the partition function, 
\bea
\chi_2(T,\mu_b) &=& \frac{1}{T}  \frac{\partial^2 \ln Z(T,\mu_b)}{\partial^2 \mu_b}.
\eea
In the classical limit, the derivative of $\chi_2(T,\mu_b)$ with respect to temperature reads
\bea
\label{eq:My1}
\left.\frac{\partial \chi_2(T,\mu_b)}{\partial T}\right|_{\mu_b=0} &=& \frac{\chi_2(T,\mu_b)}{T^2} - \frac{g}{4\pi^2}\; e^{\mu_b/T}\; T \left(\frac{m}{T}\right)^3  \left[K_1\left(\frac{m}{T}\right) + K_3\left(\frac{m}{T}\right)\right].
\eea
When implementing the result in Eq. (\ref{eq:OweP1}), we get  \cite{Tawfik:2013eua}
\bea \label{eq:Icls}
\frac{I(T,\mu_b)}{T^4} &=& \frac{I(T)}{T^4} - \frac{\chi_2(T,\mu_b)}{T^2}\, \frac{\mu_b^3}{2 T} + \frac{g}{8 \pi^2}\; e^{\mu_b/T}\; \mu_b^2 \left(\frac{m}{T}\right)^3  \left[K_1\left(\frac{m}{T}\right) + K_3\left(\frac{m}{T}\right)\right].
\eea

\begin{figure}[htb]
\centering{
\includegraphics[angle=-90,width=10cm]{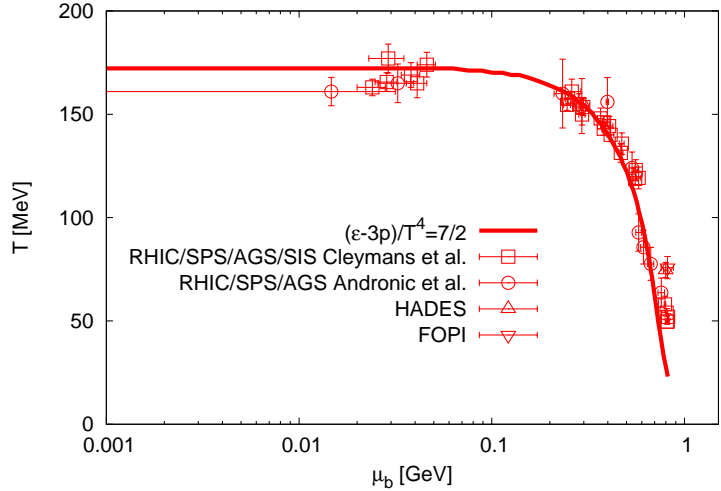}
\caption{The freeze-out parameters ($T$ vs. $\mu_b$) are compared with constant-trace anomaly calculated in HRG model (solid curve). Symbols with error bars are the phenomenologically estimated results (see text). Graph taken from Ref. \cite{Tawfik:2013eua}.}
\label{fig:e3pFO} 
}
\end{figure}

In grand canonical ensemble at  $T<T_c$
\bea
\frac{I(T,\mu_b)}{T^4} &=& \frac{I(T)}{T^4} \pm \frac{1}{2}\left(\frac{\mu_b}{T}\right)^2 \frac{\chi_{2}(T,\mu_b)}{T^2} 
+ \frac{1}{2}\left(\frac{\mu_b}{T}\right)^2 \frac{g}{2 \pi ^2} \frac{1}{T^4} \int _0^{\infty}  e^{\mu_b/T} \left[\varepsilon-\mu_b\right] \frac{F(T,\mu_b)}{\left(e^{\mu_b/T} \pm e^{\varepsilon/T}\right)^3}  p^2 dp, \hspace*{10mm} \label{eq:OweP3}
\eea
where
\bea
F(T,\mu_b) &=& \left\{ 
 \begin{array}{l l} 
 2 e^{2 \mu_b/T} + e^{(\varepsilon+\mu_b)/T} + e^{2 \varepsilon/T} & \text{for bosons}\\
  & \\
  3 e^{(\varepsilon+\mu_b)/T} - 4 e^{2 \mu_b/T} - e^{2 \varepsilon/T} & \text{for fermions} \\
 \end{array} 
\right..  \nonumber
\eea
The coefficient of $(\mu_b/T)^2$ seems to play a crucial role in estimating the chemical parameters, $T$ and $\mu_b$. The second term of Eq. (\ref{eq:OweP3}) can be decomposed into bosonic and fermionic parts
\bea
\pm\frac{1}{2}\left(\frac{\mu_b}{T}\right)^2\; \frac{\chi_{2}(T,\mu_b)}{T^2} &=& \frac{1}{2}\left(\frac{\mu_b}{T}\right)^2\; \left[ \frac{\chi^{(B)}_{2}(T)}{T^2} - \frac{\chi^{(F)}_{2}(T,\mu_b)}{T^2}\right],
\eea
revealing that the fermionic susceptibility  is to a large extend responsible for the $T-\mu_b$ curvature.

\begin{figure}[htb]
\includegraphics[angle=0,width=10.cm]{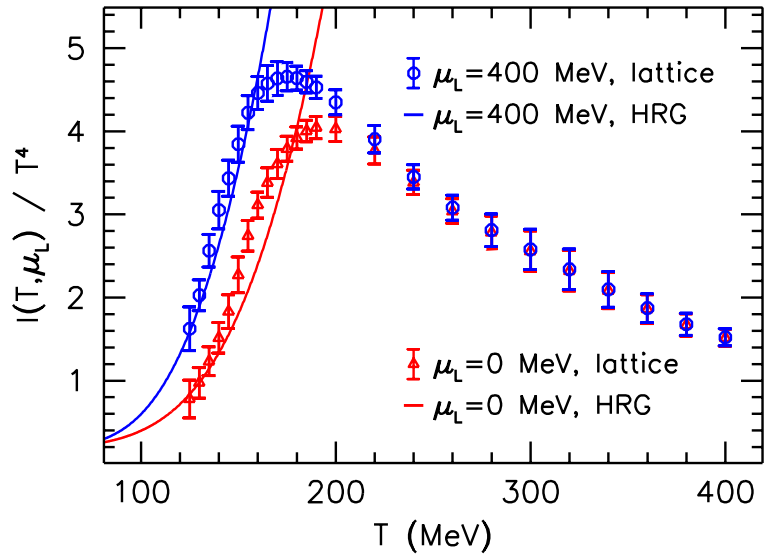}
\caption{The trace anomaly vanishing and finite light quark chemical potential $\mu_L$ as functions of $T$. Graph taken from Ref.  \cite{lgtFdr12}. }
\label{fig:lqcdI}
\end{figure}

In Fig. \ref{fig:e3pFO}, the freeze-out parameters calculated in the HRG model are plotted in a log-log graph (solid curve) \cite{Tawfik:2013eua}. The symbols with error bars represent the phenomenologically estimated parameters known as experimental data \cite{jean2006,dataCR}. They cover a center-of-mass energy ranging from couple GeV in the SchwerIonen Synchrotron (SIS) to several TeV, in the LHC. The HADES \cite{hades} and FOPI \cite{fopi} results are also illustrated.  

Fig. \ref{fig:e3pFO} presents the freeze-out diagram relating $T$ to $\mu_b$ (symbols with error bars). It is assumed that constant-trace anomaly is able to reproduce the freeze-out diagram, $T$ versus $\mu_b$. The condition that $I(T,\mu)/T^4=7/2$, Fig. \ref{fig:lqcdI} is motivated by recent lattice QCD calculations \cite{lgtFdr12}.

\subsection{Comparing with lattice QCD simulations}
\label{sec:lqcd}

As introduced on top of this section \ref{sec:hic1}, the lattice QCD simulations turn to be compatible with experiments. About ten years ago, the challenge of confronting HRG, section \ref{sec:hrg1}, with the lattice QCD results was raised. It intended to learn about the critical conditions near deconfinement.  Lattice calculations suggest that this transition is a true phase transition only in small quark mass intervals in the light and heavy quark mass regime, respectively. But, in a broad intermediate regime, the transition is not related to any singular behavior of the QCD partition function. Nonetheless, the transition temperature  is well defined through the location of maxima in response functions such as the chiral susceptibility. A collection of transition temperatures obtained in calculations with $2$ and $3$ quark flavors with degenerate masses is shown in Fig.~\ref{fig:tc_pion}.

In order to use HRG for a further comparison with lattice results, we should take into account that lattice calculations were generally performed with quark masses heavier than those realized in nature \cite{Karsch:2003vd,Karsch:2003zq,Redlich:2004gp}.
Rather than converting the bare quark masses used in lattice calculation into a renormalized mass, it is much more convenient to use directly the pion mass ($m_\pi \sim \sqrt{m_q}$), i.e. the mass of the Goldstone particle, as a control parameter for the quark mass dependence of the hadron spectrum. In section \ref{sec:krt1}, the quark mass dependence of the hadron spectrum on the lattice was examined.

\subsubsection{Thermodynamics at vanishing chemical potential}
\label{sec:QCDmu0}

\begin{figure}[htb]
\centering{
\includegraphics[width=7cm,angle=-90]{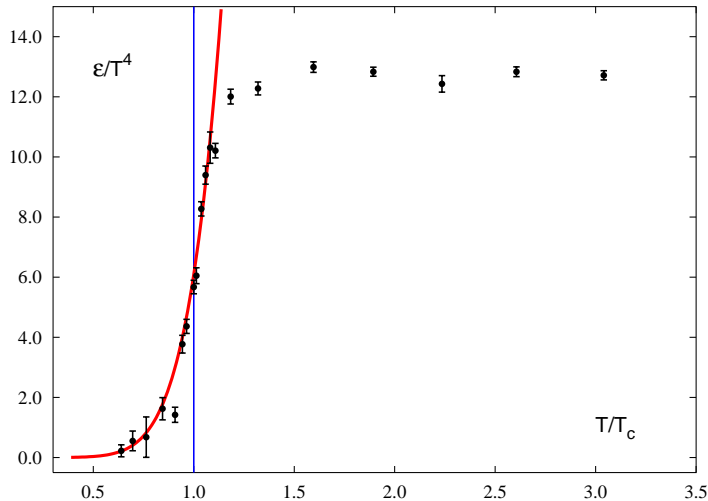}
\caption{The energy density $\epsilon$ in units of $T^4$ calculated on the lattice with (2+1) quark flavors  as a function of the $T/T_c$ ratio. The vertical lines indicate the position of the critical temperature. The full-lines are  the  results of the  hadron resonance gas model that accounts for all mesonic and baryonic resonances. The graph taken from Ref. \cite{Karsch:2003vd}.}
\label{fig:epslHRG1} 
}
\end{figure}

\begin{figure}[htb]
\centering{
\includegraphics[width=7cm,angle=-90]{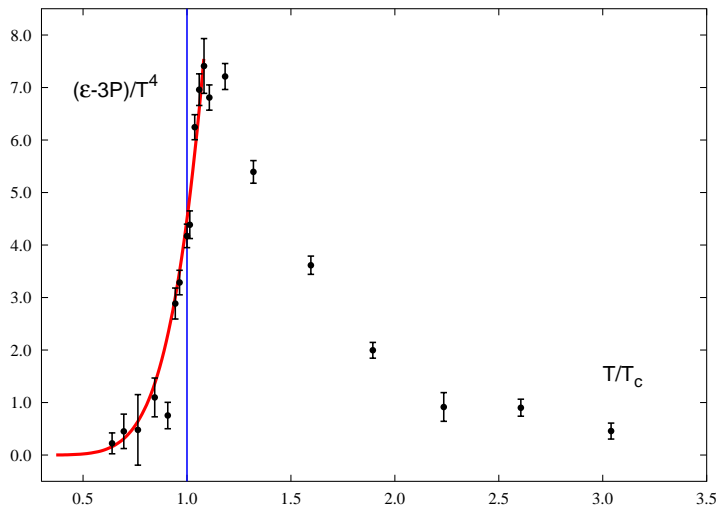}
\caption{The interaction measure  $(\epsilon -3P)/T^4$ in units of $T^4$ calculated on the lattice with (2+1) quark flavors  as a function of the $T/T_c$ ratio. The full-lines are  the  results of the  hadron resonance gas model that accounts for all mesonic and baryonic resonances. The graph taken from Ref. \cite{Karsch:2003vd}.}
\label{fig:IHRG1} 
}
\end{figure}

Long time ago, the thermodynamics of strongly interacting matter at vanishing baryon number density or chemical potential has been studied in lattice QCD calculations \cite{schladmingA,schladmingB}. As discussed in section \ref{sec:hrg1}, the basic quantity required to verify thermodynamic properties of QCD is the partition function. At vanishing chemical potential and charge neutral systems, the grand canonical partition function can be given as \cite{Karsch:2003vd}
 \be
Z(T,V) = {\rm Tr}[e^{-\beta H }] \quad ,
\label{eq3}
 \ee
where  $H$ is the Hamiltonian of the system and $\beta =1/T$ is the inverse temperature. The confined phase of QCD is to be modelled as a non-interacting gas of resonances. To do so, we use as Hamiltonian the sum of kinetic energies of relativistic Fermi and Bose particles of mass $m_i$, which contains the contributions from all resonances with masses below $\simeq2~$GeV. Summing up in Eqs.~(\ref{eqq1}), (\ref{eqq3a}) and (\ref{eqq3b}) for instance, the contributions from experimentally known hadronic states, constitutes the resonance gas model for the thermodynamics of the low temperature phase of QCD
\be
 \ln Z(T,V)=\sum_i \ln Z_i^1(T,V). \label{eqq1} 
\ee 
For particles of mass $m_i$ and  spin-isospin degeneracy factor $g_i$ the one-particle partition function
$Z^1_i$ is given by \cite{Karsch:2003vd}
\be 
\ln Z^1_i(T,V)= \frac{Vg_i}{2\pi^2} \int_0^\infty
dpp^2\eta\ln (1+\eta e^{-\beta E_i}), \label{eqq2} 
\ee 
where $E_i=\sqrt {p^2+m_i^2}$ is the dispersion relation and $\eta= -1$ for bosons and $+1$ for fermions. The energy density and pressure, respectively, reads
\bea
 \epsilon &=&\sum_i\epsilon_i^1, \label{eqq3a}\\
 P &=& \sum_i P_i^1. \label{eqq3b}
 \eea
They are also to be summed over single particle contributions $\epsilon_i^1$ and $P_i^1$, respectively,
\begin{eqnarray}
{{\epsilon_i^1}\over {T^4}} &=& \frac{g_i}{2\pi^2}\;
\sum_{k=1}^{\infty} \;(-\eta)^{k+1}\;\frac{(\beta m_i)^3}{k}
\;\left[\frac{3\;K_2(k\beta m_i)}{k\beta m} +
  \;K_1(k\beta m_i)\right]\label{eqq4} \\
\Delta_i^1&\equiv&\frac{\epsilon_i^1-3P_i^1}{T^4} = \frac{g_i}{2
\pi^2}\; \sum_{k=1}^{\infty} \;(-\eta)^{k+1}\;\frac{(\beta
m_i)^3}{k} \; K_1(k\,\beta m_i) \label{eqq5}
\end{eqnarray}
where $K_1$ and $K_2$ are modified Bessel functions.

The energy density obtained in this way starts rising rapidly at a temperature of about 160~MeV. In Figs.~\ref{fig:epslHRG1} and \ref{fig:IHRG1},  the temperature dependence of the energy density $\epsilon/T^4$, Eq.~(\ref{eqq4}), and the interaction measure $\Delta/T^4\equiv(\epsilon-3 p)/T^4$ , Eq.~(\ref{eqq5}), respectively, is compared with the lattice QCD calculations \cite{karsch1} for (2+1) quark flavors. The HRG model and the lattice data agree quite well. This indicates that for $T\leq T_c$ hadronic resonances are indeed the most important degrees of freedom in the confined phase. The energy density in the resonance gas reaches a value of 0.3~GeV/fm$^3$ at $T\simeq 155$~MeV and 1~GeV/fm$^3$ already at $T\simeq 180$~MeV. This is in good agreement with lattice calculations, which find a critical energy density of about $0.7~$GeV/fm$^3$ at $T_c\simeq 170$~MeV \cite{karsch1}. For comparison, we note that a simple pion gas would only lead to an energy density of about $0.1~$GeV/fm$^3$ at this temperature. This suggests that a more quantitative comparison between numerical results obtained from lattice calculations and the resonance gas model might indeed be meaningful.

\subsubsection{Thermodynamics at finite chemical potential}
\label{sec:QCDmun0}

Since about ten years, the first investigations of the equation of state at non-vanishing quark chemical potential, $\mu_q$, have started \cite{fodor,gavai,ejiri}. These studies of bulk thermodynamics have been performed with different lattice actions and also have used different methods; exact matrix inversion \cite{fodor} or Taylor expansion \cite{gavai,ejiri}, to extend previous calculations performed at $\mu_q=0$ into the region of $\mu_q>0$. Nonetheless, they led to qualitatively and even quantitatively similar results.

The discussion of the thermodynamics of the hadronic phase of QCD is limited in the regime of low baryon number density, $\mu_q /T \leq 1$, but high temperature, $T\sim T_c(\mu_q=0)$  \cite{Karsch:2003zq}. It intends to compare lattice QCD calculations at finite chemical potential with the predictions of HRG calculations. In lattice QCD calculations, the Taylor expansion for small $\mu_q/T$ was utilized \cite{ejiri}. Unlike the approach based on an exact inversion of the fermion determinant \cite{fodor}, the Taylor expansion, obviously, has the disadvantage of being approximate. There is, however, good reason to expect, that at least at high temperature, the contribution of terms that are beyond ${\cal O}((\mu_q/T)^4)$ order  is small. The expansion coefficients themselves provide useful information on the relevant degrees of freedom controlling the density dependence of thermodynamic quantities. It was argued that baryons and their resonances are these relevant degrees of freedom that govern thermodynamics in the confined phase at finite density \cite{Karsch:2003zq}. Tor $T\leq T_c$, it was shown that EoS at non-zero  chemical potential which has been obtained in lattice calculations can be well described by a baryonic  HRG when using the same set of approximations as used in current lattice studies \cite{Karsch:2003zq}. The importance of truncation effects in the Taylor expansion was  examined and  the influence of non-physically large quark mass values on thermodynamic  observables  and the critical conditions for deconfinement was discussed.

The distortion of the hadron mass spectrum due to non-physically large quark  masses, $m_q$ has to be deduced from lattice calculations at zero temperature \cite{Karsch:2003zq}. A generic feature of such studies is that the deviation from the physical mass value due to non-physically large values of the quark  mass becomes smaller for heavier hadronic states \cite{data1}. Moreover, it was found \cite{data3,schi} that the quark mass dependence is  well parametrized through Eq. (\ref{eq:mhparam}). The masses of only a few baryonic states constructed from $(u, d)$-quarks have been  studied in more detail on the lattice \cite{data1,data3,schi}. 

A quadratic parametrization of the quark mass dependence of baryon masses 
\bea
(m_H a)^2 &=& (m_H a)^2_{phys} + b (m_\pi a)^2, \label{eq:mhparam}
\eea
where $(m_Ha)_{phys}$ denotes the physical mass value of a hadron expressed in lattice units and $(m_Ha)$ is the value calculated on the lattice for a certain value of the quark mass or equivalently a certain value of the pion mass, $m_\pi^2 \sim m_q$ \cite{Karsch:2003zq}, 
shows at least for nucleon, delta and their parity partners   only a weak dependence on the hadron mass.
We thus take this as a general ansatz for  the parametrization of the dependence of baryon masses on the pion mass \cite{Karsch:2003zq}
\begin{equation}
{{m (m_\pi)}\over m_H}\simeq 1+A\; {{m_\pi^2 }\over {m_H^2}} \quad ,
\label{par}
\end{equation}
where $m(m_\pi)$ is the distorted hadron mass at fixed $m_\pi$ and $m_H$ is its corresponding physical value. This parametrization is consistent with the previous analysis \cite{Karsch:2003vd}, where  the MIT bag model was used in order to determine the $m_\pi$-dependence of hadron masses.

The pressure may be expanded in a power series,
\begin{equation}
{p(T,\mu_q) \over T^4} = \sum_{n=0}^{\infty} c_{2n}(T) \left( {\mu_q \over
T} \right)^{2n} \quad.\label{series}
\end{equation}
This series has been analysed up to  ${\cal O} ((\mu_q / T)^4)$ and  in addition to the density dependent change of the pressure, $\Delta p$, quantities like the quark number density, $n_q$, and quark number susceptibility, $\chi_q$, have been calculated. The latter are obtained from Eq.~(\ref{series}) through appropriate derivatives with respect to the quark chemical potential, 
\begin{eqnarray}
{\Delta p \over T^4} &=& {p(T,\mu_q) - p(T,0) \over T^4} \simeq
c_2(T) \left( {\mu_q \over T} \right)^2 +
c_4(T) \left( {\mu_q \over T} \right)^4 \quad ,  \nonumber\\
{n_q \over T^3} &=& {\partial\; p(T,\mu_q) \over \partial\; \mu_q}
\simeq 2\; c_2(T) \left( {\mu_q \over T} \right) +
4\; c_4(T) \left( {\mu_q \over T} \right)^3 \quad , \label{seriesO4} \\
{\chi_q \over T^2} &=& {\partial^2\; p(T,\mu_q) \over \partial\;
\mu_q^2} \simeq 2\; c_2(T) + 12\; c_4(T) \left( {\mu_q \over T}
\right)^2 \quad . \nonumber 
\end{eqnarray}

\begin{figure}[htb]
\centering{
\vskip -.4cm
 \includegraphics[width=10cm,height=9cm,angle=180]{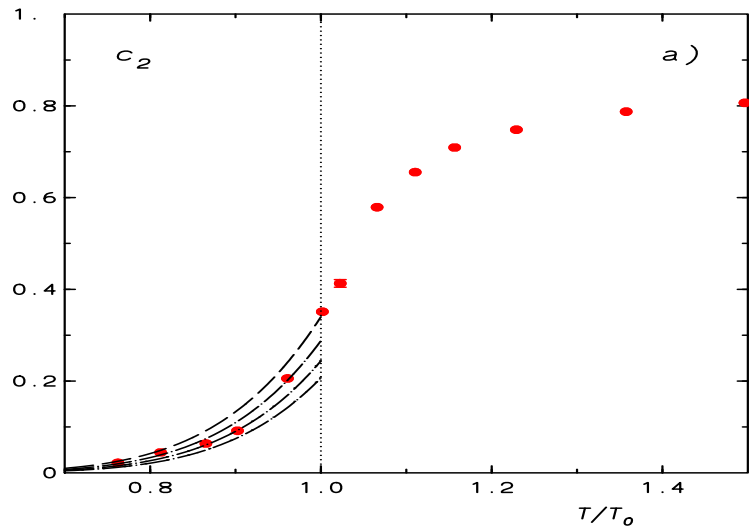} \vspace*{-1cm}
\caption{The temperature dependence of the second order expansion coefficients $c_2$. The temperature scale is given in units of the transition temperature at $\mu_q=0$, which for the quark masses used in the QCD calculation \cite{karsch1} is $T_0 \simeq 200~$MeV.  The dashed-dotted curves  show results of a resonance gas model calculation for $A=0.9, 1.0, 1.1, 1.2$ (from top to bottom). The graph taken from Ref. \cite{Karsch:2003zq}. }
\label{fig:c2}
}
\end{figure}

\begin{figure}[htb]
\centering{
\vskip -.4cm
 \includegraphics[width=10cm,height=9cm,angle=180]{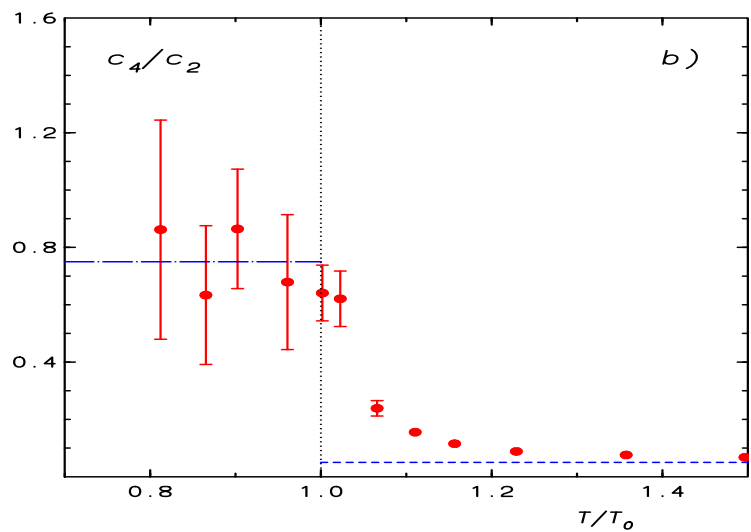} \vspace*{-1cm}
\caption{The same as in Fig. \ref{fig:c2}, but here for the temperature dependence of the ratio  $c_4(T)/c_2(T)$. The graph taken from Ref. \cite{Karsch:2003zq}.}
 \label{fig:c2c4} 
 }
\end{figure}

In the asymptotically high temperature limit, the expansion coefficients are then given by $c_2(\infty)= n_f/2$ and $c_4(\infty)/c_2(\infty)=1/2\pi^2$ respectively \cite{Karsch:2003zq}. The high temperature expansion of  $O(g^2)$  terminates also at $O(\mu_q/T)^4$. This, however, changes in the resumed $O(g^3)$ contribution. The complete expansion up to $O(g^6\ln g)$ has been presented \cite{refv}.  Thus, the ratio $c_4/c_2$ should remain small. Consequently, the leading order term dominates in the Taylor expansion for a wide range of values for $\mu_q/T$. The lattice QCD results for the expansion coefficients for $2$ quark flavors are shown in Figs.~(\ref{fig:c2})~and~(\ref{fig:c2c4}). It can be seen that  at $T\simeq 1.5 \, T_0$ the numerical results for $c_2(T)$ still deviate by about $20\%$ from the ideal gas value while the ratio $c_4/c_2$ is already close to the corresponding result expected in the infinite temperature limit. 

\begin{figure}[htb]
\centering{
\vskip -.4cm
\includegraphics[width=10cm,height=9cm,angle=180]{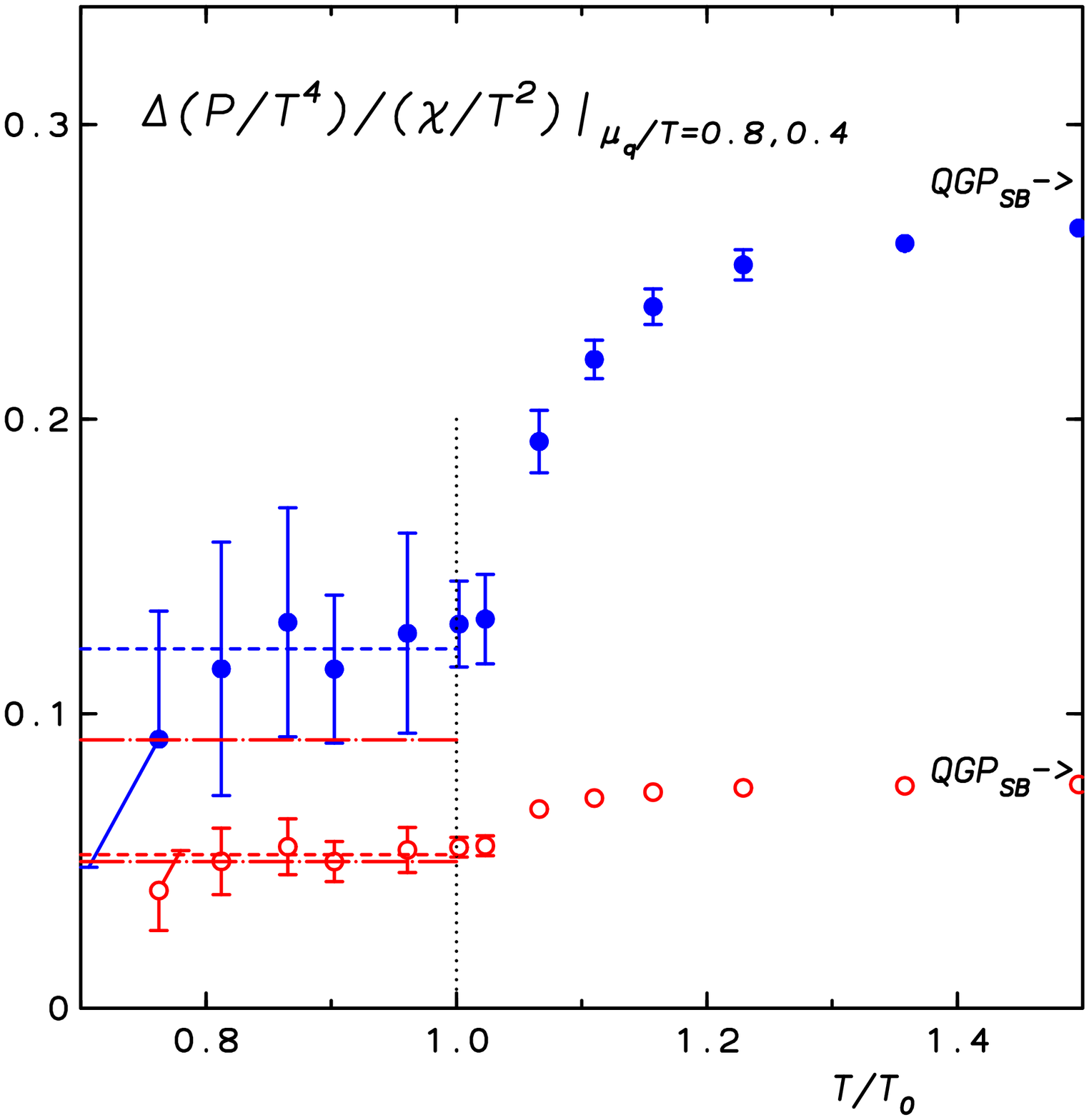}\vskip -1cm 
\caption{The ratio of pressure and quark number susceptibility versus temperature for fixed values of $\mu_q / T$. The horizontal lines are the results of HRG model. The points are the lattice calculations~\protect\cite{ejiri}. While the dashed-dotted curves represent the complete expression, the dashed curves give the result of a Taylor expansion to the
same order as that used in the lattice calculations. The graph taken from Ref. \cite{Karsch:2003zq}.}
\label{fig:ratio}
}
\end{figure}

The partition function of a resonance gas can be specified through the mass spectra for the mesonic and baryonic sectors of QCD, respectively. In a non-interacting resonance gas the partition function reads,
\be
 \ln Z(T,\mu_B, V)=\sum_{i\; \in\; {\rm mesons}} \ln Z_{m_i}^B(T,V) +
\sum_{i\; \in\; {\rm baryons}} \ln Z_{m_i}^F(T,\mu_B, V) \quad ,
\label{eqq1} 
\ee 
where $Z_{m_i}^B~(Z_{m_i}^F)$ denote single particle partition functions for bosons and fermions with mass $m_i$ and $\mu_B=3\mu_q$ is the baryon chemical potential.  Here the fermion partition function contains the contribution from a particle and its anti-particle. The total pressure of the resonance gas is built up as a sum of contributions from particles of mass $m_i$. The dependence of the pressure on the chemical potential at a fixed temperature is thus entirely due to the baryonic sector. The contribution, $p_m$, of baryons of mass $m$ to the total pressure is given by
\begin{equation}
\left.\frac{p_m}{T}\right. = \frac{d}
{2\pi^2} \int_0^\infty \!\!\!dk\,k^2
\ln\left[(1+z\exp\{-\varepsilon(k)/T\})
(1+z^{-1}\exp\{-\varepsilon(k)/T\})\right]
\label{pressure_con}
\end{equation}
where  $z\equiv\exp\{\mu_B/T\}$ is the baryonic fugacity with $\mu_B=3\mu_q$, $d$ is the spin--isospin degeneracy factor  and $\varepsilon(k)=\sqrt{k^2+m^2}$ is the relativistic single particle energy. The pressure  may be expressed in terms of a fugacity expansion as
\begin{equation}
\frac{p_m}{T^4} = \frac{d}{\pi^2} \left( \frac{m}{T} \right)^2
\sum_{\ell=1}^{\infty} (-1)^{\ell+1}\,\ell^{-2}\, K_2(\ell m/T)\,
\cosh (\ell \mu_B/T)\quad ,
\label{bessel}
\end{equation}
where $K_2$ is the Bessel function.

\begin{figure}[htb]
\centering{
\vskip -.6cm
 \includegraphics[width=10cm, height=9cm,angle=180]{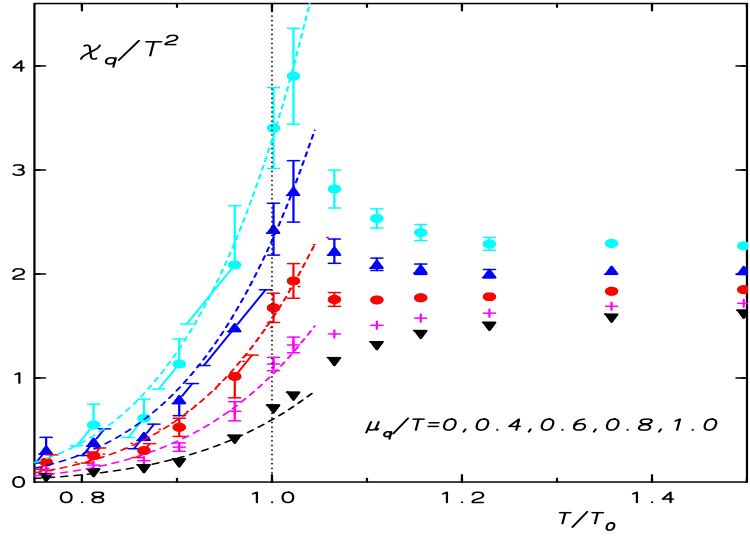}\vspace*{-1cm}
 \caption{The temperature dependence of the quark number susceptibility in 2-flavour QCD calculated in next-to-leading order Taylor expansion for different values of the quark chemical potential ~\cite{ejiri}. The lines are results from the HRG model using a distorted baryon spectrum, Eq.~(\ref{par}) with $A=1$, and treated within the same approximation as in the lattice study. The graph taken from Ref. \cite{Karsch:2003zq}.}
 \label{fig:sus}
 }
\end{figure}

The results for the change in pressure, the quark number density and quark number susceptibility are given in Eq. (\ref{pnchi}).  In order to compare the predictions of  the HRG model  with lattice results one needs to perform the Taylor expansion up to the same  order as given in  Eq.~\ref{seriesO4}. In the Boltzmann approximation, we have
\begin{eqnarray}
{\Delta p\over {T^4}} &=& F(T)[\cosh {{\mu_B}\over T}-1] \simeq
F(T) \left( \tilde{c}_2 \left( {\mu_q \over T}\right)^2
+ \tilde{c}_4 \left( {\mu_q \over T}\right)^4 \right)\quad,  \nonumber \\
{{n_{q}}\over {T^3}} &=& 3 F(T)\sinh {{\mu_B}\over T} \simeq
F(T) \left( 2 \tilde{c}_2 \left( {\mu_q \over T}\right)
+ 4 \tilde{c}_4 \left( {\mu_q \over T}\right)^3 \right) \quad,  \label{pnchi}\\
{{\chi_{q}}\over {T^2}} &=& 9 F(T)\cosh {{\mu_B}\over T} \simeq
F(T) \left( 2 \tilde{c}_2 + 12 \tilde{c}_4 \left( {\mu_q \over T}\right)^2
\right)\quad, \nonumber
\end{eqnarray}
with $\tilde{c}_2 = 9/2$ and $\tilde{c}_4=27/8$. In HRG model the expansion coefficients introduced in Eq.~\ref{seriesO4} are given by $c_{2n}= \tilde{c}_{2n} F(T)$. We note that ratios of these quantities indeed are independent of the resonance mass spectrum and only depend on the chemical potential. For fixed $\mu_q/T$ we thus expect to find that any ratio of two of the above quantities is temperature independent in the hadronic phase. Using the same order of the Taylor expansion as used in the lattice calculations such ratios only depend on $\tilde{c}_4/\tilde{c}_2 = 3/4$, {\it i.e.} the HRG model yields a temperature independent ratio $c_4/c_2$. As can be seen in  Fig.~\ref{fig:c2c4} this is indeed in good agreement with the lattice results. We note that  this result is independent of details of the hadron mass spectrum. It thus  should also   be insensitive to the change in the quark mass used in the lattice calculation. In Fig.~\ref{fig:ratio}, we show the ratio $\Delta p/T^2\chi_q$ for two values of the chemical potential. The good agreement between lattice calculations and the hadronic gas results merely reflects the  agreement found already for the ratio $c_4/c_2$. In the HRG model we can, however, provide also the complete result for this ratio,

The corresponding result for the quark number susceptibility at different values of the quark chemical potential is shown in Fig.~\ref{fig:sus} for the choice $A=1$. The agreement of  HRG model and results obtained from lattice calculations is indeed quite satisfactory. This indicates that the thermodynamics of the confined phase of QCD at finite density is to large extent governed by the baryonic resonances. The  uncertainty in the parametrization of the baryonic mass spectrum, Eq.~\ref{par}, may result in $20 \%$  error on the values of physical observables, i.e. $c_2(T)$, at $T=T_0$, Fig.~\ref{fig:c2c4}.

Fig. \ref{fig:sus} compares with lattice results \cite{ejiri} for quark number susceptibility calculated in next-to-leading order Taylor expansion for different values of the quark chemical potential. The curves are results obtained from the HRG model using a distorted baryon spectrum, Eq.~(\ref{par}), treated within the same approximation as in the lattice study.

\subsubsection{Chiral phase transition}
\label{sec:chiTr}

As introduced in section \ref{sec:krt1}, the contribution to the pressure in a free gas system due to a particle of mass $m_h$, baryon charge $B$, isospin $I_3$, strangeness $S$,  and degeneracy $g$ is given by
\begin{eqnarray}
  \label{p1}
  \Delta p=\frac{g\, m_h^2\, T^2}{2\pi^2} \,
\sum_{n=1}^\infty \,\frac{(-\eta)^{n+1}}{n^2}
\,\exp\left(n\frac{B\mu_B - I_3\mu_I - S \mu_S}{T} \right)
\, K_2\left(n\frac{m_h}{T} \right).
\end{eqnarray}
In the hadronic phase, the isospin is an almost exact symmetry~\cite{hgrTc}.   The quark-antiquark condensates are given by the derivative of pressure with respect to the constituent quark masses \cite{Tawfik:2005qh}
\begin{eqnarray}
\label{qqHRG}
<\bar{q}q>&=&<\bar{q}q>_0+ \sum_h \frac{\partial
  m_h}{\partial m_q} 
\frac{\partial \Delta p}{\partial m_h}, \nonumber \\ 
<\bar{s}s>&=&<\bar{s}s>_0+ \sum_h 
\frac{\partial m_h}{\partial m_s} \frac{\partial \Delta
  p}{\partial m_h}, 
\end{eqnarray}
where $<\bar{q}q>=<\bar{u}u>=<\bar{d}d>$ represents the light quark-antiquark condensate. $<\bar{q}q>_0$ and $<\bar{s}s>_0$ indicate the value of the light and strange quark-antiquark condensates in the vacuum, respectively. 

The computation of the quark-antiquark condensates in Eq.~(\ref{qqHRG}) requires a  modelling for two quantities \cite{Tawfik:2005qh}:
\begin{itemize}
\item The first one is the  strange quark-antiquark condensate at zero temperature and zero chemical potential, $<\bar{s}s>_0$. Using the QCD sum rules, the ratio of strange to light quark-antiquark condensates in vacuum is given by $0.8\pm0.3$~\cite{jamin} (another estimation gives $0.75\pm0.12$~\cite{narison}).  The Gell-Mann-Oakes-Renner relation connects the quark-antiquark condensates at zero temperature and zero chemical  potential to the meson masses and to their decay constants. We use the next-to-leading order result obtained in chiral perturbation theory \cite{GLsu3}: 
\begin{eqnarray}
\label{GOR}
F_K^2 m_K^2 \left(1-\kappa \frac{m_K^2}{F_\pi^2}\right) &=& \frac12 (m_q + m_s)
(<\bar{q}q>_0 + <\bar{s}s>_0),  \\
F_\pi^2 m_\pi^2 \left(1-\kappa \frac{m_\pi^2}{F_\pi^2}\right)
 &=& (m_u+m_d) <qq>_0,  
\end{eqnarray}
where $m_q$ stands for the light quark mass ($m_u=m_d$). The coefficient  $\kappa=0.021\pm0.008$ has been obtained from the low-energy coupling constants of chiral perturbation theory~\cite{GLsu3,jamin}. The coefficient  $\kappa$ does not contain any chiral logarithm. 
The kaon decay constant $F_K$ has to be known. Recently, PDG (http://pdg.lbl.gov/) published estimation for $F_{K^-}=156.1\pm 0.2\pm0.8\pm0.2~$MeV. In framework of lattice QCD, the ratio of heavy to light meson decay constants is found to be given by $F_K/F_{\pi}=1.16\pm0.04$~\cite{lattHeavy,LeutwylerRoos}. Also, PDG measured $F_{K^-}/F_{\pi^-}=1.197\pm0.002\pm0.0076\pm0.001$.
\item The second quantity is the quark mass dependence of the hadron masses. Results from lattice simulations~\cite{Karsch:2003vd,Karsch:2003zq} indicate that
\begin{eqnarray}
\label{lattMass}
\frac{\partial m_h}{\partial (m_\pi^2)}&=&\frac{A}{m_h},  
\end{eqnarray}
where $A\sim0.9 - 1.2$.  
\end{itemize}

Using these results together with the
Gell-Mann-Oakes-Renner relation, Eq.~(\ref{GOR}), one finds
that~\cite{hgrTc}  
\begin{eqnarray}
\frac{\partial m_h}{\partial m_q} &\equiv& 
\frac{\partial (m_\pi^2)}{\partial  m_q}  \frac{\partial
  m_h}{\partial (m_\pi^2)}  = 
\frac{A\;<\bar{q}q>_0 \left(1+2 \kappa \frac{m_\pi^2}{F_\pi^2}\right)}
{F_{\pi}^2\;m_h}. \label{mhmq} \\   
\frac{\partial m_h}{\partial m_s} &\equiv& \frac{\partial (m_K^2)}{\partial
  m_s} \frac{\partial (m_\pi^2)}{\partial (m_K^2)} \frac{\partial
  m_h}{\partial (m_\pi^2)} =
\frac{A\;<\bar{q}q>_0 \left(1+2 \kappa \frac{m_\pi^2}{F_\pi^2}\right)}
{F_{\pi}^2\;m_h}. \label{mhms}  
\end{eqnarray}
Relation~(\ref{mhms}) is not valid for the pion mass, since pions are almost independent of $m_s$ in chiral perturbation theory~\cite{GLsu3}.  However, Eqs. (\ref{mhmq}) and (\ref{mhms}) work reasonably well for the nucleon~\cite{Sainio}, for instance. 
The masses of a few hadrons have been shown to follow Eq.~(\ref{mhmq}) and Eq.~(\ref{mhms}) over a sizeable range of quark masses \cite{Karsch:2003vd,Karsch:2003zq}. 
Thus, the contribution of one hadron of mass $m_h$ to the light quark-antiquark condensates is given by
\begin{eqnarray}
  \label{qq}
  \frac{\Delta <\bar{q}q>}{<\bar{q}q>_0} &=& 
-\frac{g }{2\pi^2}\;T\, m_h\; \frac{A (1+2 \kappa
  \frac{m_\pi^2}{F_\pi^2})}{F_\pi^2}  \sum_{n=1}^\infty \,\frac{(-\eta)^{n+1}}{n}
\,\exp\left(n\frac{B\mu_B -I_3\mu_I-S\mu_S}{T} \right)
\, K_1\left( n\frac{m_h}{T} \right), \hspace*{1cm}
\end{eqnarray}
whereas its contribution to the strange quark-antiquark condensate is
given as
\begin{eqnarray}
  \label{ss}
  \frac{\Delta <\bar{s}s>}{<\bar{s}s>_0} &=& 
-\frac{g}{2\pi^2} \;T\, m_h\;
\frac{A (1+2 \kappa
  \frac{m_\pi^2}{F_\pi^2})}{F_\pi^2}\,\frac{<\bar{q}q>_0}{<\bar{s}s>_0} \times
\nonumber \\
& & \sum_{n=1}^\infty \,\frac{(-\eta)^{n+1}}{n}
\,\exp\left(n\frac{B\mu_B -I_3\mu_I-S\mu_S}{T} \right)
\, K_1\left(n\frac{m_h}{T} \right).
\end{eqnarray}

\begin{figure}[htb]
\centering{
\includegraphics[width=8cm]
{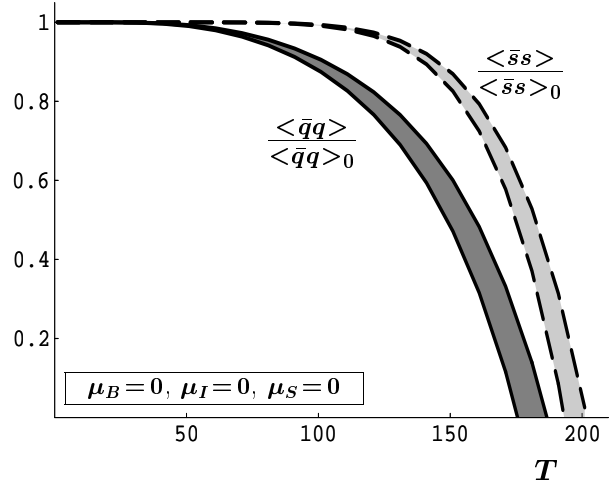}
\caption{The quark-antiquark condensates as a function of temperature $T$ at zero baryon, isospin, and strangeness chemical potentials. The light quark-antiquark condensate
  $<\bar{q}q>/<\bar{q}q>_0$ is given in dark gray and solid curve. The strange quark-antiquark condensate $<\bar{s}s>/<\bar{s}s>_0$ is illustrated in light gray and dashed curve. The graph taken from Ref. \cite{Tawfik:2005qh}.} 
  \label{fig:qqchi1}
  }
\end{figure}

\begin{figure}[htb]
\centering{
\includegraphics[width=8cm]{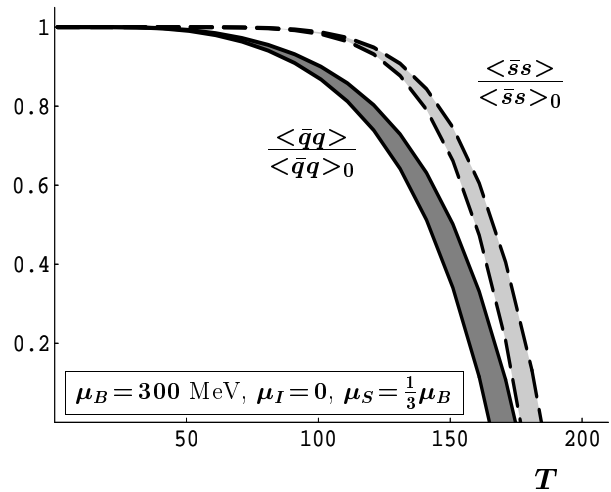}
\caption{The same as in Fig. \ref{fig:qqchi1} at baryon $\mu_B=300$~MeV and strangeness chemical potentials $\mu_S=100$~MeV.  Although the chemical potentials have been set to finite values, the critical  temperature of light quark condensate is still smaller than that of the  strange quark one. The graph taken from Ref. \cite{Tawfik:2005qh}. }
\label{fig:qqchi2}
}
\end{figure}

The quark-antiquark condensates are calculated as a function of  temperature  $T$ for various values of  baryon, isospin, and strangeness chemical potentials. The results are shown in Fig.~\ref{fig:qqchi1} and~Fig.~\ref{fig:qqchi2}, at vanishing and finite potential, respectively.  At zero chemical potential, we find that the strange quark-antiquark condensate remains large at temperatures where the light quark-antiquark condensates become small: $<\bar{s}s>/<\bar{s}s>_0=0.4\pm0.2$ where $<\bar{q}q>$ vanishes. As can be seen in Fig.~\ref{fig:qqchi2}, an increase in $\mu_B$ tends to reduce the difference between $<\bar{q}q>$ and $<\bar{s}s>$. On the other hand, we find that all condensates are equally sensitive to $\mu_B$.  

We conclude that the strange quark-antiquark condensate can be relatively large at temperatures where the light quark-antiquark condensate is very small. The  uncertainty of the value of the strange quark-antiquark condensate at critical temperature is however rather large. This is due to the uncertainty in the values of the condensates at high temperatures which we have calculated by the HRG model. This uncertainty could be reduced for the most part, if the factor $A$ in Eq.~(\ref{lattMass})
were more precisely known.

\subsubsection{Confinement-deconfinement phase diagram}
\label{sec:QCDpd}

At finite chemical potential, the fermion determinant gets complex and therefore the conventional MC~techniques are no longer applicable, the lattice configurations can no longer be generated with the probability of the Boltzmann weight. However, considerable progress has been made to overcome these problems~\cite{Fodor:2001au,deForcrand:2002ci,Allton:2002zi,D'Elia:2002gd}.  The need for effective algorithms that would overcome the sign-problem still big. The numerical studies for the equation of state at finite $\mu_B$  provides a valuable framework for understanding the experimental signatures for the phase transition from confined hadrons to QGP \cite{Rajagopal:2000wf,Tawfik:2004sw}. 

It is known from effective models such as bootstrap and Nambu-Jona-Lasinio model that the structure of the phase diagram is complex. The freeze-out curve takes a much different behavior at large chemical potential~\cite{Tawfik:2004vv,Tawfik:2004ss}. Nevertheless, one might think that for small chemical potential, $\mu_q\approx T_c$, the curvature of $T_c$-dependence upon $\mu_q$ can be fitted as a parabola, where $\mu_q$ is the quark baryon chemical potential. The situation at very large chemical potentials is not clear. One might need to take into account other effects, such as quantum effects at low temperatures~\cite{Miller:2003ha,Miller:2003hh,Miller:2003ch,Miller:2004uc,Hamieh:2004ni,Miller:2004em}, which might be able to describe the change in the correlations from confined hadrons to 
coupled quark-pairs. 

In the present review, we assume that the confinement-deconfinement phase transition is driven by a constant energy density~\cite{Tawfik:2004vv,Tawfik:2004ss}. We show that the degrees of freedom rapidly increase at $\epsilon_c(T_c,\mu_B=0)$ \cite{Karsch:2003zq,Karsch:2003vd} and $\epsilon_c(T_c,0)=\epsilon_c(T_c,\mu_B)$ \cite{Redlich:2004gp,Tawfik:2005qh}. Concretely, we propose that the existence of different transitions does not affect the assumption that $\epsilon_c$ is constant for all $\mu_B$-values. The nature of the degrees of freedom in this region is very different from that of the {\it nearly} non-interacting QGP at high $T$ and low $\mu_B$. The condition driving the QCD phase transition at finite $T$ and $\mu_B$~\cite{Karsch:2003zq,Karsch:2003vd,Redlich:2004gp} is the energy
density. Its value is not affected by the conjecture of existing of different transitions along the whole $\mu_B$-axis.  

The HRG model has been used  to determine $T_c$ corresponding to a wide range of quark (pion) masses at $\mu_B=0$~\cite{Redlich:2004gp}, section \ref{sec:TcPionm}. The masses range from the chiral to pure gauge limits. The condition of constant energy density can excellently reproduce the critical temperature $T_c$ as a function of $m_q$ and $n_f$, section \ref{sec:ecPionm}.  In lattice QCD Lagrangian, the strangeness chemical potential $\mu_s$ can have  the same value of $\mu_B/3$. 

In the HRG model, the energy density at finite chemical potential can be divided into two parts: one from the meson sector and another one from the baryon sector. For the first part, we can completely drop out the fugacity term. For symmetric numbers of light quarks, the baryon chemical potential of mesons is vanishing. But for strange mesons the strangeness chemical potential assigned to their strange quarks should be taken into account. For the baryon sector, the chemical potential is given by Eq.~(\ref{eq:mu2}).

The value of critical energy density is taken from lattice QCD simulations at $\mu_B=0$ \cite{Karsch:2000kv}. In lattice units, the dimensionless energy densities for $n_f=2$ and $2+1$ are \hbox{$\varepsilon/T^4|_{T_c}\cong 4.5\pm2$} and $\cong 6.5\pm2$, respectively. We take an average value and express it in physical units. Thus, $\epsilon_c=600\pm300\;$MeV$/$fm$^3$. It is assumed \cite{Allton:2002zi} that this value remains constant along the phase transition line; $\mu_B$-axis. The existence of different phase transitions (cross-over and first-order) and the critical endpoint, at which the transition is second-order, is assumed not to affect this assumption. For the consequences of constant $\epsilon_c(\mu_B)$  disregarding the uncertainty, we first recall the second law of thermodynamics 
\begin{eqnarray}
\partial \epsilon &=& T\,\partial s - p + \mu_B\,\partial n_B, \label{2law}
\end{eqnarray}
where $s$, $p$ and $n_B$ are the entropy, the pressure and the baryon number density, respectively. If the critical energy density $\epsilon_c$ were a decreasing function of $\mu_B$, this means that low incident energies are much more suitable to produce the phase transition from confined hadrons to deconfined QGP than the high incident energies! But the low limit should be given by the freeze-out curve~\cite{Tawfik:2004ss}, i.e. the hadronization phase diagram. We know from phenomenological observations that both freeze-out curve and phase transition line are coincident at low $\mu_B$ (high incident energy). For large $\mu_B$, the two lines are separated.  The freeze-out curve~\cite{Tawfik:2004ss} is given by $s/T^3=7$, i.e. entropy- or degrees-of-freedom-driven. Then, along freeze-out curve it is expected that $\epsilon$ slightly increase with $\mu_B$. This means the assumption that $\epsilon_c$ decreases with increasing $\mu_B$ will give a phase transition which has $T_c$ much smaller than the freeze-out temperature.

In the other case, that $\epsilon_c$ increases with increasing $\mu_B$, we expect the $\epsilon_c$ required for the phase transition gets larger with decreasing the incident energy! According to Eq.~(\ref{2law}), the phase diagram is given by $T_c=\partial \epsilon/\partial s$ at constant $n_B$ and $\mu_B= \partial \epsilon/\partial n_B$ at constant $T$. Then, for increasing $\epsilon_c(\mu_B)$, the critical temperature is expected to increase or at least remain constant. This result has the consequences
that the phase transition at large $\mu_B$ will be difficult to materialize in heavy-ion collisions or impossible. Also the phases of coupled quark-pairs (color superconductivity) will be expected for very
large $\mu_B$ or not allowed at all.

From the canonical partition function, the energy density at finite chemical potential $\mu\neq0$ can be deduced 
\begin{eqnarray}
\epsilon(T,\mu) &=& T \frac{\partial T\ln {\cal Z}(T,\mu)}{\partial T} -
  T \ln {\cal Z}(T,\mu) 
  +\mu \frac{\partial T\ln {\cal Z}(T,\mu)}{\partial
  \mu} = \frac{g}{2\pi^2}\; \int_{0}^{\infty} k^2 \;dk \;
  \frac{\varepsilon(k)}{e^{[\varepsilon(k)-\mu]/T} 
  \pm 1}. \hspace*{8mm} 
\label{eq:epslCom}
\end{eqnarray}
It is obvious that the trigonometric function included   in last expression are not truncated. To calculate this quantity in the HRG model, we sum up over all resonances we take into account.  

\begin{figure}[tb]
\includegraphics[width=10.cm,angle=-90]{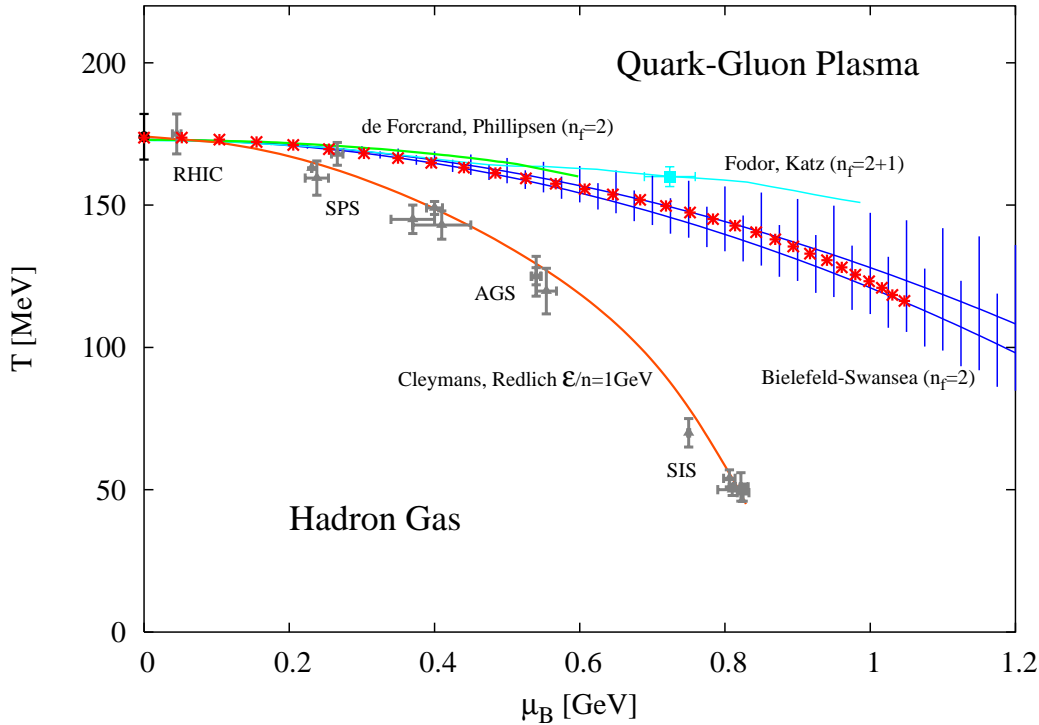} 
\caption{The $T-\mu_b$ phase diagram for $2$ and $2+1$ quark flavors is depicted. The vertical lines give the lattice results~\cite{Allton:2002zi}.  Results from $2$ \cite{deForcrand:2002ci} and $2+1$ \cite{Fodor:2001au} quark flavors are given as curves on top of the vertical lines. The HRG results are presented by astride.}
\label{Fig:ecnf22}
\end{figure}

In Fig. \ref{Fig:ecnf22}, the $T-\mu_B$ phase diagram for $2$ and $2+1$ quark flavors is drawn. The vertical lines give the lattice results~\cite{Allton:2002zi} for $2$ quark flavors. Furthermore, results from $2$ \cite{deForcrand:2002ci} and $2+1$ \cite{Fodor:2001au} quark flavors are given as curves on top of the vertical lines. The three lattice results are marked with authors. The lattice QCD simulations are performed for large quark mass. The corresponding Goldstone pion gets a mass of $770\;$MeV. The solid circles give our results with heavy quark masses. The HRG results given by astride agree well with the lattice calculations.  The transition temperature $T_c$ from hadronic matter to QGP has been determined according to a condition of constant energy density. Furthermore, we compare the transition diagram with the freeze-out one. We use the condition of constant energy per particle \cite{redlich2}, section \ref{sec:cr_en1} to determine the freeze-out curve.  Apparently, the two curves are separable at large chemical potential, while they are very close to each other at low chemical potential.

The influence of strange quark chemical potential $\mu_s$ on $T_c$ was studied \cite{Tawfik:2004vv}. For including $\mu_s$, we applied two models. In the first one, we explicitly calculated $\mu_s$ in dependence on $T$ and $\mu_B$ under the condition that the net strangeness vanishes. In the  second one, we assign, as the case in lattice QCD simulations, zero to $\mu_s$ for all $T$ and $\mu_B$. The first case, $\mu_s=f(T,\mu_B)$, is of great interest for heavy-ion collisions. The strange quantum number is entirely conserved. The current lattice results are very well reproducible. On the other hand, one can apply the second model for ultra-relativistic collisions. At RHIC and LHC energies, for instance, $\mu_B$ (and consequently $\mu_s$) should be very small.  We have shown that the proper condition that guarantees vanishing strangeness in QGP is to set  $\mu_s=\mu_q$. We did not check this explicitly. But it is obvious that $T_c(\mu_B,\mu_s=0)$ quantitatively is not very much different from $T_c(\mu_b,\mu_s(T,\mu_B))$ at small $\mu_B$.

\section{Higher order moments in statistical-thermal models}
\label{sec:hm}

The higher order moments can be studied in different physical quantities. For example, that of charged-particle multiplicity distribution have been predicted four decades ago \cite{gupta72}. The empirical relevance to the experimental measurements of the higher order moments has been proposed  \cite{endp1}. From the experimental point-of-view, we have so-far STAR measurements \cite{star1A,star1B,star1C,star2A,star2B} and lattice QCD calculations \cite{lqcd1,lqcd1a,lqcd2}. It is worthwhile to mention that the measurement of the correlation length seems to be very much crucial. On the other hand, the experimental sensitivity for the suggested signatures  \cite{star1A,star1B,star1C,star2A,star2B}, which are based on singular behavior of thermodynamical functions, should be evaluated. For the time being, we just want to mention that the experimental measurements apparently take place at the final state of the collision, which means that the signals have to survive the extreme conditions in such high energy collisions. It has been pointed out that the contribution of the critical fluctuations to the higher order moments is proportional to a positive power of $\xi$. The latter is conjectured to  diverge at the QCD critical endpoint. Such an assumption is valid in the thermodynamical limit. 

\subsection{Non-normalized higher order moments}
\label{nonnorlTc}

For the $i$-th particle, the ''first'' order moment is given by the derivative of $p=-T \partial \ln Z(T,V,\mu_i)/\partial V$ with respect to the dimensionless quantity $\mu_i/T$. When taking into account the antiparticles, we add a negative sign to the chemical potential. The first derivative describes the multiplicity distribution or an expectation operator, which is utilized to estimate the number or multiplicity density 
\bea \label{eq_m1a}
m_1(T,\mu_i ) &=& \pm \frac{g_i}{2\pi^2} T \int_{0}^{\infty}
\frac{e^{\frac{\mu_i - \varepsilon_i}{T}}\; k^2\, dk}{1\pm e^{\frac{\mu_i - \varepsilon_i}{T}}}.
\eea
The ''second'' order moment is known as the variance. It gives the susceptibility of the measurements. 
\bea \label{eq_m2a}
m_2(T,\mu_i ) &=& \pm \frac{g_i}{2\pi^2} T \int_{0}^{\infty}  \frac{e^{\frac{\mu_i -\varepsilon_i}{T}}\; k^2\, dk}{1\pm e^{\frac{\mu_i -\varepsilon_i}{T}}} 
- \frac{g_i}{2\pi^2} T  \int_{0}^{\infty} \frac{e^{2\frac{\mu_i -\varepsilon_i}{T}}\; k^2\, dk }{\left(1\pm e^{\frac{\mu_i -\varepsilon_i}{T}}\right)^2}.
\eea
The ''third'' order moment measures of the lopsidedness of the distribution. As
given in section \ref{sec:norm}, the normalization of third order moment is known as skewness or the asymmetry in the distribution. Skewness tells us about the direction of variation of the data set. 
\bea \label{eq_m3a}
m_3(T,\mu_i ) & = & \pm \frac{g_i}{2\pi^2} T \int_{0}^{\infty}
\frac{e^{\frac{\mu_i -\varepsilon_i}{T}}\; k^2\, dk}{1\pm
  e^{\frac{\mu_i -\varepsilon_i}{T}}} 
- \frac{g_i}{2\pi^2} 3T \int_{0}^{\infty}
\frac{e^{2\frac{\mu_i -\varepsilon_i}{T}}\; k^2\, dk}{\left(1\pm
  e^{\frac{\mu_i -\varepsilon_i}{T}}\right)^2} \pm  \frac{g_i}{2\pi^2} 2T \int_{0}^{\infty}  \frac{e^{3\frac{\mu_i -\varepsilon_i}{T}}\; k^2\, dk}{\left(1\pm e^{\frac{\mu_i -\varepsilon_i}{T}}\right)^3}. \hspace*{1cm}
\eea
In general, the normalization of $r-$th order moment is obtained by dividing it by $\sigma^r$, where $\sigma$ is the standard deviation. The normalization is assumed to remove the brightness dependence.

The ''fourth'' order moment compares the tallness and skinny or shortness and squatness, i.e. shape, of a certain measurement to its normal distribution. It is defined as the uncertainty is an uncertainty or {\it ''the location- and scale-free movement of probability mass from the shoulders of a distribution into its center and tails and to recognize that it can be formalized in many ways''} \cite{kurts1}  
\bea \label{eq_m4a}
m_4(T,\mu_i ) &=& 
\pm \frac{g_i}{2\pi^2} T \int_0^{\infty} \frac{e^{\frac{\mu_i -\varepsilon_i}{T}}\;
  k^2\, dk}{1\pm e^{\frac{\mu_i -\varepsilon_i}{T}}}
- \frac{g_i}{2\pi^2} 7 T \int_0^{\infty} \frac{e^{2\frac{\mu_i -\varepsilon_i}{T}}\;
  k^2\, dk}{\left(1\pm e^{\frac{\mu_i -\varepsilon_i}{T}}\right)^2} \nonumber \\
& & \pm \frac{g_i}{2\pi^2} 12 T \int_0^{\infty} \frac{e^{3\frac{\mu_i -\varepsilon_i}{T}}\; k^2\, dk}{\left(1\pm e^{\frac{\mu_i -\varepsilon_i}{T}}\right)^3} 
 - \frac{g_i}{2\pi^2} 6 T \int_0^{\infty} \frac{e^{4\frac{\mu_i -\varepsilon_i}{T}}\; k^2\, dk}{\left(1\pm e^{\frac{\mu_i -\varepsilon_i}{T}}\right)^4}. 
\eea

The ''fifth'' order moment measures the asymmetry sensitivity of the ''fourth'' order moment 
\bea \label{eq_m5a}
m_5(T,\mu_i ) &=& 
\pm  \frac{g_i}{2 \pi ^2} T \int_0^{\infty }
\frac{e^{\frac{\mu_i -\varepsilon_i}{T}} \; k^2\, dk}{1\pm e^{\frac{\mu_i -\varepsilon_i}{T}}} 
- \frac{g_i}{2 \pi ^2} 15 T \int_0^{\infty } \frac{e^{\frac{2
      (\mu_i -\varepsilon_i)}{T}} \; k^2\, dk}{\left(1\pm e^{\frac{\mu_i -\varepsilon_i}{T}}\right)^2}
 \pm \frac{g_i}{2 \pi ^2} 50 T \int_0^{\infty } \frac{e^{\frac{3
      (\mu_i -\varepsilon_i)}{T}} \; k^2\, dk}{\left(1\pm e^{\frac{\mu_i -\varepsilon_i}{T}}\right)^3} \nn \\
&-& \frac{g_i}{2 \pi ^2} 60 T \int_0^{\infty } \frac{e^{\frac{4
      (\mu_i -\varepsilon_i)}{T}} \; k^2\, dk}{\left(1\pm e^{\frac{\mu_i -\varepsilon_i}{T}}\right)^4} 
 \pm \frac{g_i}{2 \pi ^2} 24 T \int_0^{\infty } 
\frac{e^{\frac{5 (\mu_i -\varepsilon_i)}{T}}\; k^2\, dk}{\left(1\pm e^{\frac{\mu_i -\varepsilon_i}{T}}\right)^5}. 
\eea
The ''sixth'' order moment is generally associated with compound options 
\bea \label{eq_m6a}
m_6(T,\mu_i ) &=& 
\pm  \frac{g_i}{2 \pi^2} T \int_0^{\infty}
  \frac{e^{\frac{\mu_i -\varepsilon_i}{T}} \; k^2\, dk}{1\pm 
  e^{\frac{\mu_i -\varepsilon_i}{T}}}
- \frac{g_i}{2 \pi^2} 31 T \int_0^{\infty}
  \frac{e^{2\frac{\mu_i -\varepsilon_i}{T}}
  \; k^2\, dk}{\left(1\pm e^{\frac{\mu_i -\varepsilon_i}{T}}\right)^2} \nonumber \\
& & \pm \frac{g_i}{2 \pi^2} 180 T \int_0^{\infty}
  \frac{e^{3\frac{\mu_i -\varepsilon_i}{T}}
  \; k^2\, dk}{\left(1\pm e^{\frac{\mu_i -\varepsilon_i}{T}}\right)^3} 
 - \frac{g_i}{2 \pi^2} 390 T \int_0^{\infty}
  \frac{e^{4\frac{\mu_i -\varepsilon_i}{T}}
  \; k^2\, dk}{\left(1\pm e^{\frac{\mu_i -\varepsilon_i}{T}}\right)^4} \nonumber \\
& & \pm \frac{g_i}{2 \pi^2} 360 T \int_0^{\infty}
  \frac{e^{5\frac{\mu_i -\varepsilon_i}{T}}
  \; k^2\, dk}{\left(1\pm e^{\frac{\mu_i -\varepsilon_i}{T}}\right)^5} 
- \frac{g_i}{2 \pi^2} 120 T \int_0^{\infty} 
  \frac{e^{6\frac{\mu_i -\varepsilon_i}{T}}
  \; k^2\, dk}{\left(1\pm e^{\frac{\mu_i -\varepsilon_i}{T}}\right)^6}. \hspace*{7mm}
\eea

Thus, from Eqs. (\ref{eq_m1a})-(\ref{eq_m6a}), a general expression for the $r$-th order moment can be deduced  \cite{Tawfik:2012si}
\bea \label{eq:genralm1}
m_r(T,\mu_i ) &=& \frac{g_i}{2\pi^2}  T \sum_{l=1}^{r} a_{r,l} \int_0^{\infty}
\frac{e^{l \frac{\mu_i -\varepsilon_i}{T}}}{\left(1\pm e^{\frac{\mu_i -\varepsilon_i}{T}}\right)^l} \, k^2\, d k.
\eea
The coefficients read
\bea \label{coffs}
a_{r,l} &=& (\pm 1)^l (-1)^{l+1} \left[l\; a_{r-1,l} + (l-1)\; a_{r-1,l-1}\right],
\eea
where $l\leq r$ and $a_{r,l}$ vanishes $\forall\, r<1$.

\begin{figure}[htb]
\centering{
\includegraphics[angle=-90,width=10.cm]{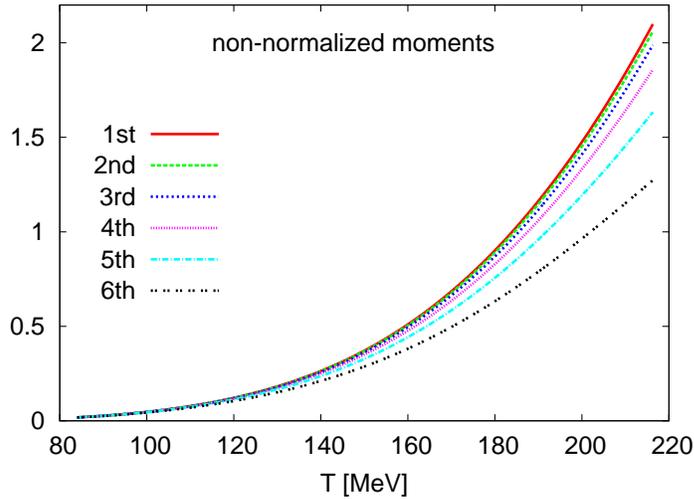}
\caption{The dimensionless non-normalized moments is given in dependence on the temperature.  The evolution of the so-called correlation parts (the integrals appearing in Eqs. (\ref{eq_12b})-(\ref{eq_m6b})) are drawn with the temperature. The graph taken from Ref.  \cite{Tawfik:2012si}. }
\label{fig:hrg1} 
}
\end{figure}

As given in Eq. (\ref{eq:genralm1}), the dependence of first six non-normalized moments on lower ones can explicitly be deduced as follows. 
\bea 
m_1 &=& \frac{g_i}{2\pi^2} T \int_{0}^{\infty}  \frac{e^{\frac{\mu_i -\varepsilon_i}{T}}\; k^2\, dk}{1 \pm e^{\frac{\mu_i -\varepsilon_i}{T}}}, \label{eq_12b} \\
m_2 &=&  m_1 - \frac{g_i}{2\pi^2} T  \int_{0}^{\infty}  \frac{e^{2\frac{\mu_i -\varepsilon_i}{T}}\; k^2\, dk}{\left(1 \pm e^{\frac{\mu_i -\varepsilon_i}{T}}\right)^2}, \label{eq_m2b} \\
m_3 &=&  -2 m_1 + 3 m_2 + 2 \frac{g_i}{2\pi^2} T \int_{0}^{\infty}  \frac{ e^{3\frac{\mu_i -\varepsilon_i}{T}}\; k^2\, dk}{\left(1 \pm e^{\frac{\mu_i -\varepsilon_i}{T}}\right)^3}, \label{eq_m3b}\\
m_4 &=& 6 m_1 -11 m_2 + 6 m_3
-6 \frac{g_i}{2\pi^2} T \int_0^{\infty} \frac{e^{4\frac{-\epsilon_i +\mu_i}{T}}\; k^2\, dk}{\left(1 \pm e^{\frac{-\epsilon_i +\mu_i }{T}}\right)^4}, \label{eq_m4b}\\
m_5 &=& -24 m_1 + 50 m_2 - 35 m_3 + 10 m_4
+ 24 \frac{g_i}{2\pi^2} T \int_0^{\infty} \frac{e^{5\frac{-\epsilon_i +\mu_i}{T}}\; k^2\, dk}{\left(1 \pm e^{\frac{-\epsilon_i +\mu_i }{T}}\right)^5}, \label{eq_m5b}\\
m_6 &=& 120 m_1 - 274 m_2 + 225 m_3 - 85 m_4 + 15 m_5 - 120 \frac{g_i}{2\pi^2} T \int_0^{\infty} \frac{e^{6\frac{-\epsilon_i +\mu_i}{T}}\; k^2\, dk}{\left(1 \pm e^{\frac{-\epsilon_i +\mu_i }{T}}\right)^6}. \label{eq_m6b} \hspace*{7mm}
\eea
Naively spoken, we conclude that raising lower order moments to higher ones is achievable through a series of all lower order moments and an additional term reflecting the correlations, themselves \cite{gupta72}. From Eqs. (\ref{eq_12b})-(\ref{eq_m6b}), the additional terms are proportional to $\langle N\rangle^r$. Apparently, they are  essential in order to judge whether going from lower to higher order moments would make the signatures of dynamical fluctuations clearer than when excluding them. 
The first and second terms can be generalized. Then, a general expression would read
\bea \label{eq:genmm}
m_r=(-)^{r-1}(r-1)c_{m_1}^{r-1} m_1 - \left[(r-1)c_{m_2}^{r-1} + (r-2)!\right] m_2 +{\cal O}(m_{>2}),
\eea
where $c_{m_{r}}$ is the coefficient of the $r$-th moment. The last term in Eq. (\ref{eq:genmm}) can be elaborated when higher order moments are calculated. The latter are essential in order to find out a clear pattern.

It is obvious that the coefficients, Eq. (\ref{coffs}), of a certain moment are to be determined from a long chain of all previous ones. Such a conclusion could be originated to about four decades \cite{gupta72}, where it has been shown that the coefficients are related to high order correlation functions. Should this assumption is proven to be valid, then expression (\ref{coffs}) gets a novel interpretation. It seems to sum up the correlation functions up to the $r$-th order. According to Ref. \cite{gupta72} and when neglecting the two-particle correlations $C_2$, then the higher order moments read
\bea
m_2 &=& \langle(\delta N)^2\rangle \approx 2 \langle N \rangle, \label{eq:mg2}\\
m_3 &=& \langle(\delta N)^3\rangle \approx 4 \langle N\rangle + C_3, \label{eq:mg3}\\
m_4 &\simeq& \langle(\delta N)^4\rangle = 6 m_3 + 3 m_2^2 - 8 m_2 + 8 C_4, \label{eq:mg4}
\eea 
where $\delta N = N-\langle N\rangle$ \cite{star1B}, 
\bea
C_3(p_1, p_2, p_3) &=& \sum_{p_1} \langle p_1p_1p_1\rangle + 3\sum_{p_1<p_2} \langle p_1p_2p_2\rangle +6\sum_{p_1<p_2<p_3} \langle p_1 p_2 p_3\rangle, \label{eq:r2}\\
C_4(p_1, p_2, p_3, p_4) &=& \sum_{p_1<p_2} \langle p_1 p_2 p_2 p_2\rangle + \sum_{p_1<p_2<p_3} \langle p_1 p_2 p_3 p_3\rangle + 8 \sum_{p_1<p_2<p_3<p_4} \langle p_1 p_2 p_3 p_4\rangle,  \label{eq:r3}
\eea
and $p_i$ is the $i-$th particle. 
To the second order moment $m_2$ we have to add the effects of the two particle correlation function, $2 \sum C_2$. The third order moment $m_3$ gets approximately three times this amount. The three and four particle correlations, Eqs. (\ref{eq:r2}) and (\ref{eq:r3}), appear first in third and fourth moment, Eq. (\ref{eq:mg4}) and (\ref{eq:r2}), respectively. 

We restrict the discussion to two particle correlations, only. It is of great interest, as it is accessible experimentally and achievable, numerically. The two particle correlations are suggested as a probe for the bulk QCD medium, energy loss, medium response, jet properties and intensity interferometry \cite{tp1,tp2,tp3,tp4,tp5,tp6,tp7A,tp7B}. In addition to this list of literature, the comprehensive review \cite{physrepA,physrepB} can be recommended. 

Taking into account the particle multiplicities, then expressions (\ref{eq:mg2}), (\ref{eq:mg3}) and (\ref{eq:mg4}) can be re-written as follows \cite{Tawfik:2012si}.
\bea
m_2 = \langle(\delta N)^2\rangle &\simeq& \langle N^2\rangle - \langle N\rangle^2, \\
m_3 = \langle(\delta N)^3\rangle &\simeq& \langle N^3\rangle - \langle N^2\rangle \langle N\rangle + 2 \langle N^3\rangle^3, \\
m_4 = \langle(\delta N)^4\rangle - 3 \langle(\delta N)^2\rangle^2 
   &\simeq & \langle\langle N\rangle^4\rangle - 2 \langle\langle N^2\rangle^2\rangle - 5 \langle\langle N^2\rangle\rangle^2 + 6 \langle\langle N\rangle^2\rangle\, \langle\langle N^2\rangle\rangle. \hspace*{7mm}
\eea
It is obvious, section \ref{nonnorlTc}, that expressions (\ref{eq_12b})-(\ref{eq_m6b}) lead to the conclusion that all moments are entirely depending on the previous ones, Fig. \ref{fig:hrg1}. Thus, the $r$-th order moment has contributions coming from all moments with orders lower than $r$.

Fig. \ref{fig:hrg1} presents the results of the higher order moments calculated in HRG model. Raising the orders of multiplicity moments results in new coefficients and new integrals. The earlier are partly characterized in Eq. (\ref{eq:genmm}). 
The integrals can be compared with the correlation functions \cite{gupta72}. We find that these integrals are related to $\langle N \rangle^r$, where $r$ is the order of the moment. It is obvious that successive moments have a difference of about one order of magnitude. Therefore, the higher order $\langle N\rangle^r$ can be disregarded with reference to their vanishing contributions.


\subsection{Normalized higher order moments}
\label{sec:norm}

\begin{figure}[htb]
\centering{
\includegraphics[angle=-90,width=10cm]{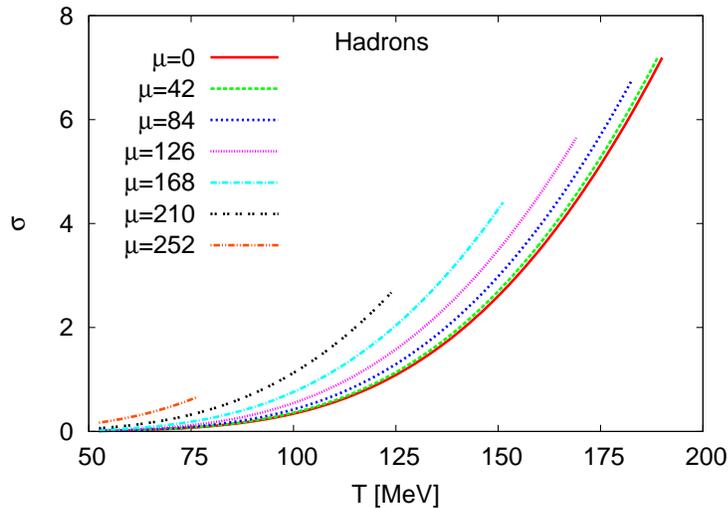}
\caption{Results for  $\sigma$ in hadronic  resonances are given in dependence on $T$ for various baryon chemical potentials (given in MeV). The graph taken from Ref.  \cite{Tawfik:2012si}. }
\label{fig:sigmaaa} 
}
\end{figure}

The normalization of higher order moments provides with a tool to relate moments with various orders to the experimental measurement. The susceptibility of the distribution give a measure for $\sigma$.  It has been shown that the susceptibility is related to $\sim\xi^2$ \cite{endp1}. The results of  $\sigma$ in hadronic resonances are calculated at different $\mu$ and given in Fig. \ref{fig:sigmaaa}.  As per the standard model, conservation of strange quantum number  is one of the global symmetries in strong interactions. The procedure of keeping strange degrees of freedom conserved in HRG was introduced \cite{Tawfik:2004sw}. This is the origin of the $\mu$-dependence. Although, baryon chemical potential $\mu$ vanishes per definition, the chemical potential associated with strange quark $\mu_S$ remains finite.  Another feature in these calculations is the assumption that the freeze-out boundary is determined by constant $s/T^3$, where $s$ is the entropy density \cite{Tawfik:2005qn}. 

\begin{figure}[htb]
\centering{
\includegraphics[angle=-90,width=10cm]{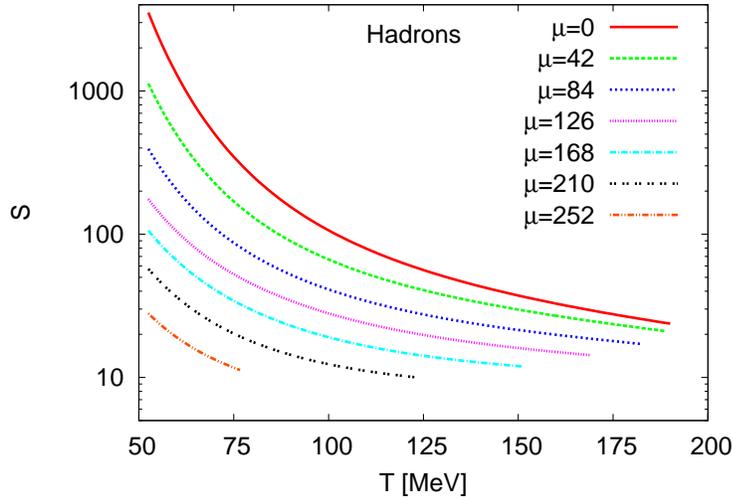}
\caption{At various baryon chemical potentials (given in MeV), the skewness $S$  for hadronic  resonance gas is given as function of $T$. The graph taken from Ref.  \cite{Tawfik:2012si}. }
\label{fig:S}
} 
\end{figure}

For standard Gaussian distribution, the skewness (''third'' order moment normalized to $\sigma^3$) is obviously vanishing. Therefore, the skewness is an ideal quantity probing the non-Gaussian fluctuation feature as expected near $T_c$. The QCD critical endpoint is conjectured to be sensitive to skewness. Experimentally, it has been shown that 
the skewness $S$ is related to $\sim\xi^{4.5}$ \cite{endp4}. The skewness for bosonic and fermionic resonance gas, respectively, reads
\bea 
S_b &=& -\frac{1}{2}\,\frac{\pi}{\sqrt{g_i}}\, T^{3/2}\, \frac{\int_0^{\infty} \,  
\text{csch}\left[\frac{\varepsilon_i - \mu_i }{2\, T}\right]^4\, \text{sinh} \left[\frac{\varepsilon_i - \mu_i }{T}\right] \; k^2 \, dk}
{\left[\int_0^{\infty} \left(\text{cosh} \left[\frac{\varepsilon_i-\mu_i }{T}\right]-1\right)^{-1}\; k^2 \, dk \right]^{3/2}}, \label{eq:Ssb}\\
S_f &=& 8\,\frac{\pi}{\sqrt{g_i}}\, T^{3/2}\, \frac{\int_0^{\infty} \,  
\text{csch}\left[\frac{\varepsilon_i - \mu_i }{T}\right]^3\, \text{sinh}
\left[\frac{\varepsilon_i - \mu_i }{2\, T}\right]^4 \; k^2 \, dk}
{\left[\int_0^{\infty} \left(\text{cosh} \left[\frac{\varepsilon_i-\mu_i }{T}\right]+1\right)^{-1}\; k^2 \, dk \right]^{3/2} }. \label{eq:Ssf}
\eea
At different chemical potentials, $S$ is calculated in dependence on $T$ and given in Fig. \ref{fig:S}.

The normalization of $4-$th order moment (kurtosis) means varying volatility or more accurately, varying variance.  The subtraction of $3$, which arises from the Gaussian distribution,  is frequently omitted \cite{kurts1,kurts2,kurts3}.  Therefore, the kurtosis is an ideal quantity for probing the non-Gaussian fluctuation feature as expected near $T_c$ and critical endpoint. A sign change of skewness or kurtosis is conjectured to indicate that the system crosses the phase boundary \cite{endp5,endp6A,endp6B}. As HRG is valid below $T_c$, the sign change is not accessible. It has been shown that kurtosis $\kappa$ is related to $\sim\xi^{7}$ \cite{endp4}
\bea
\kappa_b &=& -\frac{\pi^2}{g_i}\, T^3\, \frac{\int_0^{\infty}
  \left\{\text{cosh}\left[\frac{\varepsilon_i - \mu_i }{T}\right] + 2\right\} 
\text{csch}\left[\frac{\varepsilon_i - \mu_i }{2\, T}\right]^4 \; k^2 \, dk}
{\left[\int_0^{\infty} \left(1-\text{cosh}\left[\frac{\varepsilon_i-\mu_i }{T}\right]\right)^{-1}\; k^2 \, dk\right]^{2} } - 3,  \label{eq:Kkb}\\
\kappa_f &=& \frac{\pi^2}{g_i}\, T^3\, \frac{\int_0^{\infty}
  \left\{\text{cosh}\left[\frac{\varepsilon_i - \mu_i }{T}\right] - 2\right\} 
\text{sech}\left[\frac{\varepsilon_i - \mu_i }{2\, T}\right]^4 \; k^2 \, dk}
{\left[\int_0^{\infty} \left(\text{cosh}\left[\frac{\varepsilon_i-\mu_i }{T}\right]+1\right)^{-1}\; k^2 \, dk\right]^{2} } - 3.  \label{eq:Kkf}
\eea


\subsection{Products of higher order moments}
\label{sec:mult}

There are several techniques to scale the correlation functions. The survey system's optional statistics module represents the most common technique, i.e. Pearson or product moment correlation \cite{Tawfik:2012si}. This module includes the so-called partial correlation which seems to be useful when the relationship between two variables is to be highlighted, while effect of one or two other variables can be removed. We study the  products of higher order moments of the distributions of conserved quantities. The justification of this step is that certain products can be directly connected to the corresponding susceptibilities in lattice QCD simulation and related to long range correlations \cite{endp5,qcdlike,HM_FO3}. 

The fluctuations of conserved quantities are assumed to be sensitive to the structure of  hadronic system in its final state. As mentioned above, crossing the phase boundary or passing through critical endpoint is associated with large fluctuations. Most proposed fluctuations of observables are variations of second order moments of the distribution, such as the dynamical fluctuation of particle ratios \cite{Tawfik:2013dba,Tawfik:2010uh,Tawfik:2008ii,Tawfik:2006zh,Adamovich:2001hi} and charged dynamical measurement \cite{chargeD}. Then, the fluctuations are approximately related to $\xi^2$ \cite{corrlenght}.

\begin{figure}[htb]
\centering{
\includegraphics[angle=-90,width=10cm]{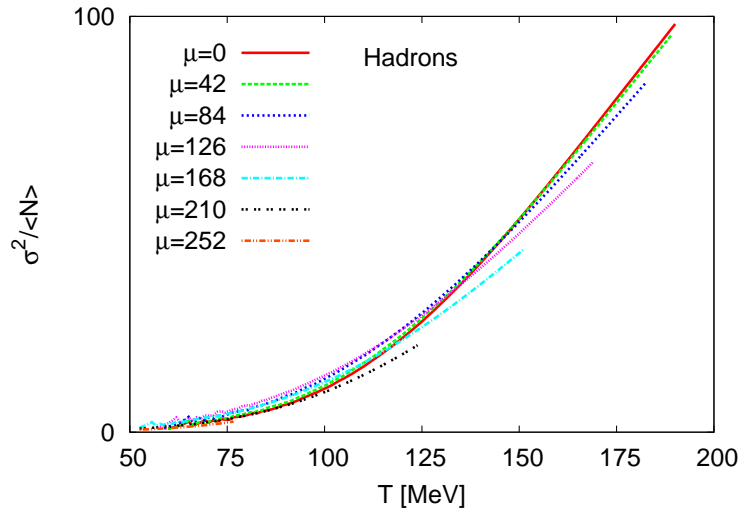}
\caption{The ratio $\sigma^2/\langle N\rangle$  is given in dependence on $T$ at different chemical potentials $\mu$ for hadronic (left), bosonic (middle) and fermionic (right) resonance gas. The graph taken from Ref.  \cite{Tawfik:2012si}. }
\label{fig:S2n} 
}
\end{figure}

The ratio of standard deviation $\sigma^2$ and the mean multiplicity $\langle N \rangle$ for fermions and bosons reads
\bea
\frac{\sigma^2}{\langle N\rangle} &=& \frac{1}{2}\, \frac{\int_0^{\infty}
  \left(1\pm \text{csch} \left[\frac{\varepsilon_i-\mu_i}{T}\right]\right)^{-1}\, k^2\, dk}{\int_0^{\infty} \left(1 \pm e^{\frac{\varepsilon_i-\mu_i }{T}}\right)^{-1}\, k^2\, dk}. \label{eq:sMb}
\eea
The results are given in Fig. \ref{fig:S2n}.

\begin{figure}[htb]
\centering{
\includegraphics[angle=-90,width=10cm]{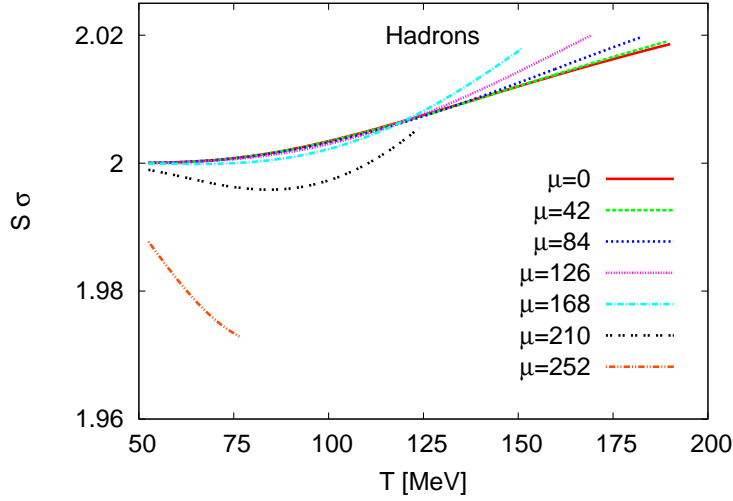}
\caption{The product $S \sigma$  is given in dependence on $T$ at different $\mu_b$ values (given in MeV). The graph taken from Ref.  \cite{Tawfik:2012si}. }
\label{fig:skewnS1} 
}
\end{figure}

The multiplication of skewness $S$ by the standard deviation $\sigma$ is directly related to the thermodynamics of the number susceptibility of the lattice QCD. In HRG, the bosonic and fermionic products read
\bea 
\left(S\; \sigma\right)_b &=& -\frac{1}{4}\; \frac{\int_0^{\infty} \, 
\text{csch}\left[\frac{\varepsilon_i -\mu_i }{T}\right]^4\; \text{sinh}\left[\frac{\varepsilon_i
      -\mu_i }{T}\right]\; k^2\,dk}{\int_0^{\infty} \left(1-\text{cosh}
  \left[\frac{\varepsilon_i -\mu_i }{T}\right]\right)^{-1}\; k^2\,dk}, \label{eq:ssigma1b} \\
\left(S\; \sigma\right)_f &=&  4 \; \frac{\int_0^{\infty} \, \text{csch}\left[\frac{\varepsilon_i
      -\mu_i }{T}\right]^3\; \text{sinh}\left[\frac{\varepsilon_i
      -\mu_i }{2\, T}\right]^4\; k^2\,dk}{\int_0^{\infty} \left(\text{cosh}
  \left[\frac{\varepsilon_i -\mu_i }{T}\right]+1\right)^{-1}\; k^2\,dk}. \label{eq:ssigma1f}
\eea
The results are given in Fig. \ref{fig:skewnS1}. It is obvious that $S\, \sigma \simeq 1$ for either bosons or fermions. Then, for hadrons, $S\, \sigma \simeq 2$. Nevertheless, the fine structure seems to reveal interesting features.


The dependence of $\sigma$ and $\sigma^2/\langle N\rangle$ on $T$ is illustrated in Figs. \ref{fig:sigmaaa} and \ref{fig:S2n}, respectively. It is obvious that both quantities have a monotonic behavior. As given in Fig. \ref{fig:skewnS1}, the product $S\, \sigma$ has a characteristic dependence on $T$.  The hadronic product makes an amazing bundle in the middle (at a characteristic $T$).  So far, we conclude that starting from the third normalized moment, a non-monotonic behavior appears. Thus, the kurtosis and its products should be expected to highlight such a non-monotonic behavior \cite{endp5,qcdlike}. 

\begin{figure}[htb]
\centering{
\includegraphics[angle=-90,width=10cm]{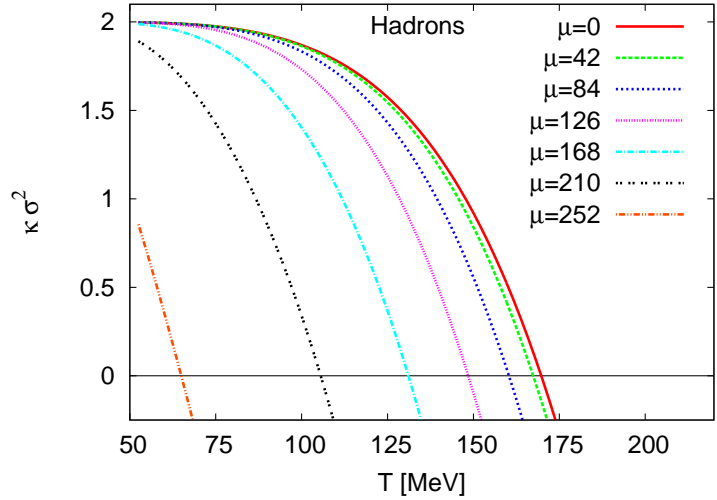}
\caption{The product ${\kappa}\,\sigma^2$ is given as a function of $T$ at various $\mu$-values for hadronic  resonance gas (right).}
\label{fig:kS2muu} 
}
\end{figure}

The multiplication of kurtosis by $\sigma^2$ called $\kappa^{eff}$ \cite{keff} is apparently equivalent to the ratio of $3$-rd order moment to $2$-nd order moment, Eq. (\ref{eq:genmm}).  In lattice QCD  and QCD-like models, $\kappa^{eff}$ is found to diverge near the critical endpoint \cite{endp5,qcdlike}. In HRG, the bosonic and fermionic products read
\bea 
\left(\kappa \sigma^2\right)_b &=& -\frac{1}{4} \frac{\int_0^{\infty}  \left\{\text{cosh}\left[\frac{\varepsilon_i
      -\mu_i }{T}\right] + 2\right\} \text{csch}\left[\frac{\varepsilon_i
      -\mu_i }{2\, T}\right]^4 k^2 dk} {\int_0^{\infty} \left(1-\text{cosh}
  \left[\frac{\varepsilon_i -\mu_i }{T}\right]\right)^{-1} k^2 dk} 
+ \frac{3 g_i}{4\, \pi^2}\, \frac{1}{T^3} \int_0^{\infty} \left(1-\text{cosh}
  \left[\frac{\varepsilon_i -\mu_i }{T}\right]\right)^{-1} k^2 dk, \hspace*{7mm}\label{eq:lsigma2b} \\
\left(\kappa \sigma^2\right)_f &=& \frac{1}{4} \frac{\int_0^{\infty}  \left\{\text{cosh}\left[\frac{\varepsilon_i
      -\mu_i }{T}\right] - 2\right\} \text{Sech}\left[\frac{\varepsilon_i
      -\mu_i }{2\, T}\right]^4 k^2 dk} {\int_0^{\infty} \left(\text{cosh}
  \left[\frac{\varepsilon_i -\mu_i }{T}\right]+1\right)^{-1} k^2 dk} 
- \frac{3 g_i}{4\, \pi^2} \frac{1}{T^3}\, \int_0^{\infty} \left(\text{cosh}
  \left[\frac{\varepsilon_i -\mu_i }{T}\right]+1\right)^{-1} k^2 dk. \label{eq:lsigma2f}
\eea
When ignoring the constant term in Eqs. (\ref{eq:Kkb}) and (\ref{eq:Kkf}), then the second terms in the previous expressions entirely disappear. The results are given in Fig. \ref{fig:kS2muu}. The thermal evolution of $\kappa\,\sigma^2$ is illustrated. We notice that increasing $T$ is accompanied with a drastic declination in $\kappa\,\sigma^2$.  The product $\kappa\, \sigma^2$ calculated at the freeze-out boundary leads to some interesting findings. \begin{itemize}
\item First, $\kappa\, \sigma^2$ almost vanishes or even flips its sign. 
\item Second, the $T$ and $\mu$ corresponding to vanishing $\kappa\, \sigma^2$ are coincident with the phenomenologically measured freeze-out parameters. 
\item Third, the freeze-out boundaries of bosons and fermions are crossing at a point located very near to the one assumed by the lattice QCD calculations to be the QCD critical endpoint \cite{Tawfik:2013dba}.
\end{itemize}

\subsection{Experimental higher order moments of particle yields}
\label{sec:HMexp}

\begin{figure}[htb]
\centering{
\includegraphics[angle=0,width=12.cm]{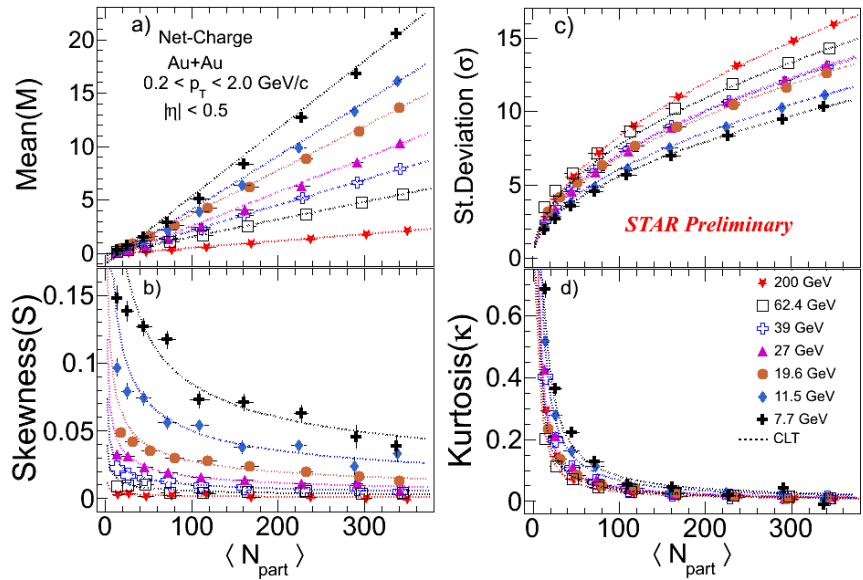}
\caption{The higher order moments mean (a), standard deviation (c), skewness (b) and
kurtosis (d) of the net-proton multiplicity are plotted with respect to $<N_{part}>$ for Au+Au collisions at various center-of-mass energies $\sqrt{s_{NN}}$. The dotted lines represent the expectation form the central limit theorem (CLT) \cite{clf}. Graph taken from Ref. \cite{Sahoo}. }
\label{fig:StarHM1} 
}
\end{figure}

In Fig. \ref{fig:StarHM1}, the first four moments of net-proton multiplicity are plotted as function of $<N_{part}>$ from STAR beam energy scan I \cite{besII}. The mean and standard deviation increase, while going from peripheral to central collisions for the given seven colliding energies, whereas $S$ and $\kappa$ decrease with the increase in the collision centrality. The evolution of these higher order moments of net-proton has been reproduced by the central limit theorem (CLT) \cite{clf}, which explains the  $<N_{part}>$  dependence of these moments.

\begin{figure}[htb]
\centering{
\includegraphics[angle=-90,width=7.cm]{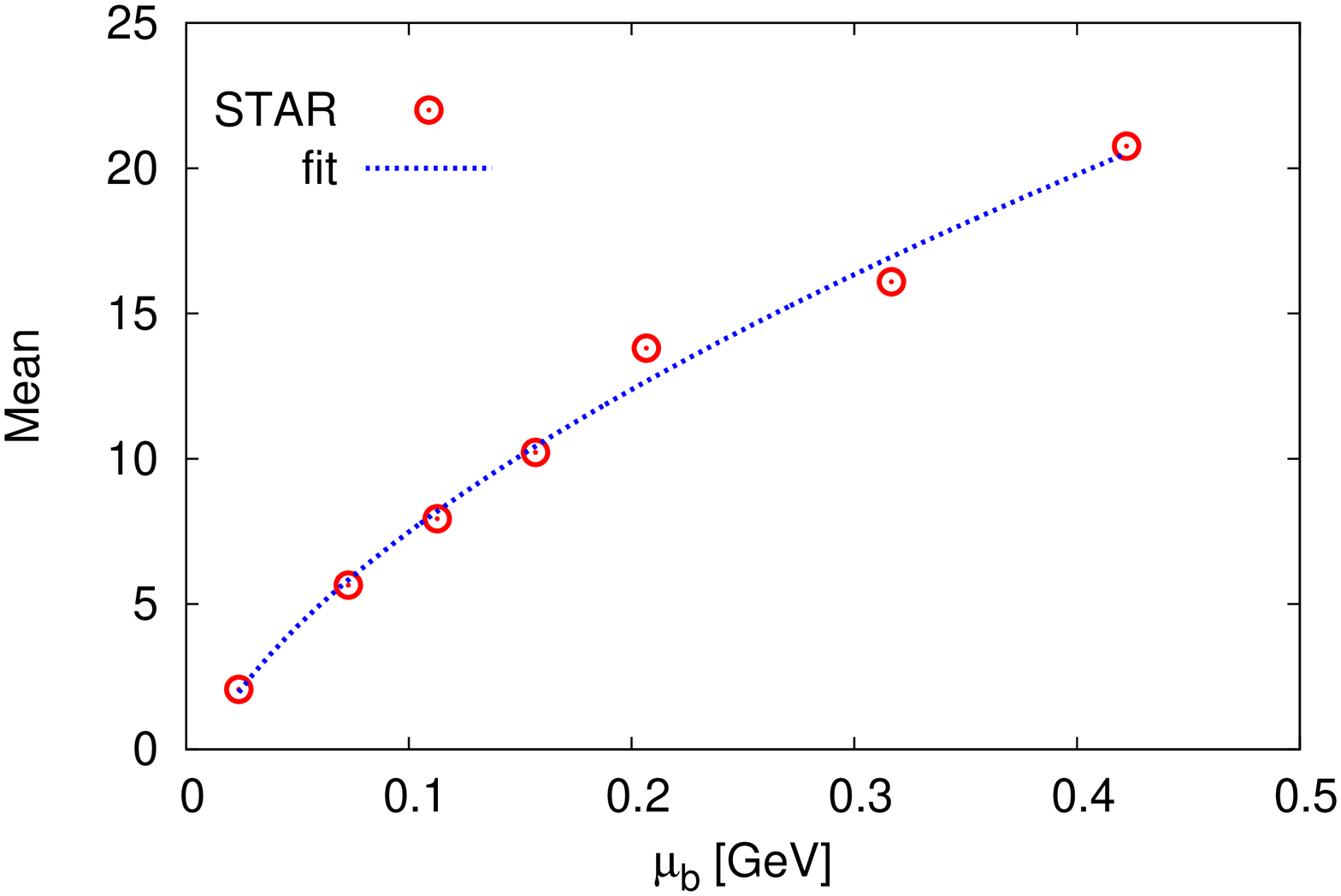}
\includegraphics[angle=-90,width=7.cm]{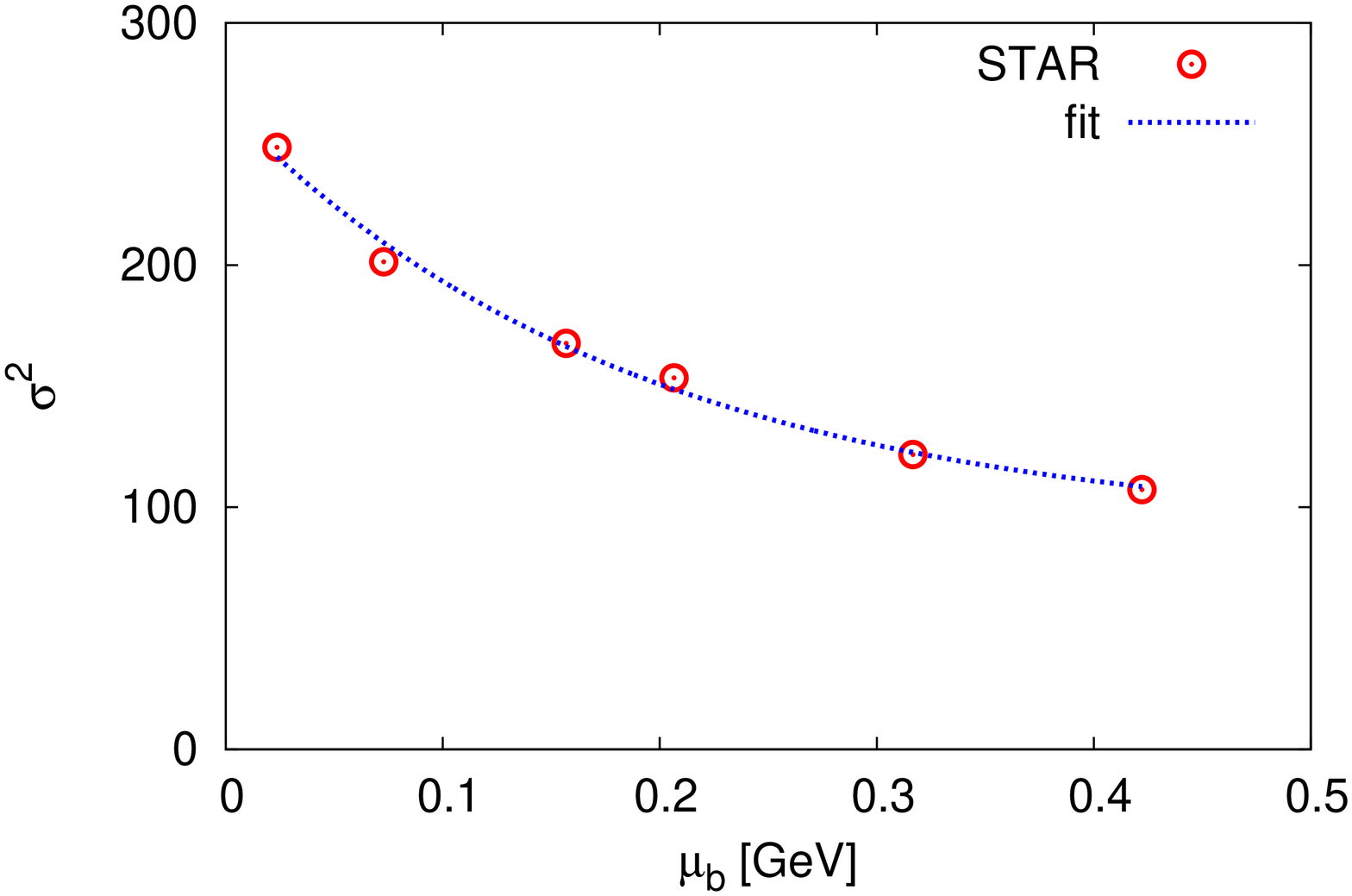} 
\caption{Mean (left panel) and standard deviation (left panel) are plotted with respect to the baryon chemical potential. The experimental data (symbols) taken from Ref. \cite{Sahoo}.}
\label{fig:myMS} 
}
\end{figure}

\begin{figure}[htb]
\centering{
\includegraphics[angle=-90,width=7.cm]{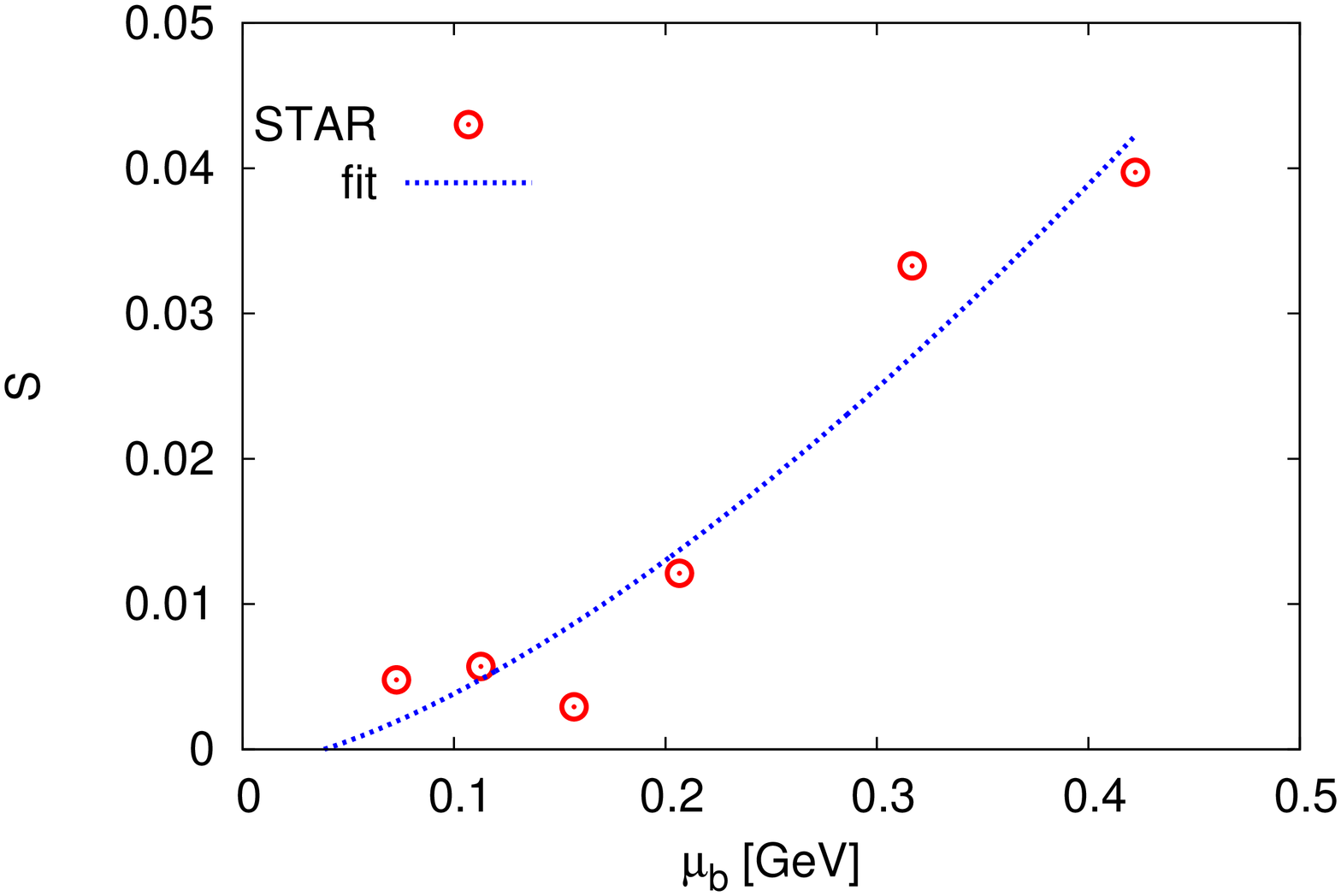}
\includegraphics[angle=-90,width=7.cm]{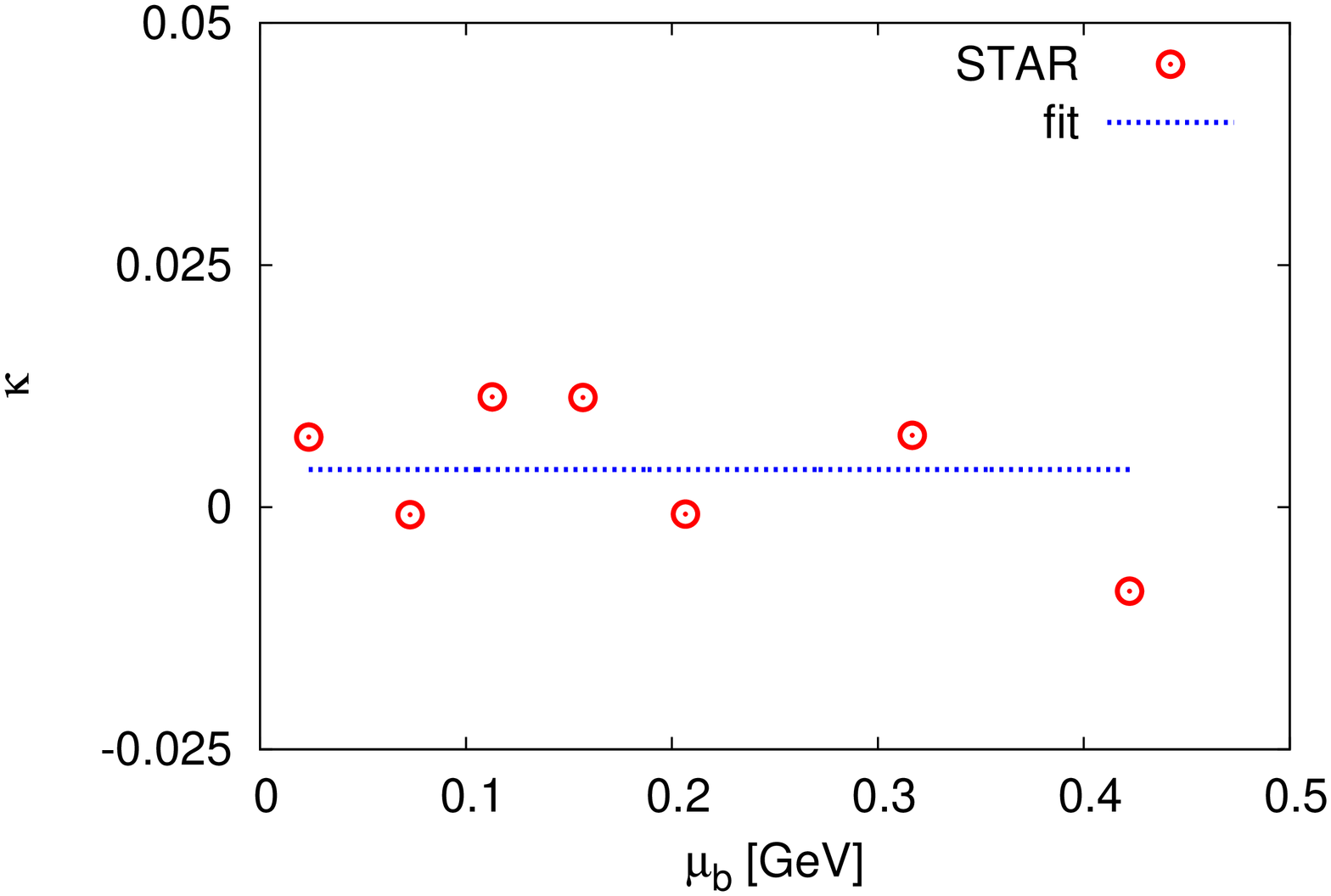}
\caption{Skewness (left panel) and Kurtosis (left panel) are plotted with respect to the baryon chemical potential. The experimental data (symbols) taken from Ref. \cite{Sahoo}. }
\label{fig:mySK} 
}
\end{figure}

\begin{figure}[htb]
\centering{
\includegraphics[angle=0,width=8.cm]{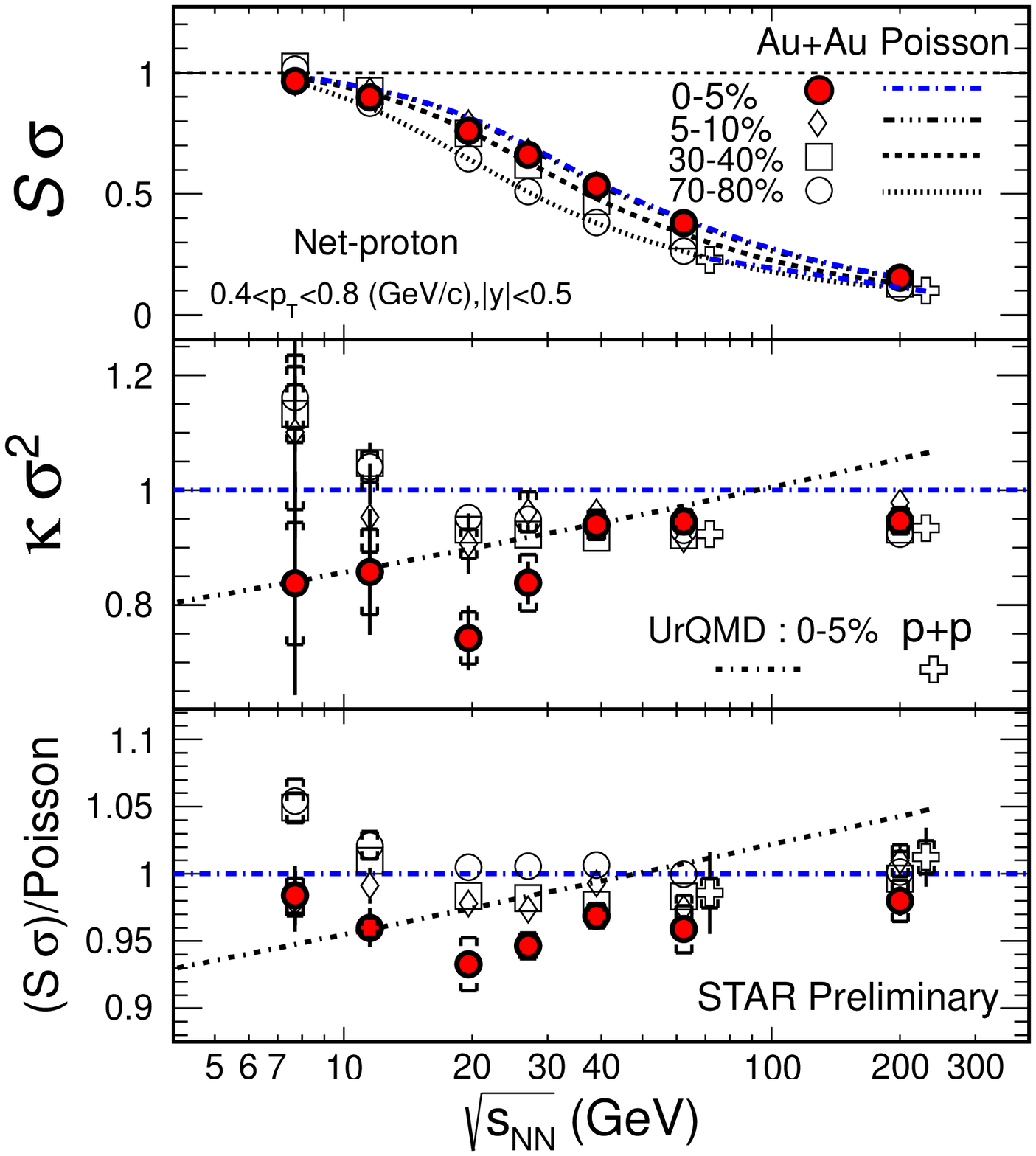}
\caption{The energy dependence of the moment products $S\,\sigma$ and $\kappa\,\sigma^2$ for net-proton distributions for different collision centralities ($0-5\%$, $5-10\%$, $30-40\%$ and $70-80\%$) measured in the STAR experiment. The results are compared to UrQMD model calculations and $pp$ collisions at $\sqrt{s_{NN}}=62.4$ and $200~$GeV. The lines in top panel are the Poisson  expectations and in the bottom panel shows the moment product $S\,\sigma$ normalized to the corresponding Poisson expectations. Graph taken from Ref. \cite{STARhm2013}. }
\label{fig:myProdHM} 
}
\end{figure}

In order to confront STAR results, Fig. \ref{fig:StarHM1}, to the statistical-thermal models, the moments measured at the highest $<N_{part}>$ are extracted, as they are related to the most-central collisions. In Figs. \ref{fig:myMS} and \ref{fig:mySK}, the first four moments of net-proton multiplicity are plotted with the baryon chemical potential $\mu_b$, which is related to the nucleon-nucleon center-of-mass energy $\sqrt{s_{NN}}$ \cite{jean2006}
\bea
\mu_b &=& \frac{1.308\pm 0.028 [\text{GeV}]}{1+(0.273\pm0.008 [\text{GeV}^{-1}]) \sqrt{s_{NN}}}.
\eea
We notice that $M$ and $S$ increase with increasing $\mu_b$, while the standard deviation decreases and $\kappa$ remains almost constant. 

Fig. \ref{fig:myProdHM} shows the energy dependence of the moment products $S\,\sigma$ and $\kappa\,\sigma^2$ of net-proton distributions for various collisions centralities ($0-5\%$, $5-10\%$, $30-40\%$ and $70-80\%$)  in Au+Au collisions measured by the STAR experiment. The bottom panel  shows the product $S\,\sigma$  normalized to the corresponding Poisson expectations. The values of $\kappa\,\sigma^2$ and normalized $S\,\sigma$ are close to the Poisson expectations for Au+Au collisions at $\sqrt{s_{NN}}=39$ , $62.4$ and $200~$GeV. There is a deviation from Poisson expectations for the $0-5\%$ central Au+Au collisions below  $\sqrt{s_{NN}}=39~$GeV. The UrQMD model~\cite{urqmd} results are also presented for $0-5\%$ centrality to understand the non-CP effects, such as baryon number conservation and hadronic scattering. The UrQMD calculations show a monotonic decrease with decreasing beam energy.

\subsection{Comparing with lattice QCD results}
\label{sec:HMlqcd}

\begin{figure}[htb]
\centering{
\includegraphics[angle=-0,width=8.cm]{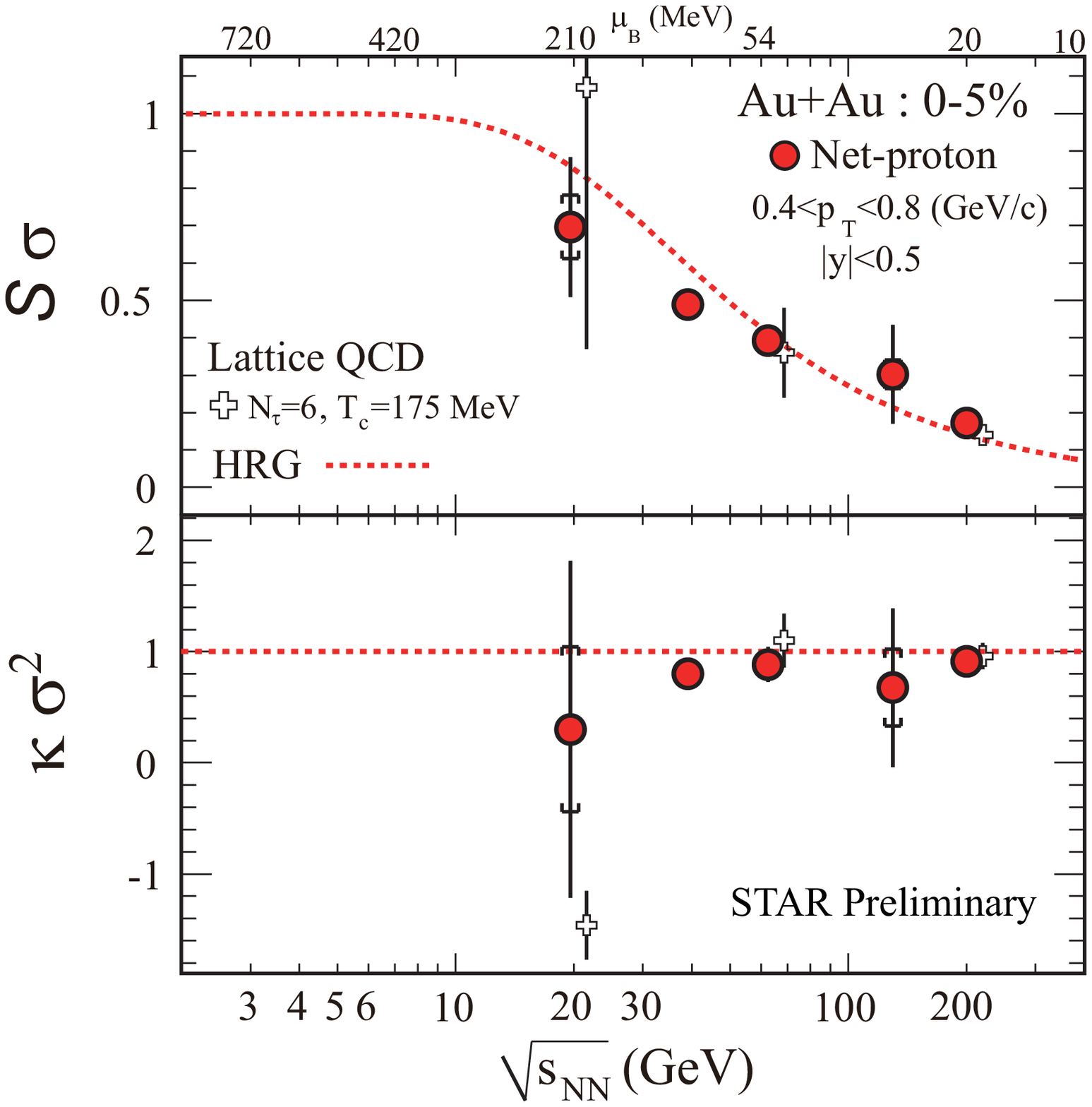}
\caption{The moment products $\kappa\,\sigma^2$ and $S\, \sigma$ of the net-proton distributions measured in central Au+Au collisions ($0-5\%$,
19.6 GeV: $0-10\%$, 130 GeV: $0-6\%$) are given as function of the center-of-mass energy. The dashed lines give
the statistical-thermal model results, while the empty markers refer to the Lattice QCD results~\cite{science}. Graph taken from Ref. \cite{star1B}. }
\label{fig:HMprodLQCD} 
}
\end{figure}

Fig. \ref{fig:HMprodLQCD} shows the moment products $S \sigma$ and $\kappa \sigma^2$ of the net-proton distributions measured in the most central Au+Au collisions ($0-5\%$, 19.6 GeV: $0-10\%$, 130 GeV: $0-6\%$) in dependence on the center-of-mass energy. The STAR measurements are compared with the lattice QCD~\cite{science} and HRG model  calculations \cite{star1B}. The lattice QCD results are obtained at temporal extension $N_{\tau}=6$. The critical temperature at $\mu_B$=0 was estimated as $T_{c}=175$ MeV. The dashed lines refer to the HRG calculations in upper panel and lower panels. They are evaluated as $S \sigma=\tanh(\mu_{B}/T)$ and $\kappa \sigma^{2}=1$, respectively. At the chemical freeze-out, the $\mu_{B}/T$ is parametrized as a function of the center-of-mass energy \cite{jean2006}. At $\sqrt{s_{NN}}$, the corresponding $\mu_B$, is illustrated in the upper band of Fig. \ref{fig:HMprodLQCD}. 

It is obvious that the moment products $\kappa \sigma^2$ and $S \sigma$ in the Au+Au collisions at $\sqrt{s_{NN}}=200$, $130$, $62.4~$GeV are consistent with the lattice QCD and the HRG model calculations, while the ones at $\sqrt{s_{NN}}=39~$GeV seems to deviates. Surprisingly, the moment product $\kappa \sigma^2$ from the lattice QCD calculations at $\sqrt{s_{NN}}=19.6~$GeV show a negative value~\cite{science}. However, due to the limited statistics, the uncertainty of the experimental measurements at $19.6$ GeV are not negligible. Such a deviation could be connected with the chiral phase-transition~\cite{chiral_HRG}. The possible existence of the QCD critical point might play a role, as well~\cite{Neg_Kurtosis}. 

Recent linear $\sigma$-model calculations demonstrate that the forth order cumulant of the fluctuations for the $\sigma$ field will be universally negative, when the QCD critical point is approached from the cross-over side~\cite{Neg_Kurtosis}. This will cause the measured $\kappa \sigma^2$ as well as kurtosis $\kappa$ of the net-proton distributions to be smaller than their Poisson expectations.

\subsection{Higher order moments of multiplicity and QCD critical endpoint}

\begin{figure}[htb]
\centering{
\includegraphics[angle=-90,width=10.cm]{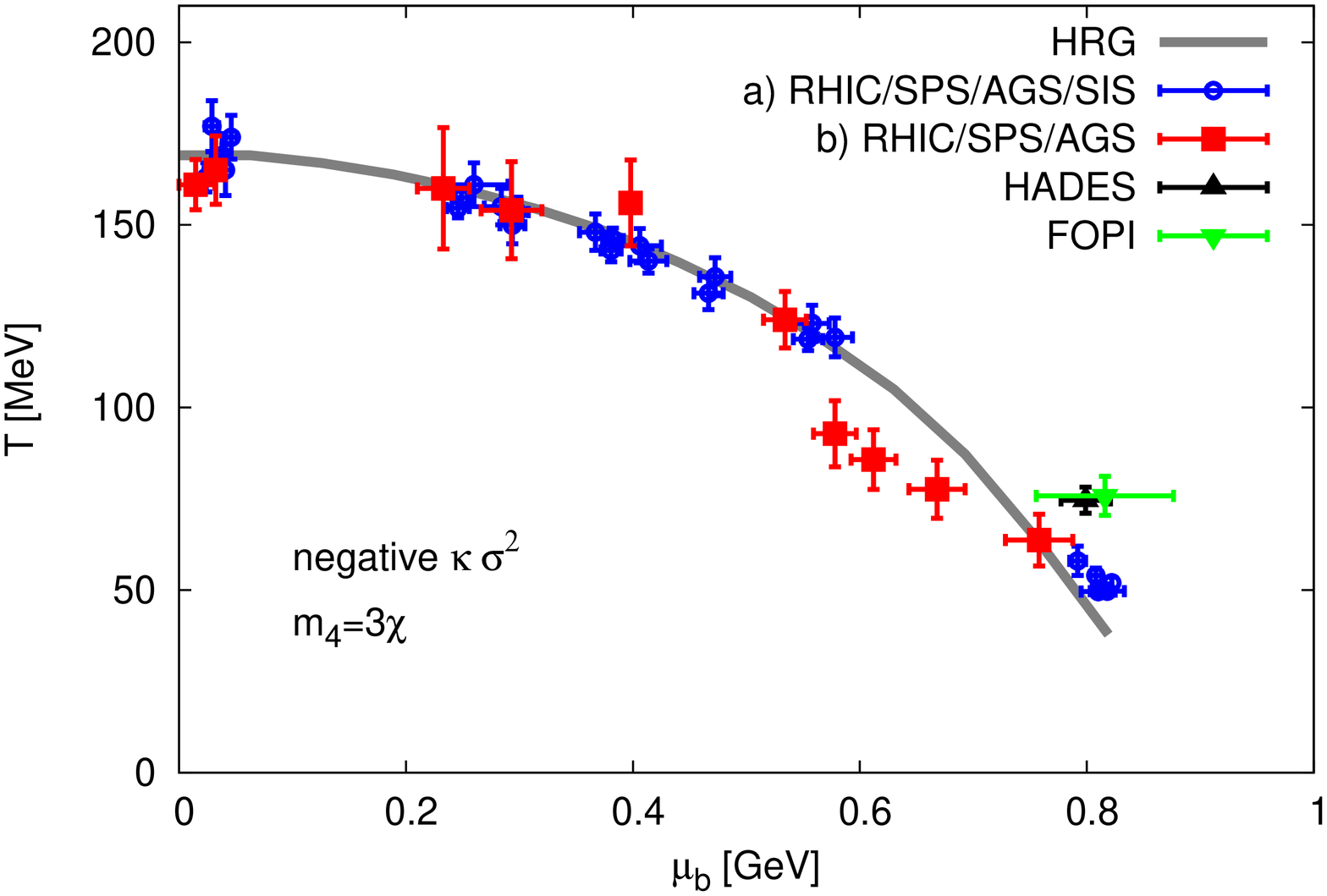}
\caption{The chemical freeze-out boundary calculated when the sign of ${\kappa}\,\sigma^2$ is flipped. The curves represent the results of hadronic (double-dotted, bosonic (dotted) and fermionic (dash-dotted) HRG. The point where the three curves intersect is coincident with the lattice QCD estimation for the critical endpoint. Graph taken from Ref. \cite{Tawfik:2013dba}. }
\label{fig:fezeout-sT3} 
}
\end{figure}

\begin{figure}[htb]
\centering{
\includegraphics[angle=-90,width=7.cm]{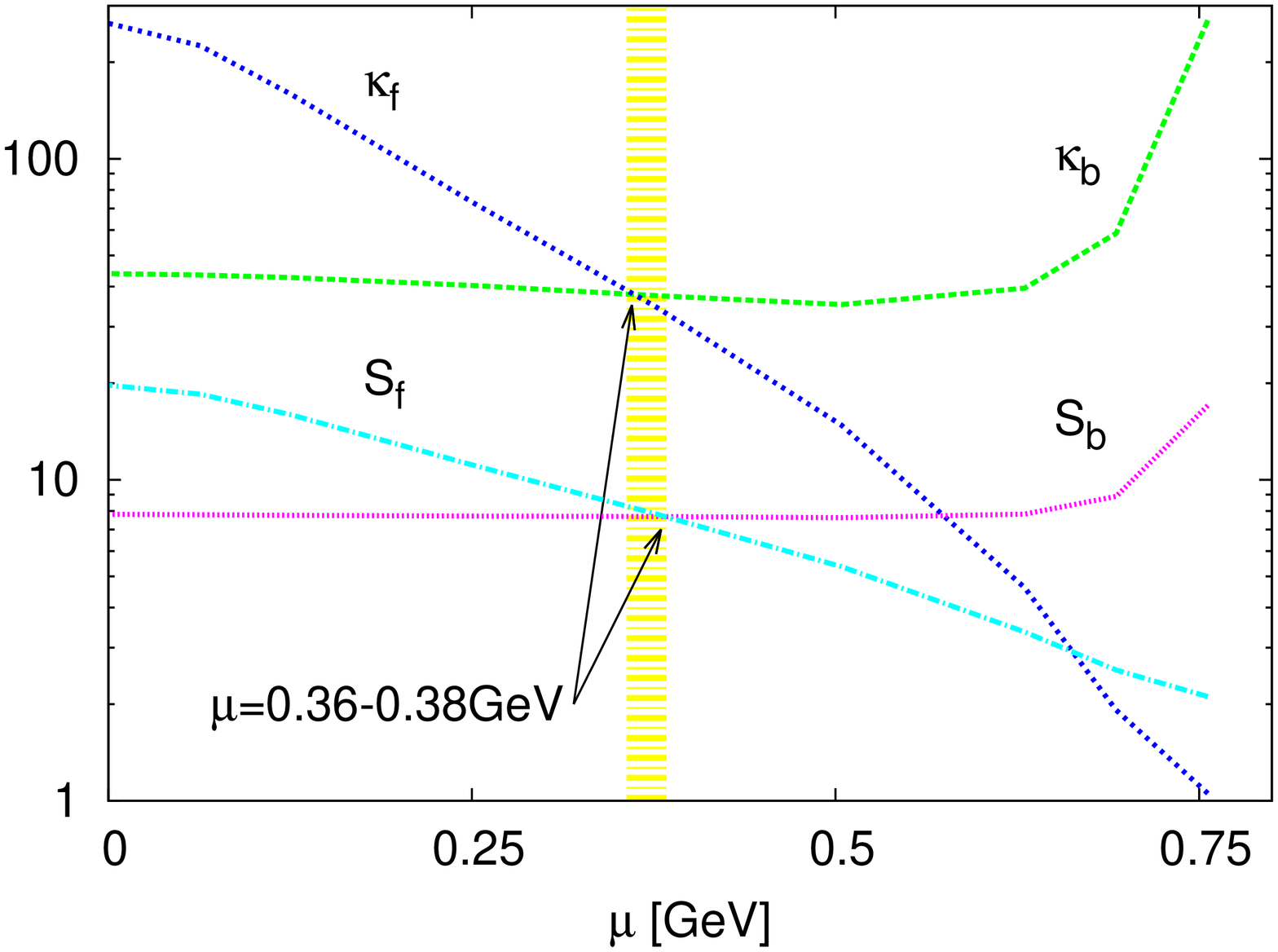}
\includegraphics[angle=-90,width=7.cm]{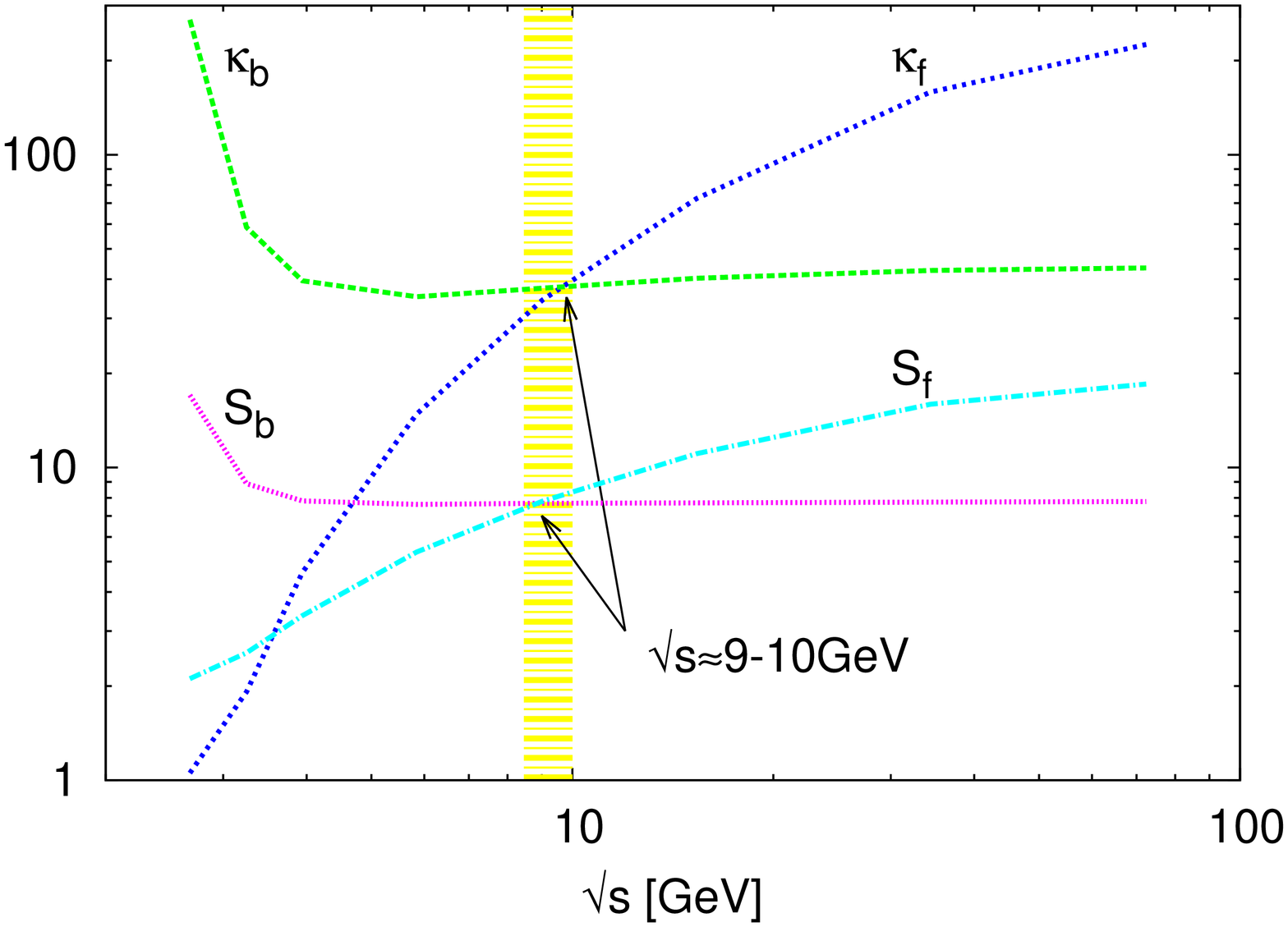}
\caption{Skewness and kurtosis calculated at the freeze-out boundary for fermions and bosons as function of chemical potential (left panel)  and center-of-mass energy (right panel). The curves of both quantities are crossing at almost the same $\mu$ value and the corresponding $T$ values are almost equal. The graph taken from Ref. \cite{Tawfik:2012si}.}
\label{fig:kSmuu} 
}
\end{figure}

In right panel of Fig. \ref{fig:fezeout-sT3}, it is interesting to notice that both fermionic and bosonic curves intersect at the hadronic curve at one point.  Furthermore, based on non-perturbative convergence radius, the critical endpoint as calculated in lattice QCD \cite{Gendp3b,Gendp3c} given by solid square  is located very near to the crossing point. It is clear that the HRG model does not contain any information on criticality related with the chiral dynamics and singularity in physical observables required to locate the critical endpoint. The HRG partition function is an approximation to a non-singular part of the free energy of QCD in the hadronic phase. It can be used only as the reference for lattice QCD calculations or heavy-ion collisions to verify the critical behavior, but it can not be used as an origin to search for the chiral critical structure in QCD medium. This is the motivation of Fig. \ref{fig:kSmuu}.

In Fig.  \ref{fig:kSmuu}, the skewness and kurtosis of bosons and fermions calculated at the freeze-out curve are given in dependence on $\mu$ (left panel) and center-of-mass energy $\sqrt{s}$ (right panel). It seems that the evolution of fermionic and bosonic skewness and kurtosis coincide at one point, marked with the vertical band. This fact would reflect the nature of the phase transition. It would be the critical endpoint connecting cross-over with the first order deconfinement phase transitions. At the QCD critical endpoint, the phase transition is conjectured to be of second order. It is worthwhile to mention the crossing point is amazingly coincident with the QCD critical endpoint measured in lattice QCD \cite{Gendp3b,Gendp3c}, Fig. \ref{fig:fezeout-sT3}, regardless its uncertainties \cite{Tawfik:2012si}.

From Eqs. (\ref{eq:Ssb})-(\ref{eq:Ssf}) and (\ref{eq:Kkb})-(\ref{eq:Kkf}), following expressions have to be solved in $\mu$, individually and/or dependently, in order to determine $\mu$ of the crossing point \cite{Tawfik:2012si}: 
\bea 
-\, \frac{\int_0^{\infty} \,  
\text{csch}\left[\frac{\varepsilon_i - \mu_i }{2\, T}\right]^4\, \text{sinh} \left[\frac{\varepsilon_i - \mu_i }{T}\right]  k^2 \, dk}
{\left[\int_0^{\infty} \left(\text{cosh} \left[\frac{\varepsilon_i-\mu_i }{T}\right]-1\right)^{-1} k^2 \, dk \right]^{3/2}} 
&=& 16\, \frac{\int_0^{\infty} \,  
\text{csch}\left[\frac{\varepsilon_i - \mu_i }{T}\right]^3\, \text{sinh}
\left[\frac{\varepsilon_i - \mu_i }{2\, T}\right]^4  k^2 \, dk}
{\left[\int_0^{\infty} \left(\text{cosh} \left[\frac{\varepsilon_i-\mu_i }{T}\right]+1\right)^{-1} k^2 \, dk \right]^{3/2} }, \label{eq:Ssf2} \\
-\frac{\int_0^{\infty}
  \left\{\text{cosh}\left[\frac{\varepsilon_i - \mu_i }{T}\right] + 2\right\} 
\text{csch}\left[\frac{\varepsilon_i - \mu_i }{2\, T}\right]^4  k^2 \, dk}
{\left[\int_0^{\infty} \left(1-\text{cosh}\left[\frac{\varepsilon_i-\mu_i }{T}\right]\right)^{-1} k^2 \, dk\right]^{2} }  
&=& \frac{\int_0^{\infty}
  \left\{\text{cosh}\left[\frac{\varepsilon_i - \mu_i }{T}\right] - 2\right\} 
\text{sech}\left[\frac{\varepsilon_i - \mu_i }{2\, T}\right]^4 k^2 \, dk}
{\left[\int_0^{\infty} \left(\text{cosh}\left[\frac{\varepsilon_i-\mu_i }{T}\right]+1\right)^{-1} k^2 \, dk\right]^{2} }.  \hspace*{8mm}\label{eq:Kkf2}
\eea
Due to the mathematical difficulties in dealing with these expressions, the integral over phase space has to be simplified. A suitable simplification is given as
\bea
\frac{p(T,\mu_q,\mu_s)}{T^4} &=& \pm \frac{1}{2\pi^2\,T^3}  \sum_{i=1}^{\infty}\,g_i\, m_i^2  \sum_{n=1}^{\infty}\, \frac{(\pm)^{n+1}}{n^2}\, \text{K}_2\left(n\frac{m_i}{T}\right)\; \exp\left[n\frac{(3 n_b+n_s)\mu_q-n_s\mu_s}{T}\right],
\eea 
where $n_b$ and $n_s$ being baryon (strange) quantum number and $\mu_q$ ($\mu_s$) is the baryon  (strange) chemical potential of light and strange quarks, respectively. The quarks chemistry is introduced in section \ref{sec:hrg1}. Accordingly, the difference between baryons and fermions is originated in the exponential function. For simplicity, we consider one fermion and one boson particle. Then, the baryon chemical potential $\mu$  at the chemical freeze-out curve, at which the fermionic and bosonic skewness (or kurtosis) curves of these two particles cross with each other can be given as \cite{Tawfik:2012si}
\bea
\mu_b = 3 n_b \mu_q &=& T \ln\left[\frac{g_b\, m_b^2\, \text{K}_2\left(\frac{m_b}{T}\right)}{g_f\, m_f^2\, \text{K}_2\left(\frac{m_f}{T}\right)}\right]. \label{eq:mubmusb}
\eea 
In the relativistic limit, $\text{K}_2(m/T)\approx 2 T^2/m^2-1/2$ while in the non-relativistic limit $\text{K}_2(m/T)\approx \sqrt{\pi T/2 m} \exp(-m/T)(1+15 T/8 m)$. It is obvious that the bosonic and fermionic degrees of freedom play an essential role in determining Eq. (\ref{eq:mubmusb}). Furthermore, it seems that the chemical potential of strange quark has no effect at the crossing point.

\begin{figure}[htb]
\centering{
\includegraphics[angle=-90,width=8.cm]{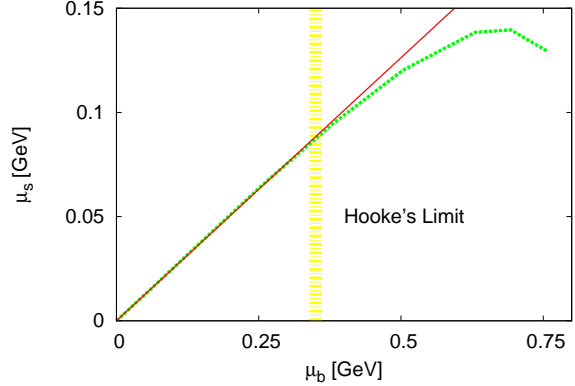}
\caption{The strangeness chemical potential $\mu_S$ as a function of the baryonic one $\mu_b$ in HRG (dotted curve). The linear fitting is given by solid curve. The graph is taken from Ref. \cite{Tawfik:2012si}.}
\label{fig:mub_mus} 
}
\end{figure}

The dependence of $\mu_S$ on $\mu_b$ as calculated in HRG is given in Fig. \ref{fig:mub_mus}. As mentioned above, $\mu_S$ is calculated to guarantee strange number conservation in heavy-ion collisions. At small $\mu_b$, $\mu_S$ has a linear dependence, $\mu_S=0.25\, \mu_b$ (Hooke's limit). At large $\mu_b$, the dependence  is no longer linear. 

As discussed in Ref. \cite{Tawfik:2012si}, the critical behavior and the existence of QCD critical endpoint can be identified by means of signatures sensitive to singular parts of the free energy, especially the ones reflecting dynamical fluctuations of conserved charges, such as baryon number and charge density \cite{cepC}. The reason that the fermionic and bosonic higher order moments are crossing on a point very near the QCD critical endpoint should be analysed, systematically \cite{Tawfik:2013dba}.


\section{Models combining hadronic ($<T_c$) and partonic ($>T_c$) phases}
\label{sec:ABTc}

We introduce three effective approaches. They are examples about other models for thermodynamics of the hadronic matter. Their analogy to the statistical-thermal models is obvious. The three models reproduce very well the lattice QCD thermodynamics below $T_c$, i.e. the effective mass and coupling seem to match that of the hadronic matter. The quasi-particle model is applicable in hadronic as well as in partonic phases, section \ref{hrg_quasiprtcl}. The linear $\sigma$-models with Polyakov-loop potentials shall be introduced in section \ref{hrg_schaefer}. Finally, we devote section \ref{hrg_koch} to the compressible bag model, in which the typical particle density at the transition requires that the size of hadrons needs to be taken into  account as it leads to a considerable suppression of the available phase-space and 
furthermore the volume of the hadrons should be allowed to vary under the effect of the pressure generated by their own thermal motion in a self-consistent way.

In all these approaches, the effective degrees of freedom is remarkably reduced. This suppression alone, however, does not explain the ideal gas behavior of lattice QCD at high temperatures, above $T_c$. Additional assumptions, should be added. For example, one can explicitly introduce a deconfined phase and then match the the two different phases. This was adopted that the same  partition function should describe both confined and deconfined phases. Appropriate dynamics in order to model the cross-over observed on the lattice should be introduced.

\subsection{Hadronic and partonic phases in quasi-particle model}
\label{hrg_quasiprtcl}

The quasi-particle model \cite{kmpf1} is constructed to reproduce the lattice QCD calculations at $T>T_c$. The free parameters are fixed for this purpose.  Accordingly, the pressure at finite $T$ and $\mu$ is given as
\bea
p &=& \sum_{i=q,g}\, p_i - B(T,\mu), \\
p_i &=& \frac{g_i}{6\, \pi^2} \int_0^{\infty} dk \frac{k^4}{E_i(k)} \left[f_i^+(k)+f_i^-(k)\right],
\eea
where the distribution function reads
\bea
f_i^{\pm}(k) &=& \frac{1}{\exp\left[\frac{E_i(k)\mp \mu}{T}\right]\pm 1}.
\eea
We will see that this model is also able to reproduce the thermodynamical quantities even in the hadronic phase. The quasi-particle dispersion relation can be approximated by the asymptotic mass shell expression near the light cone \cite{kmpf1,qp18a,qp18b},
\bea
E_i^2(k) &=& k^2 + m_i^2(T,\mu)=k^2 + \Pi_i(k; T,\mu) + (x_i\, T)^2, 
\eea
where $\Pi_i(k; T,\mu)$ is the self-energy at finite $T$ and $\mu$ and $x_i^2$ is a factor taking into account the mass scaling as using in lattice QCD simulations, section \ref{sec:krt1}. A suitable parametrization of $\Pi_i(k; T,\mu)$ is given by the hard thermal loop self-energies \cite{kmpf1,qp8}.  But, the running coupling in $\Pi_i(k; T,\mu)$ should then be replaced by an effective one, $G^2(k; T,\mu)$, which can be adjusted to reproduce the lattice QCD calculations \cite{qp17} and reflect the non-perturbative effects.

The function $B(T,\mu)$ is there to assure thermodynamical self-consistency \cite{qp8} that $n=\partial p/\partial \mu$, $s=\partial p/\partial T$ and $\partial p/\partial m_i^2=0$ \cite{qp12}. Then, the net quark number and entropy, respectively, are given as
\bea
n&=& \sum_{i=q,g}\, \frac{g_i}{2\, \pi^2} \int_0^{\infty} k^2\, dk  \left[f_i^+(k)-f_i^-(k)\right], \\
s&=& \sum_{i=q,g}\, \frac{g_i}{2\, \pi^2\, T} \int_0^{\infty} k^2\, dk \left\{\frac{\frac{4 k^2}{3}+m_i^2}{E_i(k)} \left[f_i^+(k)+f_i^-(k)\right] - \mu  \left[f_i^+(k)-f_i^-(k)\right]\right\}.
\eea

The lattice QCD simulations at finite $T$ and $\mu$ prefer the use of \cite{ejiri,Gendp3b}, 
\bea
p(T,\mu)&=&p(T,\mu=0) +\Delta p(T,\mu)= p(T,\mu=0) + T^4 \sum_{j=2}^{\infty}\, c_j \left(\frac{\mu}{T}\right)^j. \label{eq:pLQCDmu}
\eea
Then, the coefficients of Eq. (\ref{eq:pLQCDmu}) are given by the derivative at the point of vanishing chemical potential $c_j=(T^{j-4}/j!)\partial^j p/\partial \mu^j$ \cite{kmpf1}
\bea
c_2 &=& 3 \frac{n_f}{\pi^2 T^3} \int_0^{\infty} k^2 dk \frac{\exp(\omega)}{[\exp(\omega)+1]^2}, \\
c_4 &=& \frac{1}{4} \frac{n_f}{\pi^2 T^3} \int_0^{\infty} k^2 dk \frac{\exp(\omega)}{[\exp(\omega)+1]^4} \, \left\{\exp(2\omega)-4\exp(\omega)-\frac{A_2}{\omega}\left[\exp(2\omega)-1\right] +1\right\},\\
c_4 &=& \frac{3}{385} \frac{n_f}{\pi^2 T^3} \int_0^{\infty} k^2 dk \frac{\exp(\omega)}{[\exp(\omega)+1]^6} \left\{\exp(4\omega)-26\exp(3\omega)+66\exp(2\omega)-26\exp(\omega) +1 \right. \nn \\ &-& \left. \frac{10}{3}\frac{A_2}{\omega}\left[\exp(4\omega)-10\exp(3\omega)+10\exp(\omega)-1\right] \right. \nn \\
&+& \left. \frac{4}{3}\frac{A_2^2}{\omega^2}\left[\exp(4\omega)-2\exp(3\omega)-6\exp(2\omega)-2\exp(\omega)+1\right] \right.\nn \\
&+& \left. \left(\frac{5}{3}\frac{A_2^2}{\omega^3}-10T^2\frac{A_4}{\omega}\right) \left[\exp(4\omega)+2\exp(3\omega)-2\exp(\omega)-1\right]
\right\},
\eea
where the following parameters are defined at the point of vanishing chemical potential $\mu$
\bea
A_2 &=& \frac{G^2}{\pi^2} + \frac{T^2}{2} \frac{\partial^2\, G^2}{\partial \mu^2}, \\
A_4 &=& \frac{1}{\pi^2} \frac{\partial^2\, G^2}{\partial \mu^2} +  \frac{T^2}{12} \frac{\partial^4\, G^2}{\partial \mu^4}, \\
\omega &=& \frac{1}{T} \sqrt{k^2+\frac{1}{3}T^2 G^2}.
\eea
In deriving the previous expressions, flow equation \cite{qp8} was utilized 
\bea
a_{\mu} \frac{\partial\, G^2}{\partial \mu} + a_{T} \frac{\partial\, G^2}{\partial T} &=& a_{\mu T},
\eea
where the functions $a_{\mu}$, $a_{T}$ and $a_{\mu T}$ can be determined  \cite{kmpf1,qm16}, if $a_{T}(T,\mu=0)=a_{T}(T,\mu=0)$ should vanish.

Then, the derivatives of coupling $G$ at $\mu=0$ are given as  
\bea
\left.\frac{\partial^2\, G^2}{\partial \mu^2}\right|_{\mu=0} &=& \left.\frac{1}{a_{\mu}} \left(\frac{\partial\, a_{\mu T}}{\partial \mu} - \frac{\partial\, a_{T}}{\partial \mu} \frac{\partial\, G^2}{\partial T}\right)\right|_{\mu=0}, \\
\left.\frac{\partial^4\, G^2}{\partial \mu^4}\right|_{\mu=0} &=& \left.\frac{1}{a_{\mu}} \left(\frac{\partial^3\, a_{\mu T}}{\partial \mu^3} - \frac{\partial^3\, a_{T}}{\partial^3 \mu} \frac{\partial\, G^2}{\partial T} - 3 \frac{\partial^2\, a_{\mu}}{\partial \mu^2}\frac{\partial^2\, G^2}{\partial \mu^2} \right. \right.\nn \\
&-&\left. \left.\frac{3}{a_{\mu}} \frac{\partial\, a_{T}}{\partial \mu} 
\left[
  \frac{\partial^2\, a_{\mu T}}{\partial \mu \partial T} - 
  \frac{\partial^2\, a_{T}}{\partial \mu \partial T} \frac{\partial\, G^2}{\partial T} -
  \frac{\partial\, a_{T}}{\partial \mu} \frac{\partial^2\, G^2}{\partial T^2} -
  \frac{\partial\, a_{\mu}}{\partial T} \frac{\partial^2\, G^2}{\partial \mu^2}  
\right]
\right)\right|_{\mu=0}.
\eea
Accordingly,  the thermal evolution of coupling $G$ at $\mu=0$ reads
\bea
G^2(T) &=& \left\{ \begin{array}{ll}
G^2_{\text{2loop}}(T), & T\geq T_c, \\
& \\
G^2_{\text{2loop}}(T)+b\left(1-\frac{T}{T_c}\right), & T< T_c.\end{array}
\right.,
\eea
The two-loop effective coupling is given as
\bea
G^2_{\text{2loop}}(T) &=& \frac{16\, \pi^2}{\beta_0 \log(\xi^2)} \left[1-2 \frac{\beta_1}{\beta_0^2} \frac{\log(\log(\xi^2))}{\log(\xi^2)}\right],
\eea
where \cite{kmpf1}
\bea
\xi &=& \lambda \frac{T-T_s}{T_c},
\eea
and $T_s$ is a regulator at $T_c$. The parameter $\lambda$ is used to adjust the scale as found in lattice QCD simulations. The $\beta$ function \cite{betaf} depends on the coupling $g$, 
\bea
\beta &=& \frac{\partial\, g}{\partial\, \log(\Delta_{\mu})},
\eea
where $\Delta_{\mu}$ is the energy scale. 
The two-loop perturpation estimation for $\beta$ functions give
\bea
\beta_0 &=& \frac{1}{3} \left(11\, n_c - 2\, n_f\right), \\
\beta_1 &=& \frac{1}{6} \left(34\, n_c^2 - 13\, n_f\, n_c + 3\, \frac{n_f}{n_c}\right).
\eea

\subsubsection{Thermodynamics}

\begin{figure}
\begin{minipage}[t]{7cm} 
\centering
\hspace*{-8cm}\includegraphics[width=6cm,angle=90]{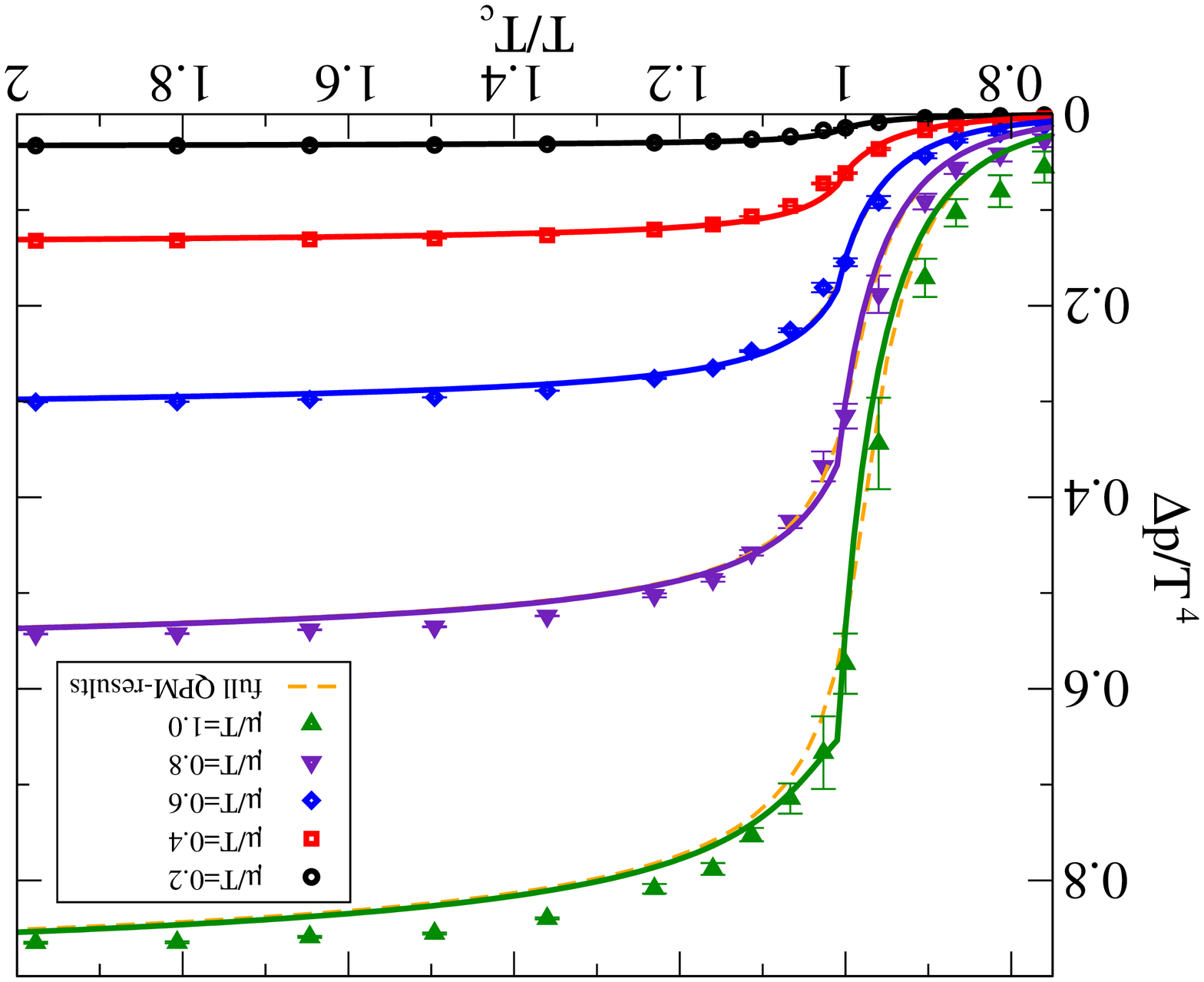}
\caption{The excess pressure  for constant $\mu / T$ of quasi-particle model is compared with lattice QCD calculations \cite{ejiri}. Graph is taken from Ref. \cite{kmpf1}. }
\label{fig:kmpf1a}
\end{minipage}
\hspace{0.5cm} 
\begin{minipage}[t]{7cm}
\centering
\hspace*{-8cm}\includegraphics[width=6cm,angle=90]{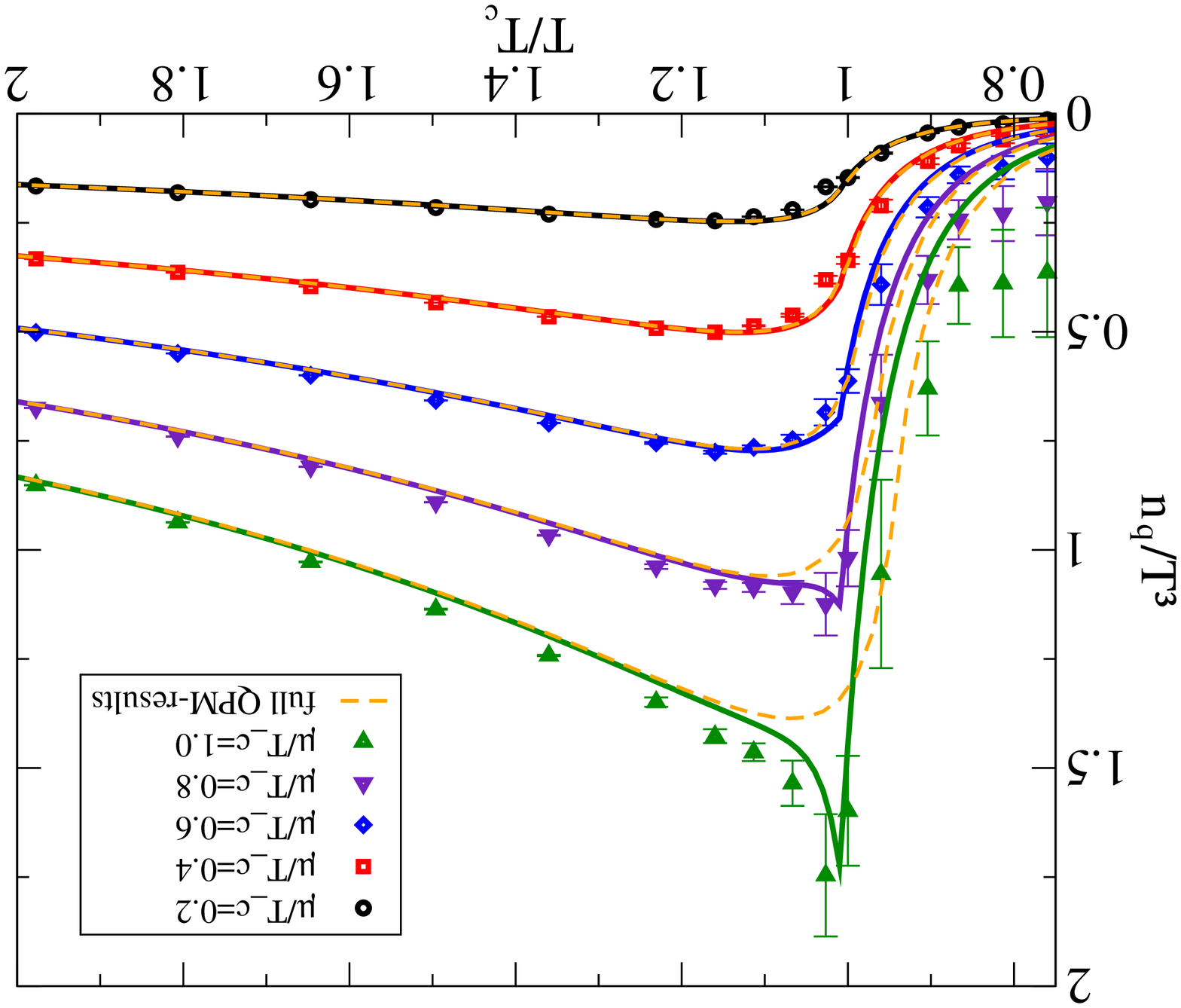}
\caption{The net quark number density for constant $\mu / T_c$ of  quasi-particle model is compared with lattice QCD calculations \cite{ejiri}. Graph is taken from Ref. \cite{kmpf1}. }
\label{fig:kmpf1b}
\end{minipage}
\end{figure}


The quasi-particle model has different free parameters, $G$, $T_s$, $\lambda$, $x_i$, etc. The procedure of fixing these parameters is elaborated in Ref. \cite{kmpf1}, for instance.
Fig. \ref{fig:kmpf1a} presents the excess pressure $\Delta p(T, \mu)/T^4$ above and even slightly below $T_c$ at small chemical potential. The individual expansion coefficients agree well with the data and turn out to depend on each other. In particular, also $p(T, \mu=0)$ follows, once $G^2(T)$ is adjusted. The normalized net quark number density is given in Fig. \ref{fig:kmpf1b}. For lattice QCD calculations (symbols),  the quasi-particle model results (solid curves) are based on the the expansion coefficients $c_j$. For comparison, the full quasi-particle model results (dashed curves) are exhibited. 

The quasi-particle model  is constructed to reproduce the lattice QCD calculations at $T>T_c$. Almost all parameters are fixed for this purpose. The figures show an excellent agreement in this region. On the other hand, the model is also able to simulate the results at $T<T_c$. Within a narrow region below $T_c$, the quasi-particle and statistical-thermal models agree, compare with section \ref{sec:lqcd}.

\subsection{Linear $\sigma$-models with Polyakov-loop potentials}
\label{hrg_schaefer}

The hiral symmetry breaking restoration (above $T_c$) can be linked with  aspects of the confinement/deconfinement transition through the linear $\sigma$-model, in which Polyakov-loop potential for three quark flavors is included \cite{schaefer09,Tawfik:2014hwa,Tawfik:2014uka}. The two-quark-flavor model \cite{Schaefer:2007pw} is combined with the chiral linear $\sigma$-model \cite{lsm1} and the Polyakov-loop potential $\Phi(\vec x)$. In temporal direction, the thermal expectation values for the color traced Wilson-loop  reads \cite{schaefer09,Tawfik:2014hwa,Tawfik:2014uka}
\begin{equation}
  \label{eq:defphi} 
  \Phi(\vec x) = \frac{1}{N_c} \langle \ensuremath{\Tr}_c \mathcal{P}
  (\vec x) \rangle,
\end{equation} 
where $A_0$ is the temporal vector field, which as given above is contained in the so-called Polyakov loop operator $\mathcal{P}(\vec x)$ in $SU(N_c)$ gauge group \cite{schaefer09},
\begin{equation}
  \label{eq:defP}
  \mathcal{P}(\vec x) = \mathrm{P} \exp \left( i \int_0^{\beta} d\tau A_0
    (\vec x, \tau ) \right),
\end{equation}
where $\mathrm{P}$ is the path ordering and $\beta$ is the inverse temperature $1/T$ \cite{Polyakov:1978vu}. It is obvious that the Polyakov loop potential is a classical variable. Details about the implementation of  Eq. (\ref{eq:defP}) are in order now. The Lagrangian apparently consists of \cite{schaefer09}:
\begin{itemize}
\item a quark-meson contribution and 
\item a Polyakov-loop potential $\mathcal{U} (\Phi[A], \Phibar[A])$, which depends on the Polyakov-loop variable $\Phi$ and its hermitian conjugate $\Phibar= \langle \ensuremath{\Tr}_c \mathcal{P}^\dagger \rangle/ N_c$, where $N_c$ gives the number of colors. 
\end{itemize}
In the quark-meson contribution, the standard derivative $\partial_\mu$ is replaced by a covariant derivative, 
\bea
  D_\mu &=& \partial_\mu - i A_\mu,\\  
  A_\mu &=& \delta_{\mu 0}  A^0.
\eea
The $A_\mu = g_s A_\mu^a \lambda^a/{2}$ includes the $SU(N_c)$ gauge coupling $g_s$. The usual Gell-Mann matrices are $\lambda^a$ with $a=1,\ldots, N_c^2-1$. 

Then, the Lagrangian can be constructed as 
\begin{equation}
  {\cal L}_{PQM} = \bar{q}\left(i \Dslash - g \phi_5 \right) q +
  \mathcal{L}_m  
  -\mathcal{U} (\Phi[A], \Phibar[A]), \label{eq:lpqm}
\end{equation}
where $q=(u, d, s)$ is the quark field for $N_f = 3$ flavors and $N_c=3$ colors.  For three quark flavors, the interaction between quarks and meson is achieved by the flavor-blind Yukawa coupling $g$. The meson matrix $\phi_5 = T_a \left(\sigma_a + i \gamma_5\, \pi_a \right)$, where $T_a=\lambda_a/2$, $a=0,\ldots, 8$ are nine generators of the $U(3)$ symmetry with $\lambda_0=\sqrt{2/3}\ \bf 1$. The generators $T_a$ are normalized to $\Tr (T_a\, T_b)=\delta_{ab}/2$ and should obey the $U(3)$ algebra $[T_a,T_b]=i\, f_{abc}\,T_c$ and $\{T_a,T_b\}=d_{abc}T_c$, respectively,  with the corresponding standard symmetric $d_{abc}$ and antisymmetric $f_{abc}$ structure constants of $SU(3)$ group and $f_{ab0}=0$ and $d_{ab0}=\sqrt{2/3} \delta_{ab}$. The nine scalar ($J^P=0^+$) mesons are labelled by the $\sigma_a$ fields and accordingly the nine pseudoscalar ($J^P = 0^-$) mesons by the $\pi_a$ fields.

The purely mesonic contribution to the Lagrangian reads 
\begin{eqnarray}
  \mathcal{L}_m &=& \Tr \left( \partial_\mu \phi^\dagger \partial^\mu
    \phi \right)
  - m^2 \Tr ( \phi^\dagger \phi) -\lambda_1 \left[\Tr (\phi^\dagger
    \phi)\right]^2 \nonumber \\
  & - &   \lambda_2 \Tr\left(\phi^\dagger \phi\right)^2
  + c   \left(\det (\phi) + \det (\phi^\dagger) \right) \label{eq:mesonL} \\
  & + & \Tr\left[H(\phi+\phi^\dagger)\right], \nonumber
\end{eqnarray}
where $\phi=T_a \phi_a=T_a \left(\sigma_a + i \pi_a\right)$ is a complex $(3\times 3)$-matrix for $N_f=3$ has been implemented. The last term in Eq. (\ref{eq:mesonL}) assures that chiral symmetry is broken, explicitly, with $H=T_a\, h_a$ is $(3\times 3)$-matrix with nine external parameters $h_a$. The $\ua$-symmetry is explicitly broken by the 't Hooft determinant term with a constant strength $c$. Further details concerning the three-flavor quark-meson part of the Polyakov-Quark-Meson (PQM) model can be found in \cite{Schaefer:2008hk}.
 
The final effective gluon field potential $\mathcal{U}$ in Eq. (\ref{eq:lpqm}) is given as function of the Polyakov-loop fields $\Phi$ and $\Phibar$. The pure Yang-Mills symmetry is guaranteed \cite{Pisarski:2000eq,Dumitru:2001xa}. The Polyakov-loop potential is motivated by the underlying QCD symmetries in pure gauge limit \cite{schaefer09}. Therefore, different ansaetze for the Polyakov-loop potential have been suggested \cite{schaefer09}:
\begin{itemize} 
\item The first one is based on Ginzburg-Landau theory \cite{Schaefer:2007pw,Ratti:2005jh}, where $Z(3)$ center symmetry is spontaneously broken in the pure Yang-Mills.
\bea
\label{eq:upoly}
  \frac{\mathcal{U}_{\text{poly}}}{T^{4}} &=& -\frac{b_2}{4}
\left(|\Phi|^2+|\bar\Phi|^2 \right)  
-\frac{b_3}{6}(\Phi^3+\Phibar^3)+\frac{b_4}{16}
\left(|\Phi|^2+|\Phibar|^2\right)^2,
\eea
with
\begin{equation}
  b_2(T) = a_0 + a_1 \left(\frac{T_0}{T}\right) + a_2
  \left(\frac{T_0}{T}\right)^2 + a_3 \left(\frac{T_0}{T}\right)^3, \nn
\end{equation}
are temperature-dependent coefficients. The pure gauge lattice results (like thermodynamics, EoS, etc.) should be reproduced. Accordingly, the potential parameters are determined. The Polyakov-loop expectation values are reproduced~\cite{Ratti:2005jh}: $a_0=6.75$, $a_1=-1.95$, $a_2=2.625$, $a_3 = -7.44$, $b_3=0.7$, $b_4=7.5$. The parameter $T_0=270~$MeV corresponds to the critical temperature. Extension Eq (\ref{eq:upoly}) to finite chemical potential was discussed \cite{Schaefer:2007pw}. At vanishing chemical potential, we find that $\bar\Phi = \Phi^\dagger$ and the polynomial expansion in $\Phi$ and $\bar\Phi$ includes the term $\Phi\bar\Phi$~\cite{Pisarski:2000eq,Ratti}.
\item The second ansatz was introduced in Ref. \cite{Roessner:2006xn} 
\bea
\label{eq:ulog}
\frac{\mathcal{U}_{\text{log}}}{T^{4}} &=& -\frac{1}{2}a(T) \Phibar \Phi 
+ b(T) \ln \left[1-6 \Phibar\Phi + 4\left(\Phi^{3}+\Phibar^{3}\right)
  - 3 \left(\Phibar \Phi\right)^{2}\right],
\eea
where 
\bea
  a(T) &=& a_0 + a_1 \left(\frac{T_0}{T}\right) + a_2 \left(\frac{T_0}{T}\right)^2, \nn \\
  b(T) &=& b_3 \left(\frac{T_0}{T}\right)^3, \nn
\eea
with  $a_0=3.51$, $a_1=-2.47$, $a_2=15.2$ and $b_3=-1.75$.
\item A third ansatz is inspired by a strong-coupling analysis \cite{Fukushima:2008wg} 
\bea
\label{eq:ufuku}
\frac{\mathcal{U}_{\text{Fuku}}}{T^4} &=& -\frac{b}{T^3}  \left[54 e^{-a/T} \Phi \Phibar 
+ \ln \left(1 - 6 \Phi \Phibar - 3 (\Phi \Phibar)^2 + 4 (\Phi^3
    + \Phibar^3)\right)\right].
\eea
We want to highlight that:
\begin{itemize}
\item the first term represents reminiscent of the nearest-neighbor interaction, 
\item the second term (logarithm) stands for the Haar measure, Eq. (\ref{eq:ulog}),
\item the parameter $a$ determines the deconfinement phase-transition, and, 
\item the parameter $b$ controls the mixing of chiral and deconfinement phase-transitions. 
\end{itemize}
Both parameters can be fixed in the pure gauge sector, $a=664~$MeV and $b= 196.2^3~$MeV$^3$. In the Polyakov-Nambo-Jena-Lasino (PNJL) model at vanishing chemical potential~\cite{Fukushima:2008wg}, a coincidence of chiral and deconfinement phase-transitions was observed at $T_c\sim200~$MeV.
\end{itemize}

\subsubsection{Thermodynamics}

In the n isospin-symmetry, the masses of light quarks are degenerated, i.e. $m_l \equiv m_u=m_d$ and the thermodynamical potential depends only on $\mu_l = (\mu_u+\mu_d)/2$ and $\mu_s$. The thermodynamical potential should be evaluated in the mean-field approximation \cite{Schaefer:2007pw,Schaefer:2008hk} in order to investigate the phase structure of this model \cite{schaefer09,Tawfik:2014hwa,Tawfik:2014uka}. Meson $U\left(\sigma_{x},\sigma_{y}\right)$, quark/antiquark $\Omega_{\bar{q}{q}}$ and Polyakov-loop fields likely contribute to the thermodynamical potential
\begin{equation}
  \label{eq:grandpot}
  \Omega = U \left(\sigma_{x},\sigma_{y}\right) +
  \Omega_{\bar{q}{q}} \left(\sigma_{x},\sigma_{y}, \Phi,\Phibar \right) +
  \mathcal{U}\left(\Phi,\Phibar\right). 
\end{equation}
The mesonic contribution reads \cite{schaefer09,Tawfik:2014hwa,Tawfik:2014uka}
\bea
  U(\sigma_{x},\sigma_{y}) &=& \frac{m^{2}}{2}\left(\sigma_{x}^{2} +
  \sigma_{y}^{2}\right) -h_{x} \sigma_{x} -h_{y} \sigma_{y}
 - \frac{c}{2 \sqrt{2}} \sigma_{x}^2 \sigma_{y} \nn \\
 & + & \frac{\lambda_{1}}{2} \sigma_{x}^{2} \sigma_{y}^{2}+
  \frac{1}{8}\left(2 \lambda_{1} +
    \lambda_{2}\right)\sigma_{x}^{4}+\frac{1}{8}\left(2 \lambda_{1} +
    2\lambda_{2}\right) \sigma_{y}^{4}. \label{eq:umeson}
\eea
The six parameters $m^{2}, c, \lambda_{1}, \lambda_{2}, h_{x}, h_{y}$ are fitted to well-known pseudoscalar meson masses $m_{\pi}, m_{K}, m^{2}_{\eta}+m^{2}_{\eta'}$ and two weak-decay constants $f_{\pi},\, f_{K}$ \cite{schaefer09}. The mass of scalar $\sigma$ meson, $\msig$, is used to complete the fit. As proposed in Ref. \cite{Schaefer:2008hk}, $\msig$ varies from $500~$MeV to $800~$MeV. This is based on the fact that the experimental measurement is not accurate,

To the fermionic part, the Polyakov loop fields are coupled \cite{schaefer09,Tawfik:2014hwa,Tawfik:2014uka}
  \bea \label{eq:Omegaqq} 
  \Omega_{\bar qq}( \sigma_{x} ,
    \sigma_{y}, \Phi,\Phibar ) &=& -2T \sum_{f=u,d,s} \int\frac{d^3p}{(2\pi)^3} \nn \\
    && 
    \left\{\ln \left[1 + 3 (\Phi + \bar \Phi
        e^{-(E_{q,f}-\mu_{f})/T})e^{-(E_{q,f}-\mu_{f})/T} +
        e^{-3(E_{q,f}-\mu_{f})/T}\right] + \right. \nn \\
    & & \left. \;\, \ln \left[1 + 3 (\bar \Phi + \Phi
        e^{-(E_{q,f}+\mu_{f})/T})e^{-(E_{q,f}+\mu_{f})/T} +
        e^{-3(E_{q,f}+\mu_{f})/T}\right] \right\}, 
\eea
where the flavor-dependent single-particle energies are $E_{q,f}=\sqrt{k^2 + m_f^2}$ and the light and strange quark masses are related to the chiral condensates $\sigma_x$ and $\sigma_y$, respectively
\bea 
m_l &=& g \sigma_x /2, \label{eq:qmassesL}\\
m_s &=& g \sigma_y /\sqrt{2}. \label{eq:qmassesS}
\eea
The Yukawa coupling $g$ is fixed to reproduce the light quark mass $m_{q} \approx 300~$MeV. The corresponding strange quark mass is $m_s \approx 433$ MeV. From Eq. (\ref{eq:Omegaqq}), one recognizes the suppressions of single-quark contributions, where $\bar\Phi, \Phi \sim 0$ (in the hadron phase). The baryon-like objects composed of $N_c=3$ quarks contribute to the potential, as well. Therefore, some confinement properties can be deduced. For $\bar\Phi, \Phi \sim 1$, the Polyakov-loop fields decoupled from the fermionic part, Eq. (\ref{eq:Omegaqq}) are similar to the quark/antiquark contributions of pure Quark-Meson (QM) model \cite{Schaefer:2008hk}.

\begin{figure}
\centering
\includegraphics[width=7.5cm,angle=0]{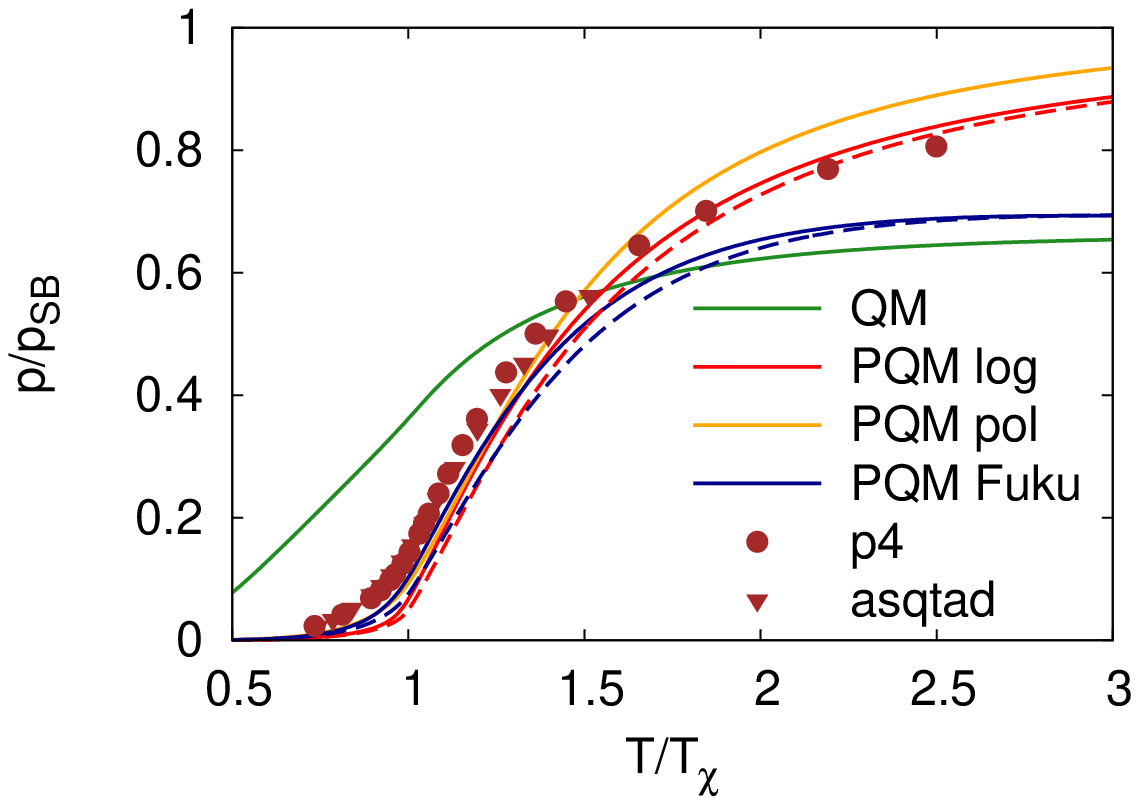}
\includegraphics[width=7.5cm,angle=0]{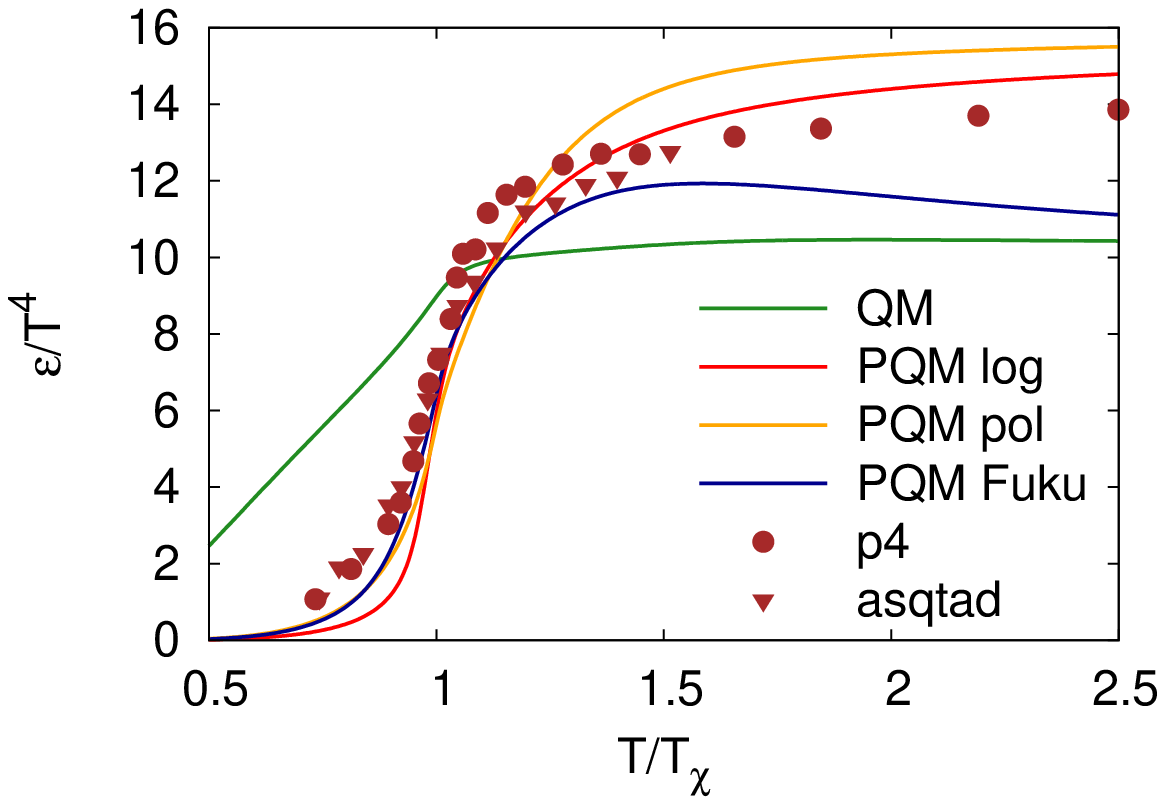}
\caption{The normalized pressure (left  panel) and the energy density (right panel) as a function of  temperature. The calculations using various Polyakov-loop potentials \cite{schaefer09} are compared wit the lattice QCD calculations~\cite{Bazavov:2009zn}  ($N_\tau=8$, p4 and asqtad actions). The solid lines correspond to larger pion and kaon masses as used in the lattice QCD simulations and the dashed lines to physical masses. Graph taken from Ref. \cite{schaefer09}.
}
\label{fig:pqmPE}
\end{figure}

\begin{figure}
\centering
\includegraphics[width=7.5cm,angle=0]{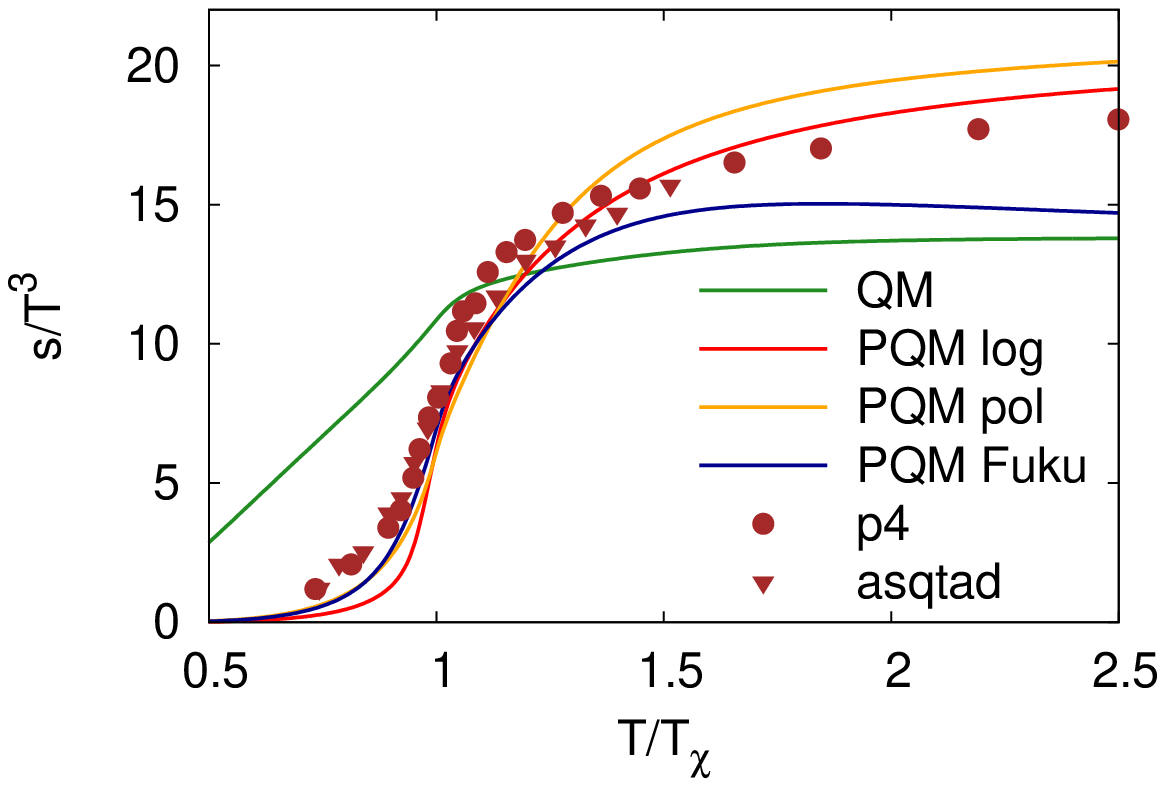}
\includegraphics[width=7.5cm,angle=0]{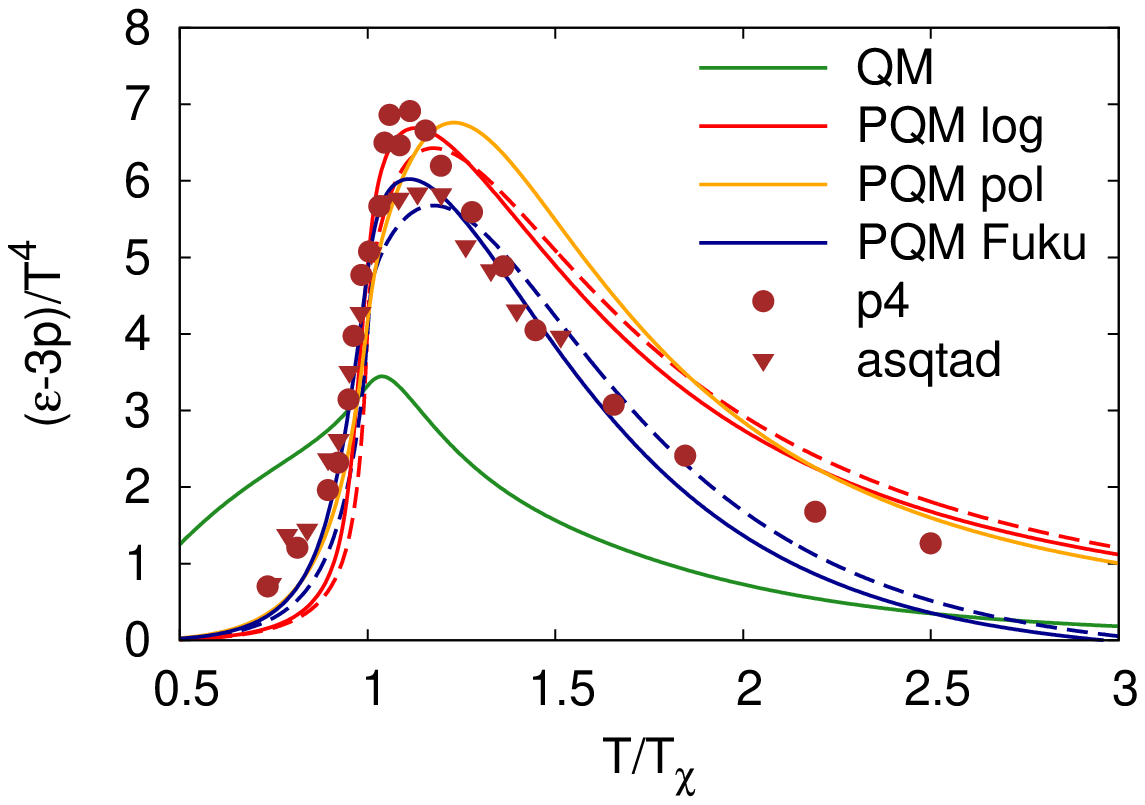}
\caption{As in Fig. \ref{fig:pqmPE}, but for the entropy (left  panel) and the interaction measure (right panel). Graph taken from Ref. \cite{schaefer09}.
}
\label{fig:pqmSIR}
\end{figure}

By using minimization conditions, the dependence of four order parameters in Eq. (\ref{eq:pqmeom}) for chiral and deconfinement phase-transition on  the temperature and the quark chemical potential  are determined as solutions of the corresponding equations of motion  \cite{schaefer09,Tawfik:2014hwa,Tawfik:2014uka}. These can be obtained by minimizing the grand potential, Eq. (\ref{eq:grandpot}), with respect to  $\vev\sigma_x$, $\vev\sigma_y$, $\vev\Phi$ and $\vev{\bar\Phi}$:
\begin{equation}
  \left.\frac{ \partial \Omega}{\partial 
      \sigma_x} = \frac{ \partial \Omega}{\partial \sigma_y}  = \frac{\partial \Omega}{\partial \Phi}=\frac{ \partial
      \Omega}{\partial \Phibar} 
  \right|_{\text{min}} = 0,   \label{eq:pqmeom}
\end{equation}
where the subscript $\text{min}$ stands for $\sigma_x=\vev{\sigma_x}, \sigma_y=\vev{\sigma_y},
\Phi=\vev{\Phi}, \bar\Phi=\vev{\bar\Phi}$ and refers a global minimum.

As an example on thermodynamic quantities, we start with the pressure, which can be obtained directly from the grand potential 
\begin{equation}
p(T,\mu_q) = - \Omega\left(T,\mu_q \right),
\end{equation}
Other thermodynamic quantities can be obtained from it by differentiation. In left-hand panel of Fig. \ref{fig:pqmPE}, the pressure for PQM and QM models,
normalized to the Stefan-Boltzmann limit (SB) of the PQM model compared to lattice QCD calculations is given as functions of the temperature. 

It is obvious that the QM model fails in describing the lattice QCD data at all temperatures, while PQM model agrees well with the lattice calculations.  The best agreement is found around $T_\chi$. The Fukushima potential achieves this good agreement but it fails at temperatures higher than $1.5\,T_\chi$. In the HotQCD lattice results, the critical temperature is as high as $T_\chi \approx 185-195~$MeV. The physical meson masses reduce the pressure in particular around $T_\chi$.

Fig. \ref{fig:pqmSIR} illustrated the entropy (left-hand panel) and interaction measure (right-hand panel). Again, we note that the QM fails to reproduce the lattice results. The logarithmic PQM describes well the entropy, while the interaction measure is well generated  in the Fukushima potential for the PQM.

\subsection{Compressible bag model}
\label{hrg_koch}

As a simple model for a gas of extended hadrons at zero chemical potential, an approach based on inspiration from the compressible bag model was proposed. The main assumptions are:
\begin{itemize}
\item at the critical temperature, the typical particle density requires that {\it volume for each hadron should be taken into consideration}. This leads to a considerable suppression (reduction) of the available phase-space and 
\item under the effect of the pressure generated by the hadrons' thermal motion, the volumes of the hadrons are allowed to vary, but  in a self-consistent way.
\end{itemize}
As a result, it is likely that the number of the effective degrees of freedom is reduced \cite{vkoch08}. For completeness, we refer to the phenomenon, in which the phase-space should be a subject of modification  \cite{Tawfik:2010kz,Tawfik:2010uh}. This suppression or reduction seems not be able to explain the lattice QCD results at high temperatures. Therefore, additional assumptions should be added, for example \cite{vkoch08}:
\begin{itemize}
\item one can explicitly adopt \cite{vkoch08} that the same partition function describes both confined and deconfined phases and
\item an appropriate dynamics should be introduced to allow observing cross-over as the case in lattice QCD.
\end{itemize}

As introduced in section \ref{sec:mitBAG}, the MIT bag model \cite{bag1} for hadrons is well suited for the purposes outlined in this section. 
\begin{itemize}
\item The MIT bag model embodies both confined and deconfined phases. Therefore, the hadrons and QGP can be described, simultaneously,  as extended bags of each other. 
\item Furthermore, the MIT bag model assumed an infinite mass spectrum of the Hagedorn type \cite{vkoch08}. 
\item Elastic interactions (an additional dynamics) between hadrons were proposed to describe the phase transition. The elastic interactions is related to the kinetic pressure. The latter in turn squeezes the bags. Other types of interactions have been discussed in section \ref{sec:intr}. 
\end{itemize}

The basic features of the MIT bag model \cite{bag1} have been introduced in section \ref{sec:mitBAG}. When $T$  approaches $T_0$, the average masses and  volume of the hadron bags rapidly increase \cite{vkoch08}. Thus, the hadrons continue occupying the whole available space. They eventually overlap with reach others. In order to avoid multiple counting of the phase space, it was proposed to exclude from the partition function the configurations with overlapping bags. For a a system with finite volume $V$, this can be achieved with an excluded volume correction, where $V$ shall be replaced by  $(V-\sum_{i=1}^N V_i)^N$, the available volume, where $V_i$ is the volume of the $i$-th particle. The constrains that total volume of all bags $\sum_{i=1}^N V_i$ should not exceed $V$ is obvious. Accordingly, the modified phase-space integral reads \cite{grons}, 
\begin{equation}
\left[\prod_{i=1}^N \frac{V}{(2 \pi)^3} \int d^3 p \right] \rightarrow 
\left[\prod_{i=1}^N \frac{1}{(2 \pi)^3} \int d^3 p \right] \left(V-\sum_{i=1}^N V_i\right)^N
\Theta{\left( V-\sum_{i=1}^N V_i \right)}.
\label{nphsp}
\end{equation} 
Accordingly, the partition function reads
\begin{eqnarray}
\label{1.7b}
Z(V,T) &=& \sum_{N=0}^{\infty} \left(\frac{T}{2 \pi}\right)^{3N/2} \frac{c_0^N}{N!}
\left[\prod_{i=1}^N \int_{0}^{\infty} d m_i \; m_i^{3/2-\alpha}  \right] \times \nonumber \\
& & 
\exp{\left[\frac{\sum_{i=1}^N m_i}{T_0}-\frac{\sum_{i=1}^N m_i}{T}\right]}\left(V-\sum_{i=1}^N V_i\right)^N
\Theta{\left(V-\sum_{i=1}^N V_i \right)}.
\end{eqnarray}
The system is supposed to pick up more energy as the temperature is increased. To allow for this, an additional dynamics should be implemented. This is the idea behind the compressibility of hadron bags~\cite{Kagiyama:2002rm}. In other words, the hadron volume is allowed to vary under the effect of the pressure, which is  generated by the thermal motion. Consequently, as $T$ and hence pressure increase, the bags will be compressed. Therefore, they acquire higher internal mass/energy density \cite{vkoch08}. This is the self-consistent dynamics added to the hadrons.

\begin{figure}
\centering
\includegraphics[width=14cm,angle=0]{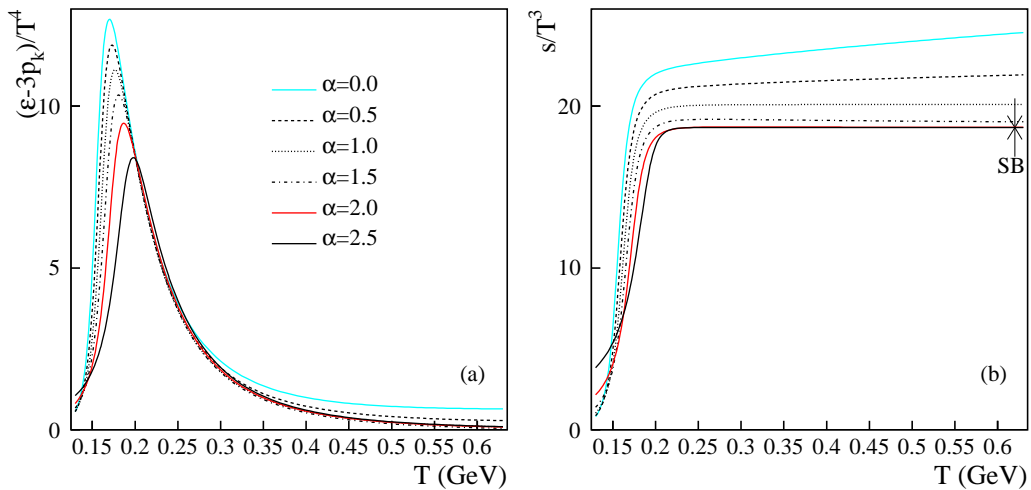}
\caption{The interaction measure $(\varepsilon-3p_k)/T^4$ (left panel) and $s/T^3$ (right panel) are given as function of temperature $T$ at different values  $\alpha$, the isobaric parameter. Graph taken from Ref. \cite{vkoch08}.
}
\label{fig:kochh4}
\end{figure}

It is apparent that at $T \ll T_0$, the system is  dilute. This means that $V \gg \sum_{i=1} V_i$ and the system behaved as a non-interacting gas with point-like  constituents. When $T$ increases, the average mass and the volume of the hadrons increases, as well. When $T \rightarrow T_0$, the critical temperature, we can apply the he dilute-gas approximation \cite{vkoch08}. Furthermore, the pressure that other particles exert increases and therefore its effect on the hadrons properties  becomes considerable. In other words, in addition to the bag pressure $B$, every hadron feels an additional thermal/kinetic pressure $p_k$. The latter is generated by the thermal motion of other hadrons  \cite{vkoch08}. It is obvious that the stability condition needed by the system, $p_r \equiv B$ should be modified. The kinetic pressure $p_k$ should be taken into account. It is worthwhile to highlight $p_k$ can be interpreted as an elastic interaction consequence. This is accounted by the infinite mass spectrum of the hadrons. The number of particles $N$ would not play any essential role..
Accordingly, when neglecting the surface effect, a simple generalization of the stability condition reads 
\begin{equation}
p_r=B+p_k(V,T).
\label{simplass}
\end{equation}  
Dependences of the volume and the mass of $i$-th hadron on $p_r$ are given as 
 \begin{eqnarray}
\label{replacements}
V_i &=& \frac{U_i}{3\, p_r}, \\
m_i &=&  U_i + B\, V_i = U_i \left(1+\frac{B}{3\, p_r} \right).
\end{eqnarray}
Then, the entropy reads
\begin{equation}
S_i=\frac{4}{3}\, \frac{U_i}{k\, (p_r)^{1/4}}. \label{entropy}
\end{equation}
This leads to the substitution that $\exp\left(\frac{m_i}{T_0}\right) \rightarrow  \exp\left(\frac{4}{3}\, \frac{U_i}{k\, (p_r)^{1/4}}\right)$,
where $\exp(m_i/T_0)$ is  exponential mass spectrum.

\subsubsection{Thermodynamics}

Based on Eq.~(\ref{1.7b}), the modified grand-canonical partition function can be written \cite{vkoch08}
\begin{eqnarray}
\label{gcpartfunUpr}
Z(V,T) &=& \sum_{N=0}^{\infty} \left( \frac{4}{3} \right)^N
\left(\frac{T}{2 \pi}\right)^{3N/2}\frac{c_0^{N}}{N!} \times \nonumber \\ 
&& \left[ \prod_{i=1}^N \int d U_i 
\left(U_i+ B \frac{U_i}{3p_r} \right)^{3/2-\alpha} 
\right] \exp{\left\{\left[ \frac{4}{3kp_r^{1/4}}-\frac{1}{T}\left(1+\frac{B}{3p_r}\right)
\right]\sum_{i=1}^N U_i\right\}} \times \nonumber \\
&&  
\left(V-\sum_{i=1}^N \frac{U_i}{3p_r} \right)^N \Theta{\left( V-\sum_{i=1}^N\frac{U_i}{3p_r}  
\right)}.
\end{eqnarray}
With the substitution, $U_i\rightarrow \eta_i=\frac{4}{3}\, U_i$, then, the partition function reads
\begin{eqnarray}
\label{partfunmpr}
Z(V,T) &=& \sum_{N=0}^{\infty} \left(\frac{T}{2 \pi}\right)^{3N/2}\frac{c_0^{N}}{N!}
\left[ \prod_{i=1}^N \int d \eta_i 
\left(\frac{3}{4} \eta_i+ B \frac{\eta_i}{4 p_r} \right)^{3/2-\alpha} 
\right] \times \nonumber \\  
&& 
\exp{\left\{\left[ \frac{1}{k p_r^{1/4}}-\frac{1}{T}\left(\frac{3}{4}+\frac{B}{4 p_r}\right)
\right]\sum_{i=1}^N \eta_i\right\}}  
\left(V-\sum_{i=1}^N \frac{\eta_i}{4p_r} \right)^N \Theta{\left( V-\sum_{i=1}^N\frac{\eta_i}{4 p_r}  
\right)}.  \hspace*{1cm}
\end{eqnarray}
It is obvious that in the dilute gas limit, $p_k\rightarrow 0$ and  $\eta_i\rightarrow m_i$. 

Fig. \ref{fig:kochh4} illustrates the dependence of the interaction measure $(\varepsilon-3p_k)/T^4$ (left panel) and $s/T^3$ (right panel)  on $T$ at different values of the isobaric parameter $\alpha$. We observe that only at $\alpha=0$ and $\alpha=0.5$, the curves do not converge to a constant value at high $T$. This is obvious in $s/T^3$ vs. $T$. Other values of $\alpha$ behave normally with $T$. The largest $\alpha$-values of very close to the SB limit.

\section*{Acknowledgement}
This work is financially supported by the World Laboratory for Cosmology And Particle Physics (WLCAPP), http://wlcapp.net/. I would like to thank David Blaschke for providing me with Heinz Koppe's paper. I am very grateful to Ludwik Turko for pointing out the work of Magaliskii and Terletskii! The last phase of this work has been accomplished in Erice-Italy. I would like to acknowledge the kind invitation and the  great hospitality of Prof. Antonino Zichichi.

\end{document}